\documentclass [preprint,12pt,authoryear]{elsarticle}
\usepackage{pdfwidgets}
\usepackage{natbib}

\usepackage[T1]{fontenc}
\usepackage{ae,aecompl}
\usepackage{tikz}

\usepackage{graphicx}	
\usepackage{hyperref}
\usepackage{amsmath}	
\usepackage{amssymb}	
\usepackage[caption=false]{subfig}
\usepackage{float}
\usepackage{lscape}
\usepackage{rotating}
\usepackage{url}
\begin{document}
\begin{frontmatter}
\title{Gravitational interaction signatures in isolated galaxy triplet systems: a photometric analysis}


\author[1,2,3]{Amira A. Tawfeek \corref{mycorrespondingauthor}} 
\cortext[mycorrespondingauthor]{Corresponding author}
\ead{amira.tawfeek@nriag.sci.eg} 

\author[4]{Saha Kanak \corref{mycorrespondingauthor}}
\ead{kanak@iucaa.in}
\author[4]{Vaghmare Kaustubh}
\author[4]{Kembhavi A. K.}
\author[1]{Ali Takey}
\author[3]{Bernardo Cervantes Sodi}
\author[3]{Jacopo Fritz}
\author[2]{Zainab Awad}
\author[1]{Gamal B. Ali}
\author[2]{Z. M. Hayman}

\address[1]{National Research Institute of Astronomy and Geophysics (NRIAG), 11421 Helwan, Cairo, Egypt}
\address[2]{Astronomy, Space Science and Meteorology Department, Faculty of Science, Cairo University, 12613 Giza, Egypt}
\address[3]{Instituto de Radioastronomia y Astrofisica, Universidad Nacional Autonoma de Mexico, Campus Morelia, A.P. 3-72, C.P. 58089 Michoacan, Mexico}
\address[4]{Inter-University Centre for Astronomy and Astrophysics (IUCAA), Pune, India}

\begin{abstract}
Galaxy triplets are interesting laboratories where we can study the formation and the evolution of small and large systems of galaxies. This study aims to investigate signs of interaction between the members of nine isolated galaxy triplet systems (27 galaxies) selected from the ``SDSS-based catalogue of Isolated Triplets'' (SIT) with members brighter than 17.0 ($m_r\le$ 17.0) in the $r-$band, and mean projected separation between the members of $r_p \leq$ 0.1 Mpc. In this work, we performed a one-dimensional (1D) fitting of the surface brightness profiles and a two-dimensional (2D) modeling of the sample galaxies. In the 1D fitting, we examined the far outer part of the light profiles of disk galaxies (22 galaxies) and categorized them into type I (simple exponential), type II (down-bending), and type III (up-bending). This fitting results showed that 55$\%$ of disk galaxies in our sample represent type III i.e are in state of interaction. In the 2D modeling, we fit smooth axisymmetric profiles to the 27 galaxies and found that 70$\%$ exhibit asymmetric features and signs of interactions in their residual images. Thus, we conclude that galaxy triplets, with projected separations ($r_p \leq$ 0.1 Mpc) between their members, are physically bounded systems that show pronounced signs of interactions.
\end{abstract}


\end{frontmatter}





\section{Introduction}

Systems of galaxy triplets are galaxy groups with the least number of galaxies and are natural laboratories that can provide insight into the formation of larger groups and perhaps even clusters \citep{Hernandez2011}. Based on the BVRI photometry of 54 triplet systems, \cite{Hernandez2011} investigated the ellipticity, position angle and $B-I$ colour maps of these systems and inferred that galaxy triplets are dominated by spirals, a population that is more representative of field galaxies. Extending to a larger sample derived from a catalogue of 1092 triplets, \cite{Omill2011} and \cite{Duplancic2013} compared physical properties such as star-formation rate (SFR), stellar population age, colour indices, compactness and concentration indices of galaxy triplets and compact groups. Based on their similarity, they found that galaxy triplets are a natural extension of compact groups, later confirming this result through the analysis of their dynamical parameters, finding that the member galaxies of these triplets belong to the same dark matter halo, which leads to a co-evaluation of the member galaxies \citep{Duplancic2015}. Interestingly, many of the galaxy triplets are found in the outer parts of clusters, filaments, or walls and are different from the void population of galaxies \citep{Fernandez2015}. 

More recently, using spectral synthesis and emission line measurements, \cite{Costa2016} studied the 
stellar mass and age of triplet galaxies and suggested that most of the galaxies in the triplet systems are 
passive or quenched. This is intriguing. How do galaxies in a triplet become passive or quench 
their star-formation when their environment is not favorable?  On the other hand, it is well understood 
that the relaxation time of compact group and galaxy triplet systems would be much shorter 
compared to the Hubble time \citep{Binney1987} and one would expect the members of the triplet system to undergo tidal interaction since their formation. However, it remains unclear at what extent galaxy triplet systems show any signs of interaction and if there is a way to quantify such interactions. 

The most popular optical ways used to identify the state of interaction between galaxies consists in investigating their morphologies, star formation region properties and their colour distribution \citep{Veilleux2002,Hernandez2011,Kilerci2014}. In addition, the analysis of hydrodynamical simulation and panchromatic spectral energy distribution play a key role in the identification of signs of interactions between galaxies \citep{Steinmetz1999, Springel2000, Springel2005(a), Hopkins2006, Costa2016}. 

The goal of this paper is to understand the state of gravitational interaction in  galaxy triplet systems by deeply investigating their morphologies. In our previous paper, \cite{Tawfeek2019} applied a statistical study on the ``SDSS-based catalogue of Isolated Triplets'' (SIT) compiled by \cite{Fernandez2015}, containing 315 isolated triplet systems, to estimate their physical and
dynamical parameters and hence, understand the nature of galaxy
triplet systems and their evolution.
In this paper, we selected a sample of nine galaxy triplet systems, from the same catalog (SIT), with small projected separation between their members ($r_p$ $<$ 0.1 Mpc, as estimated by \cite{Tawfeek2019}) and apparent $r-$band magnitude $m_r\le$ 17.0. 
We first extract the surface brightness distribution of galaxies in our sample using IRAF- $Ellipse$ task \citep{Jedrzejewski1987, Milvang1999}, written by the National Optical Astronomy Observatories (NOAO) in Tucson-Arizona-USA, and apply 1D modeling to these profiles using one or a combination of various S$\acute{e}$rsic profiles. 
In general, the stellar disk of most disk galaxies is well defined by exponential profile \citep{deVaucouleurs1959}. However, it was found that in many cases this profile breaks down at large radius, in the outer region of the disk \citep{Freeman1970, Barnaby1992,Pohlen2002,Erwin2005,Pohlen2006}.

According to the study made by \cite{Freeman1970}, the profiles of disk galaxies are classified into two basic types: type I, in which the disk shows a simple exponential; and type II, where the outer part of the disk is down-bending. Then, a third type of up-bending manner in the outer part of the disk, was added by \cite{Erwin2005, Erwin2008} denoted by type III profile.   
These breaks have been studied by a number of authors who found a general agreement in interpreting such features as caused mainly by star formation thresholds \citep{Schaye2004, Erwin2005, Naab2006, Boissier2007}
In particular, type II is generated by star formation processes where the gas density changes at large radii \citep{Schaye2004, Elmegreen2006} while, type III profiles are probably produced by minor merger \citep{Younger2007, Erwin2008, Laine2014} where disk galaxies are found within a hierarchical universe, in which mergers occurs frequently \citep{Somerville1999,Somerville2000}. Moreover, \citet{Younger2007} used a hydrodynamical simulations of minor mergers of galaxies and investigated the nature of antitruncated (type III) stellar disk. They found that antitruncations are dominated by progenitor stars that have been transferred from the primary disk to larger radius via interaction.   

Beside 1D modeling, we performed a 2D surface brightness decomposition of the galaxy images using GALFIT \citep{Peng2002} and derive the 
structural components that not only become useful to understand these systems but also to gauge the kind of interactions or merger processes these galaxies in triplets have gone through \citep{Simmons2004, Weinzirl2007, Gabor2009, Peng2010, Hernandez2011}.

This paper is organized as follows: in Section~\ref{s:Sample} we introduce the process of sample selection. Section ~\ref{s:Methodology} describes the data processing and the 1D fitting and 2D decomposition. The main results of the photometric analysis and signs of interactions are presented in Section~\ref{Sec:results}. Our summary and conclusions are given in Section~\ref{s:conclusions}. The detailed analysis of our sample is presented in the Appendix.

\section{Galaxy triplets sample}
\label{s:Sample}

In this study, we drew a sample from the ``SDSS-based catalogue of Isolated Triplets'' (SIT) published by \cite{Fernandez2015} and based on the Sloan Digital Sky Survey- Tenth Data Release (SDSS-) \citep{Ahn2014}. This catalogue comprises 315 isolated triplets with spectroscopic information, $r-$band model magnitudes ($m_r$) in the range 14.5 $\le m_r \le$ 17.7 over a redshift (z) range 0.006 $\le$ z $\le$ 0.080. They used spectroscopic information instead of the photometric one to avoid the uncertainty and biasing produced by projection effects in case of photometric redshift data. \cite{Fernandez2015} based their isolation criterion on galaxies with no neighbours within radial velocity difference $\Delta$ $v$ $\le$ 500 $km~s^{-1}$ and the three members are found within 1 Mpc radius. They also performed a star-galaxy separation in order to prevent classifying a star as a galaxy. In addition, they removed multiple identifications by choosing unique systems where the brightest galaxy is the primary one and have two neighbours according to the isolation criteria. In these systems, galaxies are considered to be physically bound to the primary galaxy (G1, the brightest one in the system), if $\Delta$ $v$ $\le$ 160 $km~s^{-1}$ and the projected separation ($r_p$) is smaller than 450 kpc, as mentioned by \citet{Fernandez2014}.

For gravitational interaction investigations we select our sample under two restrictions,

\begin{enumerate}

 \item The mean projected separation between the members of each system is as small as possible ($r_p \leq$ 0.1 Mpc), where tidal strength increases as projected separation between the members of triplets decreases \citep{Fernandez2015,Tawfeek2019}, i.e. gravitational force more likely affect the photometric and the morphological characteristic of galaxies in triplet systems.
 
 \item The $r-$band apparent magnitude ($m_r$) of the three galaxies in each system should be brighter than 17 ($m_r\le$ 17), and hence all features from the center of the galaxy until its very far and faint edges, where gravitational interaction between galaxies appears, could be investigated.

Under these conditions, we obtained nine systems (i.e. 27 galaxy members), after discarding three systems with high noise in their SDSS images. For the nine triplet systems, we downloaded their corrected and reduced $r-$band fits images from the SDSS-DR12, one of the most recent data release of the SDSS-III that contains observations up to July 2014 \citep{DR12}. 
\end{enumerate}

Before applying our analysis, the background stars and the projected objects (e.g. nearby galaxies) have been masked from all images by using $imedit$ task in IRAF.
In all images, the north direction is adjusted to be upwards and the east direction is at the left. The basic parameters of our sample are reported in Table~\ref{tbl:basic}. According to SIMBAD and LEDA, the morphological classification of the sample reveals 13 spirals (S), 10 ellipticals (E), three lenticulars (S0), and one irregular ($I_{rr}$) galaxy.

\begin{table*}
  \begin{center} 
   \tabcolsep 5.8pt
   \scriptsize
    \caption{The basic parameters of the nine selected galaxy triplets taken from the SIT catalogue. Column (1) is the index of the triplet system, column (2) indicates the SIT-ID and a rank of galaxies in each system. The third (3) and the fourth (4) columns are the right ascension and the declination of each galaxy, respectively. The fifth (5) column is the galaxy redshift, column (6) represents the morphological type, and the last two columns (7) and (8) reveals the apparent magnitude in the $r-$band and the mean projected separation ($\overline r_p$) between the members of each triplet system, respectively. The morphological type is obtained from literature. The triplets parameters are updated from the SDSS-DR14 (as stated in our previous paper, \cite{Tawfeek2019})}. 
    \label{tbl:basic}\\
    
\begin{tabular}{|l|l|l|l|l|l|l|r|}
\hline
  \multicolumn{1}{|c|}{Index} &
  \multicolumn{1}{|c|}{SIT-ID} &
  \multicolumn{1}{c|}{RA} &
  \multicolumn{1}{c|}{Dec} &
  \multicolumn{1}{c|}{z} &
  \multicolumn{1}{c|}{Type} &
  \multicolumn{1}{c|}{m$_r$}&
  \multicolumn{1}{c|}{$\overline r_p$}\\
  \multicolumn{1}{|c|}{} &
  \multicolumn{1}{|c|}{} &
  \multicolumn{1}{c|}{(deg)} &
  \multicolumn{1}{c|}{(deg)} &
  \multicolumn{1}{c|}{} &
  \multicolumn{1}{c|}{} &
  \multicolumn{1}{c|}{} &
  \multicolumn{1}{c|}{(Mpc)}\\
  \multicolumn{1}{|c|}{(1)} &
  \multicolumn{1}{|c|}{(2)} &
  \multicolumn{1}{c|}{(3)} &
  \multicolumn{1}{c|}{(4)} &
  \multicolumn{1}{c|}{(5)} &
  \multicolumn{1}{c|}{(6)} &
  \multicolumn{1}{c|}{(7)} &
  \multicolumn{1}{c|}{(8)}\\
\hline
   & 30G1 & 348.122 & 13.942 & 0.03411 & Sb & 14.52&  \\
 1 & 30G2 & 348.102 & 13.951 & 0.03412 & Sb & 15.04 & 0.08 \\
   & 30G3 & 348.108 & 13.914 & 0.03389 & S0 & 15.32 & \\
   &101G1 & 206.035 & 45.341 & 0.03848 & S0 & 15.38 & \\
 2 &101G2 & 206.017 & 45.310 & 0.03865 & SBab & 15.76&0.10  \\
   &101G3 & 205.974 & 45.331 & 0.03838 & SBc & 16.81& \\
   &104G1 & 128.212 & 30.879 & 0.06579 & E & 15.60 & \\
 3 &104G2 & 128.232 & 30.893 & 0.06606 & E & 16.56&0.09 \\
   &104G3 & 128.213 & 30.897 & 0.06624 & SB & 16.83& \\
   &125G1 & 209.305 & 12.021 & 0.02074 & S0 & 13.07 &\\
 4 &125G2 & 209.308 & 11.998 & 0.02050 & SABa & 14.01 &0.05\\
  &125G3 & 209.330 & 11.976 & 0.02036 & Sc & 14.83&\\
  &197G1 & 122.560 & 51.873 & 0.06055 & E & 15.55&\\
 5&197G2 & 122.537 & 51.881 & 0.06053 & E & 15.91 &0.08\\
  &197G3 & 122.571 & 51.887 & 0.06058 & E & 16.30 &\\
  &217G1 & 133.510 & 17.343 & 0.06678 & E & 15.59 &\\
 6&217G2 & 133.488 & 17.329 & 0.06671 & Sc & 16.40 & 0.10 \\
  &217G3 & 133.485 & 17.348 & 0.06696 & E & 16.97 &\\
  &263G1 & 229.252 & 15.566 & 0.04810 & SBb & 14.69& \\
 7 &263G2 & 229.237 & 15.544 & 0.04800 & S & 15.19& 0.08 \\
  &263G3 & 229.256 & 15.573 & 0.04789 & S & 15.72 &\\
  &264G1 & 199.061 & 41.494 & 0.02075 & Sc & 14.29 &\\
 8 &264G2 & 199.026 & 41.501 & 0.02088 & I & 14.73 & 0.05 \\
  &264G3 & 198.997 & 41.499 & 0.02101 & E & 16.78 &\\
  &280G1 & 128.417 & 9.9637 & 0.03076 & E & 14.14 &\\
 9 &280G2 & 128.396 & 10.015 & 0.03055 & E & 15.57& 0.10 \\
  &280G3 & 128.432 & 9.9904 & 0.03062 & SBc & 16.62& \\
\hline
\end{tabular}
 \end{center}
\end{table*}

\section{Photometric analysis: surface brightness profiles and decomposition of triplet galaxies}
\label{s:Methodology}  
 
In this section we describe the main techniques applied throughout this study. We used both IRAF's $ELLIPSE$ task and GALFIT programs to investigate 1D fitting of surface brightness profiles and 2D fitting of galaxy images, respectively.  

\subsection{1D modelling} 

We applied IRAF's $ELLIPSE$ task on $r-$band SDSS images of disk galaxies (Lenticulars and spirals) in our sample, after discarding elliptical galaxies, to investigate the distribution of light along the surface of each galaxy by fitting its isophotes into ellipse along the semi-major-axis, where intensity, position angle, ellipticity, and centroids can be extracted  \citep{Jedrzejewski1987, Milvang1999}. 

The intensity profiles of disk galaxies are then categorized into type I (simple exponential), type II (down-bending), and type III (up-bending) according to the classification made by \cite{Erwin2005, Pohlen2006, Erwin2008}. Type II is subdivided into two classes depending on the location of the breaking point. If this point is located near or at the bar region, then we call it ``inner'' break (type II.i). If it is outside the bar region, then it is known as an ``outer'' break (type II.o). Finally, if a galaxy does not have a bar and the intensity profile is down-bending, it is denoted with a plain ``type II'' profile. Type III is also subdivided into type III-d and type III-s. In type III-d, the outer part of the profile, beyond the break, is part of the main galaxy disk. In this case, the outer light appears to have the same ellipticity as the inner light. Oppositely, type III-s is when the outer region is a part of spheroid where the isophotes become rounder at larger radii \citep{Erwin2008}. 

\begin{figure*}
\centering
\includegraphics[scale=0.75, bb=80 20 500 100]{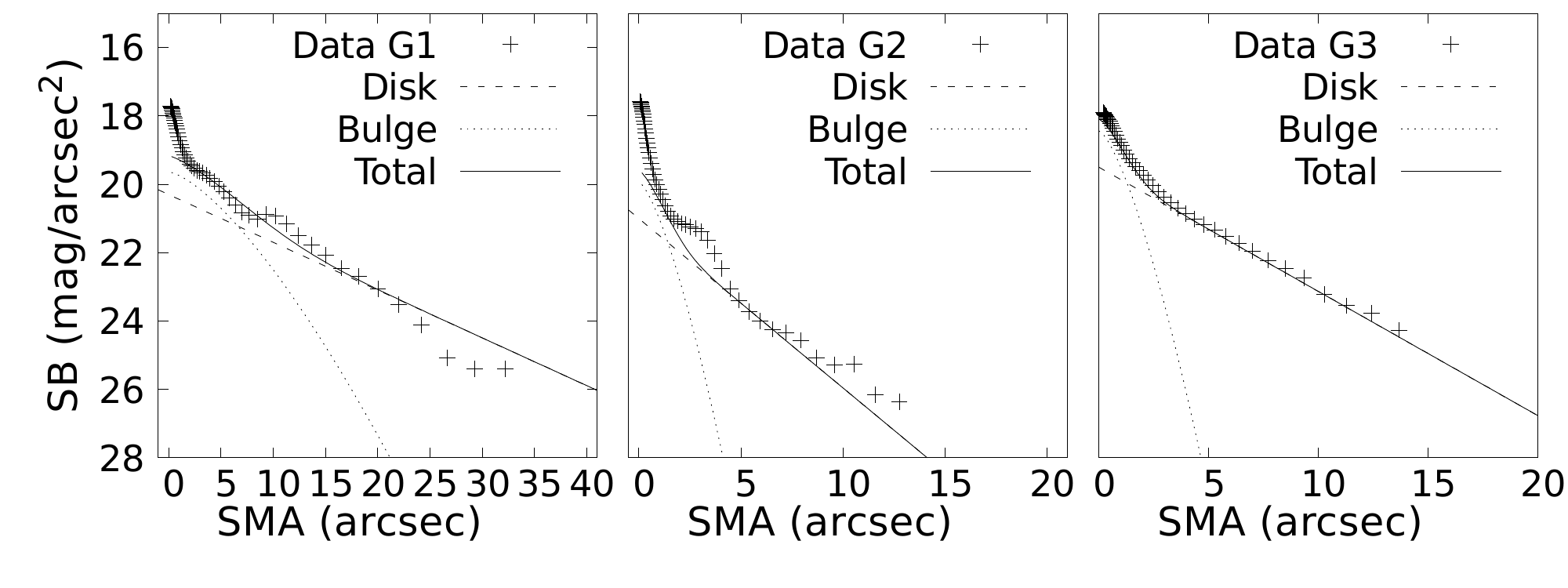}
\caption{An example of 1D fitting of the intensity profiles of galaxies in one of our triplet sample (SIT 30) showing type II break at the upper left panel, type III break in the middle panel, and type I profile in the right panel. The profiles were fitted via GNUPLOT program.}
\label{fig:1D_30analysis}
\end{figure*}

Figure~\ref{fig:1D_30analysis} shows the three different types of intensity profile of the triplet system SIT-30 (one of our sample) after fitting the intensity profiles of its members.

This classification is useful in explaining the ongoing activities, evolution and star formation in galaxies. In addition, it may help in identifying the state of interaction between galaxies in groups (e.g. galaxy triplet systems), where simulations found that type II profiles are caused by a drop in star formation rate due to the reduced amount of cooled gas \citep{Roska2008}. Meanwhile, \citet{Azzollini2008} found that the colour profile of type II become redder after the break radius that could be due to change in the star formation around the break. On the other hand, further studies and cosmological simulations show that type III profiles could be a result of minor mergers \citep{Younger2007, Erwin2008, Laine2014}, where mild gravitational interaction cause significant perturbation in the disks of the involved galaxies.  

To identify the break type in each intensity profile, we used a S$\acute{e}$rsic profile ($Ser$) to fit the bulge and the disk components of each profile by using GNUPLOT \footnote{\url{http://www.gnuplot.info/docs_5.0/gnuplot.pdf}}. S$\acute{e}$rsic profile can describe the brightness profile through the S$\acute{e}$rsic index ``$n$'' with 1.5 $< n <$ 10 usually representing a bulge, with the particular value of $n=4$ for the deVaucouleur profile, $n =$ 1 representing the disk component with an exponential profile and $n \sim$ 0.5 representing a bar. This profile is given by the following equation:
 
 \begin{equation}
  I_{b}(r) = I_e \exp[-b((\frac{r}{r_e})^{1/n} -1)],
  \label{eq:bulge function}
 \end{equation}

 where $I_{b}(r)$ is the bulge intensity at radius $r$, $I_e$ is the intensity at $r_e$, $r_e$ is the effective radius of the S$\acute{e}$rsic component that contain half the total intensity of the bulge, $n$ is the S$\acute{e}$rsic index, a parameter measuring the concentration of light towards the center. The factor $b$ is a normalization constant determined by $n$ \citep{Phillipps2005}. 
 
 For the disk component, $n=1$, and Equation~\ref{eq:bulge function} reduces to;
 \begin{equation}
  I_{d}(r) = I_s \exp[-(\frac{r}{r_s})].
  \label{eq:disk function}
 \end{equation}
 
The total intensity profile is described by the sum of Equation~\ref{eq:bulge function} and Equation~\ref{eq:disk function}. 

For an appropriate classification, we ignored the breaks in the inner region that could be due to the existence of bar or an inner ring structure.

Finally, we estimate the break radius ($R_{br}$) of type II and III profiles and normalized it with respect to the scale length ($R_S$) of the corresponding galaxy (computed by GALFIT in the 2D decomposition) to investigate its correlation with the tidal strength ($Q$) of each system. Information about $Q$ and how we compute it are given in Section~\ref{ss:R&Q}.

\subsection{2D modelling} 
 
For 2D decomposition, we used GALFIT (3.0) software  to fit and decompose the fine structures of galaxies \citep{Peng2002}. This software depends on parametric fitting using the Levenberg–Marquadt algorithm to minimize the residuals between the data image (DATA) and modeled (MODEL) image describing as follows,

\begin{equation}
 \chi^2_{\nu}= \frac{1}{N} \Sigma_x \Sigma_y \frac{[DATA(x,y)-MODEL(x,y)]^2}{\sigma(x,y)^2},
\end{equation}

The sum is taken over the image pixels,
and $\sigma ( x , y )$ is the sigma image that indicates the statistical uncertainty of each pixel. Sigma image can be extracted manually or automatically via GALFIT.
To obtain an appropriate sigma image we allowed GALFIT to create an automatic one by using the RDNOISE, GAIN, and NCOMBINE parameters in the header of the fits image. The model image consists of a sum of model components, i.e., for bulge, disk, bar etc., created by the radial profiles functions and the image point-spread function (PSF image).
The PSF image is the most important input for GALFIT to perform a proper fitting process.
An ideal PSF image should be created with high signal-to-noise (infinite), a flat and zero background, match the detailed shape of a stellar image, and centered in the image. 
In this study, we used Gaussian profile to create an appropriate PSF image as mentioned by \citet{Peng2002}.
 
Finally, the radial profile functions were chosen according to the morphological type of each galaxy as follows;
Elliptical galaxies were modeled using a S$\acute{e}$rsic ($Ser$) profile or its specific case de Vaucouleurs ($DeV$) where the S$\acute{e}$rsic index $n$ is fixed to 4. Spiral and lenticular galaxies are fitted by using $Ser$ plus an exponential disk ($Exp$) model. In some cases we used two models ($Ser$ + $Exp$) to decompose elliptical galaxies. Our measurement accuracy depends on the value of $\chi^2_{\nu}$ which, when the model properly represents the data, should be $\sim$ 1. 

The most important output from GALFIT is the residual image where signs of interaction can be identified in the form of; dust lanes, asymmetries, rings, bridges, tidal tails, tidal arms, fan arms, or offset bars \citep{Hernandez2011}. We also quantified the difference between the center of the bulge and the disk in case of two component fitting, where an offset between them might be interpreted as an evidence for an ongoing tidal interaction \citep{Knapen2010,Valenzuela2014,Bonita2015}.

\begin{figure*}
\centering
\subfloat{\includegraphics[scale=0.22, bb=2 2 540 400]{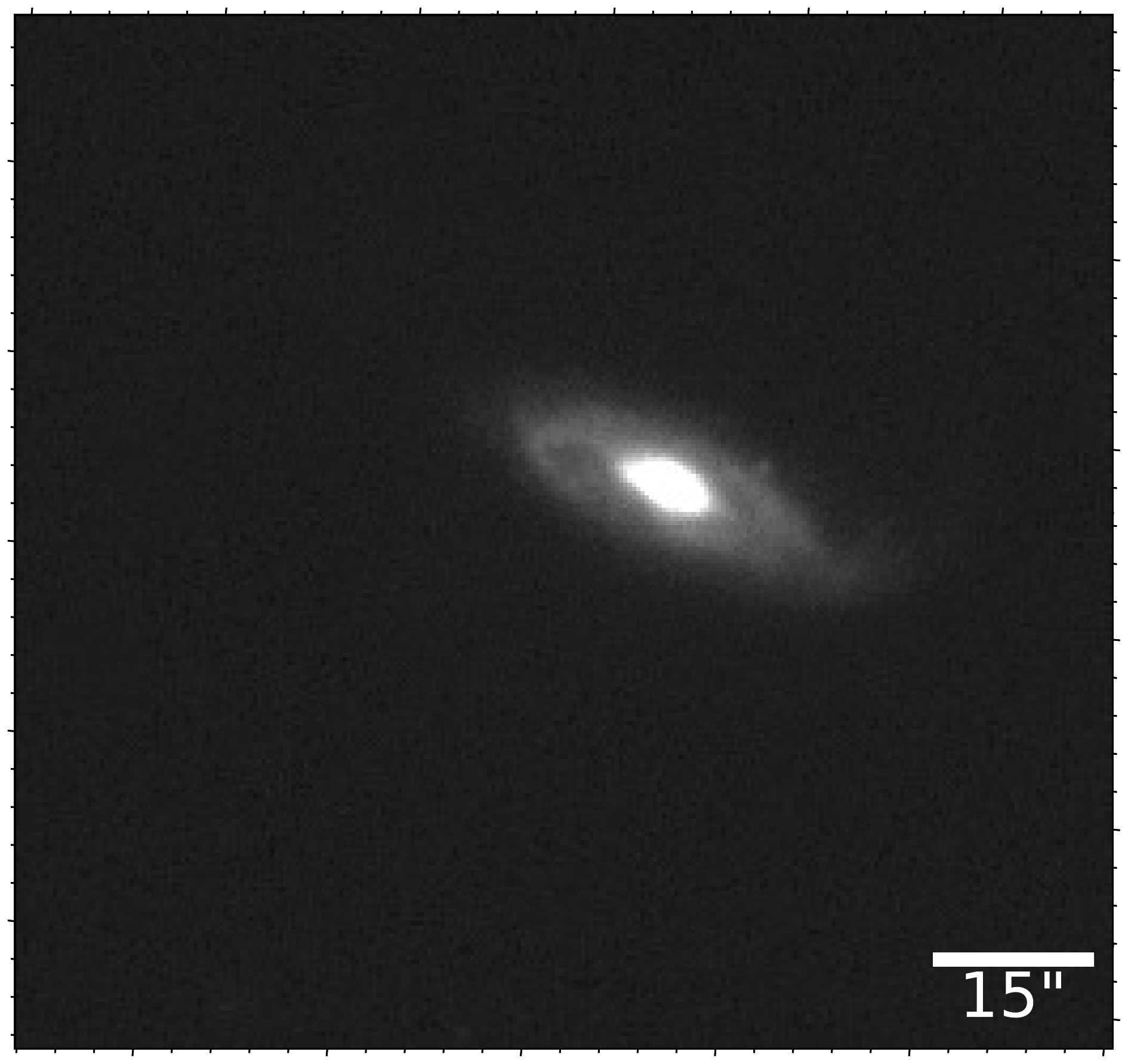}\label{125G2_original}}
 \hspace{0.2cm}
\subfloat{\includegraphics[scale=0.22, bb=2 2 540 400]{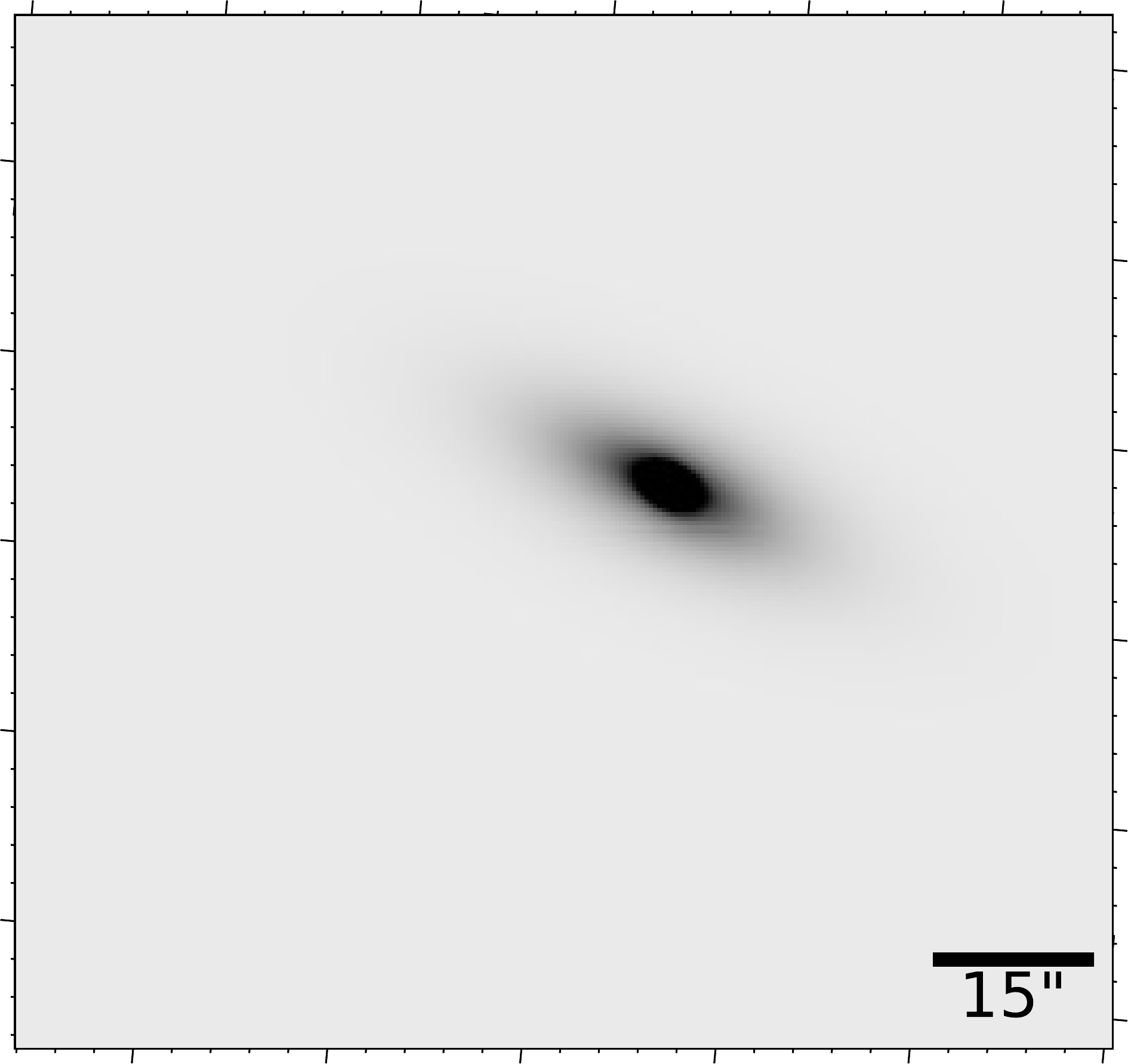} \label{125G2_model}}
\hspace{0.15cm}
\subfloat{\includegraphics[scale=0.22, bb=2 2 540 400]{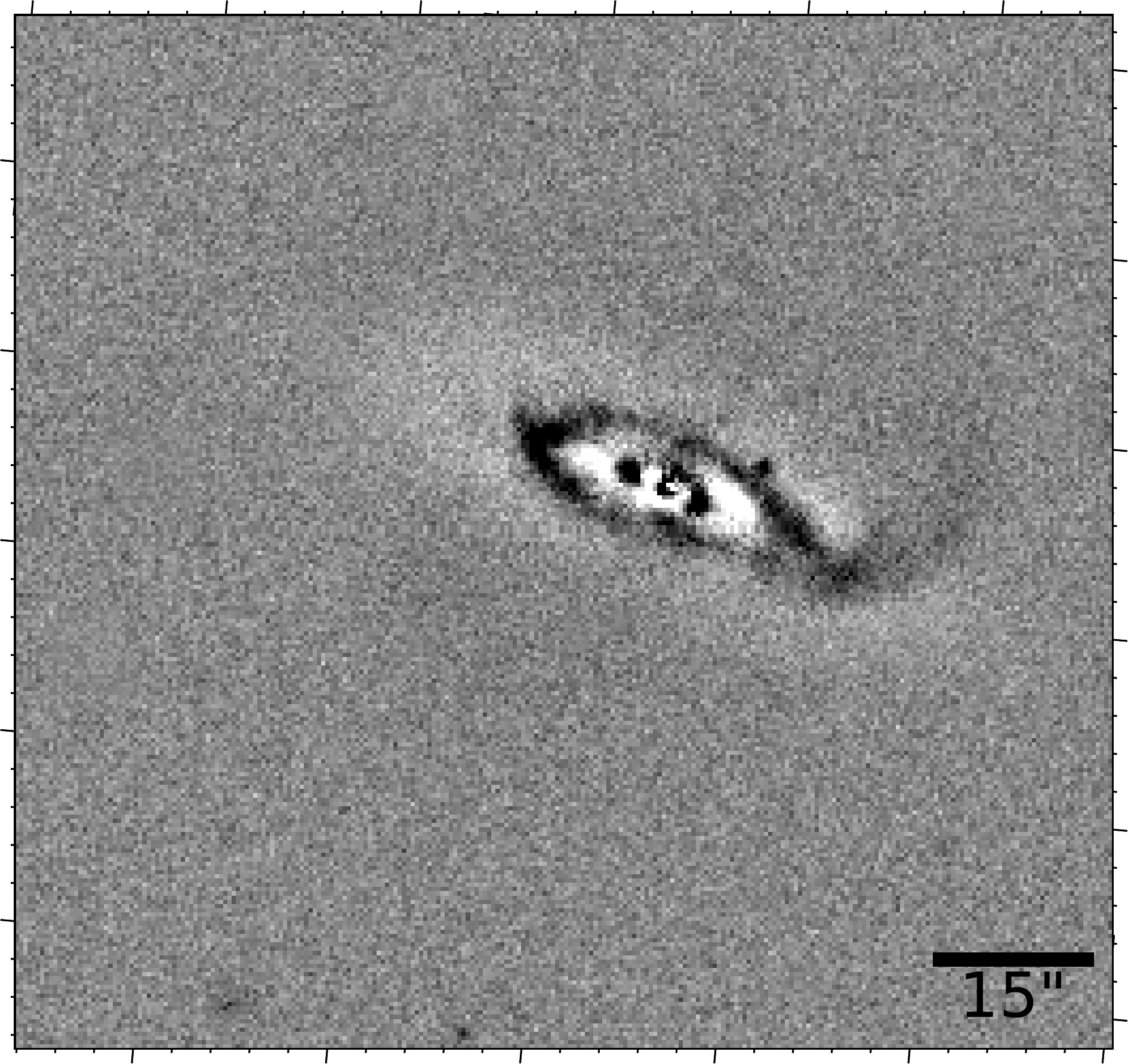} \label{125G2_resid2}}\\
\caption{An example of GALFIT output images for the 2D modeling of SIT 125G2 (one of our galaxy sample); at the left is the original cleaned image in the $r-$band, the middle panel is the model image and at the right is the residual image that results from subtracting the model image from the original one.}
\label{fig:2Dmodeling}
\end{figure*}

Figure~\ref{fig:2Dmodeling} shows an example of GALFIT output images, after fitting SIT 125G2 (a galaxy in our sample) using two components, including the original image (Figure~\ref{125G2_original}), after removing the projected objects, the model image (Figure~\ref{125G2_model}), and the residual image (Figure~\ref{125G2_resid2}) where a tidal tail appears clearly in the East-direction.

A detailed description of the 1D and 2D modelling analysis of all galaxies in our sample is describe in Sec~\ref{appendix}. In 2D decomposition we discussed the result of using two components including the SDSS image, the model image and the residual one.

\section{Results}
\label{Sec:results}

\subsection{Breaks in disk galaxies}

A break in the luminosity profile of disk galaxies is defined when a change in the slope of the exponential scalelength occurs. The break radius is thus determined at the intersection point between the straight line derived by 1D fitting representing the exponential profile and the data representing the distribution of light in the target disk galaxy \citep{Erwin2005, Martin2012, Laine2014}. In this study we ignored the changes in the exponential profile that arise in the shape of humps and$/$ or fluctuations in the inner region of disk galaxies, which are usually caused by bars or asymmetric structures in the galaxy core. For that reason, we define the break radius in our sample as a continuous change in the slope of the outer region in the luminosity profiles of disk galaxies.

To identify disk galaxies in our sample we did not depend mainly on the morphological classification mentioned in Table~\ref{tbl:basic}. Instead, we used the 1D fitting and 2D decomposition to separate disk galaxies from elliptical ones. An elliptical galaxy was re-classified as a lenticular or early-type spiral if a disk component was found. Surprisingly, from the original classification in Table~\ref{tbl:basic} we found five galaxies that are more likely distinguished as disk galaxies, four of which were classified as ellipticals in literature, and the fifth was classified as irregular galaxy. A detailed discussion for our re-classification of each case is included in the ~\ref{appendix}. Following our modified classification, we obtained 22 disk galaxies in our sample. 

In this section, we will mention the recorded fraction of each break type in the 22 disk galaxies in our sample, along with the distribution of the break radius. In addition, the correlation between the break radius and the tidal strength of each galaxy will be presented.

\subsubsection{Break type fraction}
We classified the light profiles of the 22 disk galaxies according to the observed breaks in the outer region into type I (no breaks), type II (down-bending), and type III (up-bending) as stated by \citet{Erwin2005}. We found that type III covers 55$\%$ (12 galaxies) of disk galaxies in our sample, while type II represents 36$\%$ (8 galaxies), one galaxy had a complex profile of type III+II, and two galaxies show type I profile. This indicates that the majority of disk galaxies in our sample are type III, i.e in state of interaction/minor merger \citep{Younger2007, Erwin2008, Laine2014}. 
The profile type of each galaxy and its corresponding break radius, diameter and tidal strength are presented in Table~\ref{tbl:1D_results}.

Moreover, we found that 83$\%$ of type III are strongly correlated with signs of interaction, clarified later in the 2D decomposition, such as; ring, tidal tails, bridge and tidal arms.

\subsubsection{The distribution of the normalized break radius ($R_{br}/R_S$)}

\begin{figure*}
 \centering
 \includegraphics[scale=0.8, bb=45 22 420 200 ]{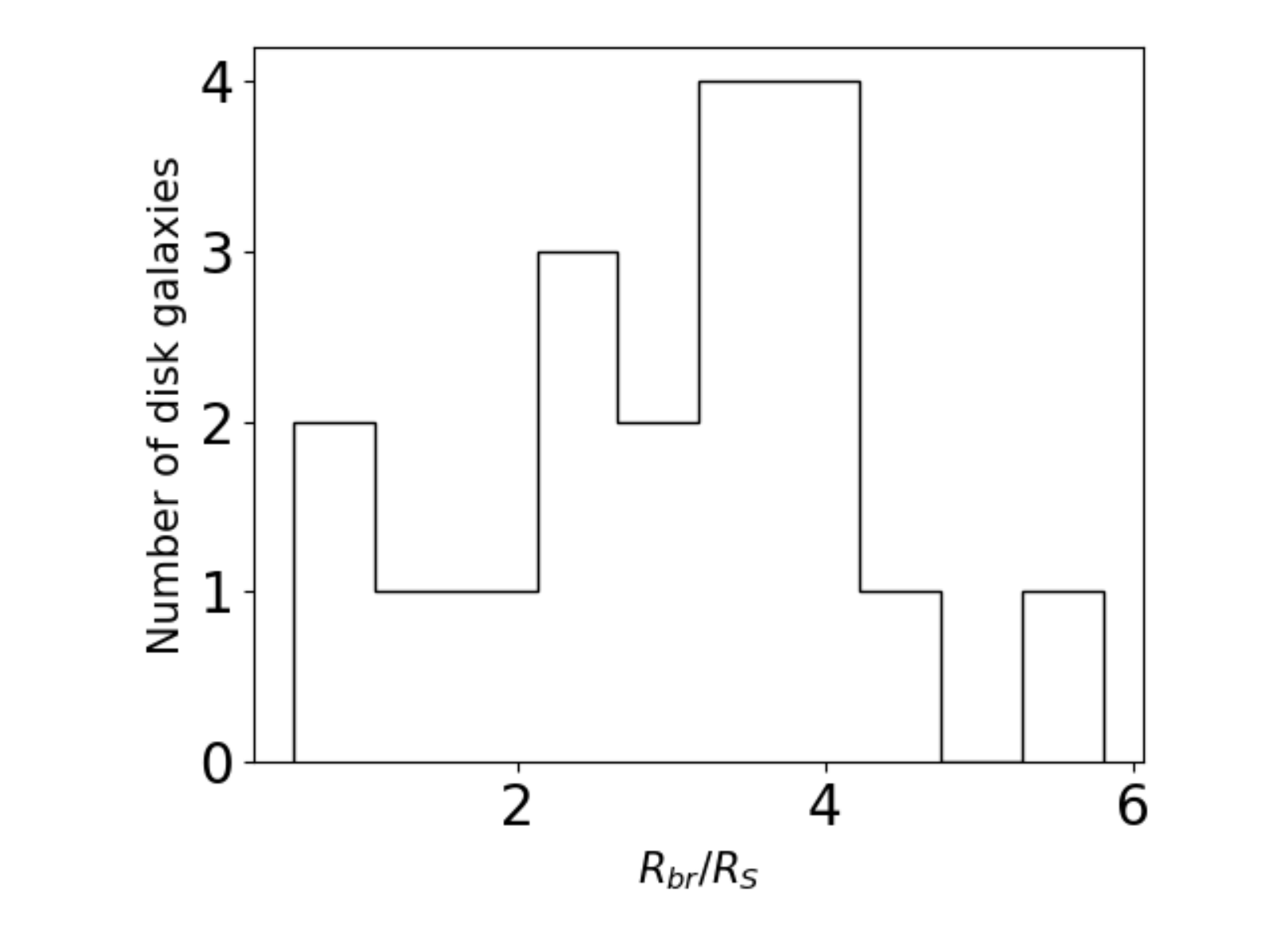} 
 \caption{The distribution of the normalized break radius (type II and III) of disk galaxies in our sample.}
 \label{fig:R_b}
\end{figure*}

In addition to the break type, we quantified the break radius $R_{br}$ in the outer region of the disk with profile type II and III, and normalized it with respect to the scale length of the disk $R_S$ (computed by GALFIT). In general, we found that the breaks in our sample appear mostly in the outer region between 2.3 $<$ $R_{br}/R_S$ $<$ 4.3 and only two galaxies show a break in the inner region at $R_{br}/R_S$ $<$ 1, as shown in Figure~\ref{fig:R_b}. This probably suggests that our triplet systems are in a weak state of interaction where the effect of gravitational interaction appears mainly in the outer region of the galaxies.

\subsubsection{The correlation between the normalized break radius ($R_{br}/R_S$) and tidal strength ($Q$)}
\label{ss:R&Q}

In order to clarify if the breaks in the luminosity profile is related to the star formation process or the gravitational interaction between galaxies in triplets, we investigated the correlation between the normalized break radius and the tidal strength (Dahari parameter) between members of galaxy triplets. 

The Dahari parameter ($Q$) \citep{Dahari1984} is used to identify the tidal strength between the primary galaxy and its companions within 1Mpc. This parameter depends mainly on the diameter of each galaxy ($D$) and the projected distance between them ($r_p$) \citep{Laine2014}. Following \citet{Laine2014}, the Dahari parameter can be computed by using the following equation;

\begin{equation}
    Q_i=\frac{(D_p D_c)^{1.5}}{(r_p)^3} ,
    \label{eqn:Dahari}
\end{equation}

where $D_p$ and $D_c$ are the diameters of the primary and the companion galaxy, respectively, and $r_p$ is the projected separation between them. 

For triplet systems, the Dahari parameter will be the logarithm of the sum of $Q_i$;

\begin{equation}
    Q=log_{10}(\sum\limits_{i=1}^n Q_i).
    \label{eqn:Dahari_sum}
\end{equation}

\begin{figure*}
 \centering
 \includegraphics[scale=0.8, bb=45 22 420 200]{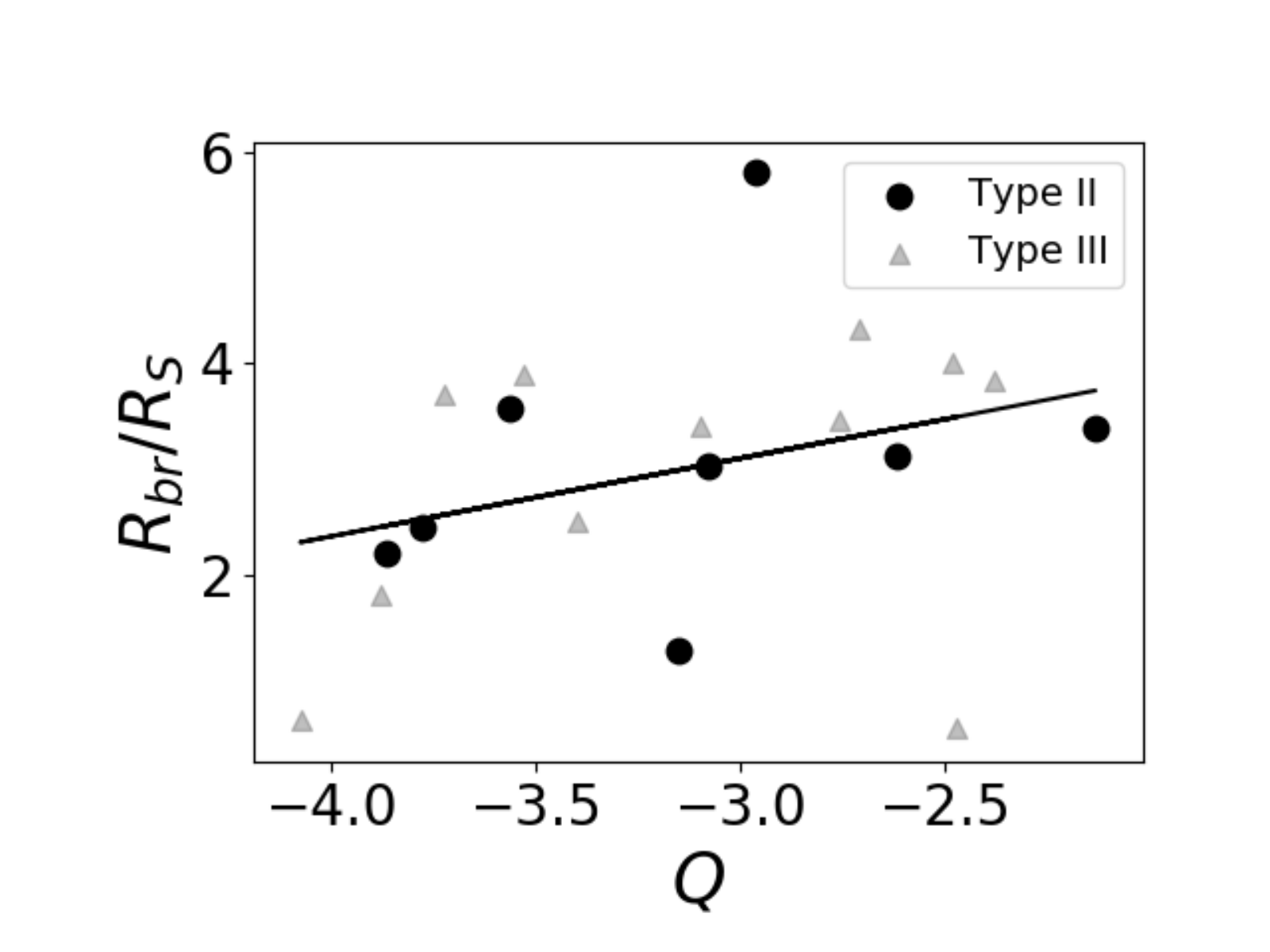} 
 \caption{The normalized break radius versus the Dahari parameter for type II profiles (denoted by circle) and type III profiles (triangle shape), showing a weak positive correlation with both types. The Dahari parameter is computed using Eq.~\ref{eqn:Dahari_sum} according to \citet{Laine2014}}
 \label{fig:R_Q}
\end{figure*}

Figure~\ref{fig:R_Q} shows a positive trend between the Dahari parameter and the normalized break radius of disk galaxies in our sample, with a correlation coefficient of 0.332. This finding is in a good agreement with the one published by \citet{Laine2014} who found that type III profiles are positively correlated with the tidal strength. It is also clear that the tidal strength appears to be stronger in the outskirts of disk galaxies where the gravitational interaction is more disruptive. 

\subsection{Distribution of SIT sample on the colour Magnitude Diagram (CMD)}

The colour-magnitude Diagram (CMD) is a useful tool to study the stellar population and evolution of galaxies \citep{Dey2000,Gusev2000,Pozzetti2000, Eisenhardt2006, Strazzullo2010}. This diagram is subdivided into two main regions: the `red sequence' of early type galaxies with dominant bulge component, more massive, little star forming, passive galaxies, and the `blue cloud' of late type galaxies with dominant disk, less massive, with active star formation \citep{Blanton2003, Driver2006,Wyder2007}. 

\begin{figure*}
 \centering
 \includegraphics[scale=0.8, bb=45 22 420 200]{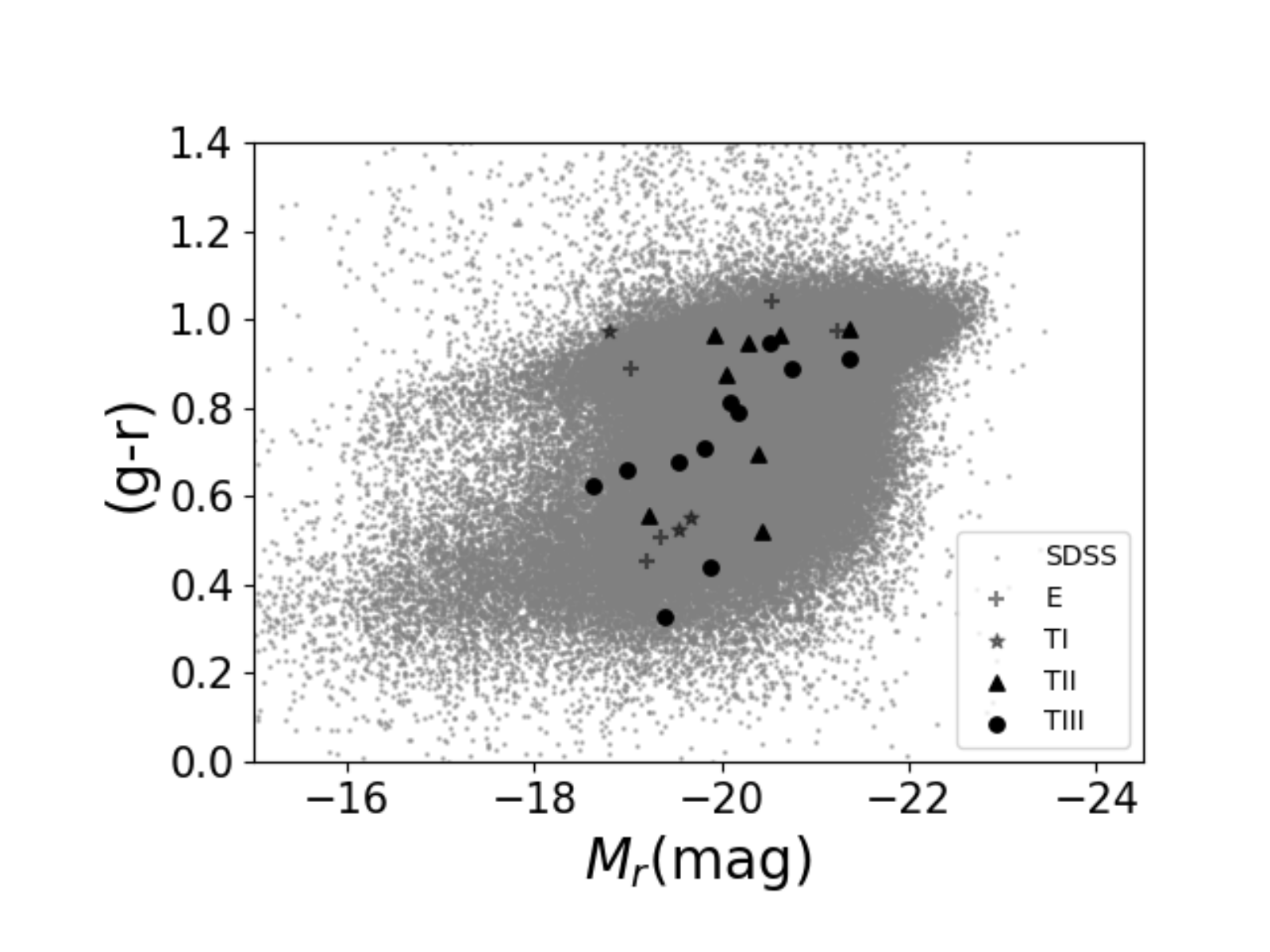} 
 \caption{The colour-magnitude diagram (CMD) of SIT sample over plotted over the KIAS-VAGC SDSS galaxies sample. M$_r$ is the absolute magnitude in $r$-band and (g-r) is the colour difference between the model magnitude in $g$- and $r$- bands. E indicates elliptical galaxies represented by $+$ symbol, TI are disk galaxies with simple exponential profiles (type I) with $*$ symbol, TII are disk galaxies with type II breaks denoted by a triangle and TIII are disk galaxies with type III breaks denoted by a circle.}
 \label{fig:CMD}
\end{figure*}

For this investigation, we overplotted our sample on the KIAS Value-Added Galaxy Catalog (KIAS-VAGC) that includes the data of 114,303 galaxies in the SDSS, based on Data Release 7 (DR7), corrected from galaxy extinction and k-correction \citet{Choi2010}. Although disk galaxies are mostly distinguished as late, blue galaxies, we found some disk galaxies in our sample fall on the red sequence, as shown in  Figure~\ref{fig:CMD}. The red colour of SIT disk galaxies might indicate the presence of dust in the galactic plane that generate redder colour and change the actual colour of these galaxies \citep{Martin2012}. Generally, it is clear that both type II (triangular shape) and type III galaxies (circular shape) are randomly scattered in the red sequence and the blue cloud, i.e. type III galaxies do not belong to the blue cloud where star formation is active. This might indicates that the up-bending in disk galaxies of SIT occurred long ago. This explains the weak positive correlation between the normalized break radius and the tidal strength (Dahari parameter) as stated in the previous section (see Figure~\ref{fig:R_Q}). In addition, we noticed that two elliptical galaxies fall in the blue cloud one of which show a clear stellar ring (SIT 280G1) in the residual image of the 2D decomposition. This might be explained by the existence of star formation caused by gravitational interaction between member of SIT systems.   

\begin{table*}
  \begin{center} 
   \tabcolsep 5.8pt
    \caption{Disk-profile parameters of 1D modeling for galaxies that show breaks of type II or type III.(1) galaxy ID, (2) and (3) are the break radius in $kpc$ and its normalization, respectively. (4) The isophotal diameter of each galaxy as extracted from NED, (5) the projected separation in $kpc$ as computed by \citet{Tawfeek2019}, (6) is the Dahari parameter, (7) is the break type; 2 for type II break and 3 represents type III break}.
    \label{tbl:1D_results}\\
  
\begin{tabular}{|l|r|r|r|r|r|r|}
\hline
  \multicolumn{1}{|c|}{ID} &
  \multicolumn{1}{c|}{R$_{br}$(kpc)} &
  \multicolumn{1}{c|}{R$_{br}$/R$_{d}$} &
  \multicolumn{1}{c|}{D(kpc)} &
  \multicolumn{1}{c|}{r$_{p}$(kpc)} &
  \multicolumn{1}{c|}{Q} &
  \multicolumn{1}{c|}{Ptype} \\
  \multicolumn{1}{|c|}{(1)} &
  \multicolumn{1}{c|}{(2)} &
  \multicolumn{1}{c|}{(3)} &
  \multicolumn{1}{c|}{(4)} &
  \multicolumn{1}{c|}{(5)} &
  \multicolumn{1}{c|}{(6)} &
  \multicolumn{1}{c|}{(7)} \\
\hline
  30G1 & 13.59 & 3.39 & 9.15 & 5.363 & -2.13 & 2\\
  30G2 & 10.19 & 3.846 & 6.19 & 9.621 & -2.374 & 3\\
  101G1 & 6.86 & 1.8 & 3.93 & 9.632 & -3.875 & 3\\
  101G2 & 7.66 & 3.704 & 4.85 & 10.506 & -3.72 & 3\\
  101G3 & 6.85 & 2.195 & 4.82 & 12.706 & -3.862 & 2\\
  104G1 & 12.62 & 3.571 & 4.5 & 10.869 & -3.564 & 2\\
  104G3 & 8.89 & 3.889 & 3.02 & 8.837 & -3.53 & 3\\
  125G1 & 7.56 & 3.462 & 3.33 & 3.683 & -2.757 & 3\\
  125G2 & 8.3 & 3.125 & 5.75 & 4.667 & -2.616 & 2\\
  125G3 & 6.19 & 3.409 & 3.1 & 7.893 & -3.093 & 3\\
  217G2 & 12.78 & 3.03 & 7.09 & 10.038 & -3.078 & 2\\
  217G3 & 5.13 & 2.5 & 3.56 & 12.256 & -3.396 & 3\\
  263G1 & 7.54 & 0.544 & 4.85 & 9.606 & -2.468 & 3\\
  263G2 & 11.28 & 2.449 & 4.84 & 12.454 & -3.775 & 2\\
  263G3 & 5.63 & 4.0 & 3.72 & 2.858 & -2.479 & 3\\
  264G1 & 7.56 & 5.806 & 4.99 & 4.27 & -2.958 & 2\\
  264G2 & 6.76 & 4.324 & 3.51 & 3.431 & -2.708 & 3\\
  280G2 & 3.67 & 0.612 & 2.42 & 9.768 & -4.074 & 3\\
  280G3 & 3.67 & 1.277 & 4.37 & 6.982 & -3.151 & 2\\
\hline\end{tabular}
\end{center}
\end{table*}

\subsection{Identifying signs of interaction/merger in our sample}

We based the identification of an interaction or merger in our sample according to the study by \cite{Hernandez2011}, where late-type galaxies are pinpointed as interacting (I) if they are associated with faint bridge, warps, asymmetries, or tidal arms. On the other hand, early-type galaxies are likely interacting if they show fan-like arms, tail-like features, rings, or shells. We include those considerations to the diagnose of type III-d profile in either early- or late-type galaxies as a state of suspected interaction (SI).

Meanwhile, \cite{Hernandez2011} classified early-type galaxies with evidence of a merger (M) or possible merger (PM) if they reveal a double nuclei, or prominent dust lanes, while late-type galaxies are distinguished as merger if they show double nuclei, strong tidal tails, bridge, or strong morphological transformations. Otherwise, galaxies that do not show any features from the above mentioned are denoted by (NI) i.e, no evidence for interaction.

\begin{figure*}
\centering
\subfloat{\includegraphics[scale=0.2, bb=2 2 540 400]{Figures_pdf/125G2_residue2-eps-converted-to.pdf} \label{125G2_tidal tail}}
\hspace{0.12cm}
\subfloat{\includegraphics[scale=0.21, bb=2 2 540 400]{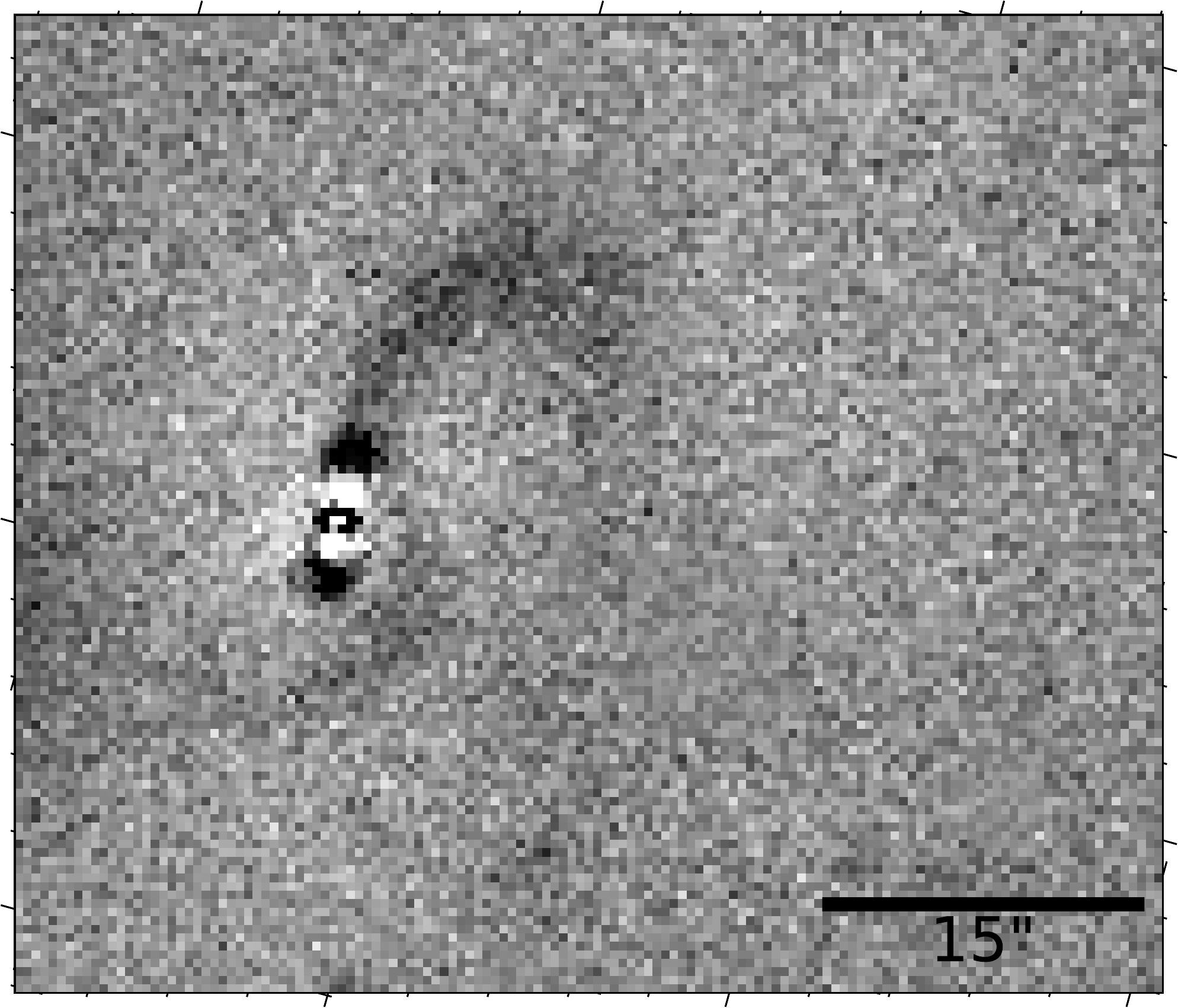} 
\label{263G3_tidal arm}}
\hspace{0.12cm}
\subfloat{\includegraphics[scale=0.24, bb=2 2 540 400]{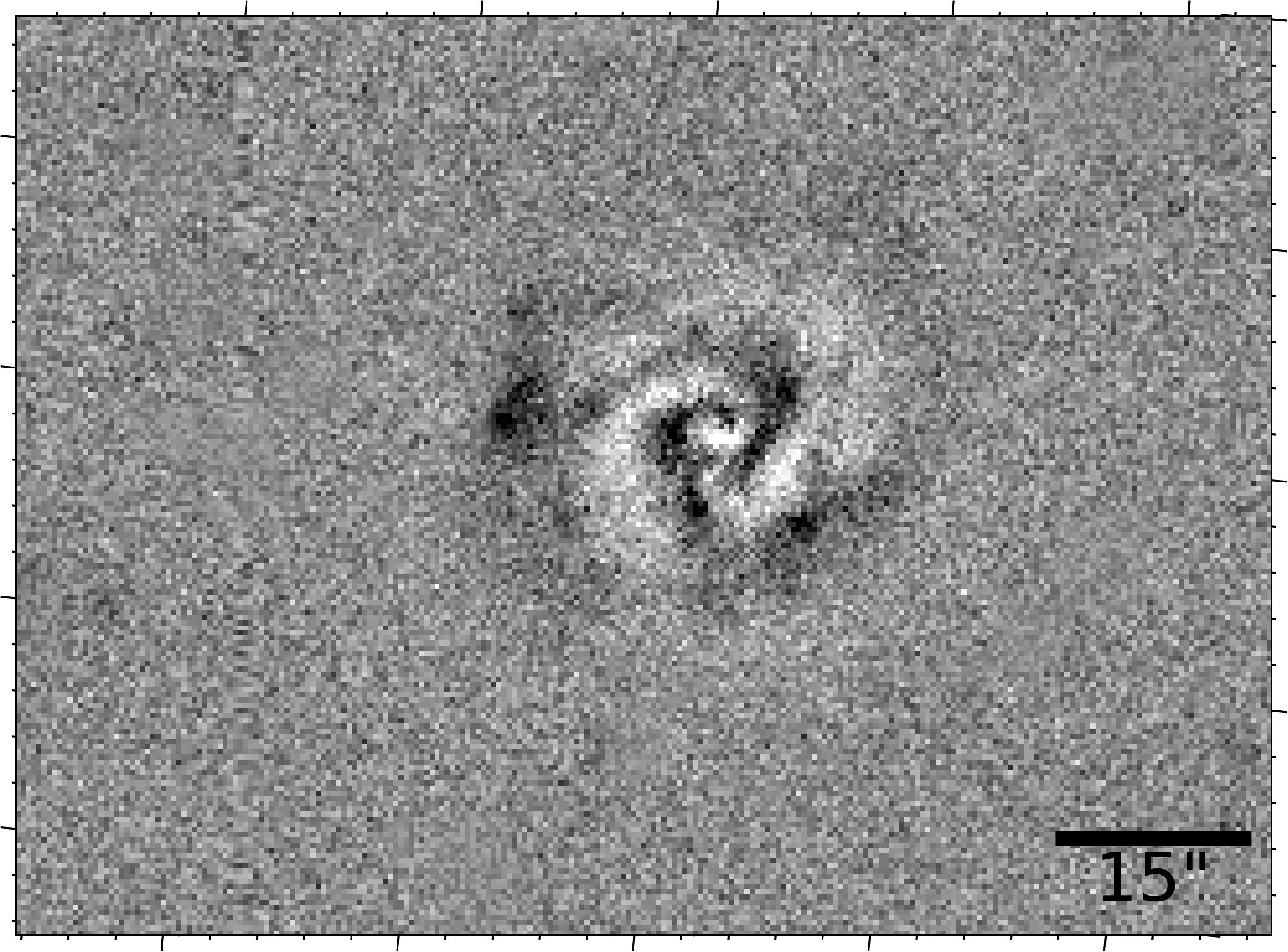} \label{125G3_fan arms}}\\
\vspace{0.12cm}
\subfloat{\includegraphics[scale=0.23, bb=2 2 540 400]{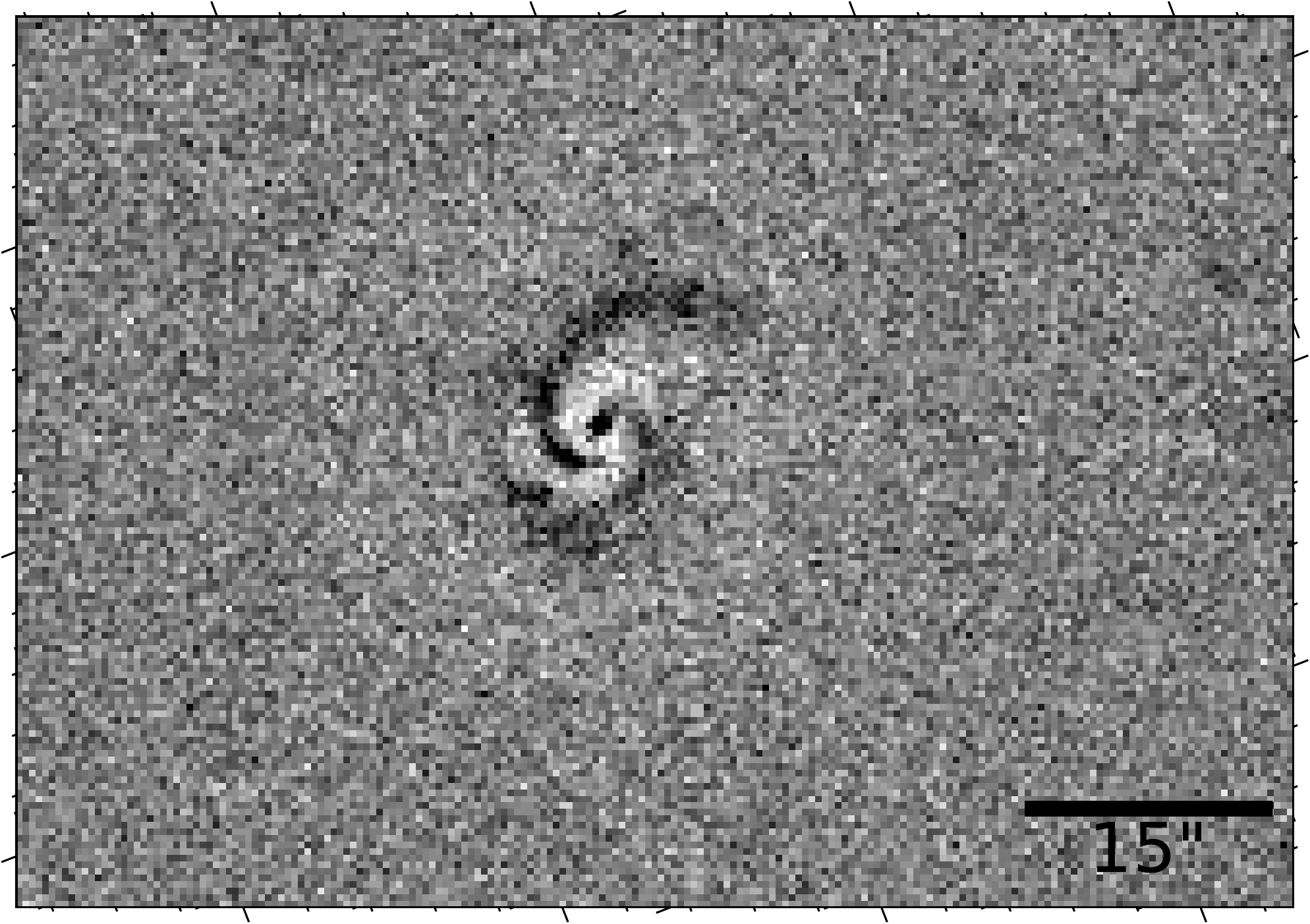}
\label{217G2_asymmetric arms}}
\hspace{0.2cm}
\subfloat{\includegraphics[scale=0.22, bb=2 2 540 400]{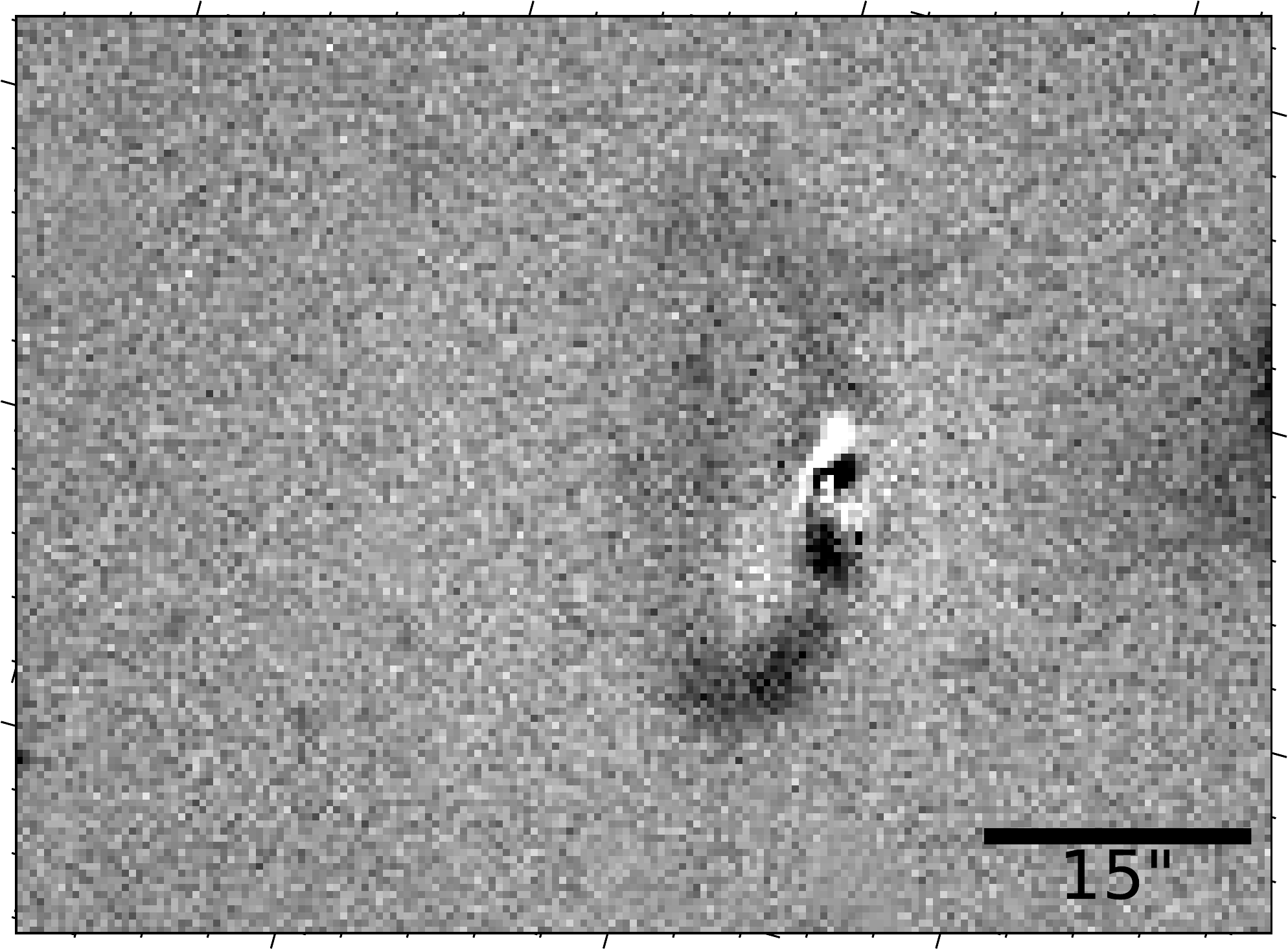}
\label{263G1_bridge}}\\
\subfloat{\includegraphics[scale=0.22, bb=2 2 540 400]{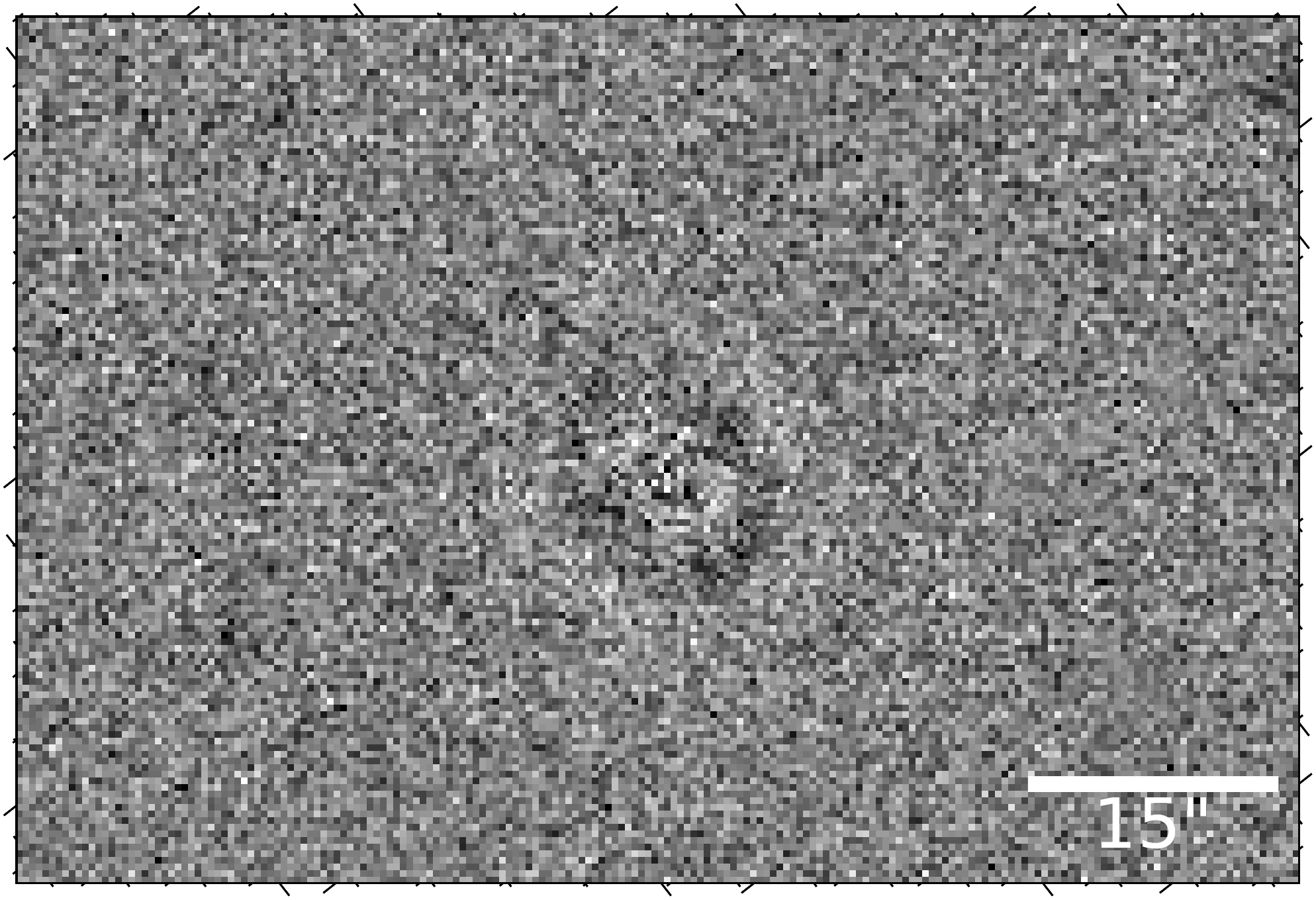}\label{104G1_ring}}
 \hspace{0.2cm}
\subfloat{\includegraphics[scale=0.17, bb=2 2 540 400]{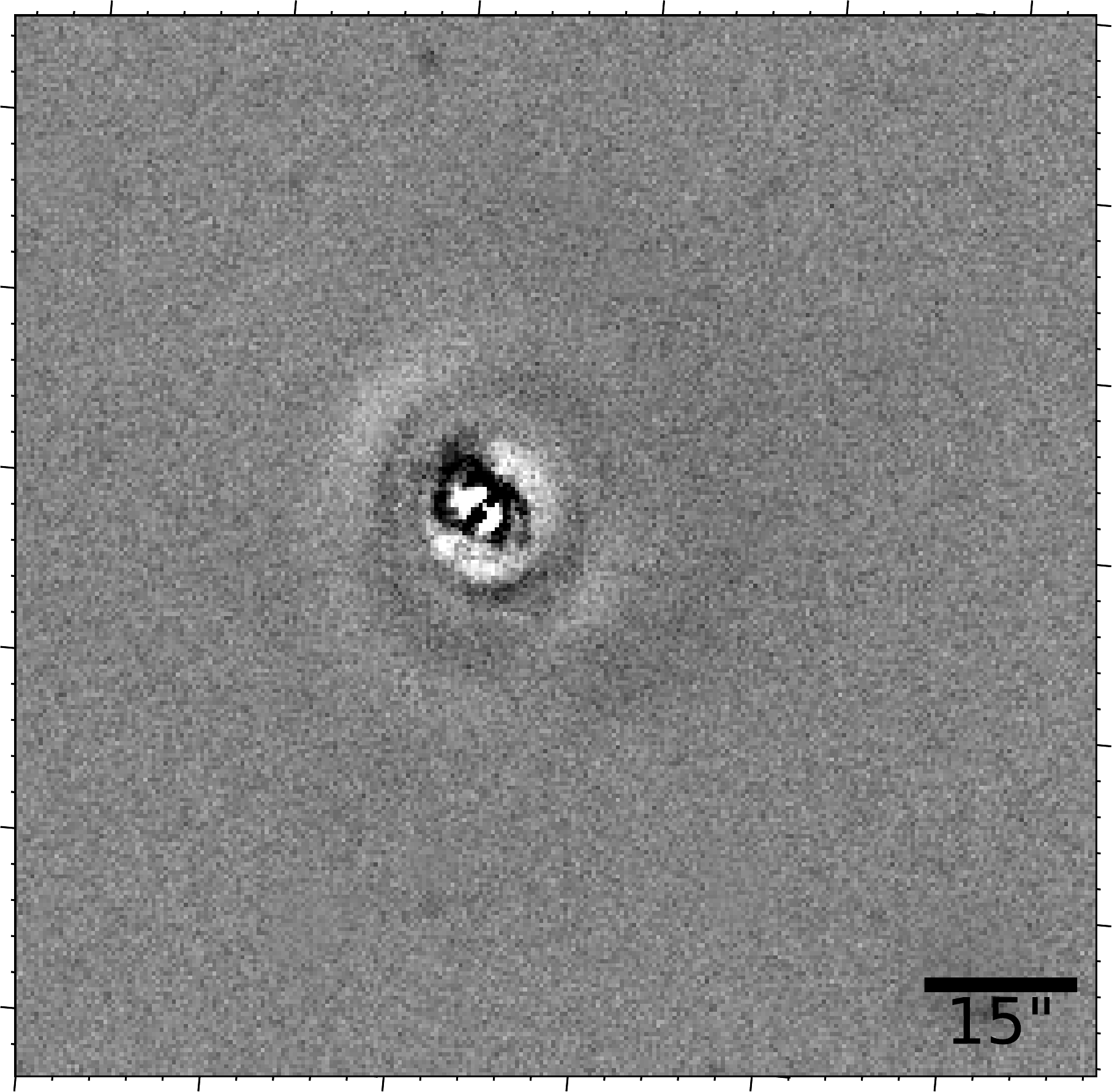}\label{125G1_ring}}
\hspace{0.15cm}
\subfloat{\includegraphics[scale=0.23, bb=2 2 540 400]{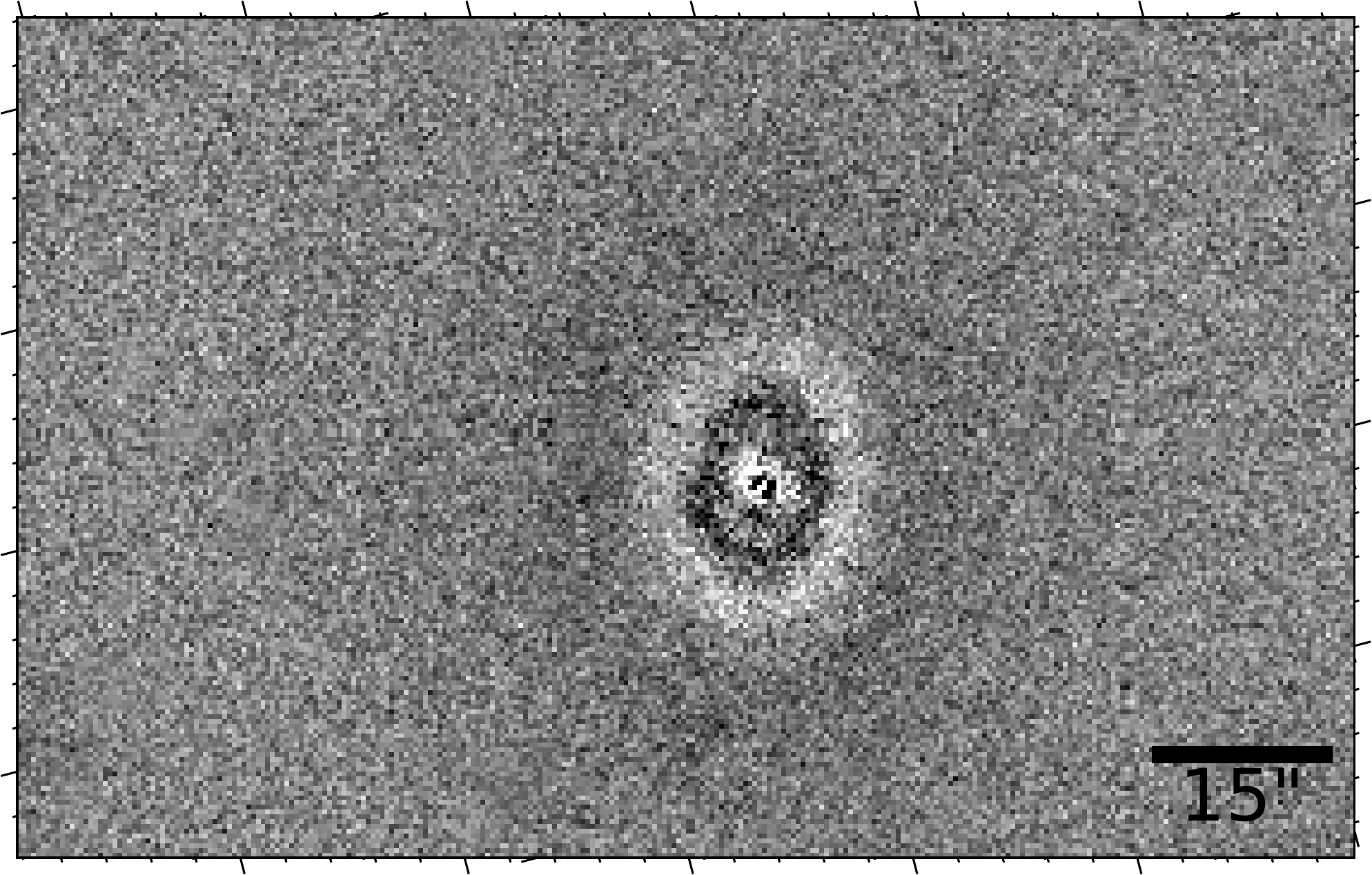}\label{280G1_ring}}
\caption{Some examples of GALFIT's output residual images clarifying signs of interaction in our sample; tidal tail in SIT 125-G2 in the top right panel, tidal arms in SIT 263-G3 in the top middle panel, fan arms in SIT 125-G3 in the top left panel, asymmetric arms in SIT 217-G2 in the middle left panel, bridge in SIT 263-G1 in the middle right panel and at the bottom are residuals images showing a faint ring in SIT 104-G1 at the left, shells in SIT 125-G1 at the middle and a clear ring in SIT 280-G1 at the right.}
\label{fig:2dsignsofinteraction}
\end{figure*}

Following this strategy we found that three galaxies (11$\%$) fall in a state of possible merger (PM), 16 (60$\%$) reveal interaction signs (I), three (11$\%$) with suspected interaction (SI), and only five galaxies (18$\%$) do not show any evidence of interaction (NI). Figure~\ref{fig:2dsignsofinteraction} illustrates some examples of signs of interactions (e.g. tidal tail, fan arms, bridge, asymmetric arms) clarified in the residual images of our SIT sample produced by GALFIT.

\subsection{Rings and shells in the residual images}

Rings (inner or outer) and shells are directly related to the internal dynamics and evolution of galaxies. They are considered to be a signature of interaction, accretion, and merger sequences \citep{Weil1993,Pop2018,Rampazzo2020}. Inner rings are old diffuse red rings formed with ancient stars, while outer rings are composed mainly of massive, relatively young blue stars \citep{Mutlu2017}. They are associated with active star formation that caused mainly by galaxy interaction, collision or merger \citep{Elmegreen1992,Buta1996, Bekki1997,Elmegreen2006(a), Hernandez2011}.

Shells are more common in early-type galaxies. They are defined as faint scattered stellar tidal with large open angel. Conflicting to outer rings, \citet{Atkinson2013} found that shells are more common in red than blue galaxies despite being correlated with minor and major merger \citep{Weil1993, Pop2018, Rampazzo2020}.  

As a result of the 2D decomposition we found that 59$\%$ (16 galaxies) of our sample reveals rings and/or shell structures. Ring structure were found in 10 galaxies; five (3 early-type and 2 late-type galaxies) of them declare a blue (g-r) in the CMD. The remaining six galaxies (5 late-type and 1 early-type) exhibit a shell structure and fall on the red sequence in the CMD in agreement with \citet{Atkinson2013} who found that shells are usually found in red galaxies i.e. spirals (late-type galaxies).

In addition, we found that 62$\%$ (10 from 16) of galaxies that exhibit ring and/or shell are associated with other signs of interaction (e.g. type III break, tidal arms, tails, etc) in their luminosity profiles and residual images. Moreover, we noticed that G1 (the brightest galaxy in each triplet system) is usually associated with a ring and/or shell structure. In our sample we found a ring in 78$\%$ of G1 (seven systems, five are early-type galaxies and two are late-type) with only two systems (22$\%$) lacking ring structures in their central galaxies (G1),  indicating that the probability of finding rings and/or shells in early-type galaxies is twice as the one found in late-type galaxies. Since early-type galaxies are considered to be the final product of galaxy interaction and galaxy evolution (\citet{Rampazzo2020}), finding rings in such galaxies confirm the strong correlation between rings and galaxy interaction. 

Figures~\ref{104G1_ring}, ~\ref{125G1_ring} and ~\ref{280G1_ring} clarifies some examples of residual images from our sample showing a ring and/or shell structure.

\section{Summary and conclusions}
\label{s:conclusions}

The key objective of this study is to understand and identify the state of gravitational interaction in galaxy triplet systems. To achieve this goal we used two different techniques, to identify the signs of interaction, on a sample of nine galaxy triplet systems (27 galaxies) with $r_p$ $\leq$ 0.10 Mpc, and $m_r$ $<$ 17.0, selected from the ``SDSS-based catalogue of Isolated Triplets'' (SIT)
\citep{Fernandez2015}.

The first technique consisted in fitting the one-dimensional (1D) light profile of the galaxies classified as ``disky" in our sample of 27 galaxies (22 out of the total). Like this, we were able to exploit the breaks in these profiles to identify signs of gravitational interaction, where type I represents a simple exponential profile without breaks, type II presents a down-bending break, and type III shown an up-bending break as previously done by \cite{Erwin2008}. Accordingly, we found that;
\begin{itemize}
      \item Type III covers 55$\%$ (12 galaxies) of disk galaxies in our sample, while type II represents 36$\%$ (8 galaxies), one galaxy had a complex profile of type III+II and only two galaxies show type I profile. This indicates that the majority of disk galaxies in our sample are in state of interaction/minor merger (type III profiles) \citep{Younger2007, Erwin2008, Laine2014}.
      \item Comparing with the study by \cite{Erwin2008}, who analyse the surface brightness profile of 66 early-type barred galaxies, we found that; in both studies type I profiles are less frequent than type II and type III ;15$\%$ (2 galaxies) in our study versus 27$\%$ (16 galaxies) in Erwin study. In our study, type III presents the highest percent; 55$\%$ (12 galaxies) versus 30$\%$ (19 galaxies) in Erwin study, while in \cite{Erwin2008} type II is the most common class; 48$\%$ (31 galaxies) versus 19$\%$ (8 galaxies) in our study. This suggests that signs of gravitational interaction are more frequent in our sample i.e. SIT triplet systems are in state of interaction.
      \item The distribution of the normalized break radius ($R_{br}$/$R_S$) indicated that the majority of our sample shows a break in the outer region (2.3 $<$ $R_{br}$/$R_S$ $<$ 4.3). This might suggest that triplets in our sample are in a weak state of interaction that affects the outer region of the galaxies only. 
      \item Type III profiles show a weak positive correlation with the tidal strength ($Q$) between the members of SIT systems and a random scattered between the red sequence and the blue cloud in the colour-magnitude diagram (CMD). This might probably reveals that up-bending in disk galaxies (type III) of SIT occurred long time ago.
\end{itemize}

The second approach we adopted in the current study is the two dimensional (2D) decomposition based on decompose the fine entire structures of galaxies that are not visually clear in their observed images. From this analysis, signs of interaction (rings, tidal arms, fan arms, tidal tails, bridges, or bi-nuclei) can be identified by subtracting the model image from the original image of each galaxy (i.e. the residual image) via GALFIT. Following this methodology we found the following;
\begin{itemize}
    \item The majority of galaxies in our sample (70$\%$) show asymmetric features and/or signs of interactions in their residual images. A small fraction (11$\%$) has model residuals that are compatible with the galaxy being symmetric and hence not showing any detectable signs of disturbance; only one galaxy was successfully fitted by using a single $Ser$ component.
    \item Galaxies that reveal signs of interactions in their residual images are strongly correlated with type III profiles (up-bending break).
    \item Ring and/or shell structures were found in 59$\%$ (10 with ring and 6 with shell) of galaxies in our sample, 62$\%$ of them are associated with other signs of interactions such as tidal arms, type III break and tidal tails. These features are considered signatures of past interaction or merger.
    \item 50$\%$ of galaxies with rings (5 out of 10) are located in the blue cloud in the CMD that might suggest a recent active star formation caused by merging events. While, four out of six galaxies with shell structure have a red (g-r) color profiles and the remaining two galaxies are located in the green valley.
    \item Seven of nine central galaxies (78$\%$, five early-type and 2 late-type galaxies) reveals a ring structure while only one elliptical (SIT 197G1) and one sprial (SIT 264G1) lack such structure. This confirms that rings play a big role in galaxy interaction. 
    \item The classification of six galaxies (see Table~\ref{tbl:results}) has been modified from the published one after applying the 2D decomposition, hence, 2D decomposition can be used to perform a finer morphological classification.
    \item Three galaxies (11$\%$) are in state of possible merger (PM), 16 (60$\%$) reveal interaction signs (I), three (11$\%$) with suspected interaction (SI), and only five galaxies(18$\%$) do not show any evidence of interaction (NI). 
    \item In agreement with \cite{Hernandez2011} we found that spirals are dominant in triplet systems with different stage of interaction, where our sample's modified classification is composed of 16 (59$\%$) spirals and 11 (40$\%$) elliptical and lenticulars. In addition, we confirm that bars are almost twice common in late-type more than in early-type spirals, where we recorded seven barred spiral galaxies, five are late-type and two are early-type. Finally, we affirm that ellipticals represent the lowest fraction in triplet systems (5 galaxies, 18$\%$), while late-type spirals are relatively higher (10 galaxies, 37$\%$).
    
\end{itemize}

The output parameters of the above mentioned methodologies are presented in Table~\ref{tbl:results}.

In sum, we can conclude that using 1D fitting and 2D decomposition reveal the state of interaction in 81$\%$ of galaxies in our sample with $r_p$ $\leq$ 0.10 Mpc and $m_r$ $<$ 17.0.

\newpage
\begin{landscape}
\begin{table}
\begin{center}
\scriptsize
    \caption{\scriptsize{Summary of the results of the 1D fitting and 2D decomposition of the 27 studied galaxies. Column (1) is the galaxy ID, the second column represents the published morphological classification of each galaxy, while the third one referees to the modified classification found through this study with the indication of a ring (R) or bar (B), if found. Column (4) is the break type in the intensity profile of each galaxy. Columns (5), (6), (7) and (8) are GALFIT output parameters by using one $Ser$ component. Columns (9) to (14) are GALFIT output parameters by using two components. The last column identify the main feature(s) of interaction with a flag indicating if the galaxy is in a state of interaction (I), suspected interaction (SI), possible merger (PM), or without evidence of interaction (NI).  NF refers to undefined values.}} 
    \label{tbl:results}
\begin{tabular}{|l|l|l|l|l|l|l|l|l|l|l|l|l|l|l|l|l|l|} 
\hline
  \multicolumn{1}{|c|}{ID} &
  \multicolumn{1}{c|}{Published} &
  \multicolumn{1}{c|}{Modified} &
  \multicolumn{1}{c|}{Intensity } &
  \multicolumn{4}{c|}{$Ser$ 1comp} &
  \multicolumn{3}{c|}{$Ser$ 2comps} &
  \multicolumn{3}{c|}{Disk} &
  \multicolumn{1}{c|}{Sign(s) and state }\\
  \multicolumn{1}{|c|}{} &
  \multicolumn{1}{c|}{G-type} &
  \multicolumn{1}{c|}{R$/$B} &
  \multicolumn{1}{c|}{Type} &
  \multicolumn{1}{c|}{mag} &
  \multicolumn{1}{c|}{n} &
  \multicolumn{1}{c|}{$R_e$($''$)} &
  \multicolumn{1}{c|}{$\chi^2_{\nu}$} &
  \multicolumn{1}{c|}{mag} &
  \multicolumn{1}{c|}{n} &
  \multicolumn{1}{c|}{$R_e$($''$)} &
  \multicolumn{1}{c|}{mag} &
  \multicolumn{1}{c|}{$R_S$($''$)} &
  \multicolumn{1}{c|}{$\chi^2_{\nu}$} &
  \multicolumn{1}{c|}{of interaction}\\
   \multicolumn{1}{|c|}{(1)} &
  \multicolumn{1}{c|}{(2)} &
  \multicolumn{1}{c|}{(3)} &
  \multicolumn{1}{c|}{(4)} &
  \multicolumn{1}{c|}{(5)} &
  \multicolumn{1}{c|}{(6)} &
  \multicolumn{1}{c|}{(7)} &
  \multicolumn{1}{c|}{(8)} &
  \multicolumn{1}{c|}{(9)} &
  \multicolumn{1}{c|}{(10)} &
  \multicolumn{1}{c|}{(11)} &
  \multicolumn{1}{c|}{(12)} &
  \multicolumn{1}{c|}{(13)} &
  \multicolumn{1}{c|}{(14)} &
  \multicolumn{1}{c|}{(15)}\\
\hline
  30G1 & Sb & Sb/R  & II.o   &15.4&2.3&9.4  &2.79 & 16.5 & 4.0& 5.2&  16.01&5.9& 1.7 & shell, ring(I)\\ 
  
  30G2 & Sb & SBb/R & III-d  &16.0&2.7&4.2  &2.37  & 17.6 & 0.79 &0.8 &16.30&3.9& 1.15 &ring, III-d(I)\\
  
  30G3 & S0 & S0 &  I  & 16.37 & 1.95 & 3.4 &1.30& 17.9&0.95&1.0&  16.71&2.9 & 1.05 &(NI)\\ 
  
  101G1 & S0 & S0/R &  III-d &16.43&2.07&3.0&1.18&17.2& 1.13&1.4& 16.85&5.0 & 0.97 & ring, III-d(I)\\
  
  101G2 & SBab & SBab & III-d &16.57&2.79&4.7&1.58 &18.9&0.56&0.6& 16.97&2.7 & 1.15 & III-d (SI) \\ 
  
  101G3 & SBc & SBc &  II.i & 17.47&1.43&7.2&1.08 & 20.7 & 0.57 &1.9 &17.58 &4.14& 1.05 &tidal arms, asymmetries(I)\\
  
  104G1 & E & E/S0/R &  II & 16.64&1.85&4.1& 1.14 &18.9 & 0.63 & 0.9&16.85 &2.76& 1.06 & ring(I) \\
  
  104G2 & E & S0 & I & 17.54&1.08&1.9&1.23 & 19.6 & 0.39 & 0.8& 17.69&1.4 &1.12 &(NI)\\ 
  
  104G3 & SB & S & III-s & 17.8&1.4&1.9&1.12 &19.2 & 0.63 & 0.8& 18.08&1.8 & 1.06 & (NI)\\
  
  125G1 & S0 & S0/R &  III-d & 14.02&1.66&4.8&1.52 & 14.9 & 1.0 &2.5&14.56 &5.2& 1.08 & shell, ring, III-d(I)\\
  
  125G2 & SABa & SABa/R &  II.i+II.o & 14.8&2.5&7.8 &2.34& 16.6&0.82& 1.5& 15.18&6.4& 1.43 & tidal like, asymmetries(PM) \\
  
  125G3 & Sc & Sc & II-o & 15.7&0.99 &7.2&1.07&18.8& 4.0&34.9&15.71&4.4 & 1.06 & tidal fan arms, asymmetries(I)\\
  
  197G1 & E & E & NF  & 16.7&1.7&2.6 &1.05&  17.99& 0.82& 1.0&16.99&2.6& 0.80 & (NI)\\
  
  197G2 & E & E & NF  & 16.9&1.7&2.2 & 1.24& 18.4& 0.67 &0.8&17.21&2.05& 0.91 & (NI)\\ 
  
  197G3 & E & E/R &NF  & 17.36&1.31&2.3&1.13& 17.4&1.3& 2.2&18.79& NF& 1.12 & ring (I)\\
  
  217G1 & E & E/R & NF & 16.4&2.5&3.5& 17.9& 0.93 &0.9&1.24&16.73&7.92& 0.9 & ring (I)\\ 
  
  217G2 & Sc & SBc & II.o & 17.1&1.36&5.7& 1.17& 17.1& 1.36& 5.6 & 17.32&3.3& 1.17  & tidal arms, asymmetries (I) \\
  
  217G3 & E & Sa & III-d & 17.3&4.0&4.2&1.59& 19.1&0.69& 0.6&18.32&1.65 & 1.08 &  III-d (SI) \\ 
  
  263G1 & SBb & SBb & III-d & 15.6&2.84&8.3& 0.87& 16.4& 1.88& 3.6& 15.48&14.7& 0.87 & bridge, tidal tails,  III-d (PM)\\
  
  263G2 & S & S & III-d & 16.3&1.53&4.6& 1.47& 17.3 & 1.18& 2.6& 16.74&4.9& 1.23  &tail arm, shell, asymmetries (I)\\ 
  
  263G3 & S & S/R & III-d & 16.8& 1.37&1.7&3.47& 16.9&1.17&1.5& 16.99&1.5& 0.85  & tidal arm, ring,  III-d (PM)\\
  
  264G1 & Sc & Sc & III-d+II.o &15.1&2.21&10.8& 1.95& 17.2&1.46&3.1& 17.16&3.1& 1.63 &tidal arms, shell, III-d (I)\\ 
  
  264G2 & I & Sc & III-d & 15.6&0.79&5.5& 2.26&15.7&0.96&5.2& 15.85&3.7& 1.41 &tidal arms, shell, III-d (I)\\
  
  264G3 & E & S & II.o & 18.57&0.85&4.5& 1.06& & & & & & & (NI) \\
  
  280G1 & E & E/R & NF & 15.1&2.35&6.3 &0.98& 16.99& 1.15& 1.4& 15.49&4.2& 0.98  & ring (I)\\ 
  
  280G2 & E & S0/Sa & III-d & 16.5&2.19&3.8&1.10& 16.99 & 2.05 & 3.4& 17.31&9.8& 1.07  &shell, III-d (I)\\ 
  
  280G3 & SBc & SBc & II.o & 17.4&1.05&5.0& 1.94& 17.6&0.94&4.7& 15.17&360.7*& 1.91  &warp, tidal arm (I) \\
\hline
\end{tabular}
\end{center}
\end{table}
\end{landscape}

\section{Acknowledgements}

The authors thank the thorough reading of the original manuscript by the anonymous referee and his/her insightful comments that helped to improve the quality of the paper and clarify the results.

Amira A. Tawfeek thanks the Inter-University Centre for Astronomy and Astrophysics (IUCAA), Pune, India and the researchers there for their immense knowledge and great effort in teaching me essential programs used in this work.

Amira A. Tawfeek and B. Cervantes Sodi acknowledge financial support through PAPIIT project IA103520, from DGAPA-UNAM, México.

Funding for SDSS-III has been provided by the Alfred P. Sloan Foundation, the Participating Institutions, the National Science Foundation, and the U.S. Department of Energy Office of Science. The SDSS-III web site is http://www.sdss3.org/.

SDSS-III is managed by the Astrophysical Research Consortium for the Participating Institutions of the SDSS-III Collaboration including the University of Arizona, the Brazilian Participation Group, Brookhaven National Laboratory, Carnegie Mellon University, University of Florida, the French Participation Group, the German Participation Group, Harvard University, the Instituto de Astrofisica de Canarias, the Michigan State/ Notre Dame/ JINA Participation Group, Johns Hopkins University, Lawrence Berkeley National Laboratory, Max Planck Institute for Astrophysics, Max Planck Institute for Extraterrestrial Physics, New Mexico State University, New York University, Ohio State University, Pennsylvania State University, University of Portsmouth, Princeton University, the Spanish Participation Group, University of Tokyo, University of Utah, Vanderbilt University, University of Virginia, University of Washington, and Yale University.


\bibliographystyle{elsarticle-num-names}  
\bibliography{refbib_SIT} 


\newpage

\appendix

\section{Detailed analysis of triplet systems in our sample}
\label{appendix}

In this section, the detailed analysis of our sample using 1D and 2D decomposition is presented. In 2D decomposition we discussed the result of using two components, where almost all galaxies in our sample require more than one profile to be fitted. We identified the galaxies in every system as G1, G2 and G3 according to their brightness, where G1 represents the brightest galaxy and G3 the faintest one. In all Figures, the SDSS image of each triplet system in r-band is included in the first panel, the 1D fitting of G1, G2, and G3 are illustrated in the second panel(from left to right) with a fixed scalebar length of 15$''$ in all images. The last three panels represent the 2D decomposition of G1, G2, G3 (from top to bottom) by using two profiles, including the cleaned image (at the left), the model image (at the middle) and the residuals (at the right).

\subsection{The triplet system SIT 30}
\label{ss:SIT 30}

\begin{figure*}
\centering
\subfloat{\includegraphics[scale=0.5, bb=130 -150 100 100]{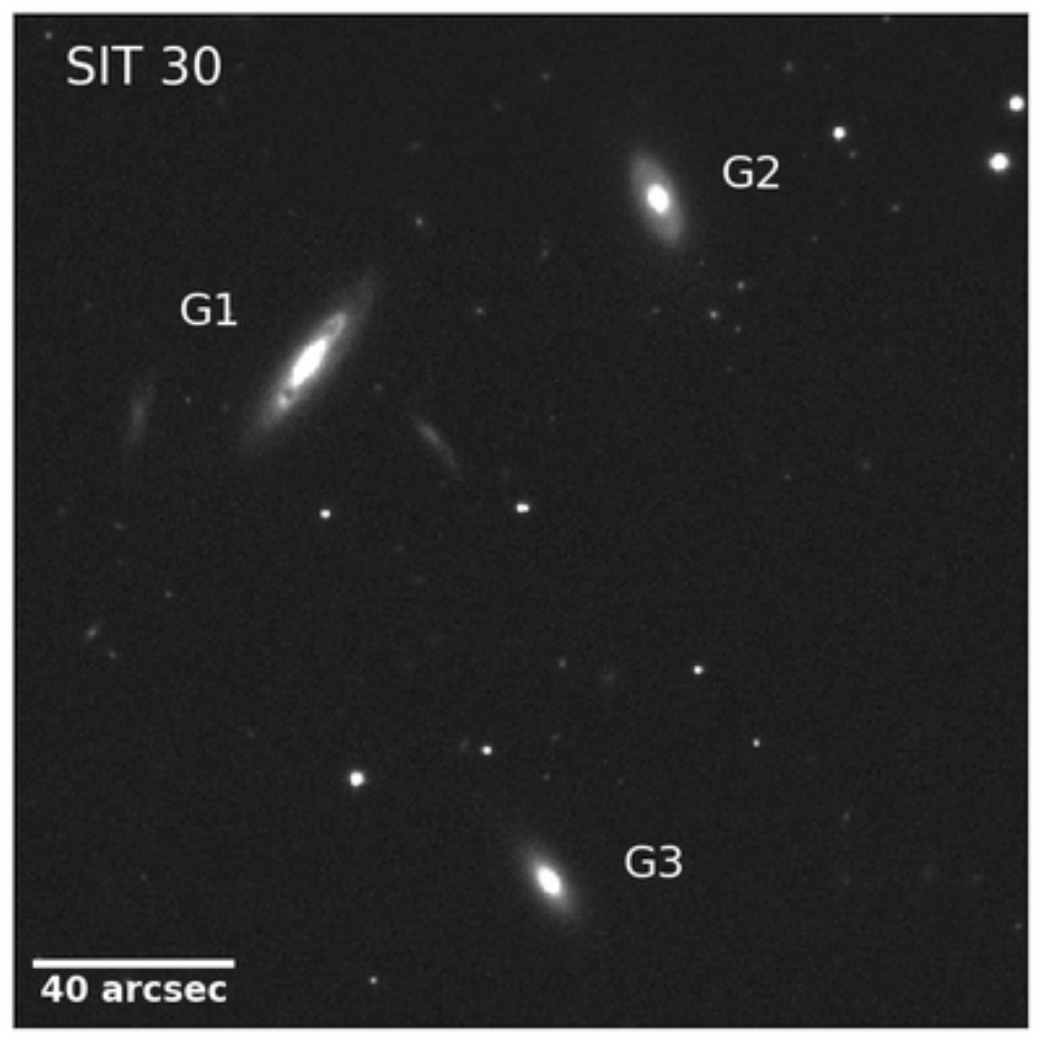}
\label{fig:SIT30}}\\
\subfloat{\includegraphics[scale=0.7, bb=80 20 500 100]{Figures_pdf/30_SB-eps-converted-to.pdf}\label{30_SB}}\\
\subfloat{\includegraphics[scale=0.2, bb=2 2 540 400]{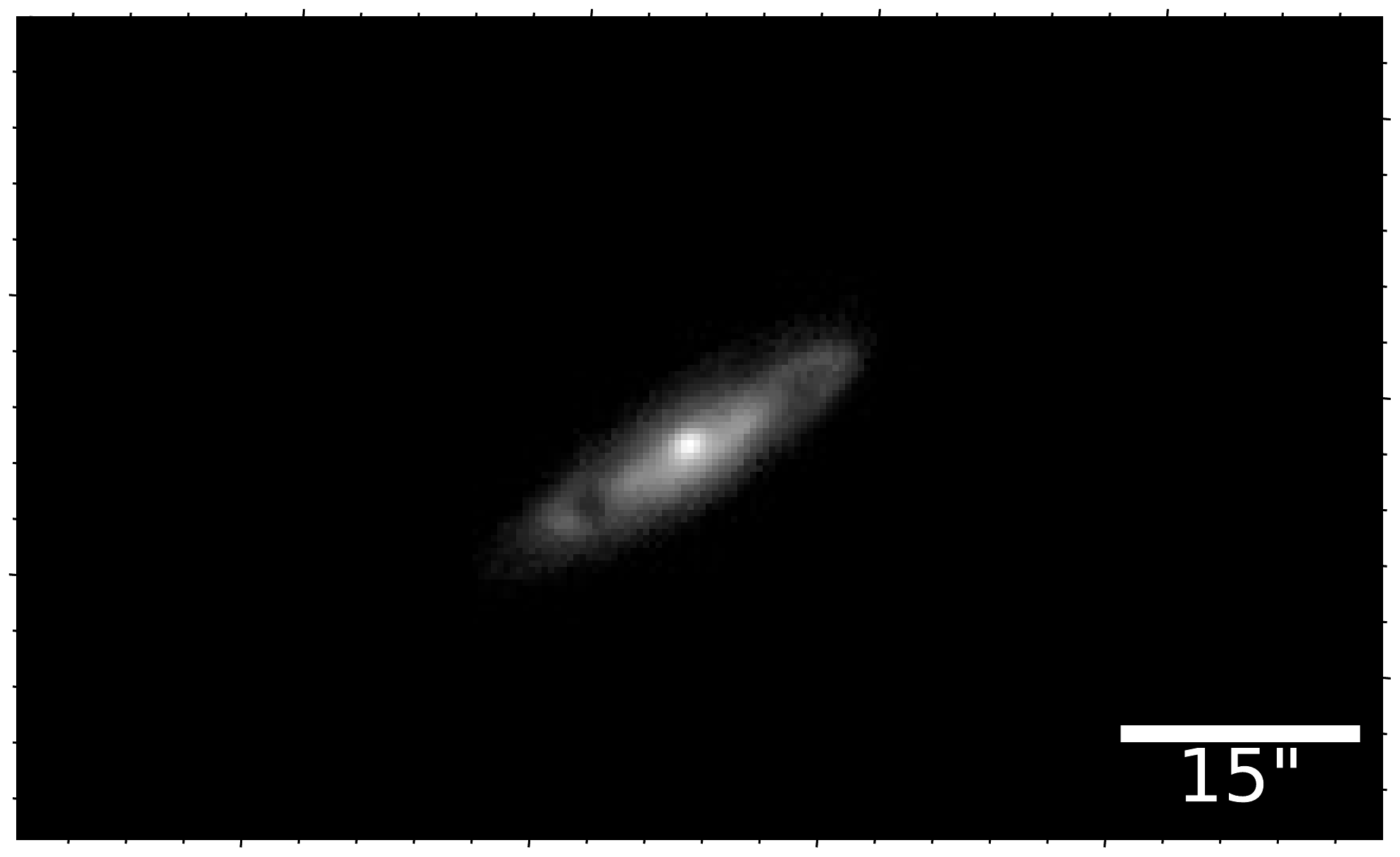}\label{30G1original}}
 \hspace{0.3cm}
\subfloat{\includegraphics[scale=0.2, bb=2 2 540 400]{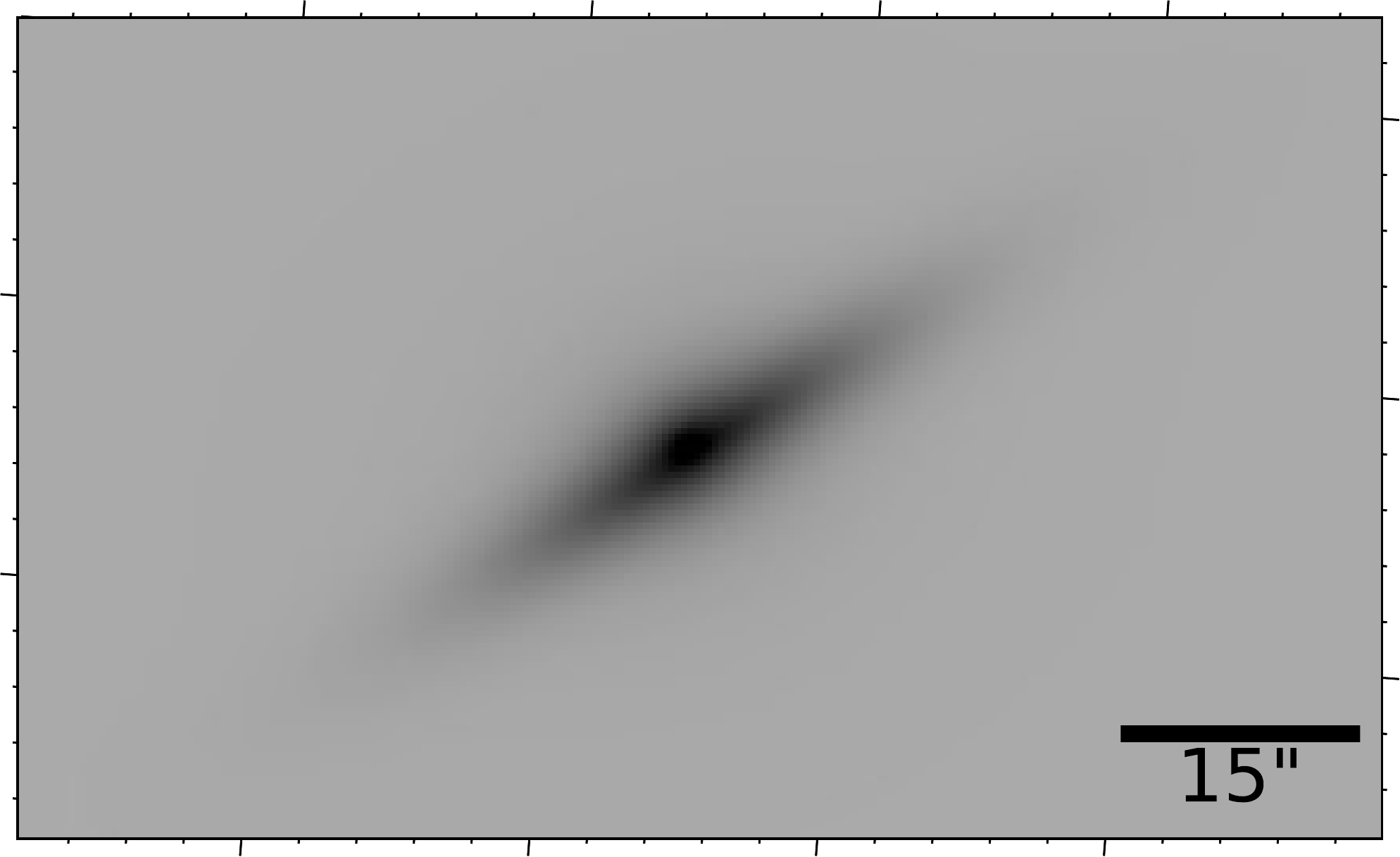} \label{30G1resid1}}
 \hspace{0.3cm}
\subfloat{\includegraphics[scale=0.2, bb=2 2 540 400]{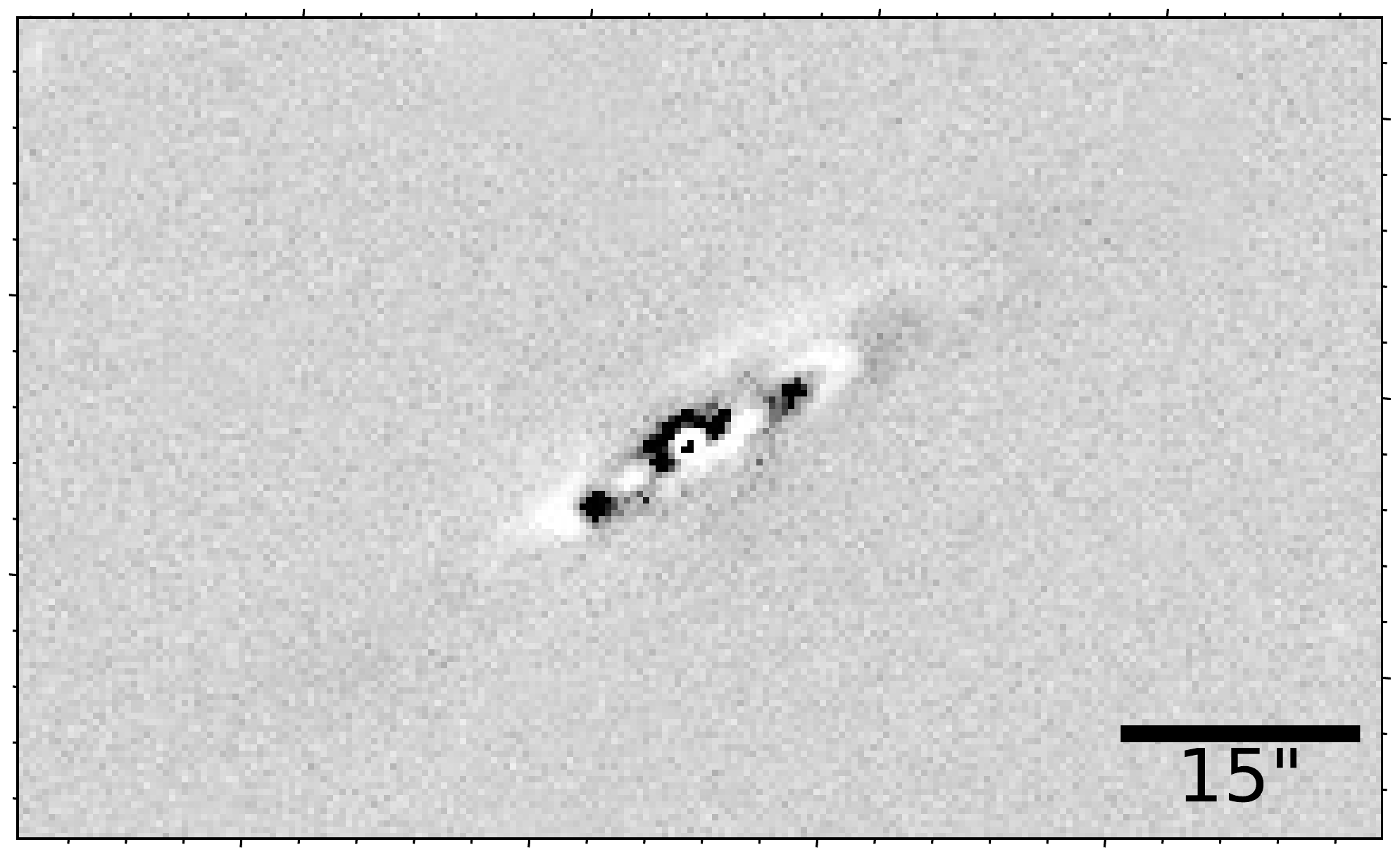} \label{30G1resid2}}\\
\subfloat{\includegraphics[scale=0.2, bb=2 2 540 400]{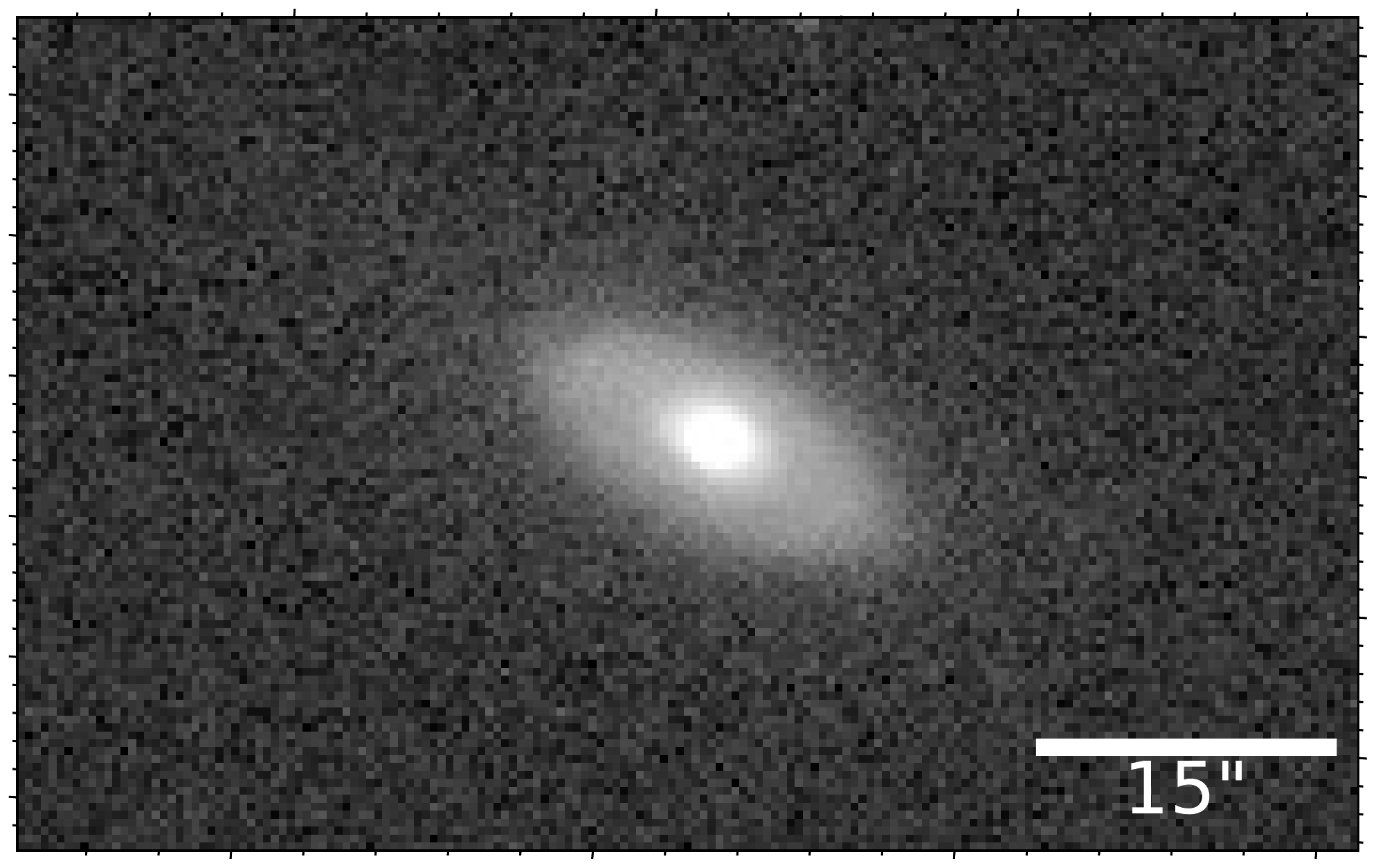}\label{30G2original}}
 \hspace{0.3cm}
\subfloat{\includegraphics[scale=0.2, bb=2 2 540 400]{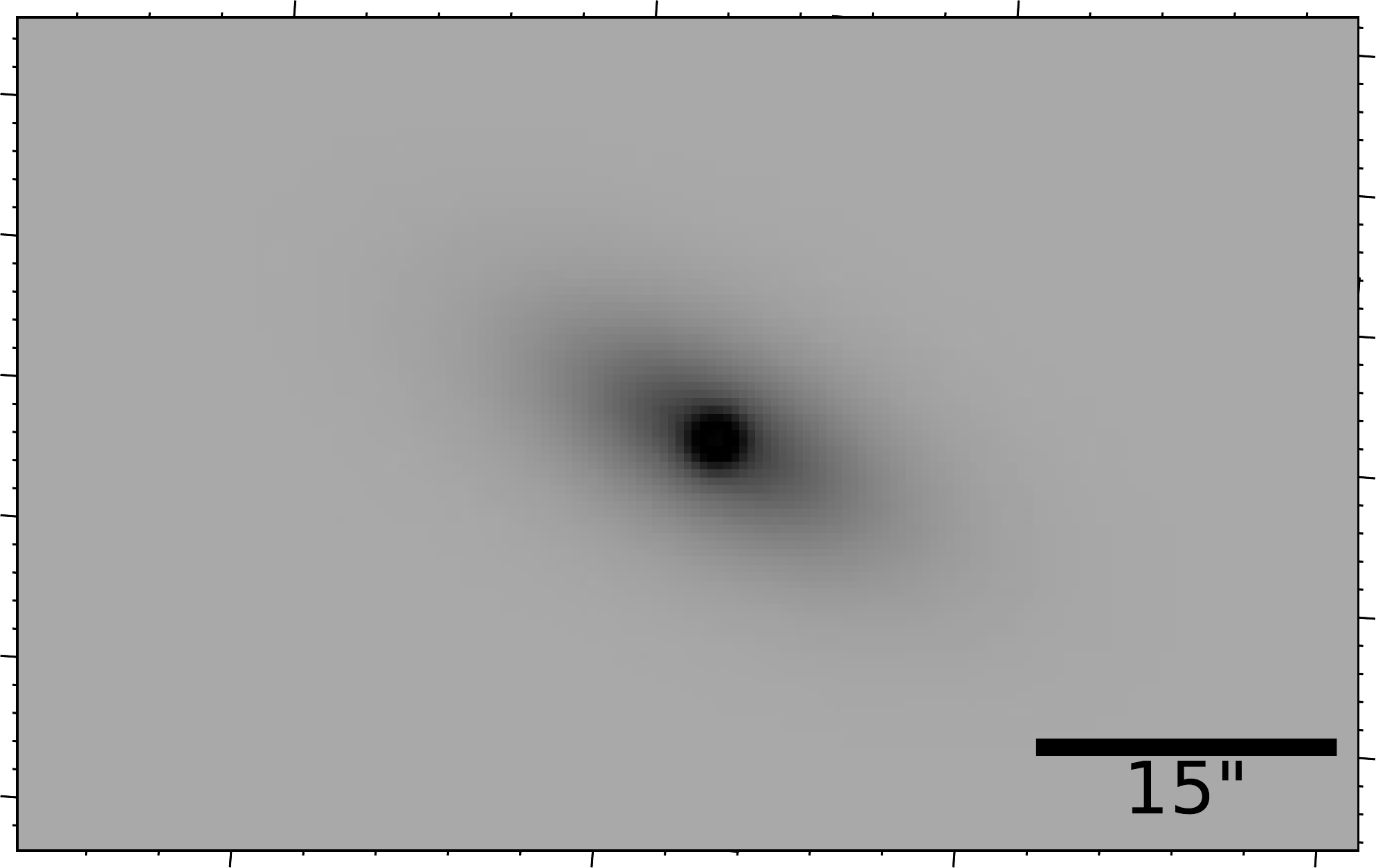} \label{30G2resid1}}
\hspace{0.3cm}
\subfloat{\includegraphics[scale=0.2, bb=2 2 540 400]{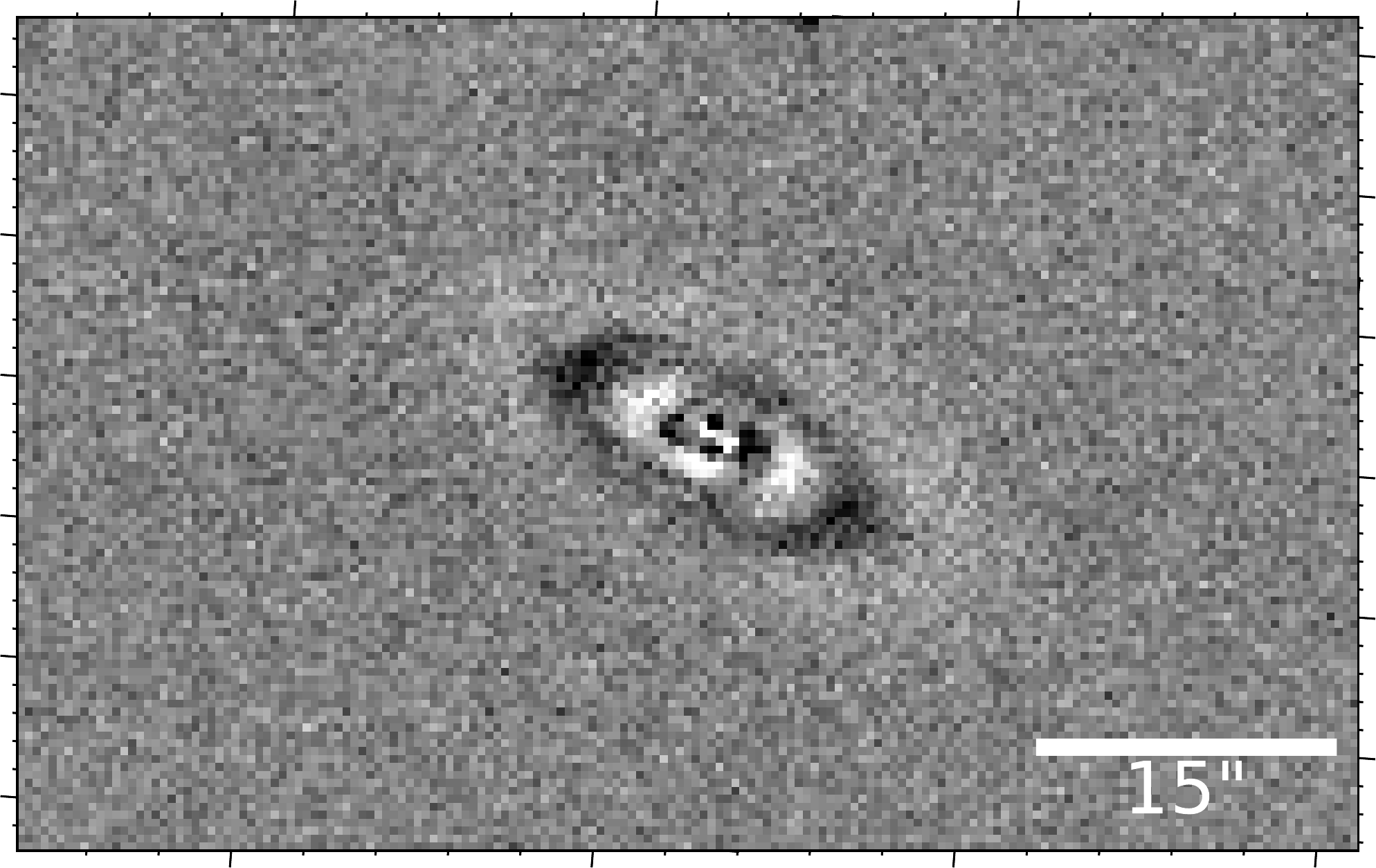} \label{30G2resid2}}\\
\subfloat{\includegraphics[scale=0.2, bb=2 2 540 400]{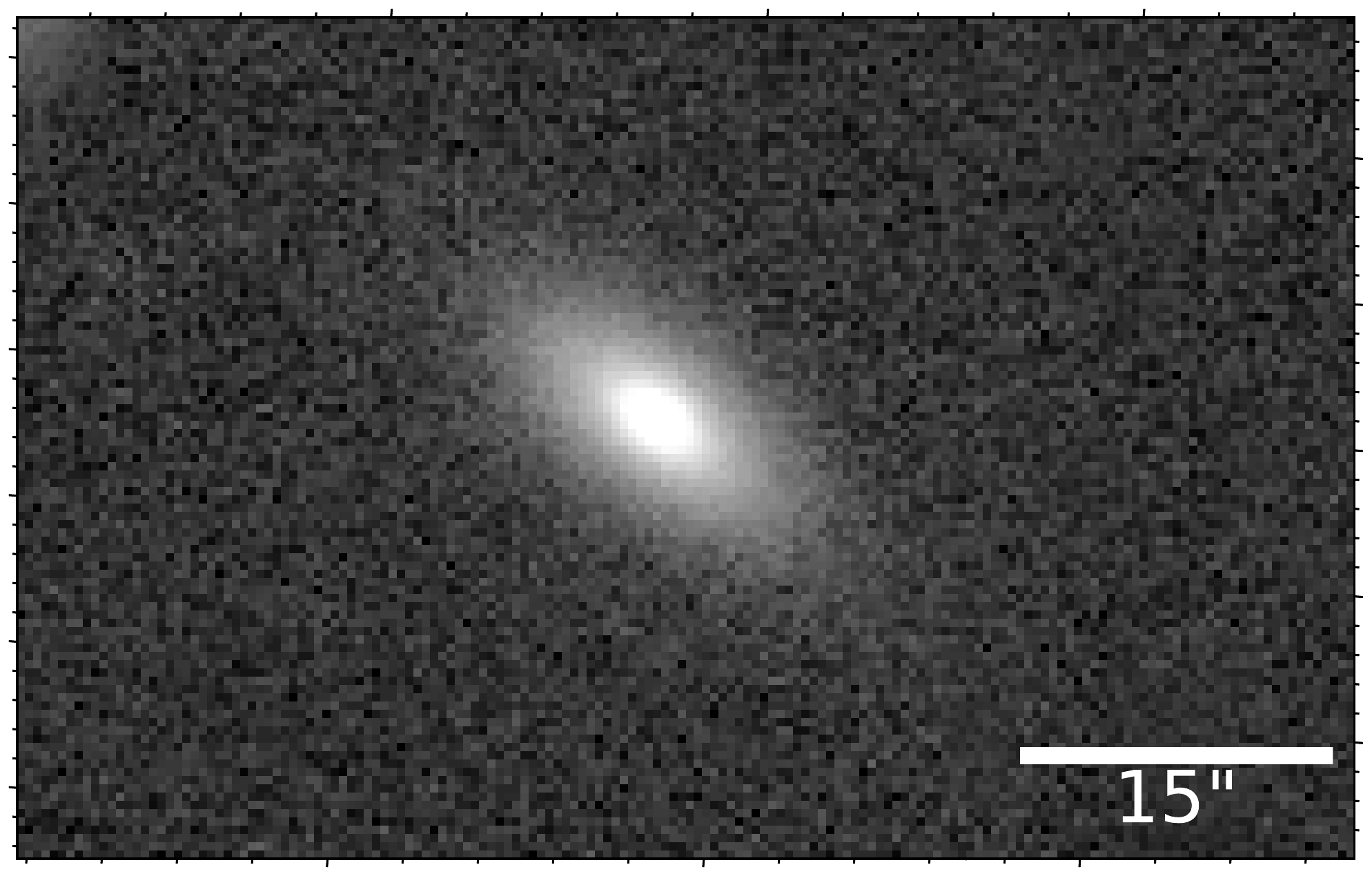}\label{30G3original}}
\hspace{0.3cm}
\subfloat{\includegraphics[scale=0.2, bb=2 2 540 400]{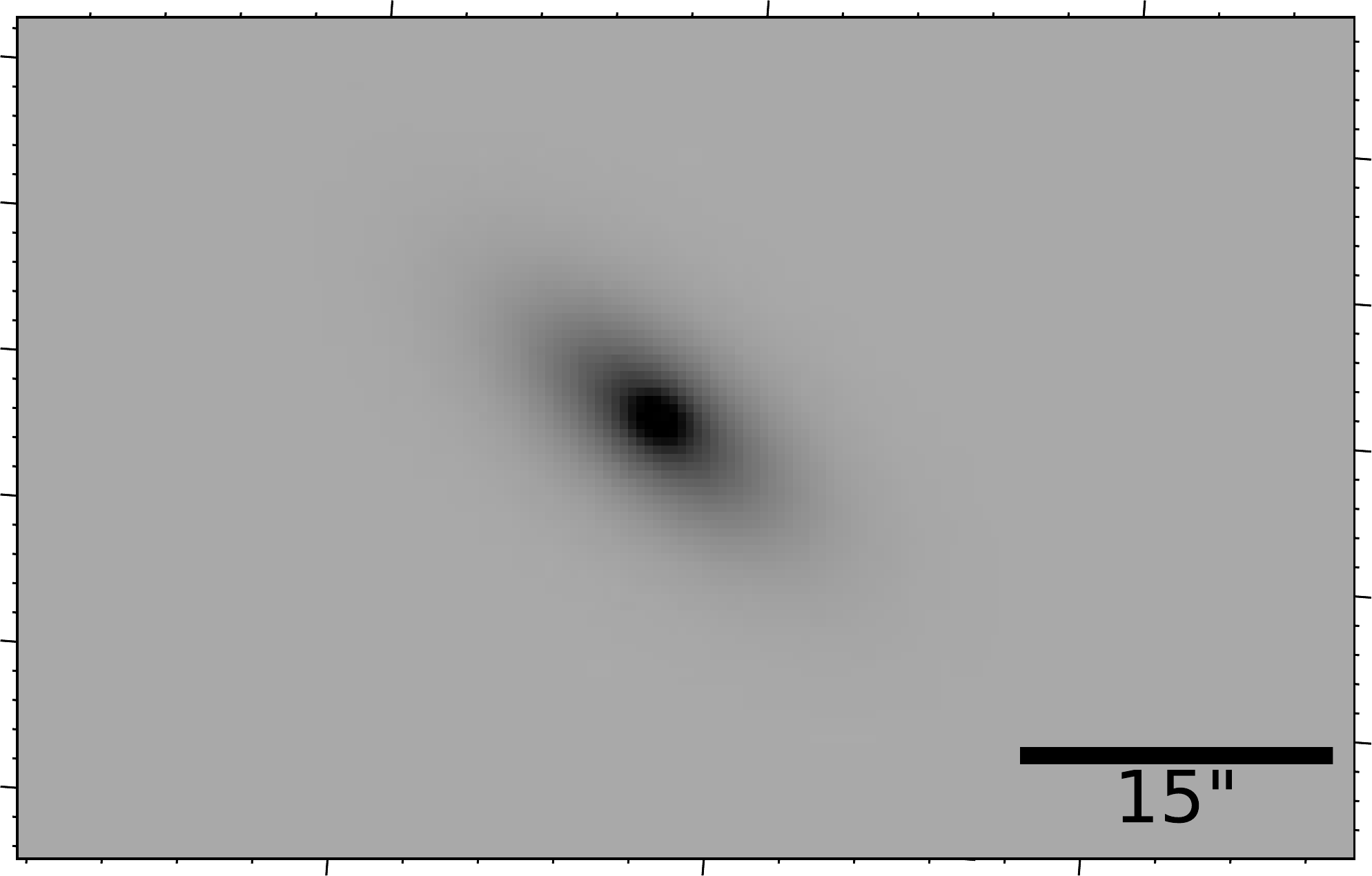} \label{30G3resid1}}
\hspace{0.3cm}
\subfloat{\includegraphics[scale=0.2, bb=2 2 540 400]{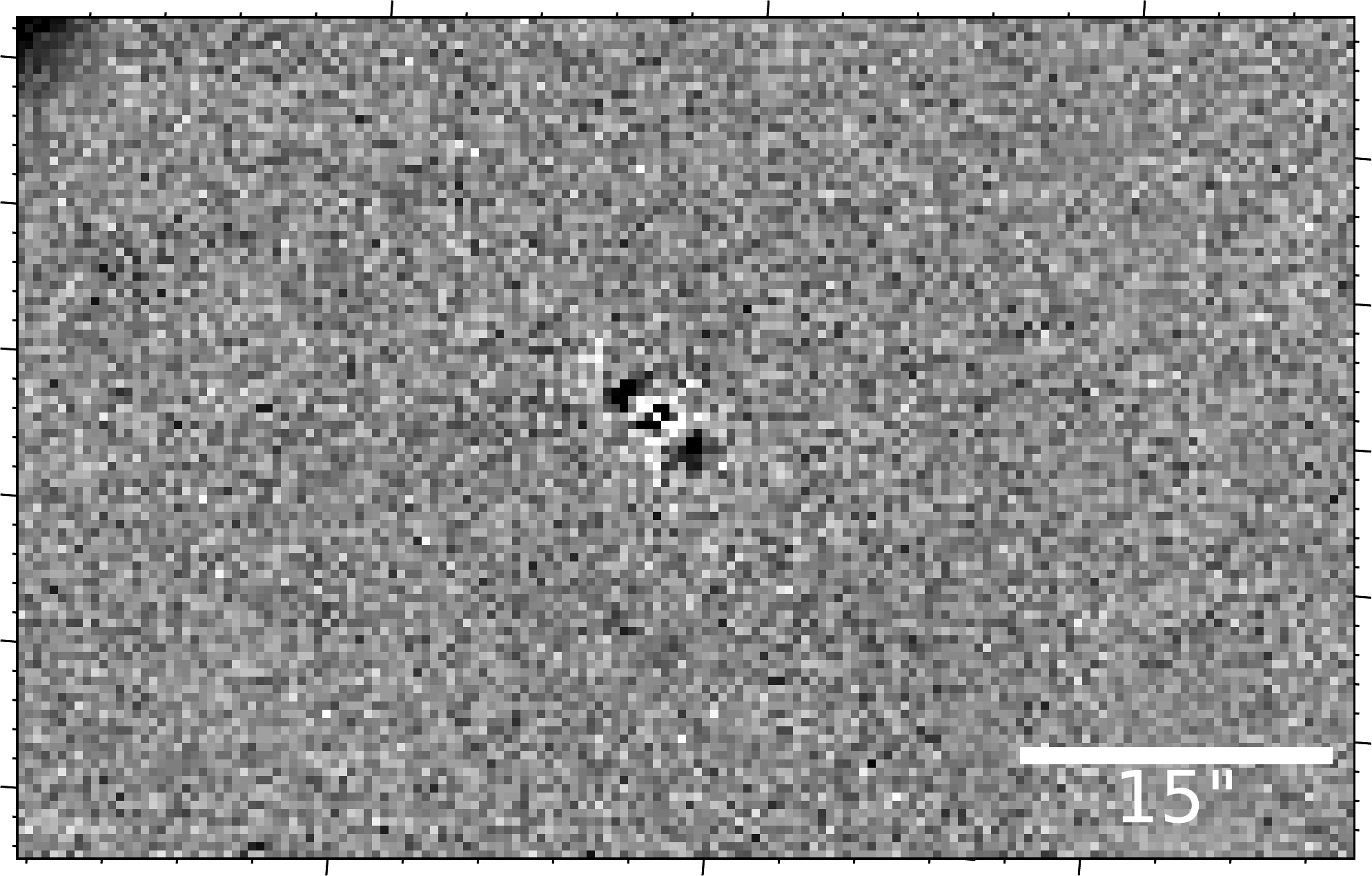} \label{30G3resid2}}\\
\caption{Photometric analysis of the triplet system SIT 30. The upper panel is the SDSS image of the triplet system in r-band. The second panel is the 1D fitting of G1, G2 and G3 from left to right. The last three panels are the cleaned, the model, and the residual image of G1, G2 and G3 using two profiles, respectively, from top to bottom.}
\label{fig:30analysis}
\end{figure*}

SIT 30 is a triplet system of three luminous galaxies (Figure~\ref{fig:SIT30}). \textbf{G1} and \textbf{G2} are two spiral galaxies (Sb) with a prominent luminous bulge and a huge halo surrounding each of them \citep{Oh2013}. The image of \textbf{G1} shows two symmetric spiral arms closed to its center, while the spiral arms of \textbf{G2} are not clearly observed. \textbf{G3} is a lenticular galaxy with evidence of a disk and a weak elongated component \citep{Oh2013}. 

The surface brightness profile of \textbf{G1} (Figure~\ref{30_SB}, left panel) shows multiple breaks at large radii. The excess ``shoulder''  between Semi-Major Axis (SMA) $7''<$ SMA $<$ 11 $''$ indicates the existence of a bar in this region. The down-bending break beyond 24$''$ represents type II.o profile.
Although \textbf{G2} is classified as Sb, an evidence of a bar is remarked in its intensity profile between 2$''<$ SMA $<$ 3.5$''$. Beyond the bar region the profile represents a perfect example of type III-d with a shallower exponential profile (Figure~\ref{30_SB}, middle panel).
The intensity profile of \textbf{G3} (Figure~\ref{30_SB}, right panel) shows a simple exponential type I profile without breaks.

A two-dimensional decomposition of \textbf{G1} with de Vaucouleur ($Dev$) plus $Exp$ leaves residuals of spiral arms with $\chi^2_{\nu}$ = 1.70 (Figure~\ref{30G1resid2}). 

Decomposition of \textbf{G2} using $Ser$ and $Exp$ profiles produce a residual image with ``peanut-shaped'' halo extending out to 7.9$''$ from the center of the galaxy and a $\chi^2_{\nu}$ = 1.15 (Figure~\ref{30G2resid2}).

Figure~\ref{30G3resid2} shows the residuals of the fit for \textbf{G3} using a Ser + Exp profile with only faint structures visible and a $\chi^2_{\nu}$ of 1.05. 

\subsection{The triplet system SIT 101} 

\begin{figure*}
\centering
\subfloat{\includegraphics[scale=0.5, bb=130 -150 100 100]{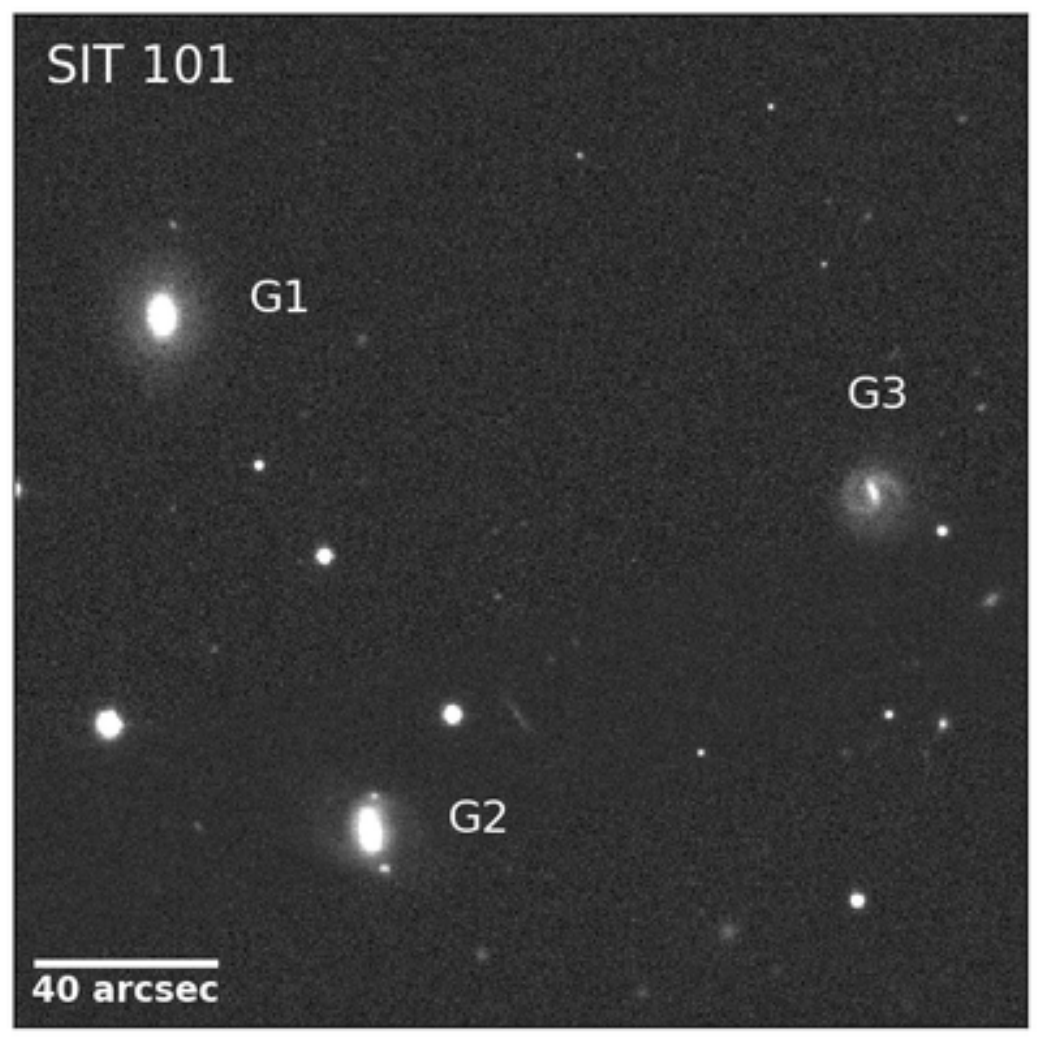} \label{fig:SIT101}}\\
\subfloat{\includegraphics[scale=0.7, bb=80 20 500 100]{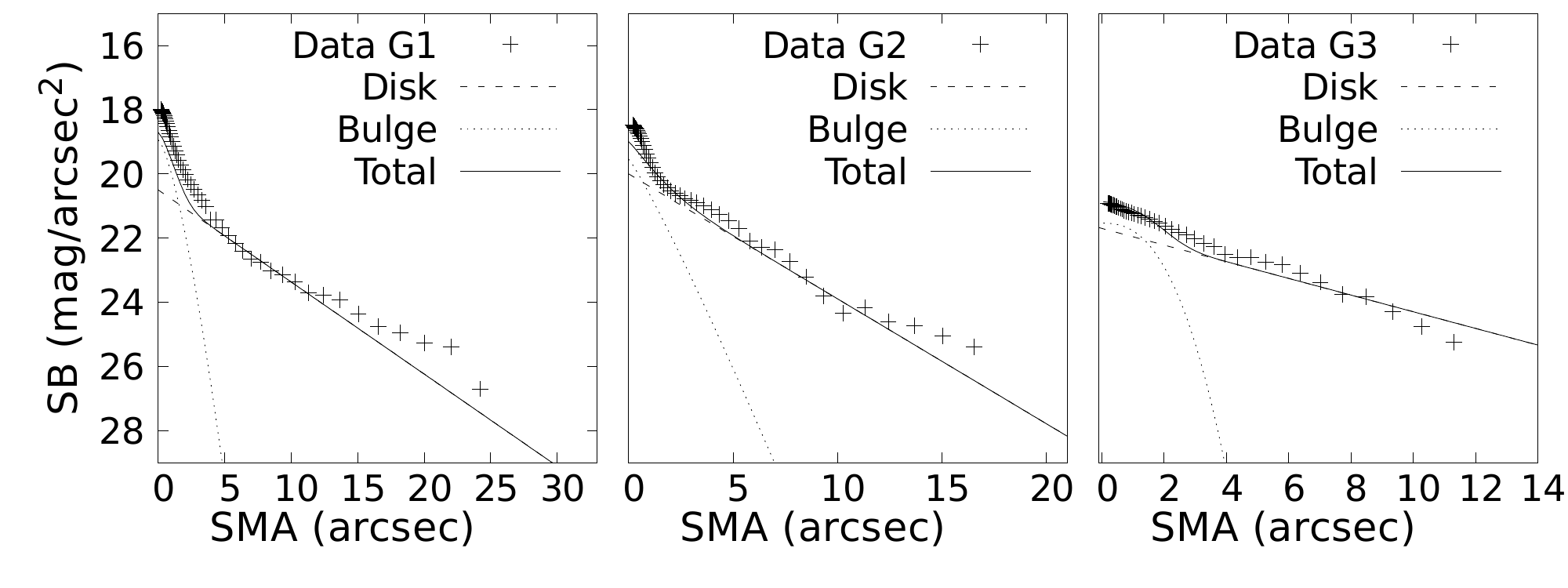}\label{101_SB}}\\
\subfloat{\includegraphics[scale=0.2, bb=2 2 540 400]{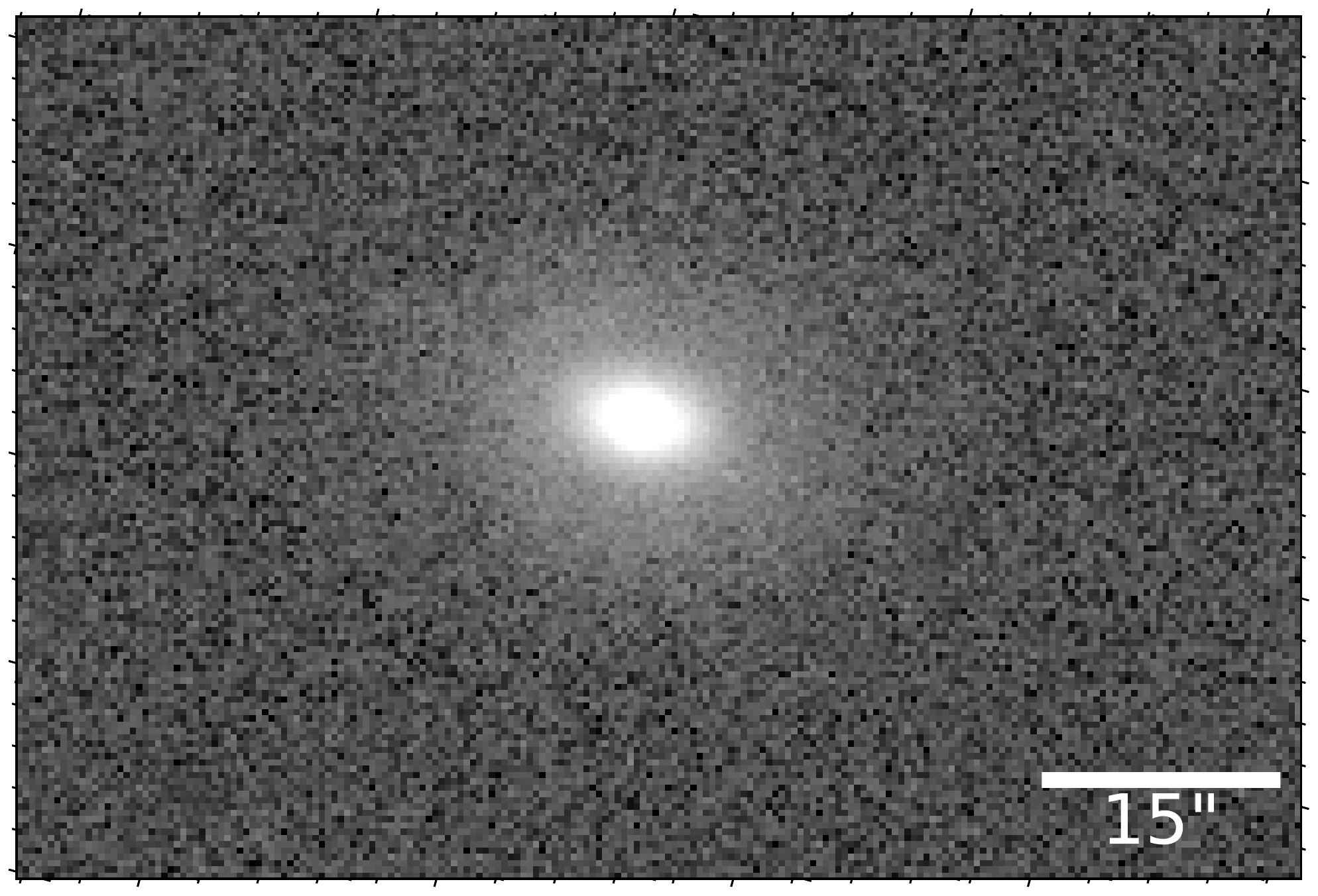}}
\hspace{0.3cm}
\subfloat{\includegraphics[scale=0.2, bb=2 2 540 400]{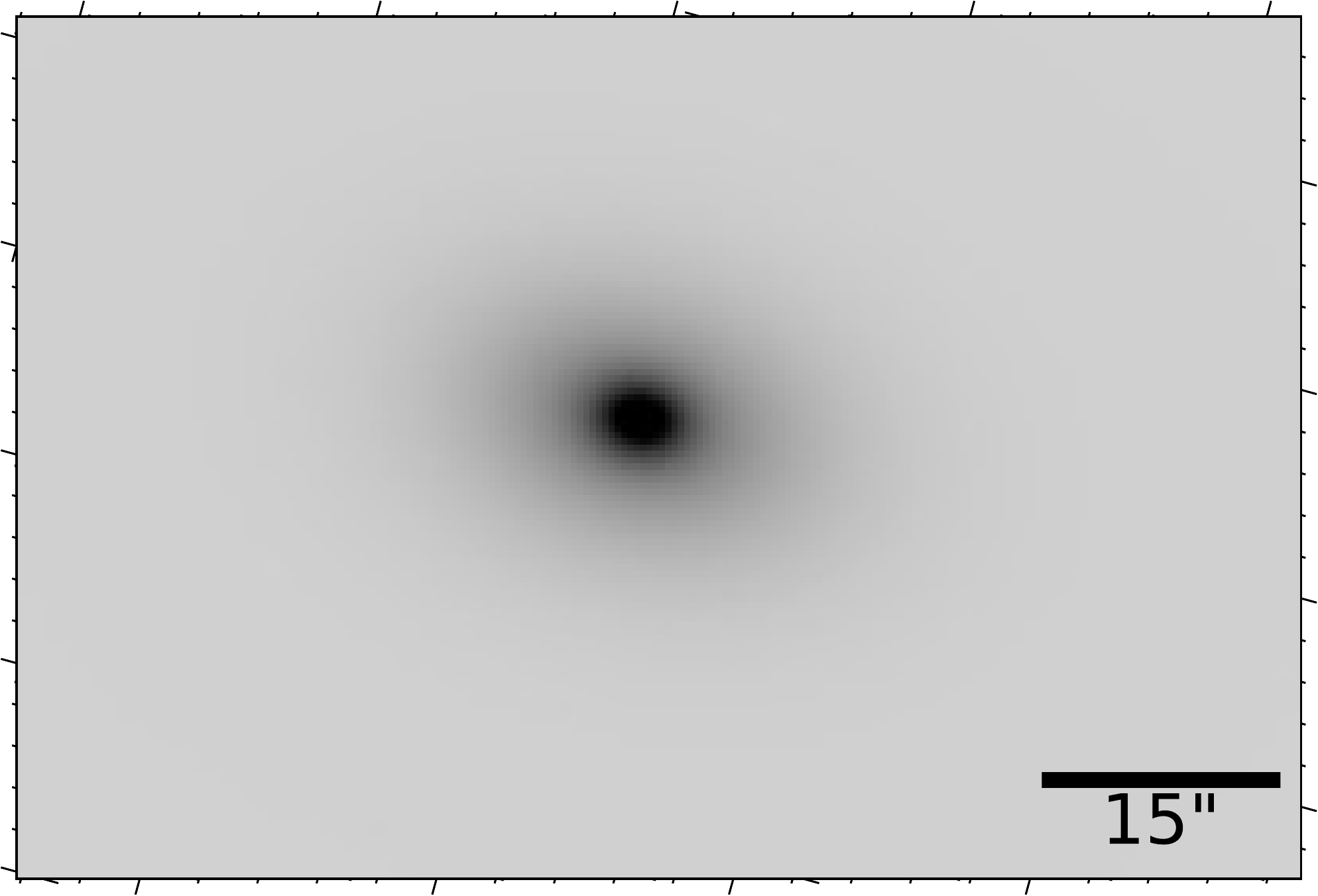}\label{101G1resid1}}
\hspace{0.3cm}
\subfloat{\includegraphics[scale=0.2, bb=2 2 540 400]{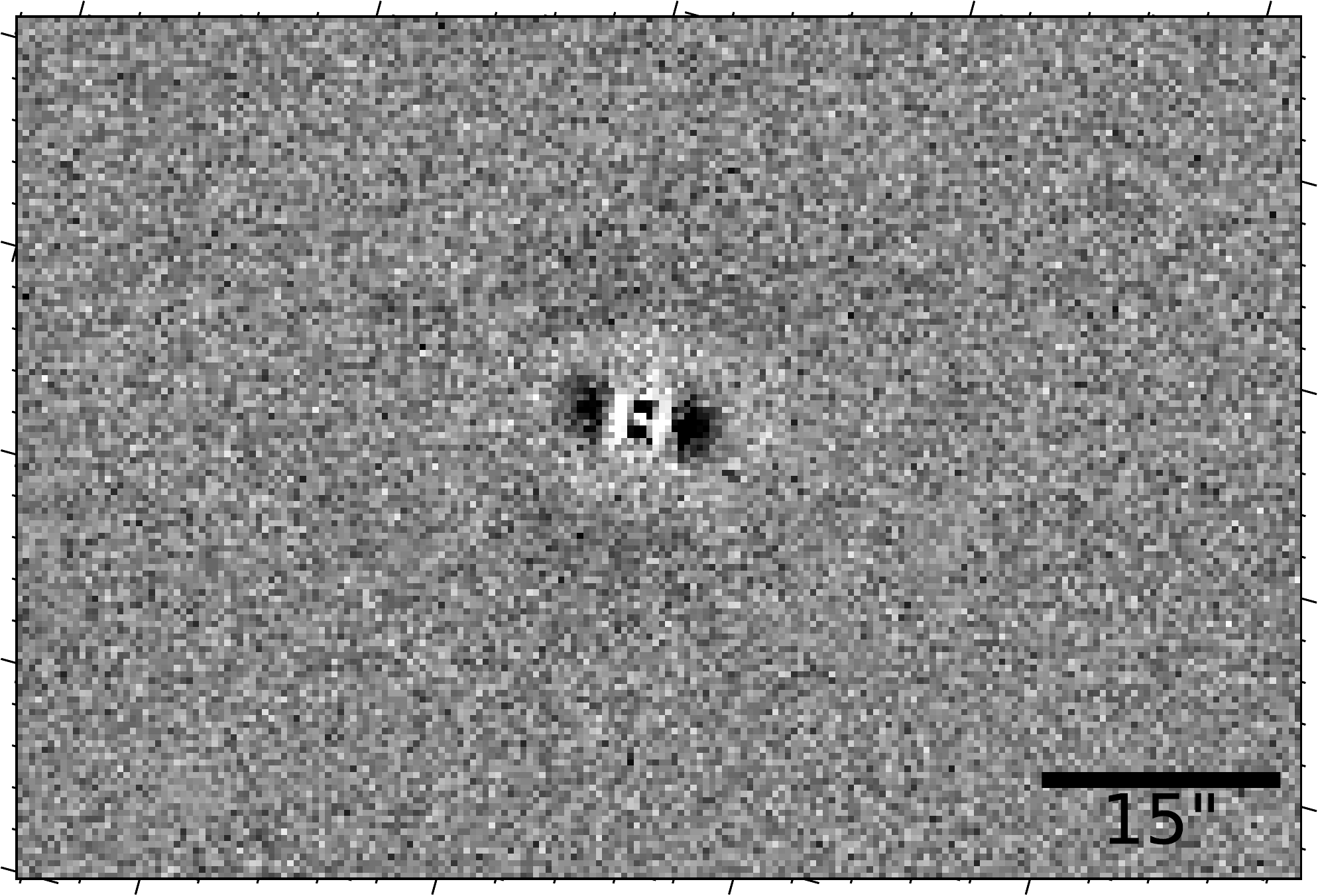}\label{101G1resid2}}\\
\subfloat{\includegraphics[scale=0.2, bb=9 9 400 500]{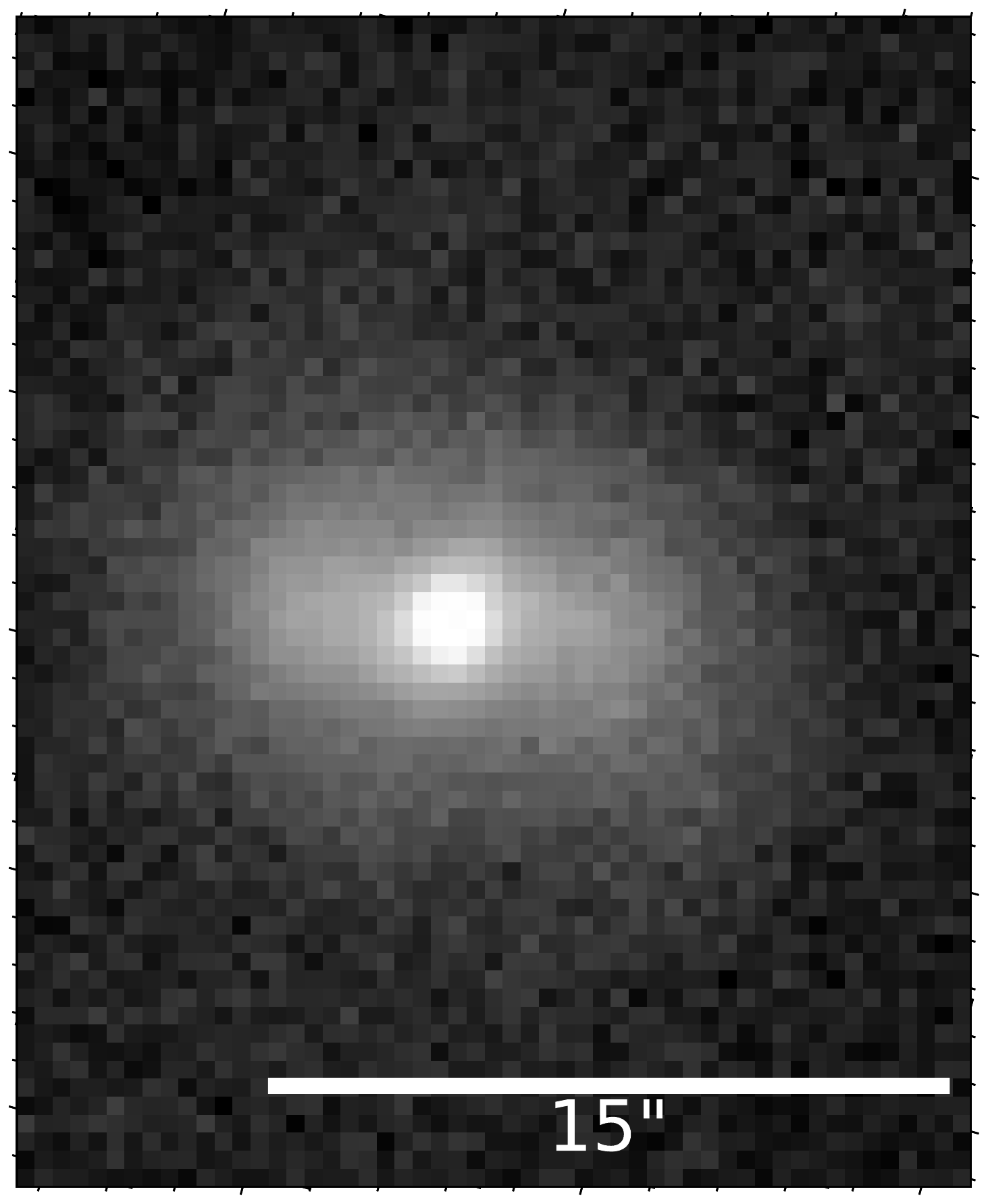}\label{101G2original}}
 \hspace{0.3cm}
\subfloat{\includegraphics[scale=0.2, bb=9 9 400 500]{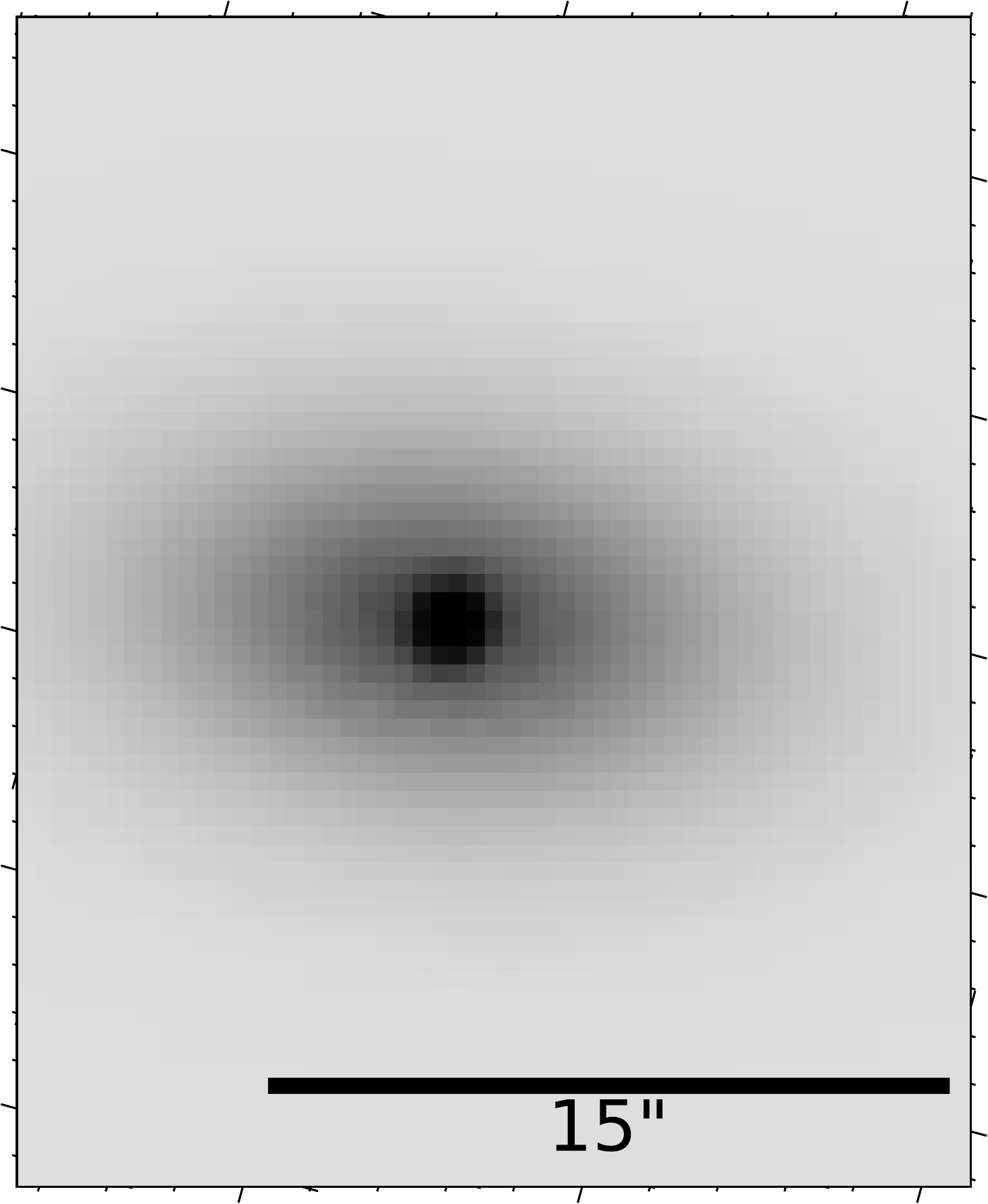} \label{101G2resid1}}
\hspace{0.3cm}
\subfloat{\includegraphics[scale=0.2, bb=9 9 400 500]{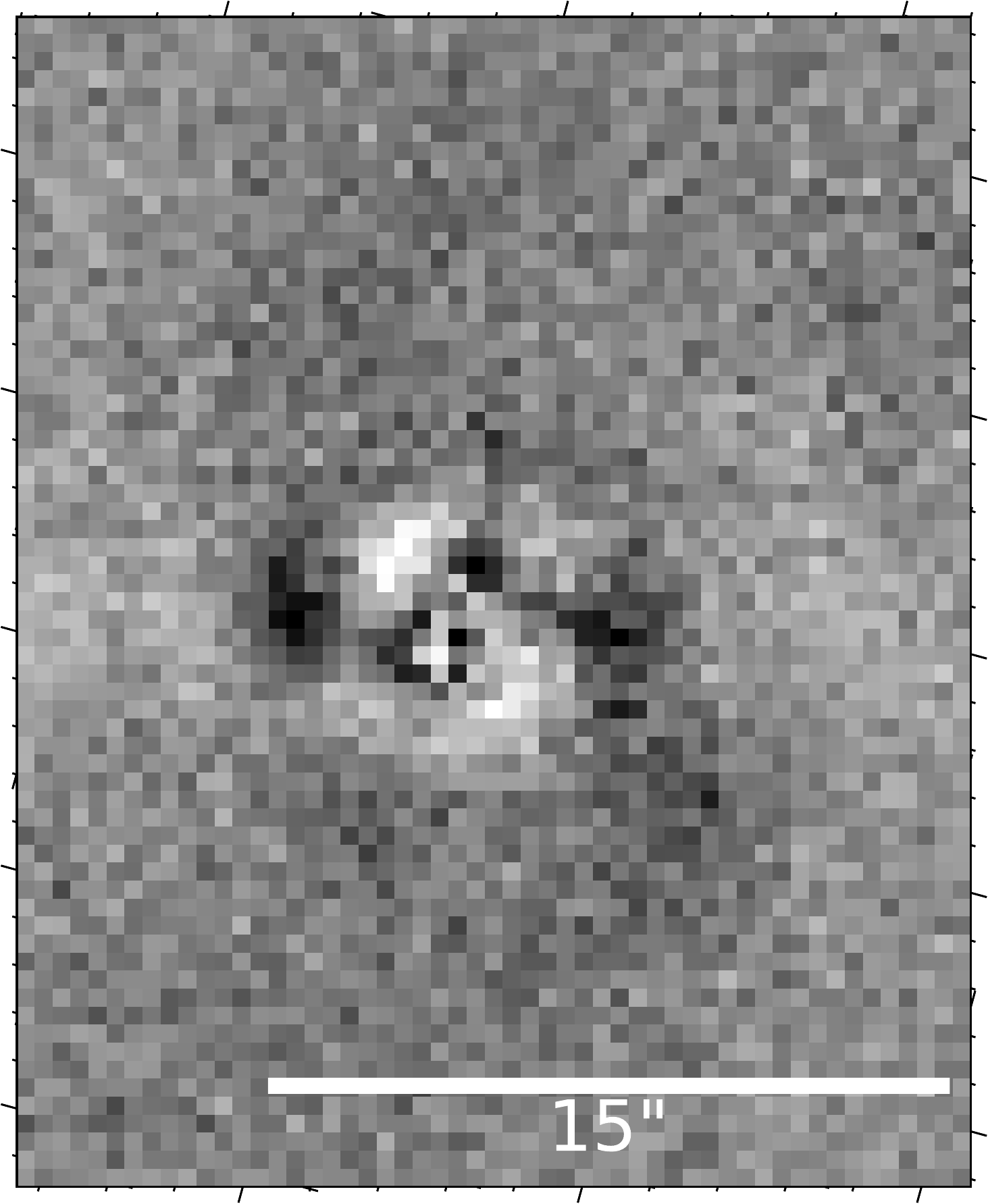} \label{101G2resid2}}\\
\subfloat{\includegraphics[scale=0.2, bb=2 2 540 400]{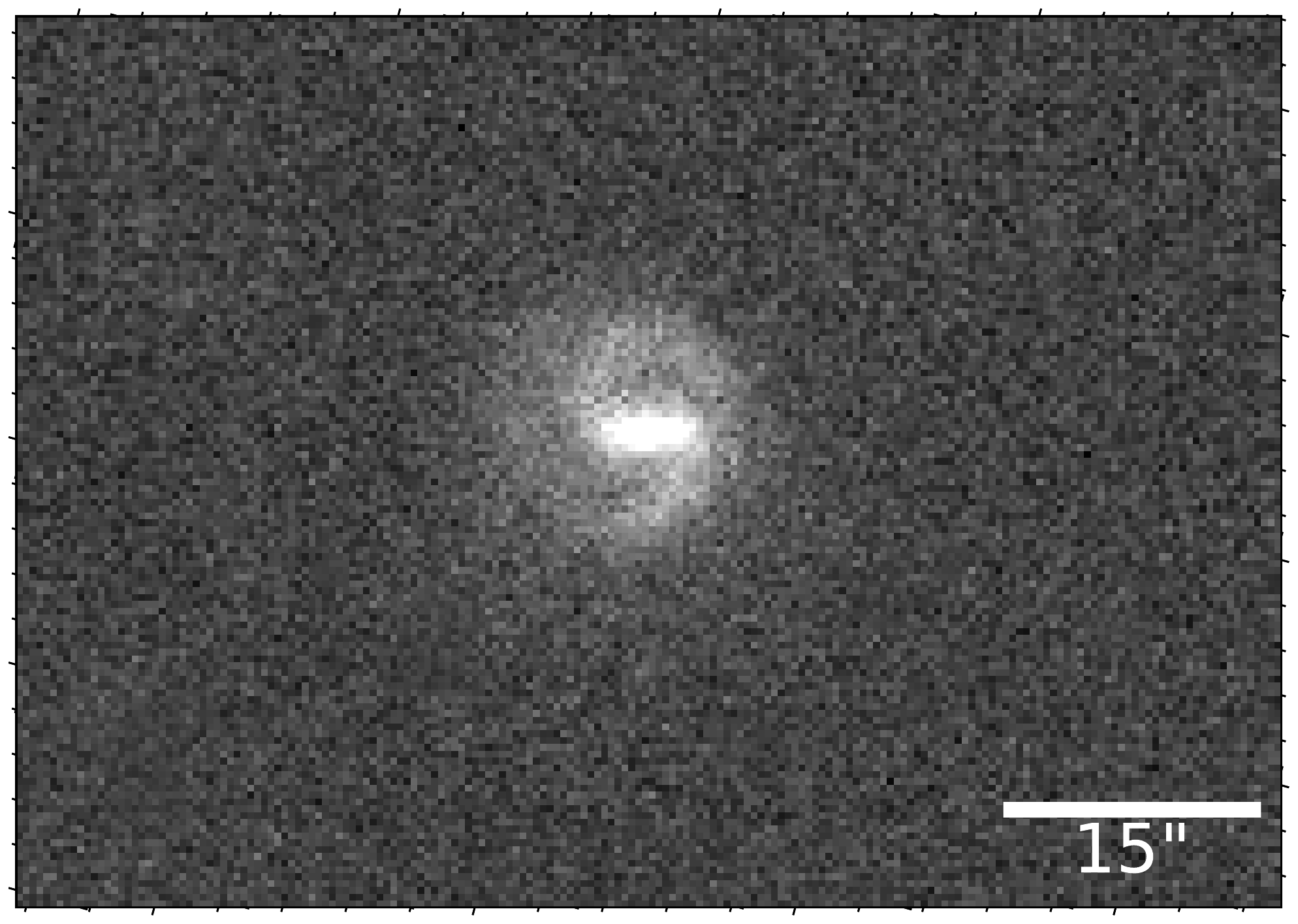}\label{101G3original}}
\hspace{0.3cm}
\subfloat{\includegraphics[scale=0.2, bb=2 2 540 400]{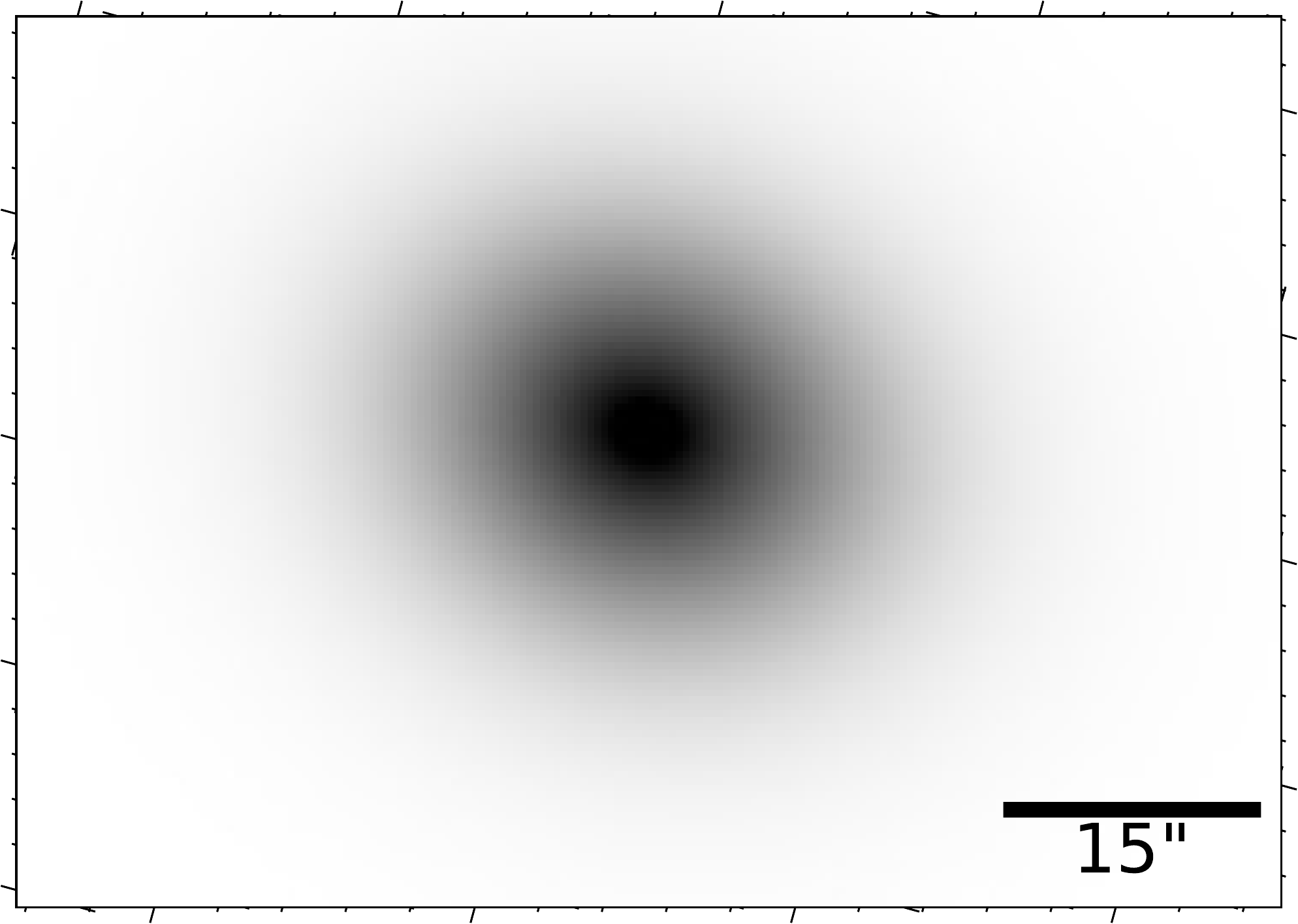} \label{101G3resid1}}
\hspace{0.3cm}
\subfloat{\includegraphics[scale=0.2, bb=2 2 540 400]{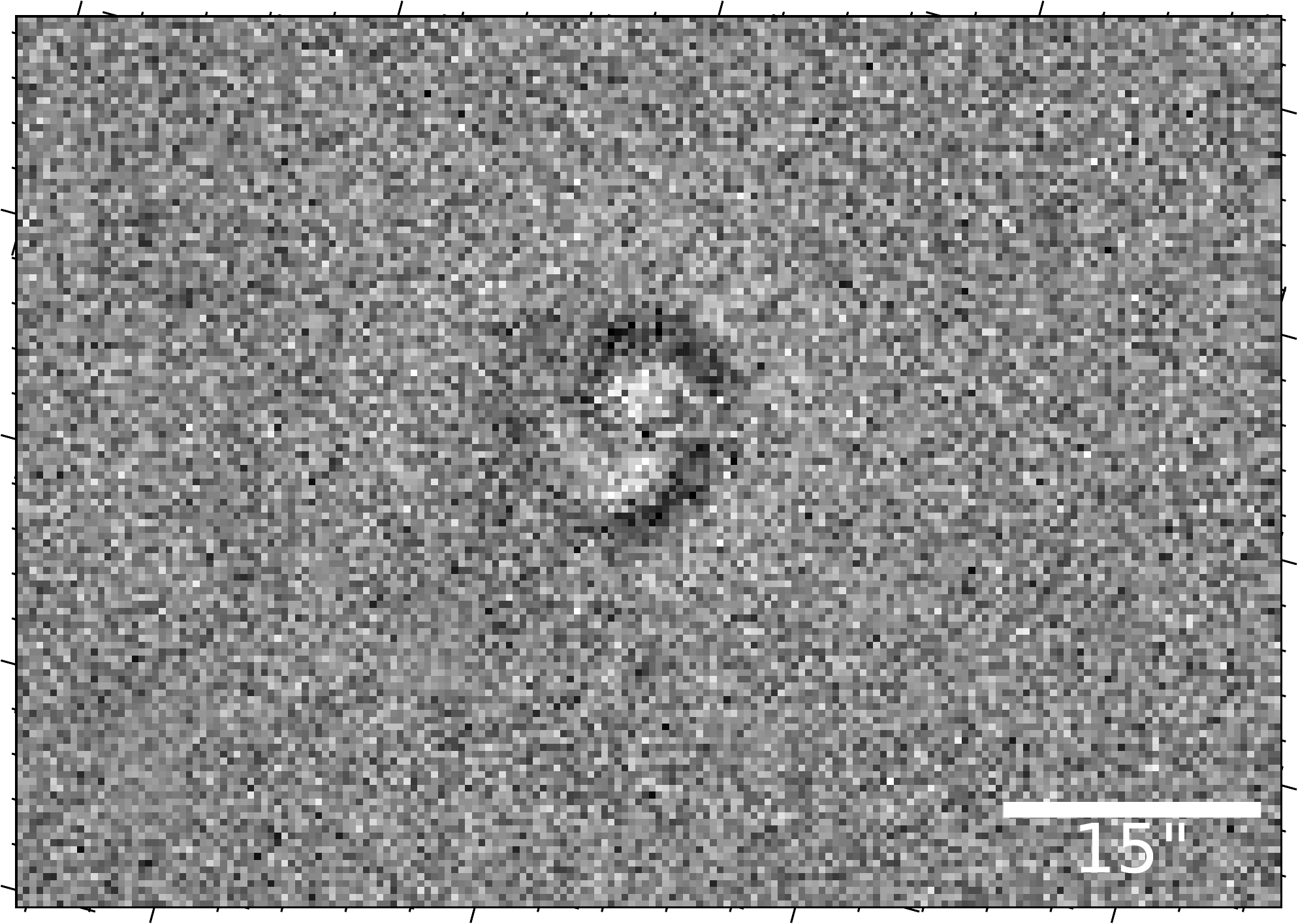} \label{101G3resid2}}\\

\caption{Photometric analysis of the triplet system SIT 101. See caption Fig.~\ref{fig:30analysis}}
\label{fig:101analysis}
\end{figure*}

The triplet system SIT 101 (Figure~\ref{fig:SIT101}) includes three disk galaxies. \textbf{G1} is a lenticular galaxy  with a slightly elongated central disk without definite substructure along it \citep{Khim2015}. \textbf{G2} and \textbf{G3} are barred spiral galaxies with a set of two faint arms, emerging from the end of their bars \citep{Willett2013}.

The intensity profiles of \textbf{G1 and G2} (Figures~\ref{101_SB}, left and middel panels) reveal anti-truncations of type III-d with a break at SMA = 11$''$ and 10.5$''$, respectively, while \textbf{G3} represents a type II.i profile with truncation at 8.3$''$ (Figure~\ref{101_SB}, right panel).

The decomposition of \textbf{G1} using $Ser$ plus $Exp$ profiles leave two small positive nodes in both sides of the galaxy as residuals with a $\chi^2_{\nu}$ of 0.97, as shown in Figure~\ref{101G1resid2}. 

Using $dev$ + $Exp$ profiles upon decomposing \textbf{G2} reveals a residual image (Figure~\ref{101G2resid2}) with an S-Shape structure around the nucleus that is presumably caused by a circum-nuclear dust ring with $\chi^2_{\nu}$ = 1.15. Furthermore, the residuals shows the trace of two faint spiral arms extended in the East and West directions.

Decomposition of \textbf{G3} using $Ser$ + $Exp$ profiles show, in Figure~\ref{101G3resid2}), a prominent bar associated with two symmetric spiral arms in the East and the West directions and a $\chi^2_{\nu}$ of 1.05. This bar was fitted by fixing the $Ser$ index ($n$) to 0.5.

\subsection{The triplet system SIT 104}

\begin{figure*}
\centering
\subfloat{\includegraphics[scale=0.5, bb=130 -150 100 100]{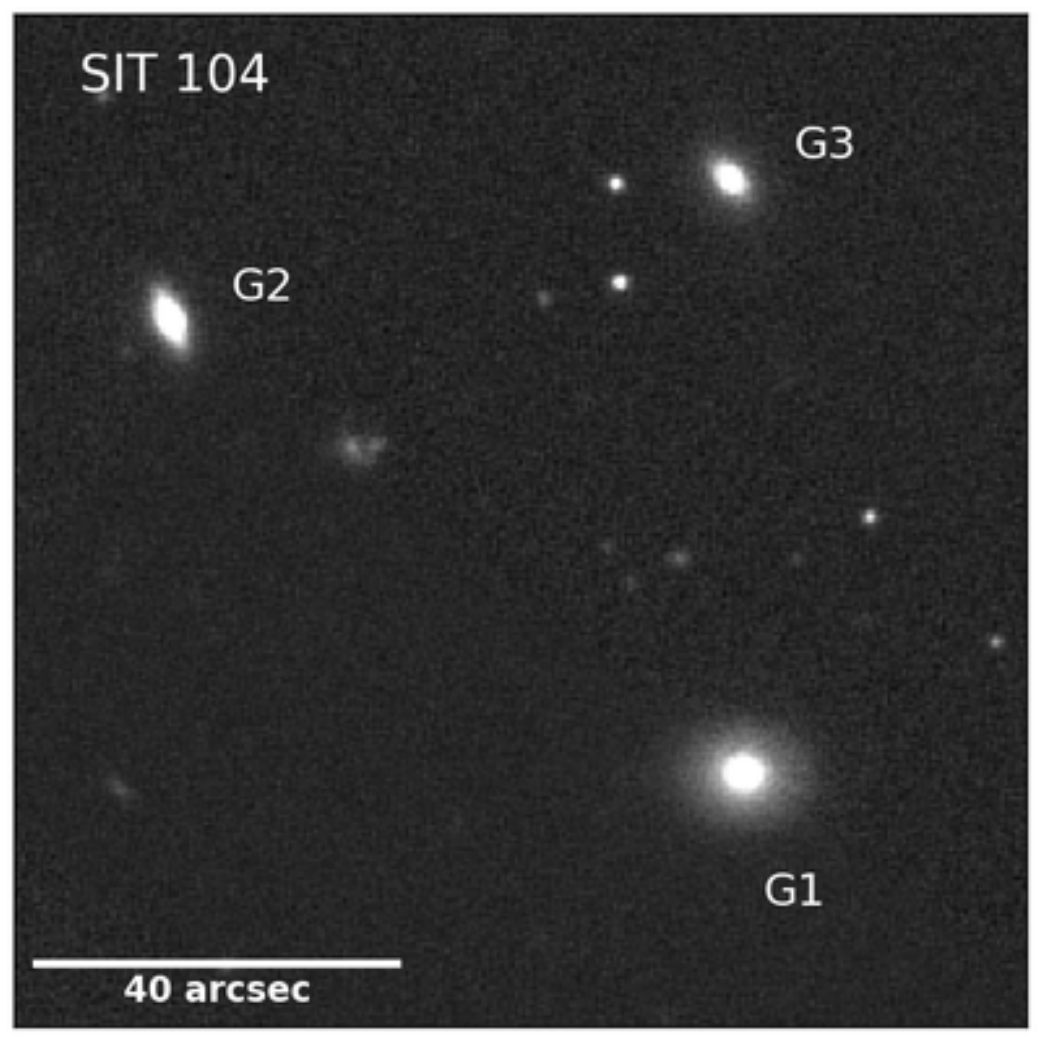}
 \label{fig:SIT104}}\\
\subfloat{\includegraphics[scale=0.7, bb=80 20 500 100]{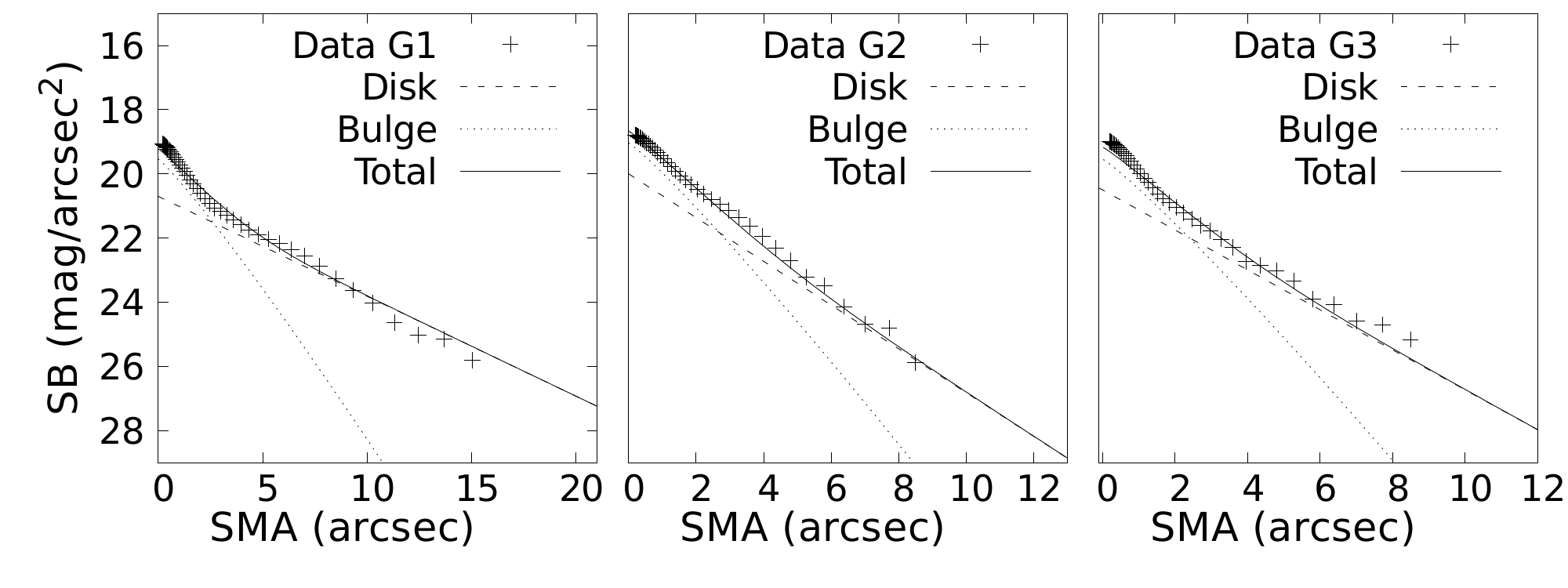}\label{104_SB}}\\
\subfloat{\includegraphics[scale=0.2, bb=2 2 540 400]{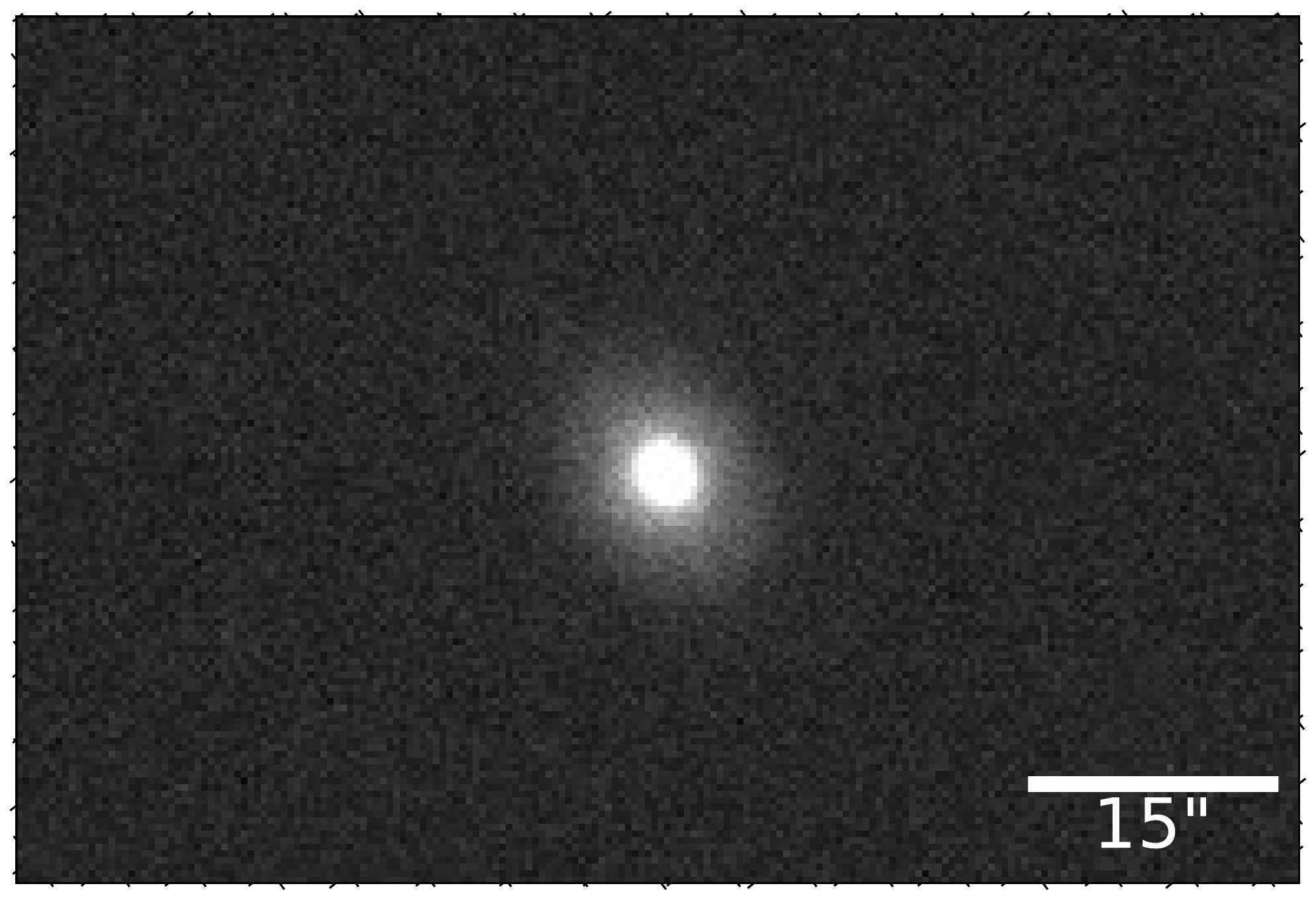}\label{104G1original}}
 \hspace{0.3cm}
\subfloat{\includegraphics[scale=0.2, bb=2 2 540 400]{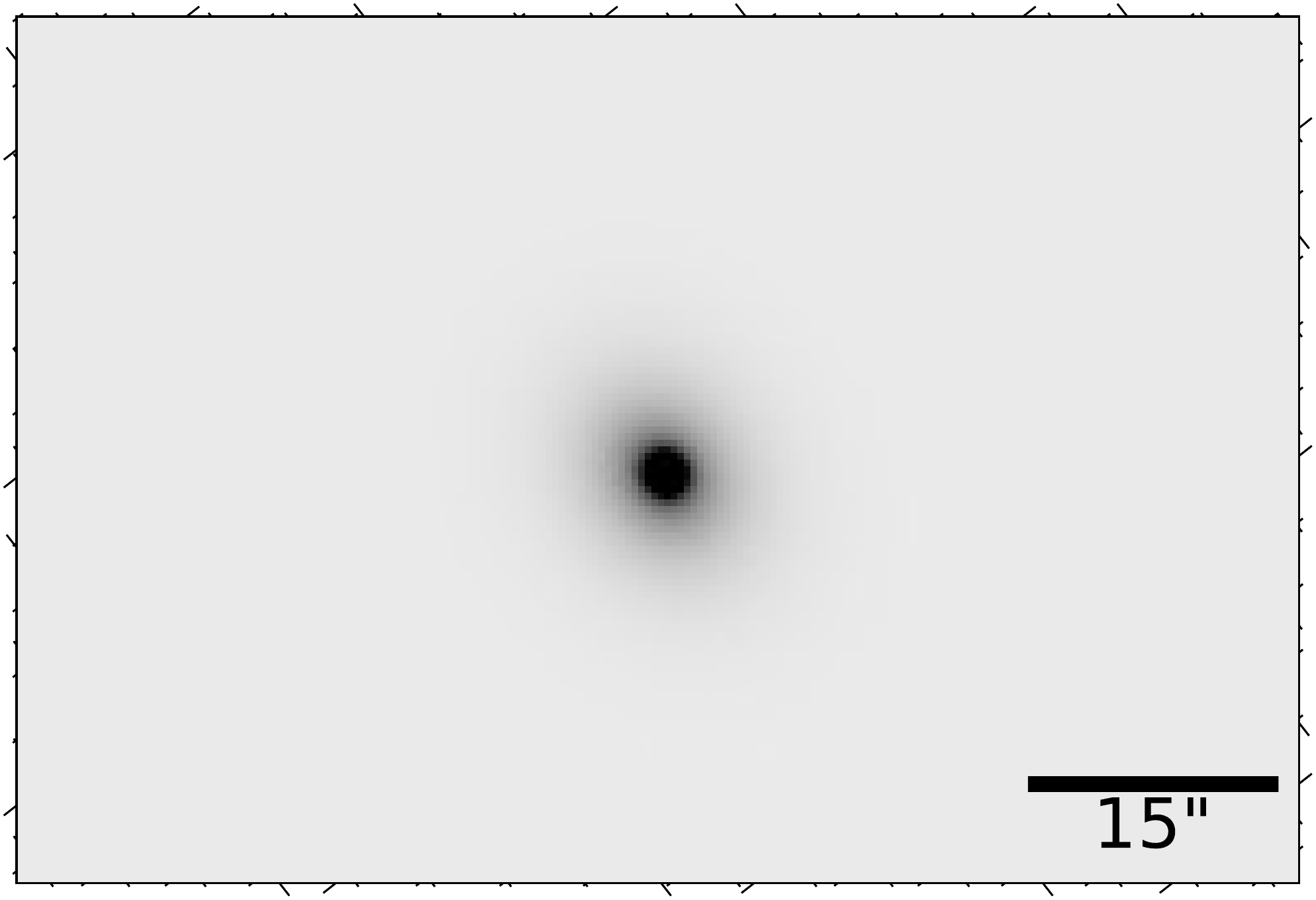} \label{104G1resid1}}
\hspace{0.3cm}
\subfloat{\includegraphics[scale=0.2, bb=2 2 540 400]{Figures_pdf/104G1_residue2-eps-converted-to.pdf} \label{104G1resid2}}\\
\subfloat{\includegraphics[scale=0.2, bb=2 2 540 400]{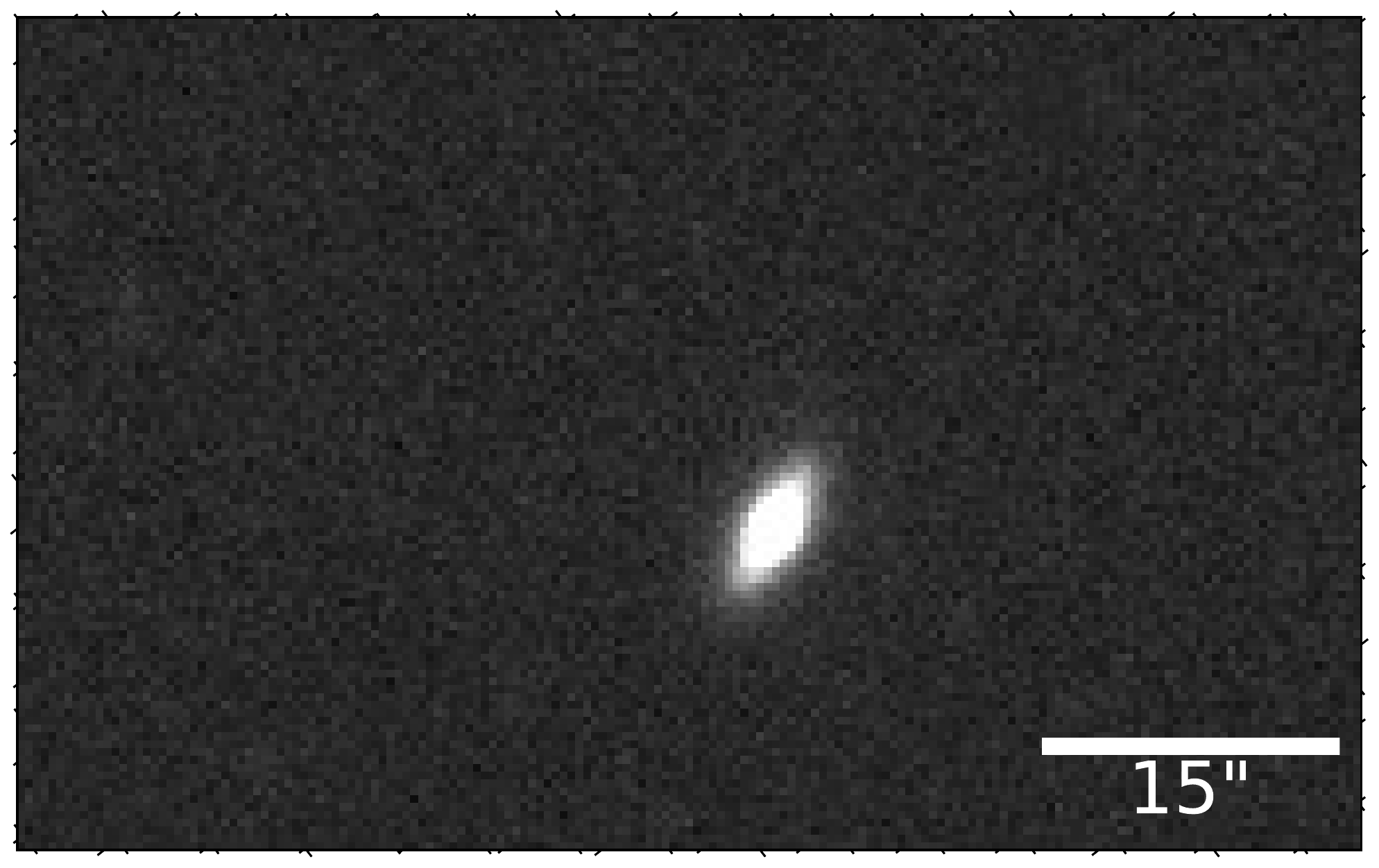}\label{104G2original}}
\hspace{0.3cm}
\subfloat{\includegraphics[scale=0.2, bb=2 2 540 400]{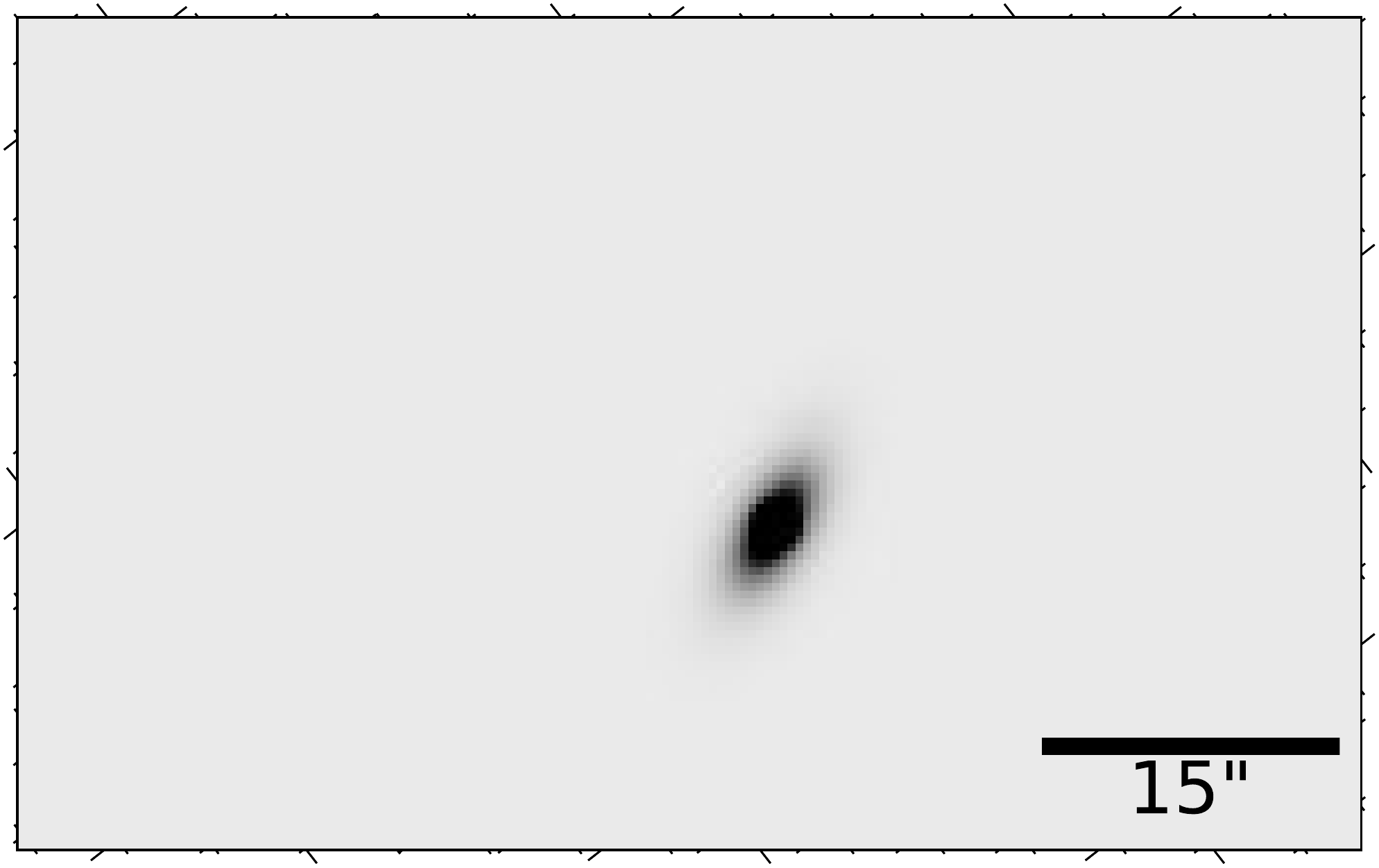} \label{104G2resid1}}
\hspace{0.3cm}
\subfloat{\includegraphics[scale=0.2, bb=2 2 540 400]{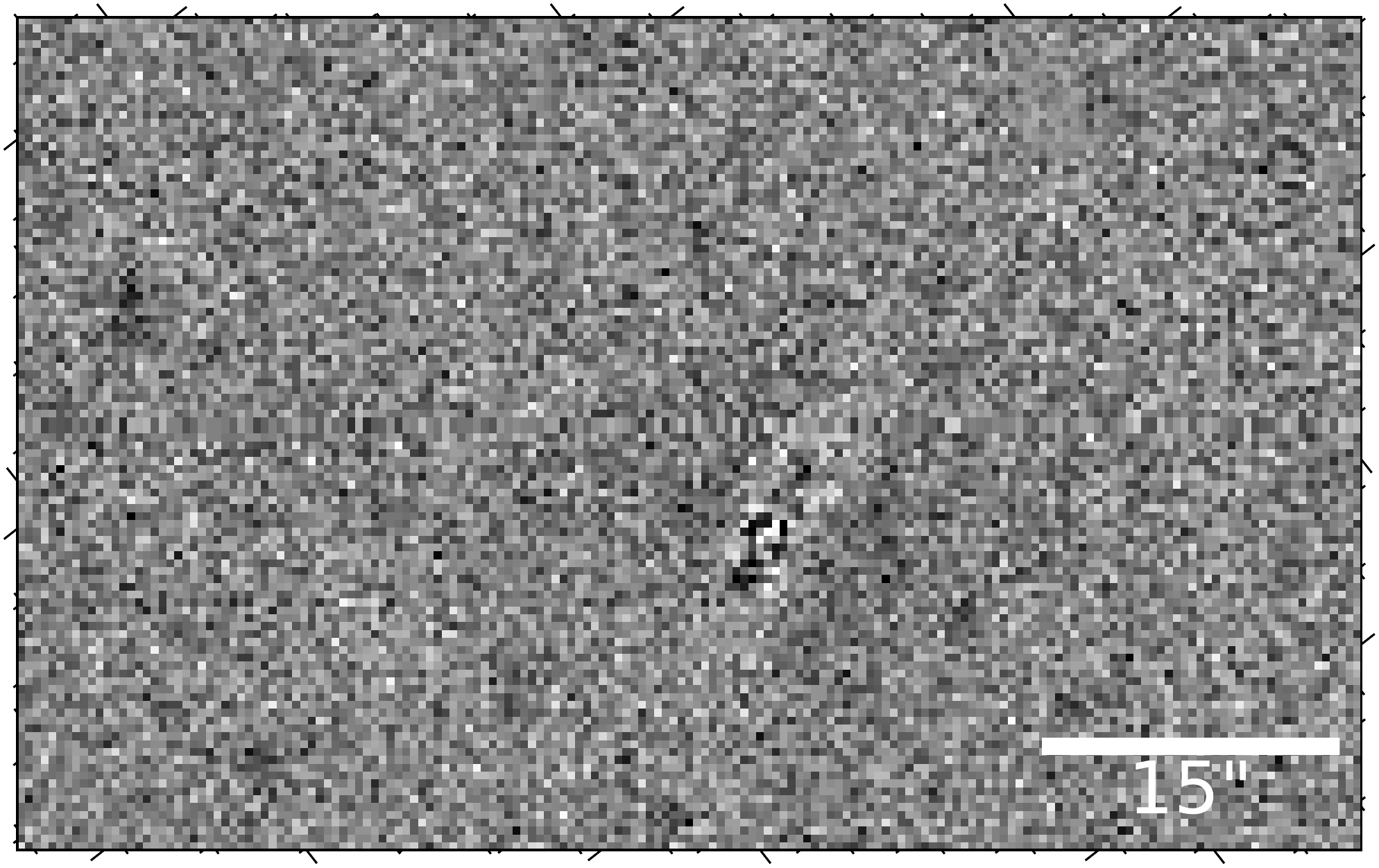} \label{104G2resid2}}\\
\subfloat{\includegraphics[scale=0.2, bb=2 2 540 400]{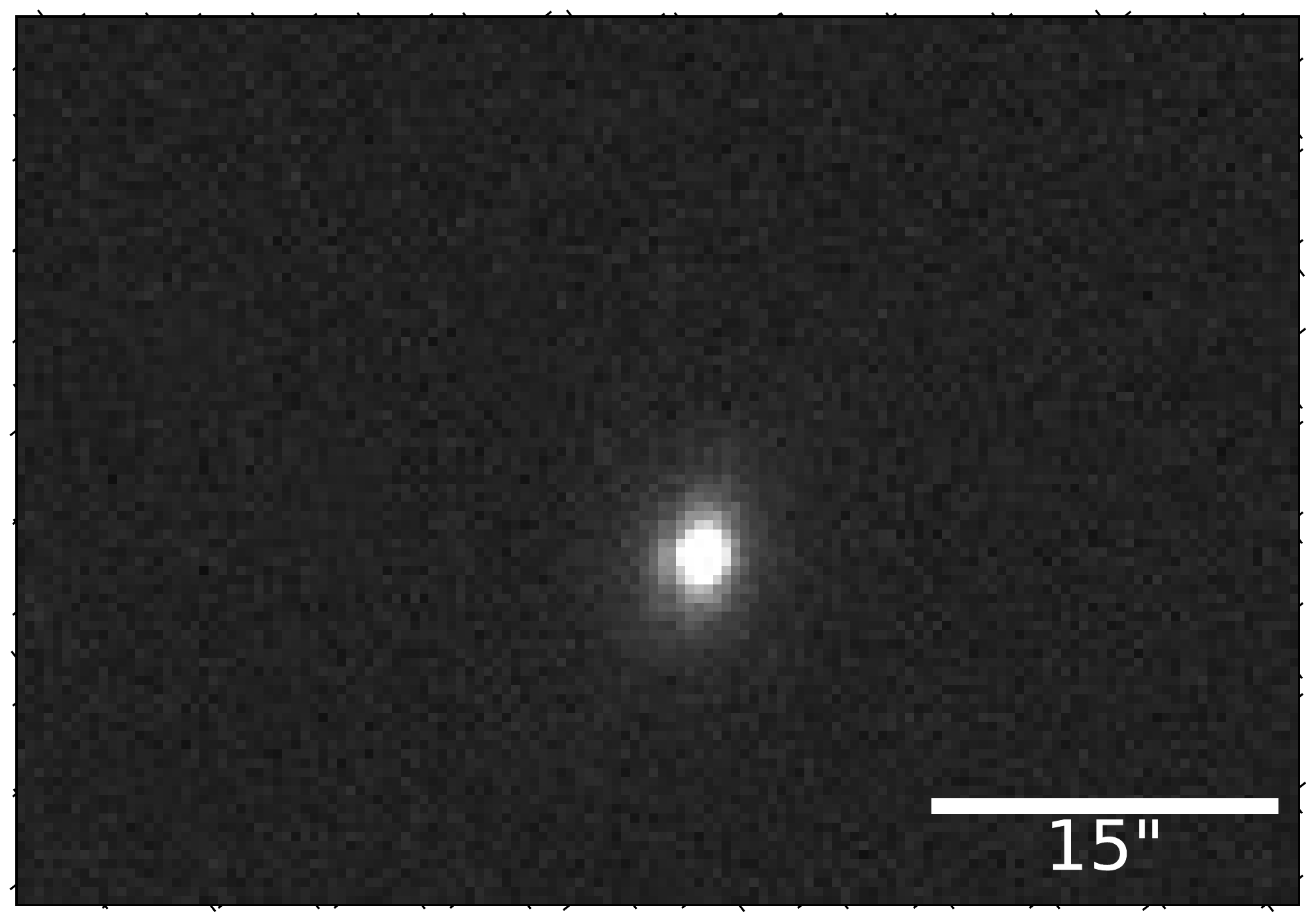}\label{104G3original}}
 \hspace{0.3cm}
\subfloat{\includegraphics[scale=0.2, bb=2 2 540 400]{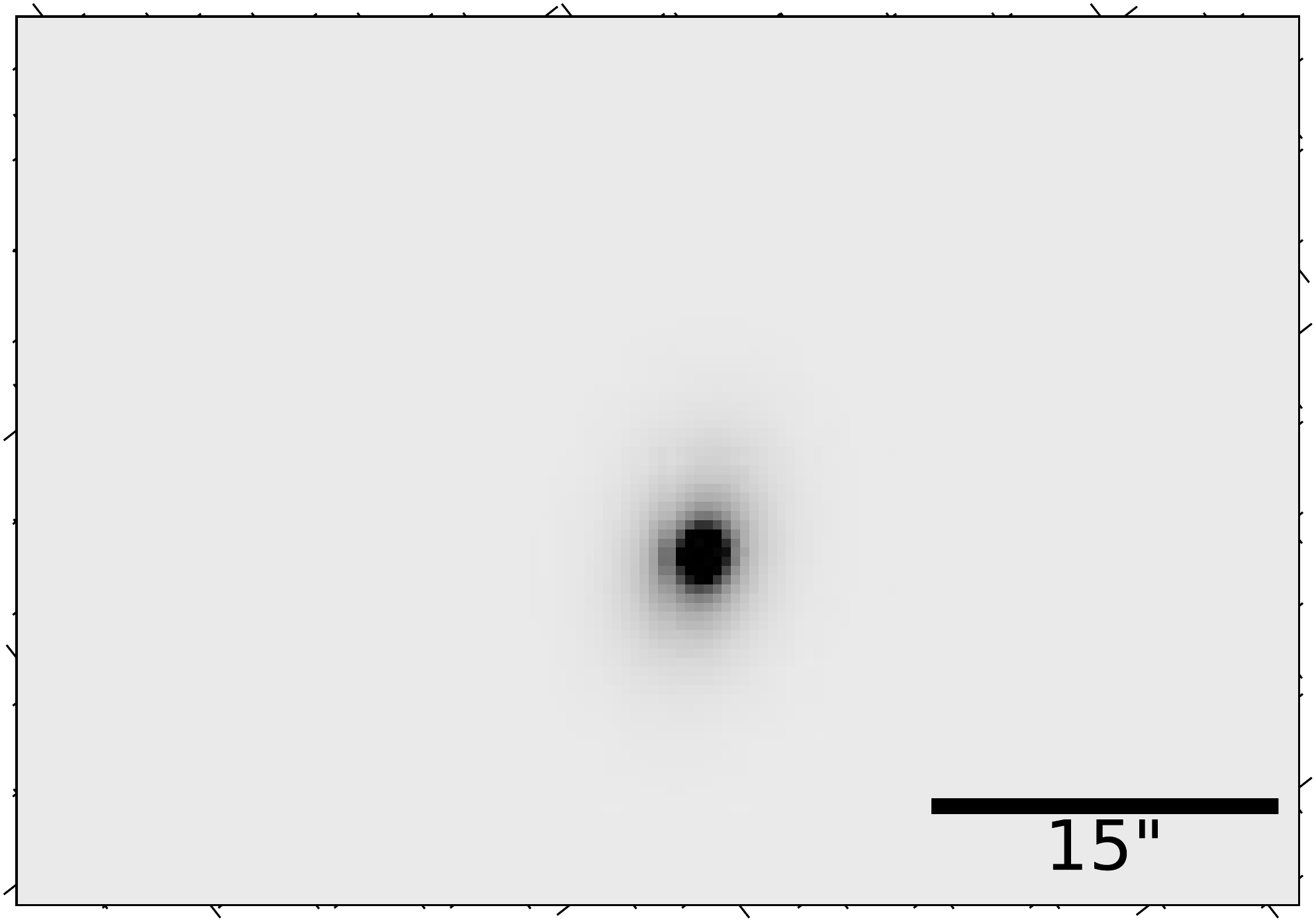} \label{104G3resid1}}
\hspace{0.3cm}
\subfloat{\includegraphics[scale=0.2, bb=2 2 540 400]{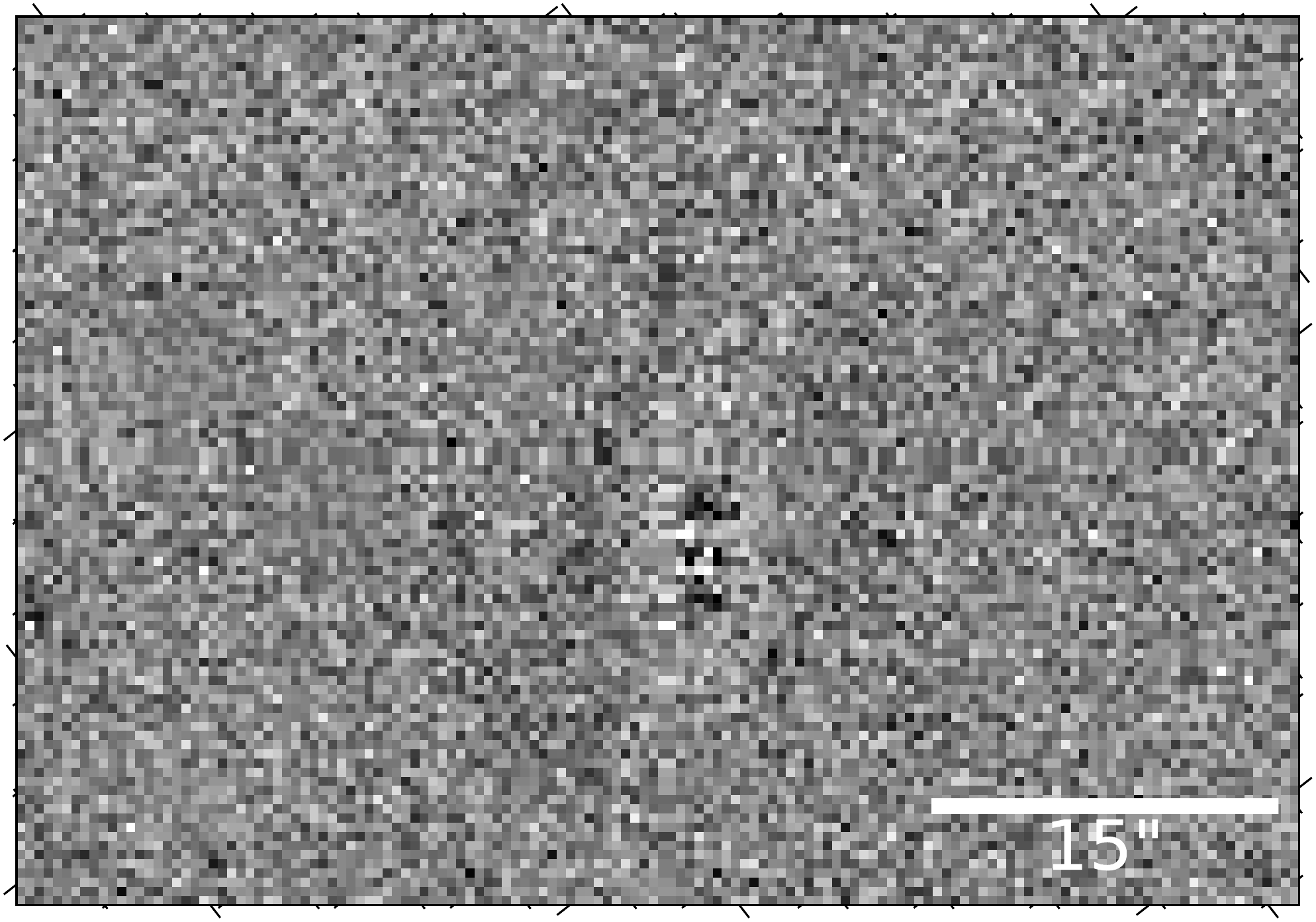} \label{104G3resid2}}\\

\caption{Photometric analysis of the triplet system SIT 104. See caption Fig.~\ref{fig:30analysis}}
\label{fig:104analysis}
\end{figure*}

The triplet system SIT 104 has two elliptical galaxies (\textbf{G1} and \textbf{G2}) and one barred spiral galaxy (\textbf{G3}) \citep{Willett2013} (Figure~\ref{fig:SIT104}). \textbf{G1} is an elliptical galaxy with cuspy core without evidence of a disk component, while the elliptical galaxy \textbf{G2} shows a central elongated bulge structure with an evidence of disk component. This galaxy is classified as S0 by \cite{Nair2010}. The optical image of the spiral barred galaxy \textbf{G3} does not reveal the existence of neither spiral arms nor bar component on its elongated disk component. 

The 1D fitting of \textbf{G1} (Figure~\ref{104_SB}, left panel) shows a down-bending truncation at SMA $=$ 9.5$''$ of type II profile with an outer ring. An E/S0 classification might be more appropriate for this galaxy.
The intensity profile of \textbf{G2} is clearly distinguished as type I profile (Figure~\ref{104_SB}, middel panel), while the intensity profile of \textbf{G3} resembles a type III-s profile with a break at SMA $=$ 6.5$''$ (Figure~\ref{104_SB}, right panel).

Decomposition of \textbf{G1} using $Ser$ + $Exp$ confirms the presence of a nuclear stellar disk in \textbf{G1} with a semi-major axis of $SMA$ = 2.76$''$ in the residual image with a $\chi^2_{\nu}$ of 1.06 (Figure~\ref{104G1resid2}). An E/S0 classification of this galaxy is more reasonable. 

Decomposition of \textbf{G2} with $Ser$ + $Exp$ profile admits a clear fit without residuals (see Figure~\ref{104G2resid2}), and the $\chi^2_{\nu}$ of 1.12. This result confirm the classification of \textbf{G2} as an S0 galaxy \citep{Nair2010}.

Figure~\ref{104G3resid2} illustrate the decomposition of \textbf{G3} using $Ser$ + $Exp$ profiles, which covers the whole components of the galaxy and a $\chi^2_{\nu}$ of 1.06 were obtained. This galaxy could be classified as an unbarred spiral rather than SB galaxy, where the bar structure could not be identified in neither the 1D and 2D decomposition, nor the position angle profile of this galaxy.

\subsection{The triplet system SIT 125}

\begin{figure*}
\centering
\subfloat{\includegraphics[scale=0.5, bb=130 -150 100 100]{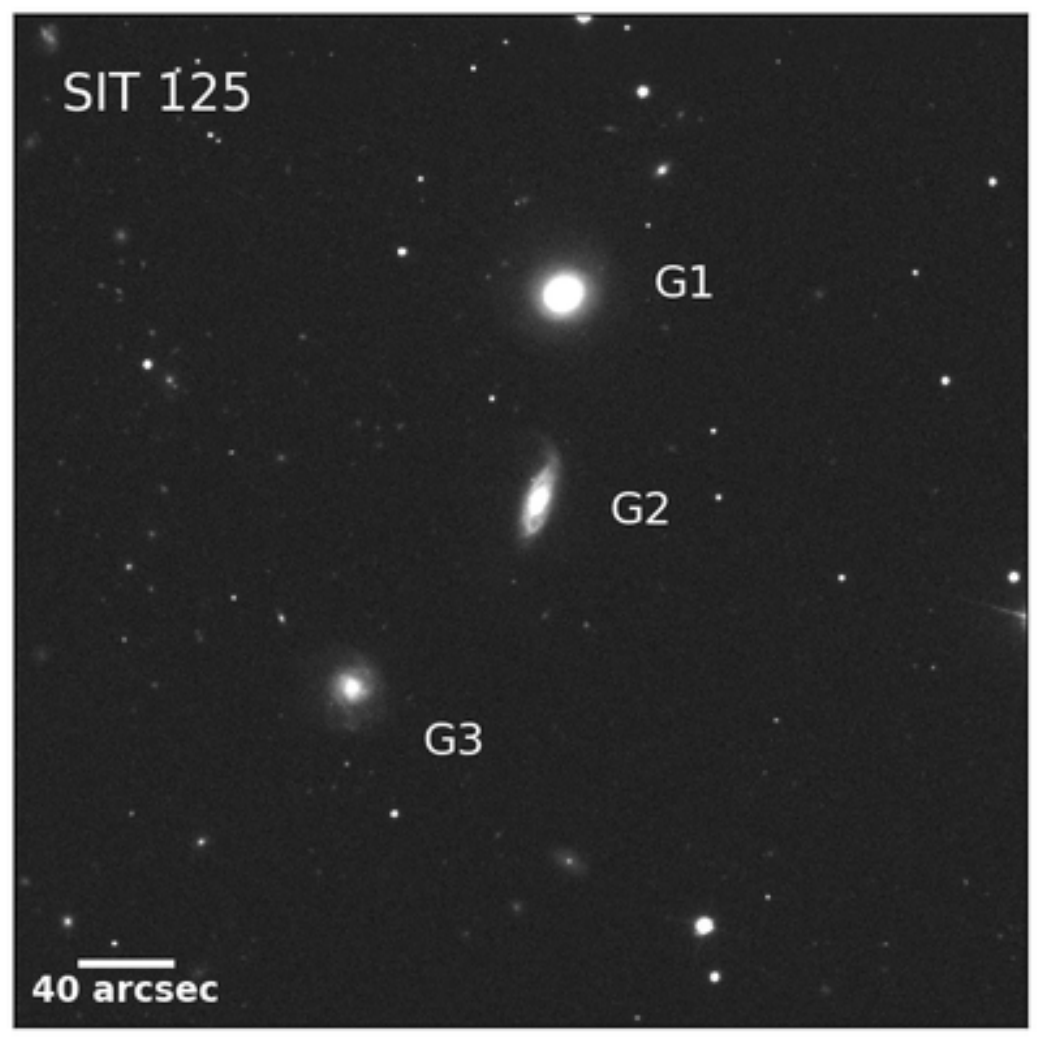}
 \label{fig:SIT125}}\\
\subfloat{\includegraphics[scale=0.7, bb=80 20 500 100]{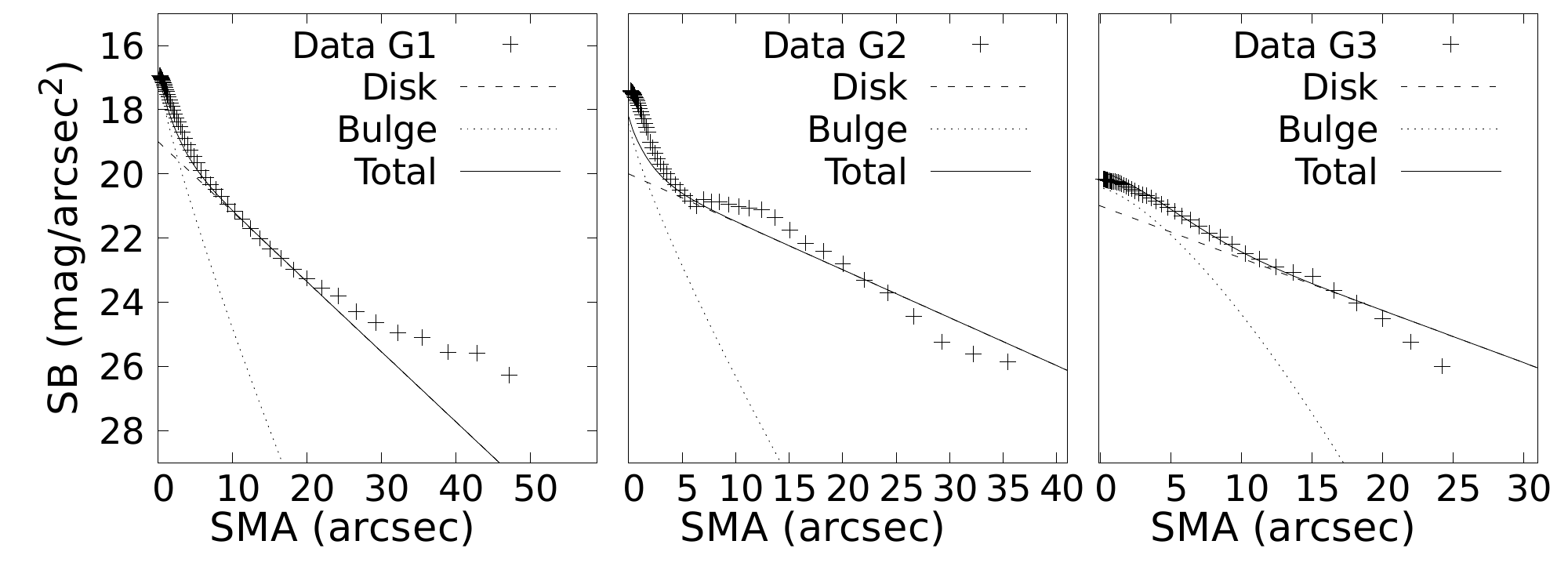}\label{125_SB}}\\
\subfloat{\includegraphics[scale=0.2, bb=5 8 515 504]{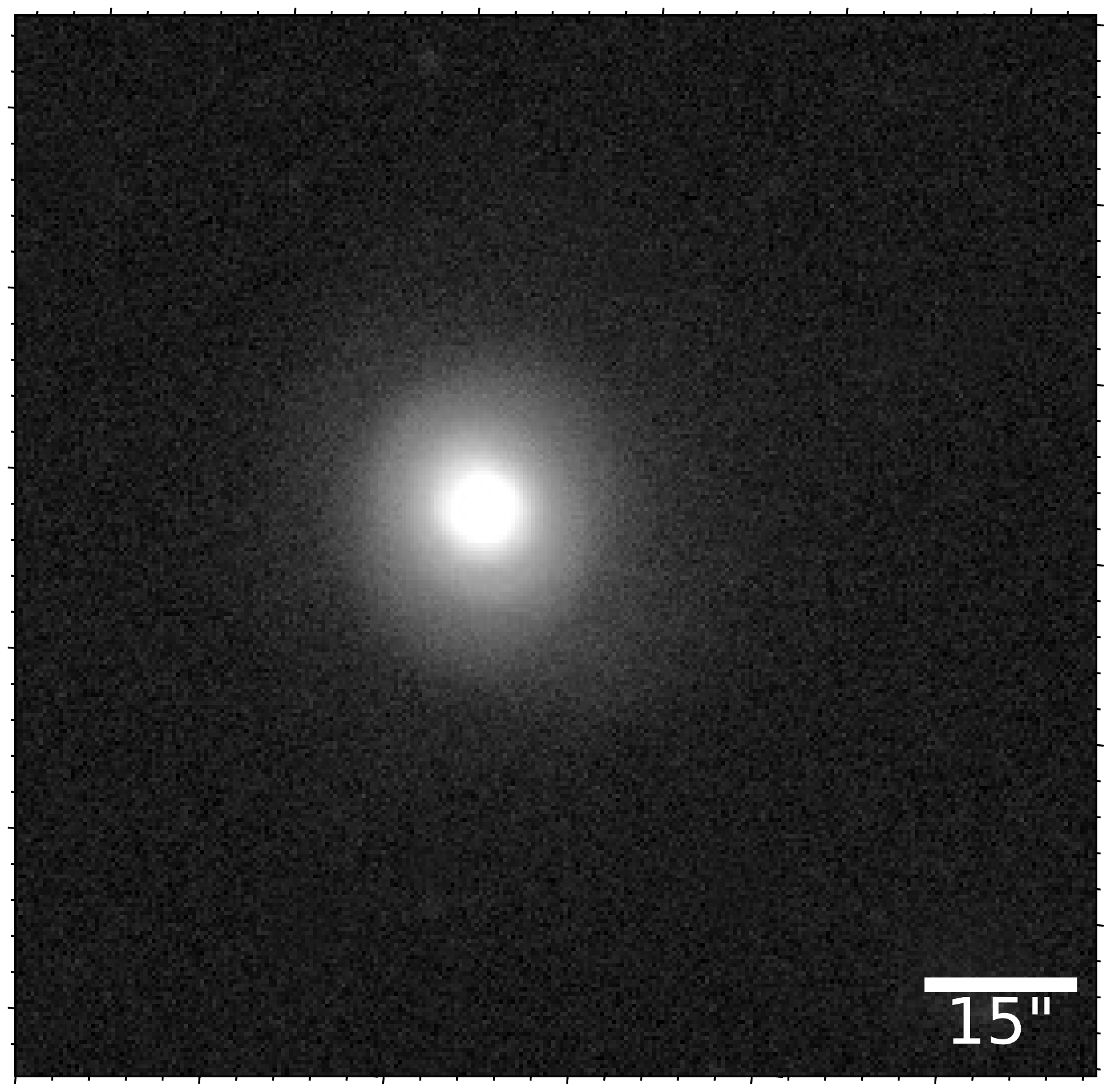}\label{125G1original}}
\hspace{0.3cm}
\subfloat{\includegraphics[scale=0.2, bb=5 8 515 504]{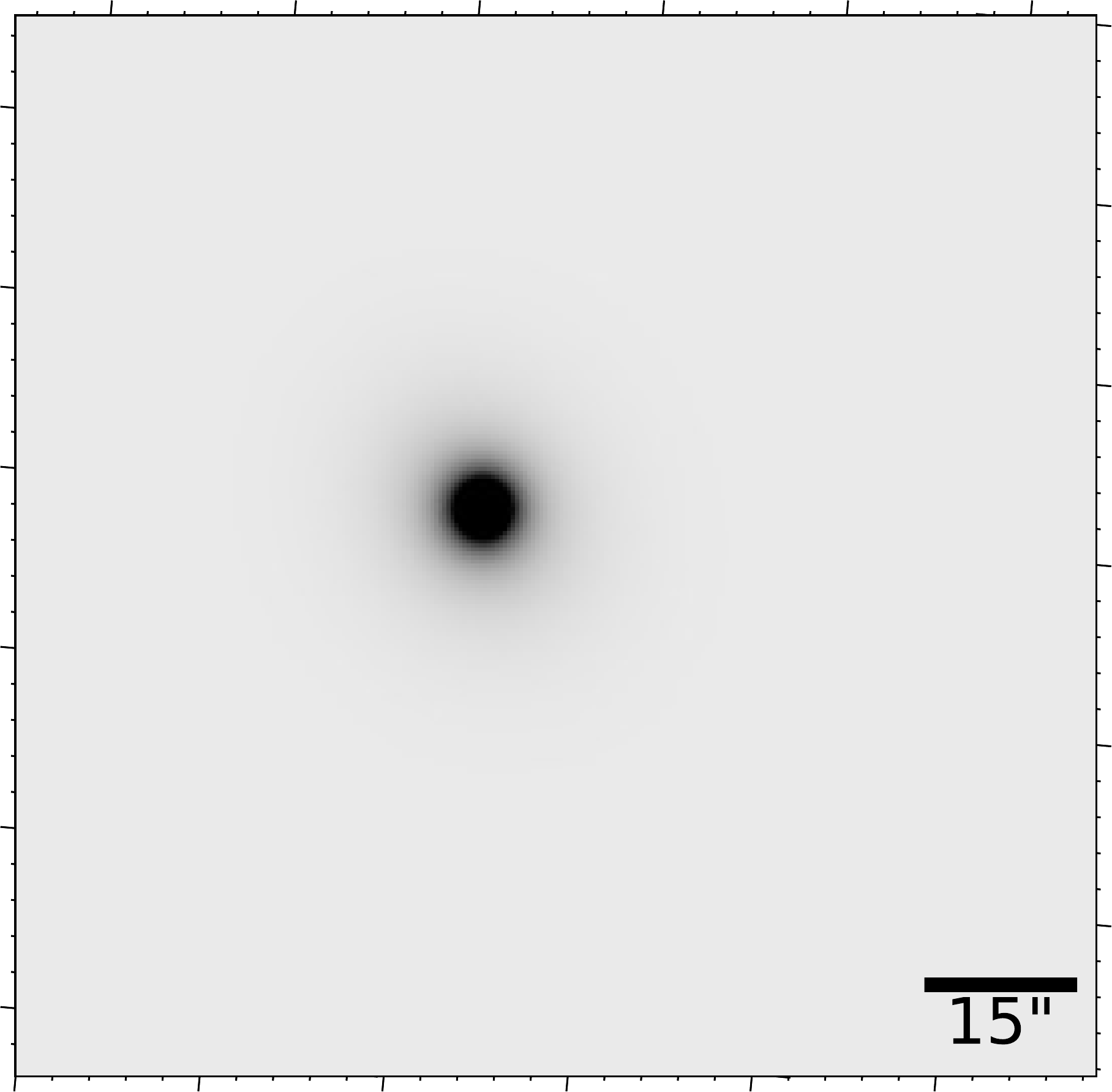} \label{125G1resid1}}
\hspace{0.3cm}
\subfloat{\includegraphics[scale=0.2, bb=5 8 515 504]{Figures_pdf/125G1_residue2-eps-converted-to.pdf} \label{125G1resid2}}\\
\subfloat{\includegraphics[scale=0.2, bb=8 8 540 506]{Figures_pdf/125G2_original-eps-converted-to.pdf}\label{125G2original}}
 \hspace{0.3cm}
\subfloat{\includegraphics[scale=0.2, bb=8 8 540 506]{Figures_pdf/125G2_model2-eps-converted-to.pdf} \label{125G2resid1}}
\hspace{0.3cm}
\subfloat{\includegraphics[scale=0.2, bb=8 8 540 506]{Figures_pdf/125G2_residue2-eps-converted-to.pdf} \label{125G2resid2}}\\
\subfloat{\includegraphics[scale=0.2, bb=8 8 560 416]{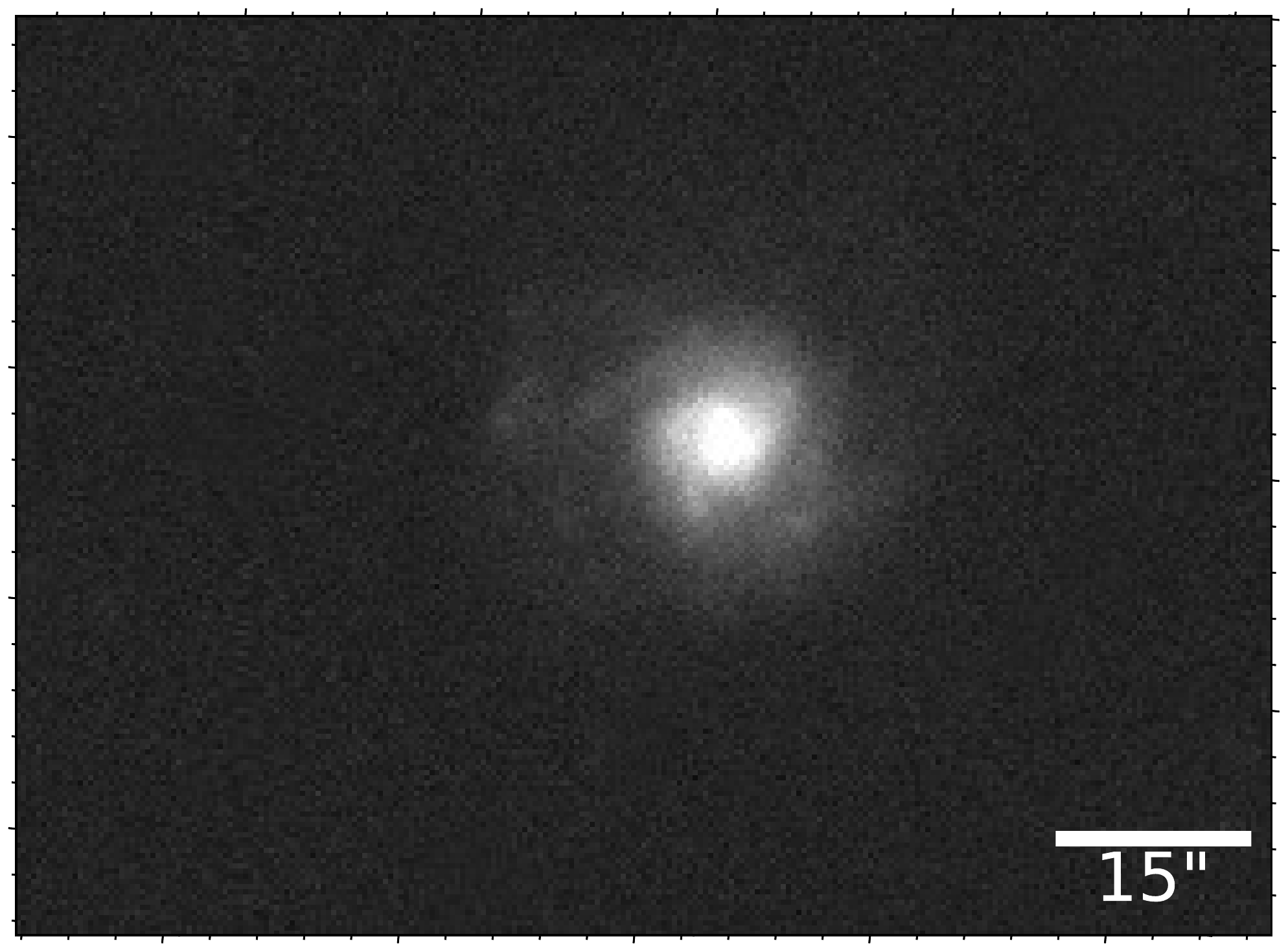}\label{125G3original}}
\hspace{0.3cm}
\subfloat{\includegraphics[scale=0.2, bb=8 8 560 416]{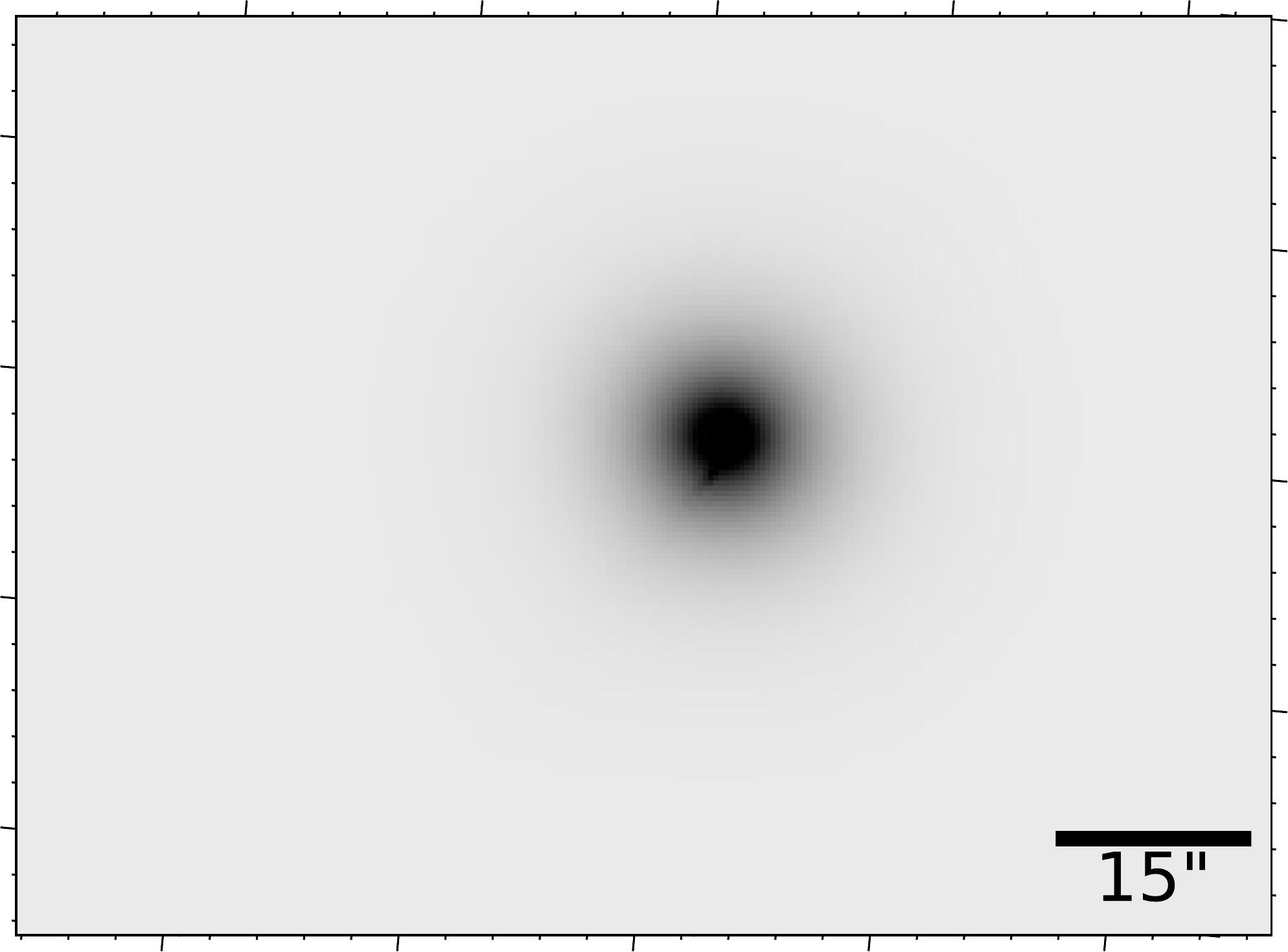} \label{125G3resid1}}
\hspace{0.3cm}
\subfloat{\includegraphics[scale=0.2, bb=8 8 560 416]{Figures_pdf/125G3_residue2-eps-converted-to.pdf} \label{125G3resid2}}

\caption{Photometric analysis of the triplet system SIT 125. See caption Fig.~\ref{fig:30analysis}}
\label{fig:125analysis}
\end{figure*}

SIT 125 (Figure~\ref{fig:SIT125}) is a hierarchical triplet where \textbf{G1} and \textbf{G2} are close neighbors (with angular separation of 1.44 arcmin) and \textbf{G3} is separated from \textbf{G1} by 3.09 arcmin \citep{Tawfeek2019}. \textbf{G1} is an S0 galaxy with an extremely bright nucleus and a clear circular disk around it \citep{Nair2010}. \textbf{G2} is classified  as SABa galaxy \citep{Nair2010} with two asymmetric arms at the N and the S directions. The northern arm forms a bridge extending toward \textbf{G1} while the southern arm is faint and short. \textbf{G3} is an Sc galaxy with a set of fuzzy bluish arms that bifurcate outwards \citep{Nair2010}. At the outer most regions, the arms winded clockwise enclosing the galaxy and forming an asymmetric structure, which might be of tidal origin.

\textbf{G1} represents a type III-d profile with an up-bending break at SMA $=$ 22$''$ (Figure~\ref{125_SB}).
The intensity profile of \textbf{G2} (Figure~\ref{125_SB}, middle panel) shows two breaks at SMA $\sim$ 7$''$ and SMA $=$ 24.5$''$. The first one classified as type II.i, while the second one is of type II-o. 
The intensity profile of \textbf{G3} matches well a type II-o with a down-bending truncation at SMA $\sim$ 16.5x$''$ (Figure~\ref{125_SB}). 

Decomposition of \textbf{G1} using a single $Ser$ and $Exp$ profile reveals a significant bulge with an outer ring at 12.6$''$ and a $\chi^2_{\nu}$ of 1.08 (Figure~\ref{125G1resid2}. 

Decomposition of \textbf{G2}, using $Ser$ + $Exp$,  reveals a bi-nuclei in the residual image with a strong tidal tail at the Northern side of the galaxy and a $\chi^2_{\nu}$ of 1.43 (Figure~\ref{125G2resid2}). These residuals affirm the state of interaction between \textbf{G2} and \textbf{G1}.

Fitting \textbf{G3} with $Ser$ and $Exp$ profile leaves the spiral arms as residuals revealing a pure exponential unperturbed d
isk galaxy and a $\chi^2_{\nu}$ of 1.06 has been achieved (Figure~\ref{125G3resid2}). 

\subsection{The triplet system SIT 197}

\begin{figure*}
\centering
\subfloat{\includegraphics[scale=0.5, bb=130 -150 100 100]{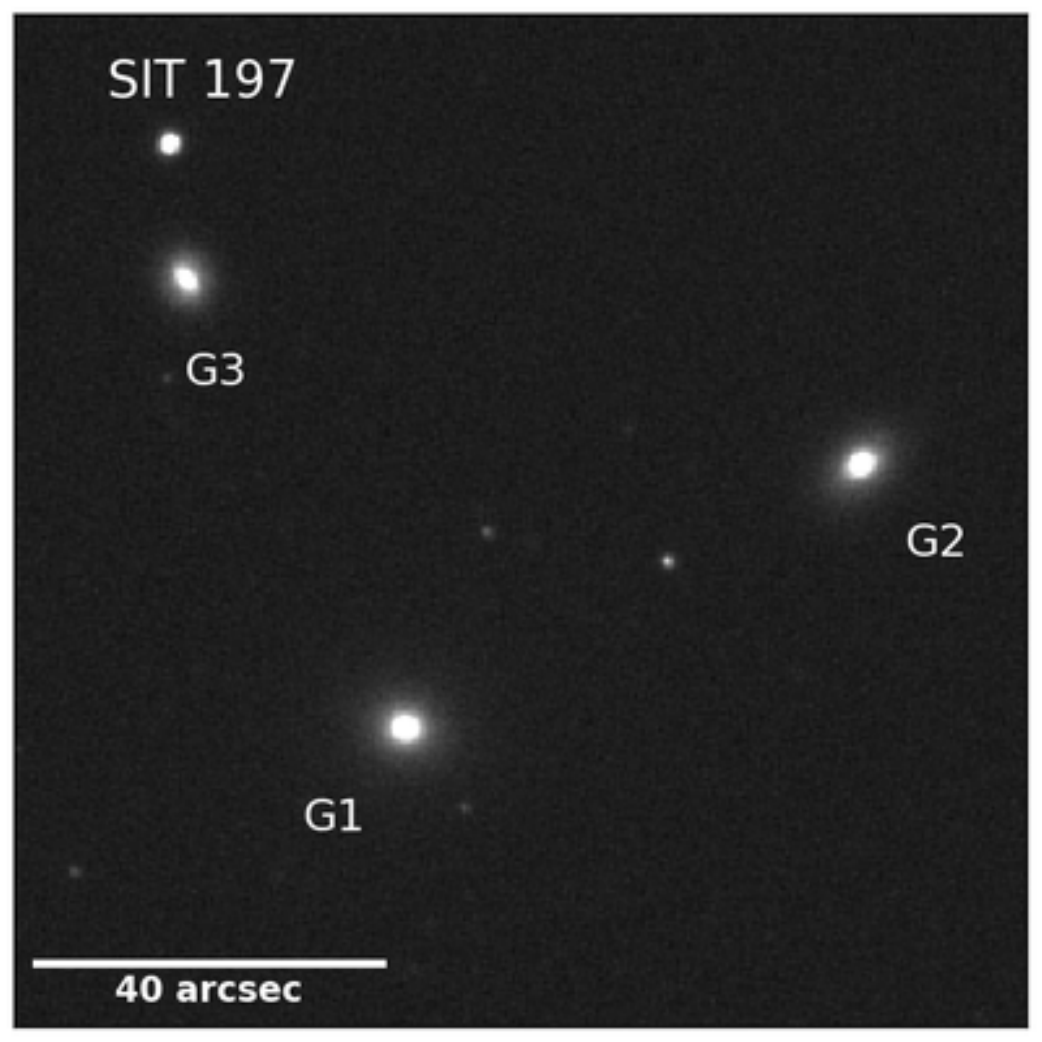}
 \label{fig:SIT197}}\\
\subfloat{\includegraphics[scale=0.7, bb=80 20 500 100]{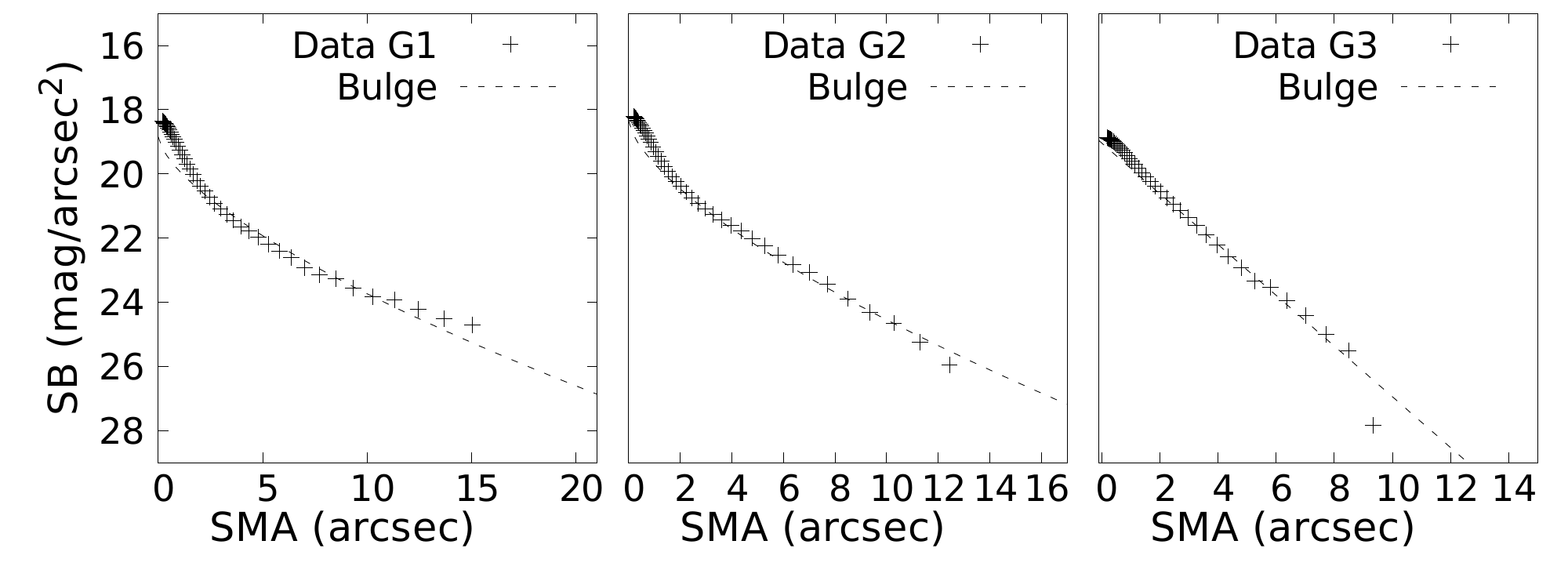}\label{197_SB}}\\
\subfloat{\includegraphics[scale=0.2, bb=8 8 560 416]{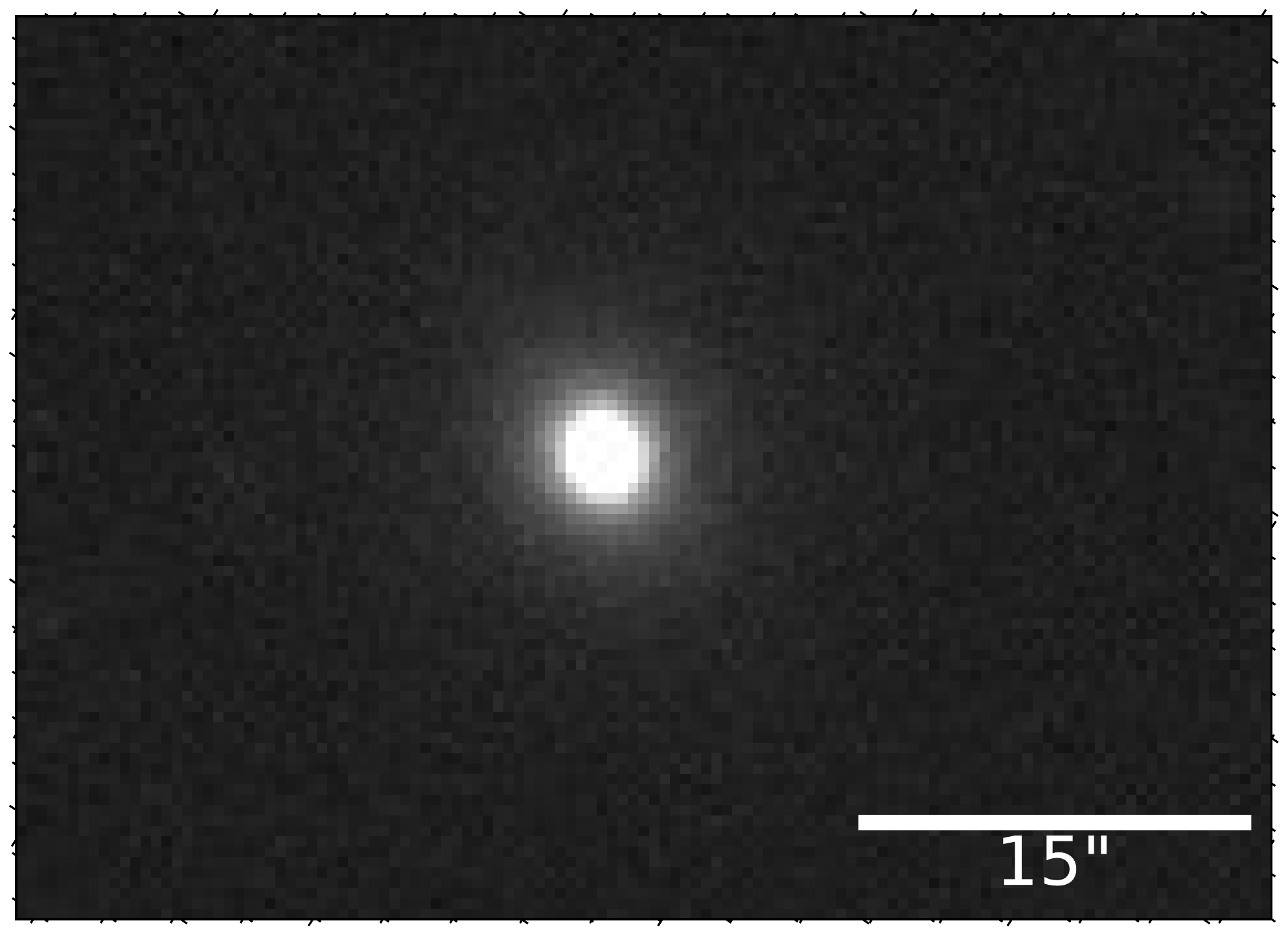}\label{197G1original}}
 \hspace{0.3cm}
\subfloat{\includegraphics[scale=0.2, bb=8 8 560 416]{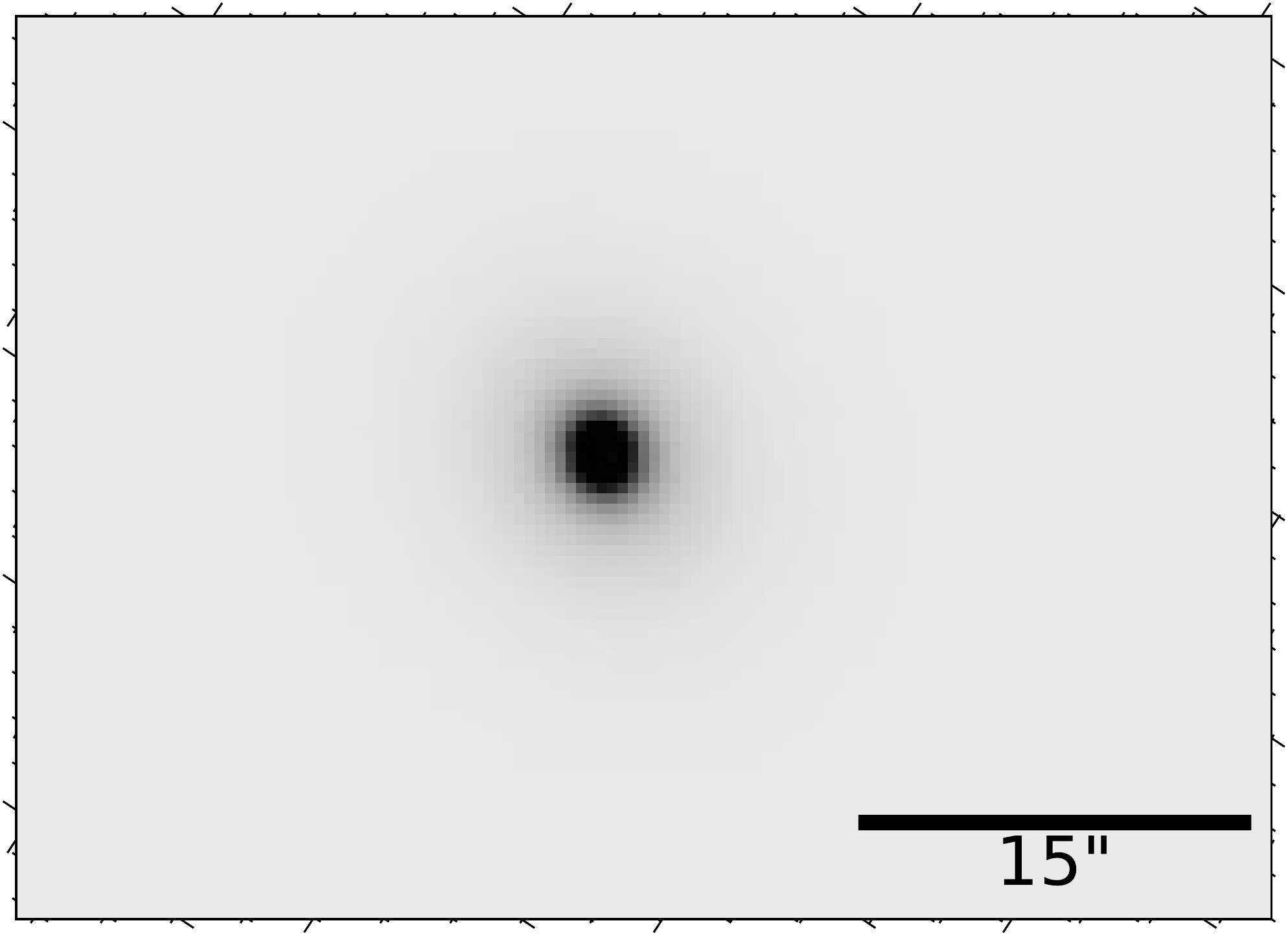} \label{197G1resid1}}
\hspace{0.3cm}
\subfloat{\includegraphics[scale=0.2, bb=8 8 560 416]{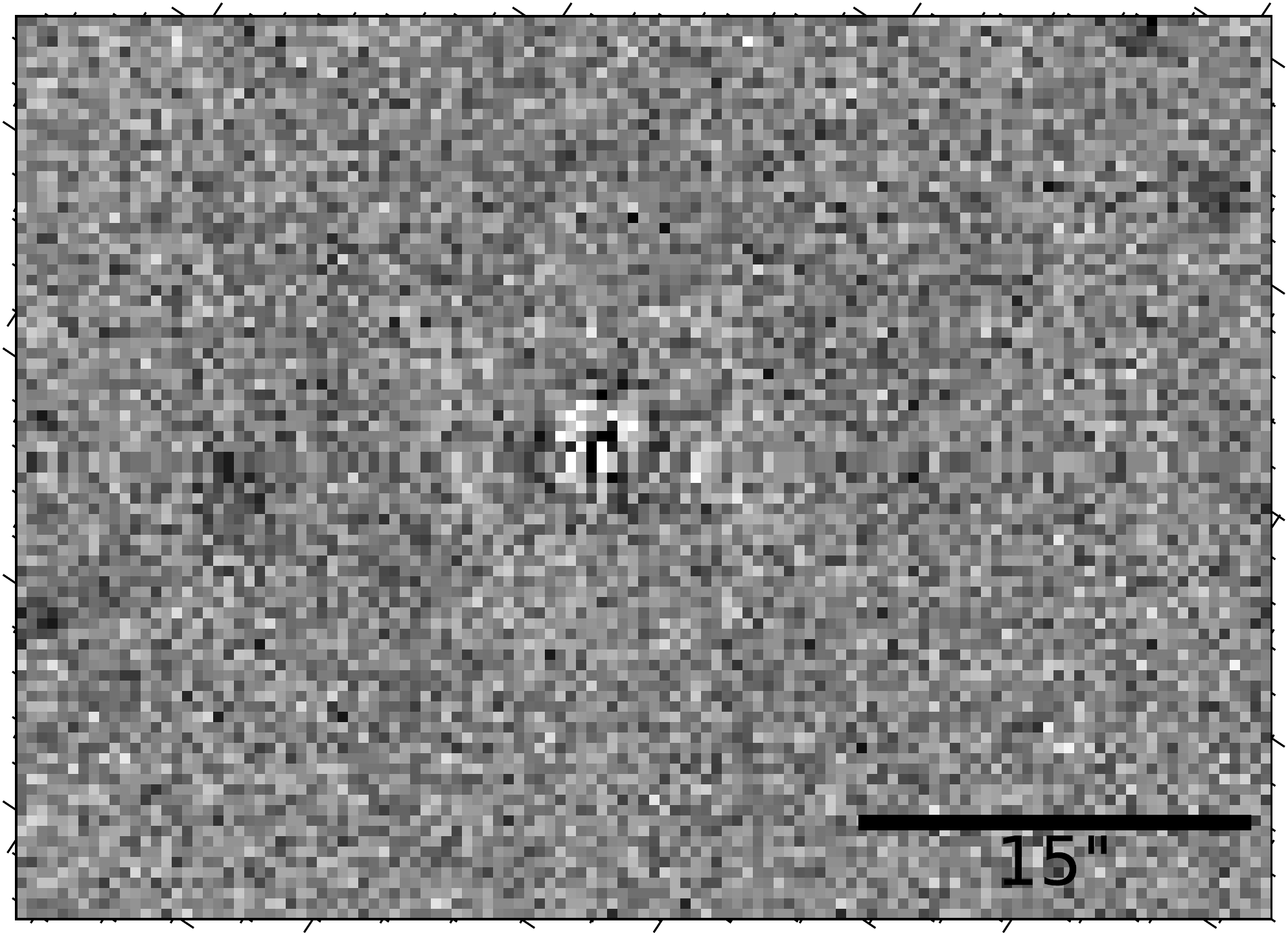} \label{197G1resid2}}\\
\subfloat{\includegraphics[scale=0.2, bb=8 8 560 470]{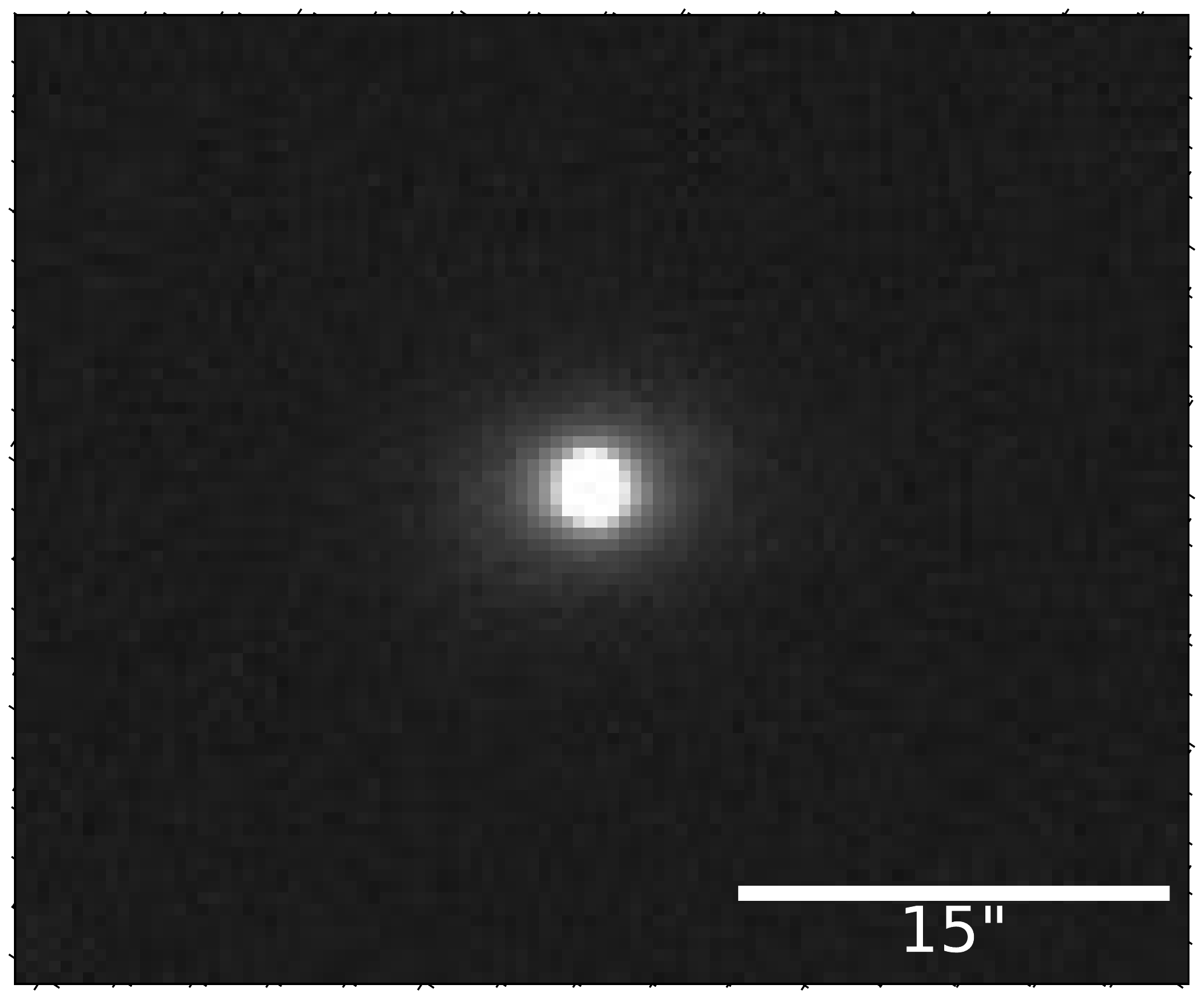}\label{197G2original}}
 \hspace{0.3cm}
\subfloat{\includegraphics[scale=0.2, bb=8 8 560 470]{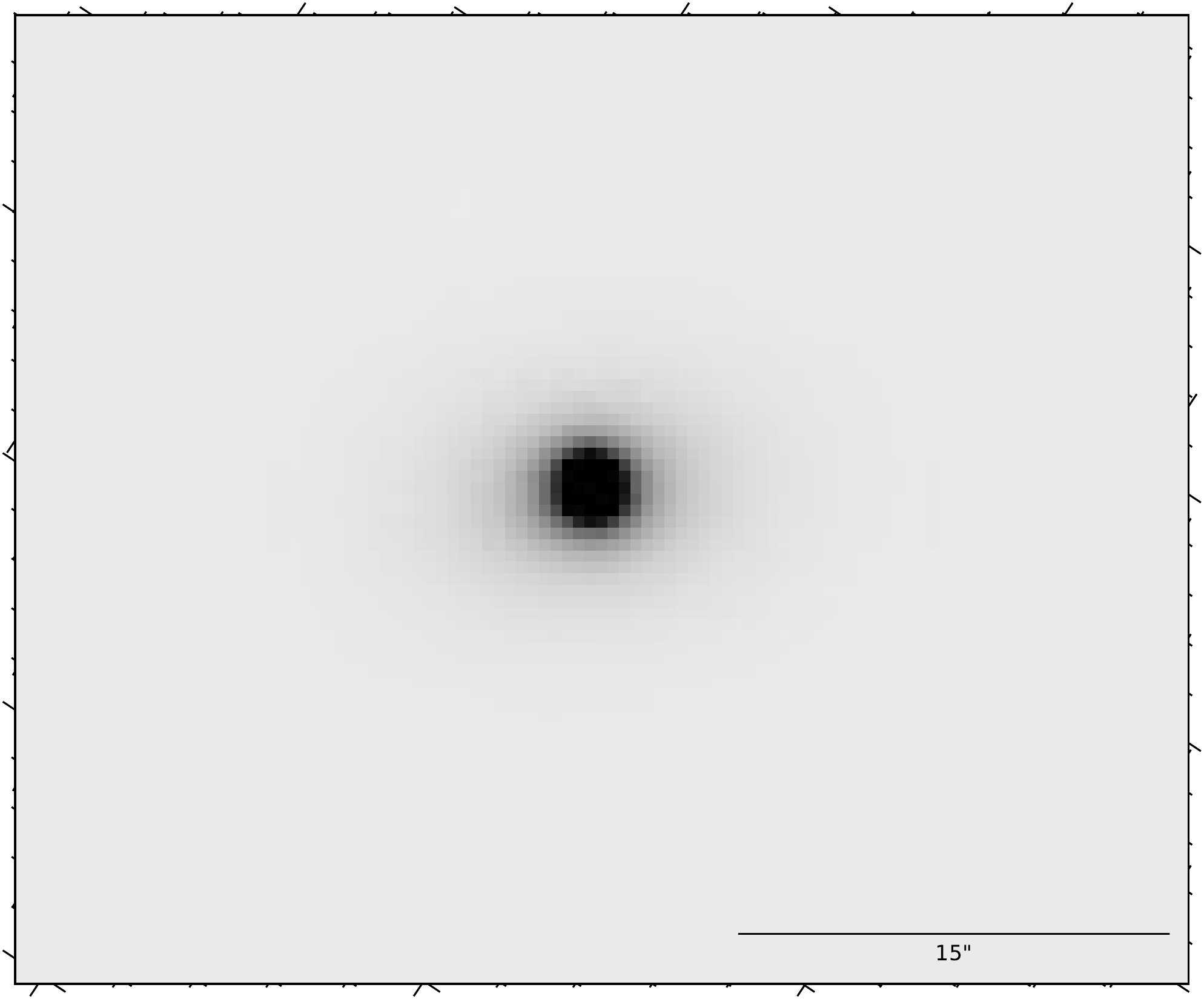} \label{197G2resid1}}
\hspace{0.3cm}
\subfloat{\includegraphics[scale=0.2, bb=8 8 560 470]{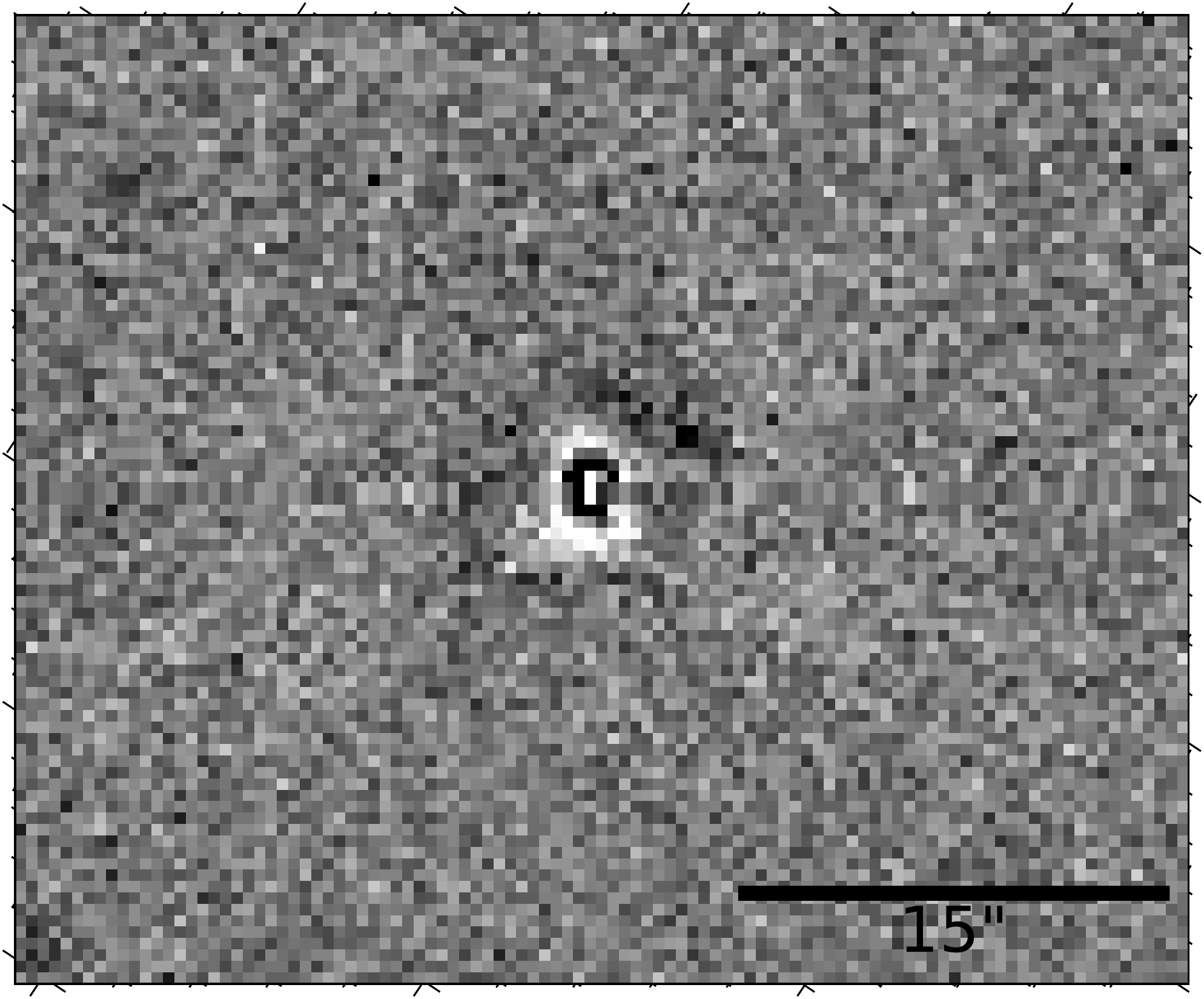} \label{197G2resid2}}\\
\subfloat{\includegraphics[scale=0.2, bb=8 8 560 416]{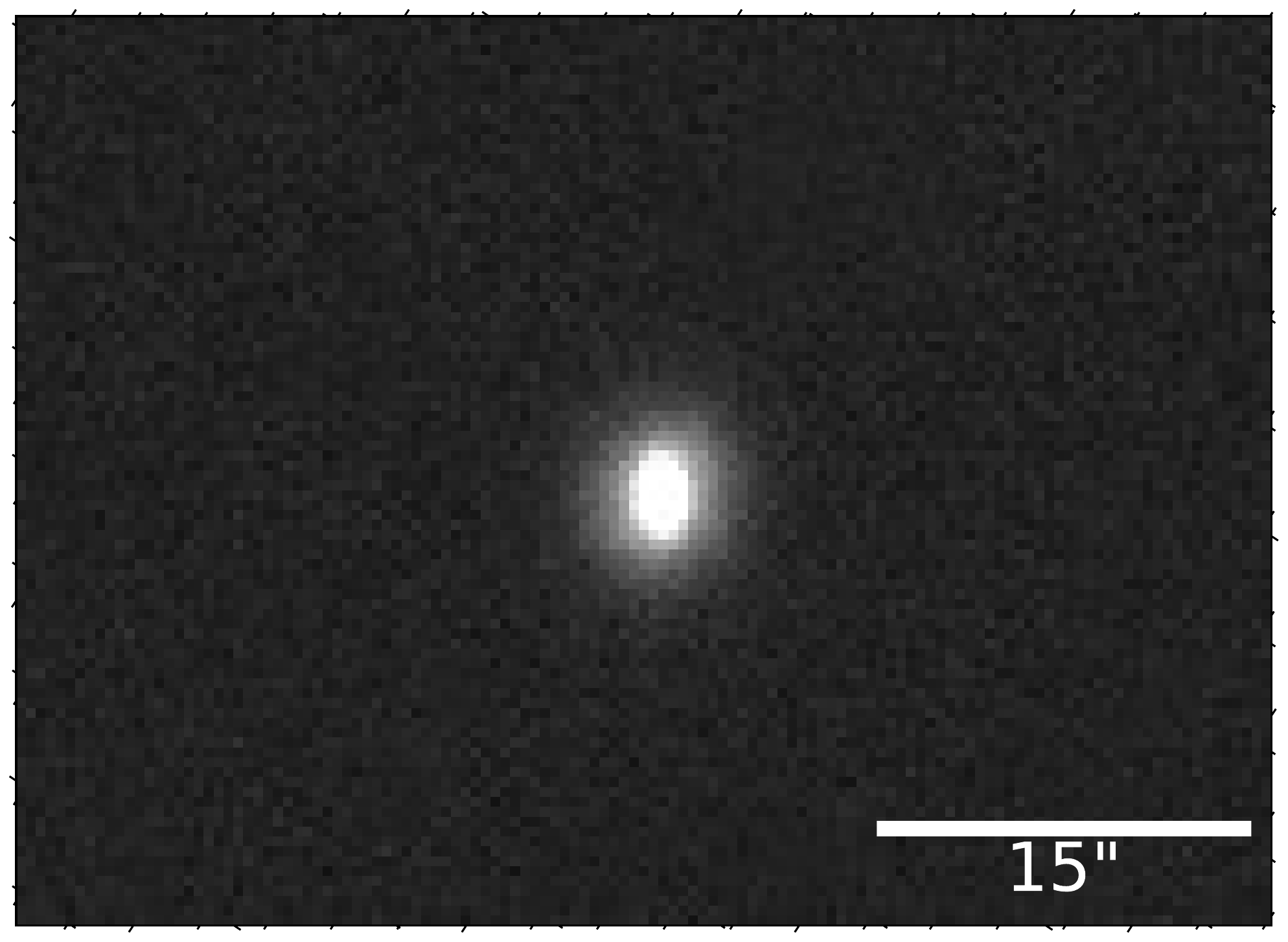}\label{197G3original}}
\hspace{0.3cm}
\subfloat{\includegraphics[scale=0.2, bb=8 8 560 416]{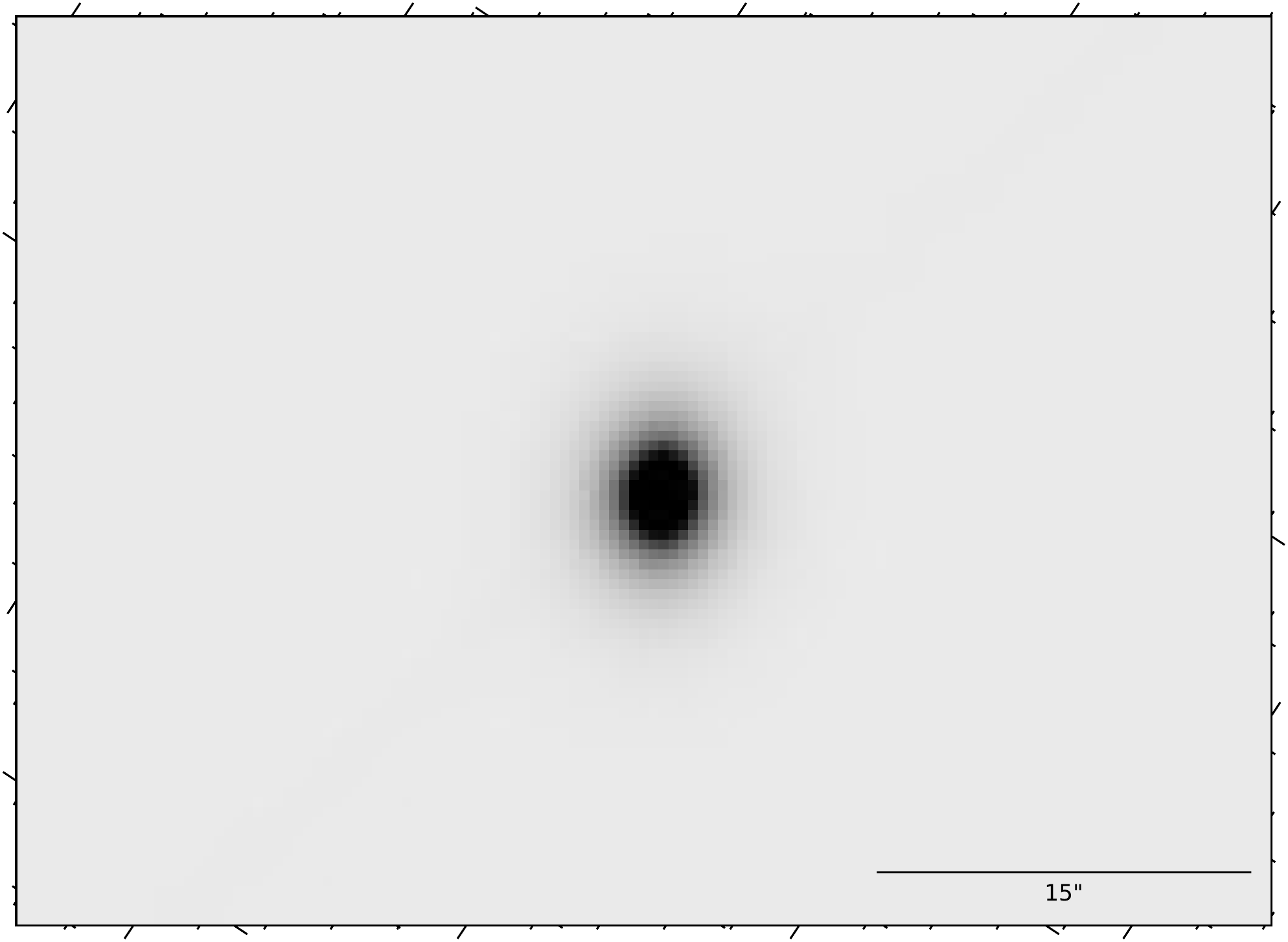} \label{197G3resid1}}
\hspace{0.3cm}
\subfloat{\includegraphics[scale=0.2, bb=8 8 560 416]{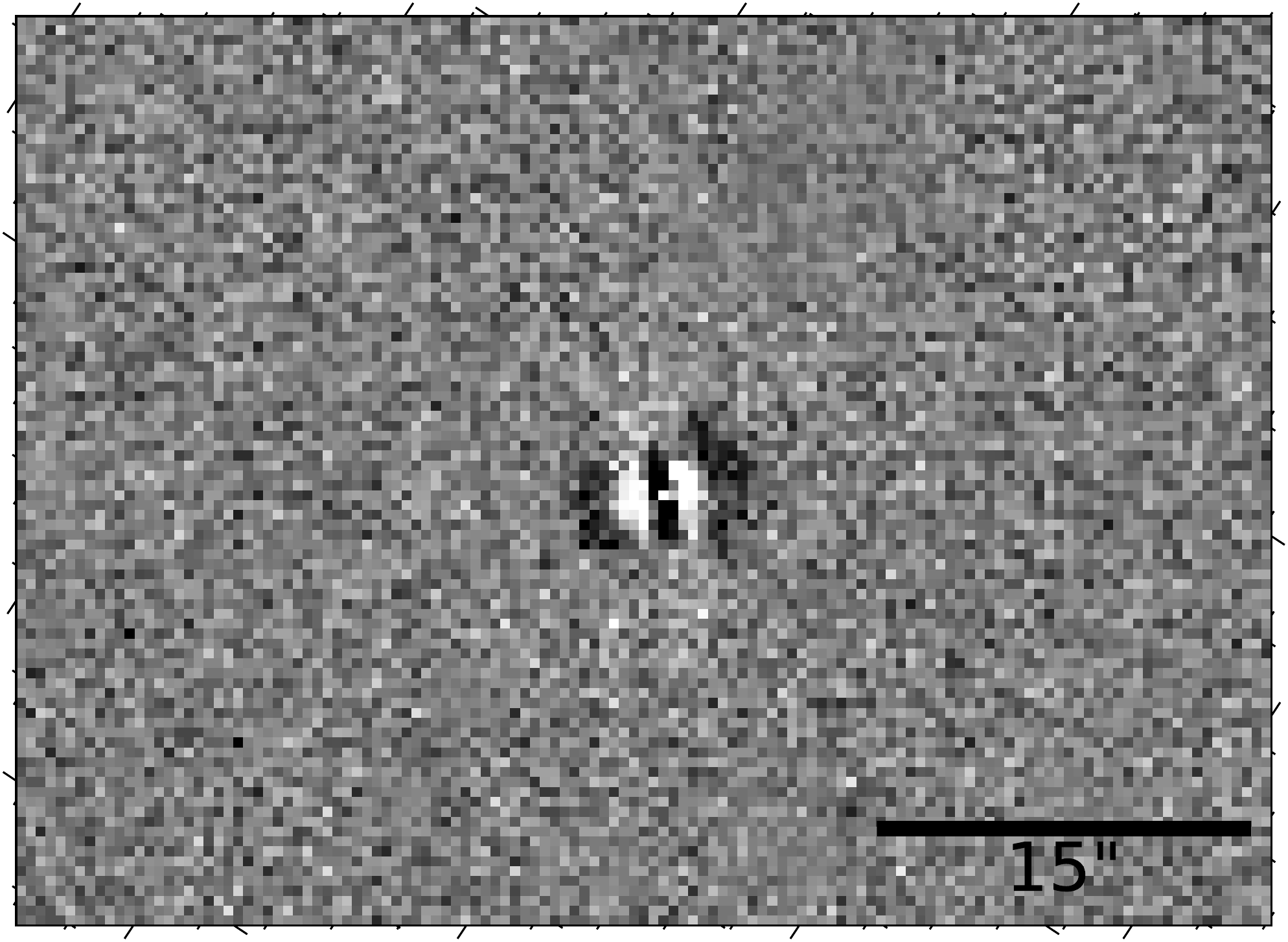} \label{197G3resid2}}\\

\caption{Photometric analysis of the triplet system SIT 197. See caption Fig.~\ref{fig:30analysis}}
\label{fig:197analysis}
\end{figure*}

SIT 197 (Figure~\ref{fig:SIT197}) is a unique system in our sample, it is built of three elliptical galaxies \citep{Willett2013}. \textbf{G1} and \textbf{G2} have nearly circular isophotes with a bright cuspy core at the center while, \textbf{G3} shows elongated isophotes, with higher ellipticity.

The intensity profiles of the three elliptical galaxies were fitted by using single $Ser$ component with $Ser$ index ($n$)of 1.5, 1.5 and 1, respectively  (Figures~\ref{197_SB}). An E/S0 classification might be more suitable for \textbf{G3}.

Despite \textbf{G1} being an elliptical galaxy, two profiles ($Ser$+ $Exp$) were required to fit the galaxy components with a $\chi^2_{\nu}$ of 0.80 (Figure~\ref{197G1resid2}).

Similarly, the decomposition of \textbf{G2} using $Ser$ + $Exp$ covers nearly all the fine structures of the galaxy and a $\chi^2_{\nu}$ of 0.91 was achieved  (Figure~\ref{197G2resid2}). 

Decomposition of \textbf{G3} using two models ($Ser$ + $Exp$) reveals a central dark bipolar patches as a residuals with $\chi^2_{\nu}$ of 1.12 (Figure~\ref{197G3resid2}).

\subsection{The triplet system SIT 217}

\begin{figure*}
\centering
\subfloat{\includegraphics[scale=0.5, bb=130 -150 100 100]{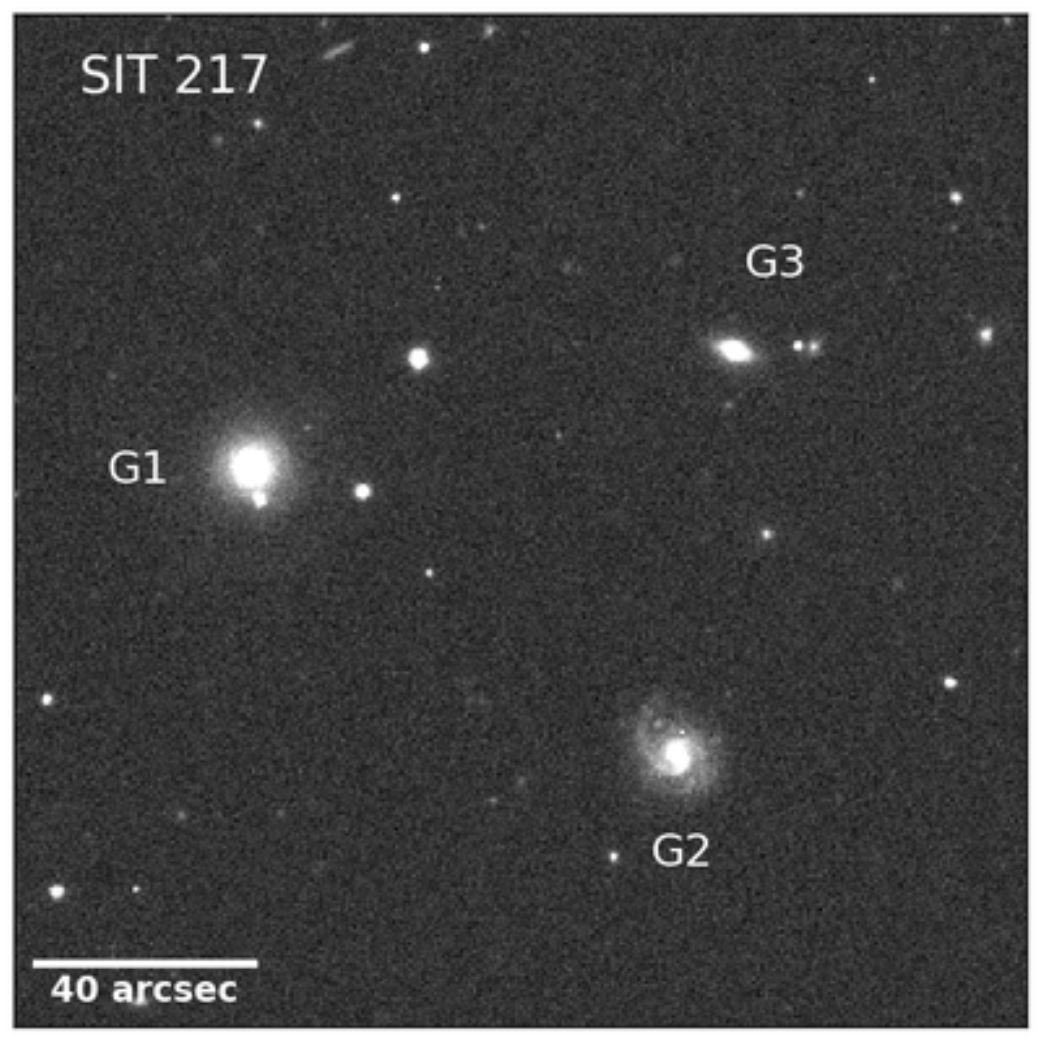}
 \label{fig:SIT217}}\\
\subfloat{\includegraphics[scale=0.7, bb=80 20 500 100]{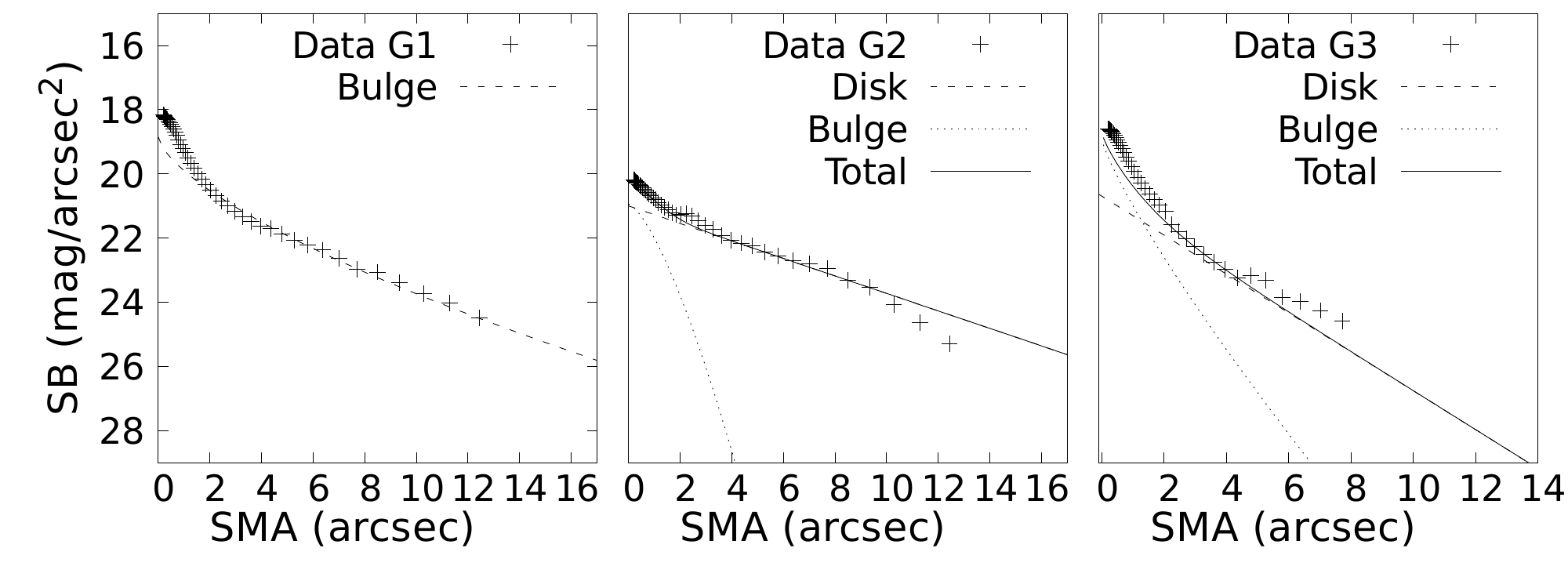}\label{217_SB}}\\
\subfloat{\includegraphics[scale=0.2, bb=8 8 560 416]{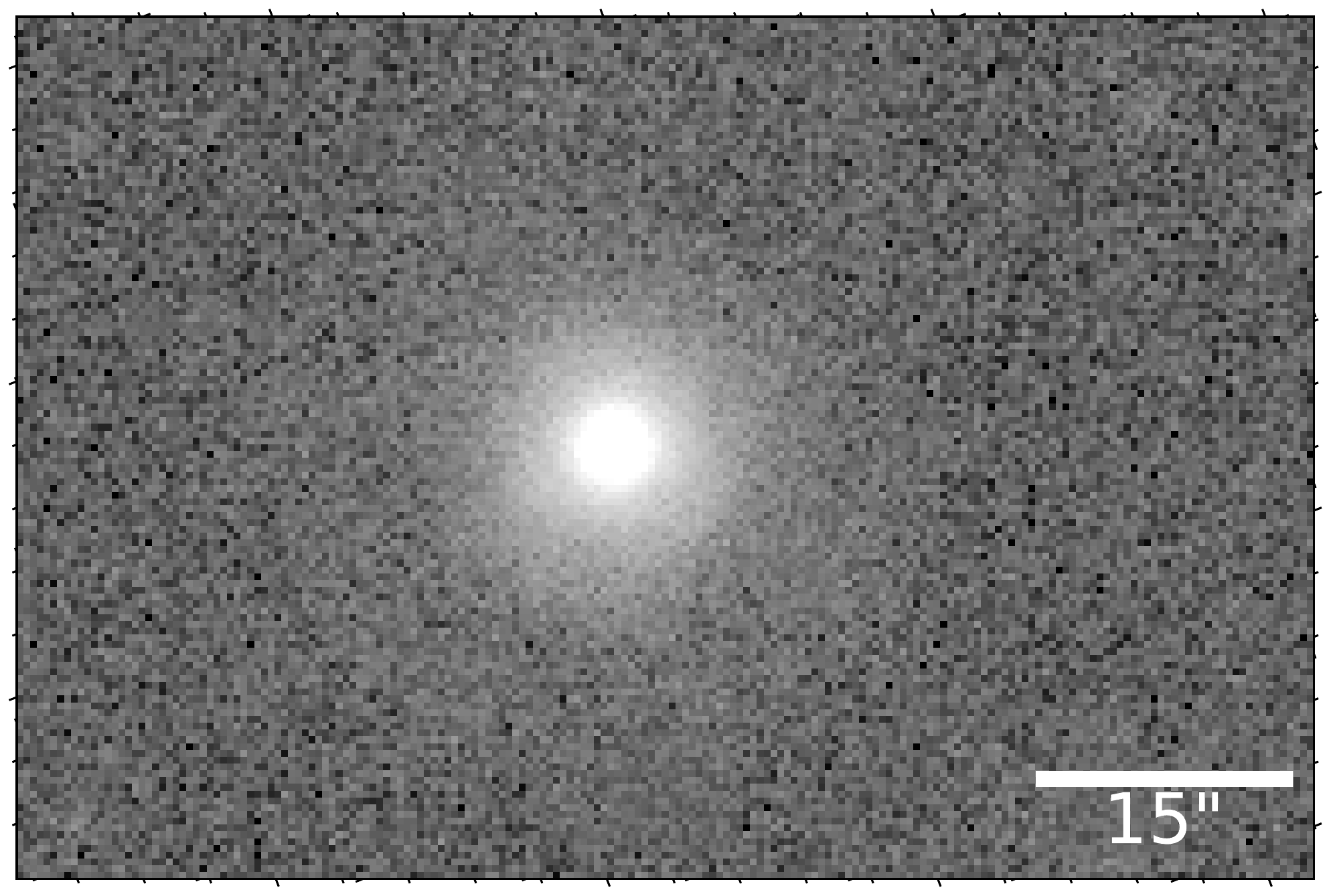}\label{217G1original}}
 \hspace{0.3cm}
\subfloat{\includegraphics[scale=0.2, bb=8 8 560 416]{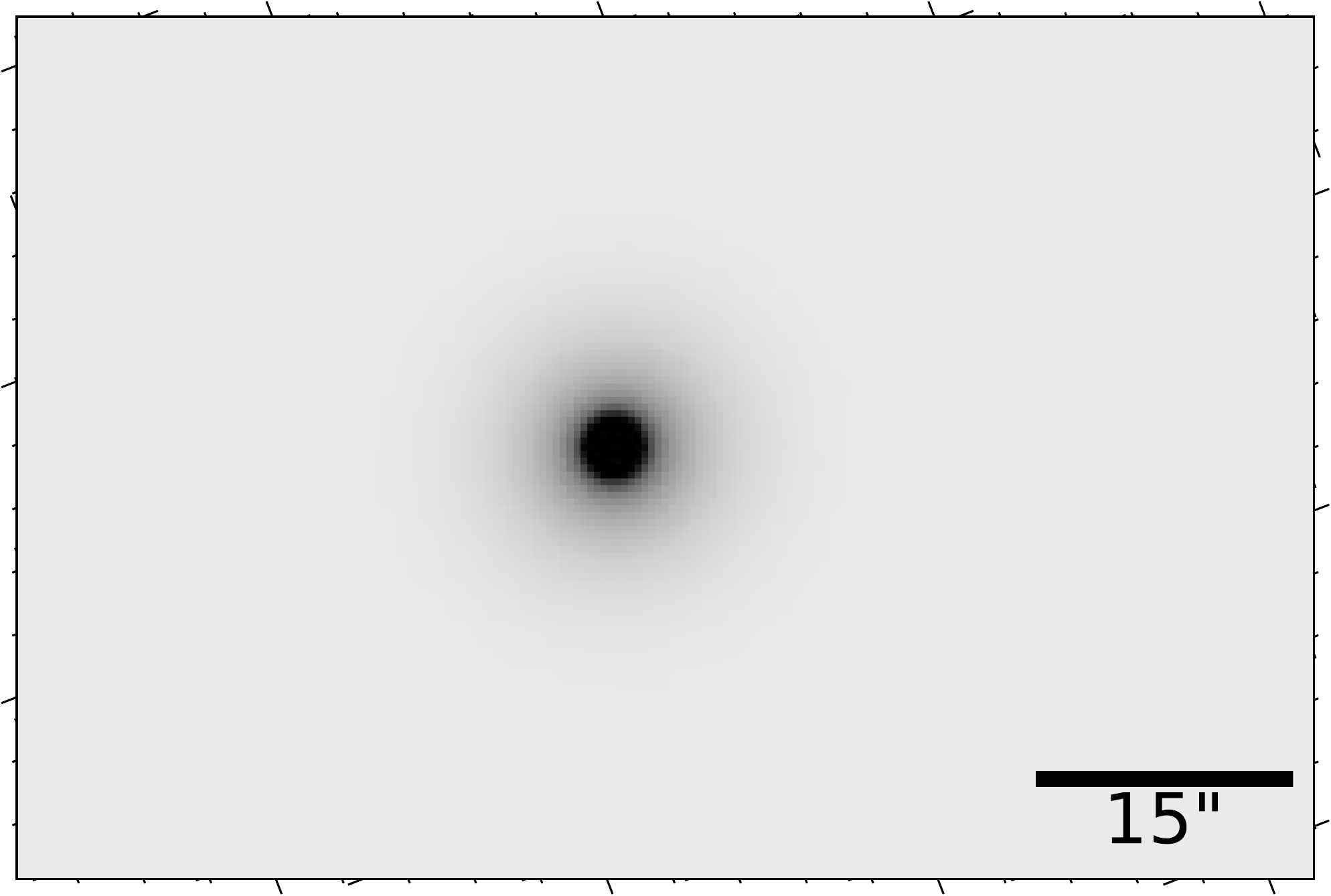} \label{217G1resid1}}
\hspace{0.3cm}
\subfloat{\includegraphics[scale=0.2, bb=8 8 560 416]{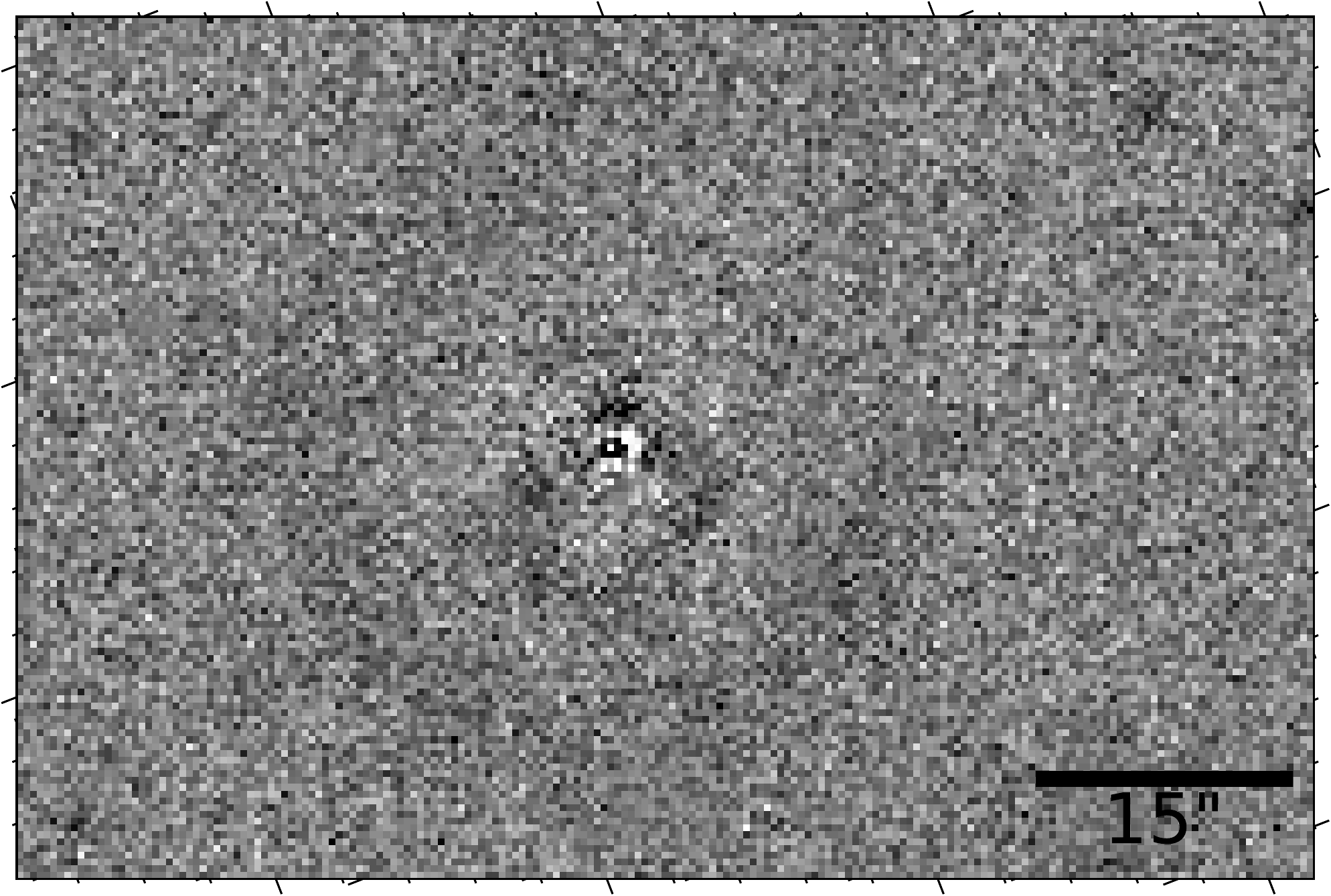} \label{217G1resid2}}\\
\subfloat{\includegraphics[scale=0.2, bb=8 8 560 416]{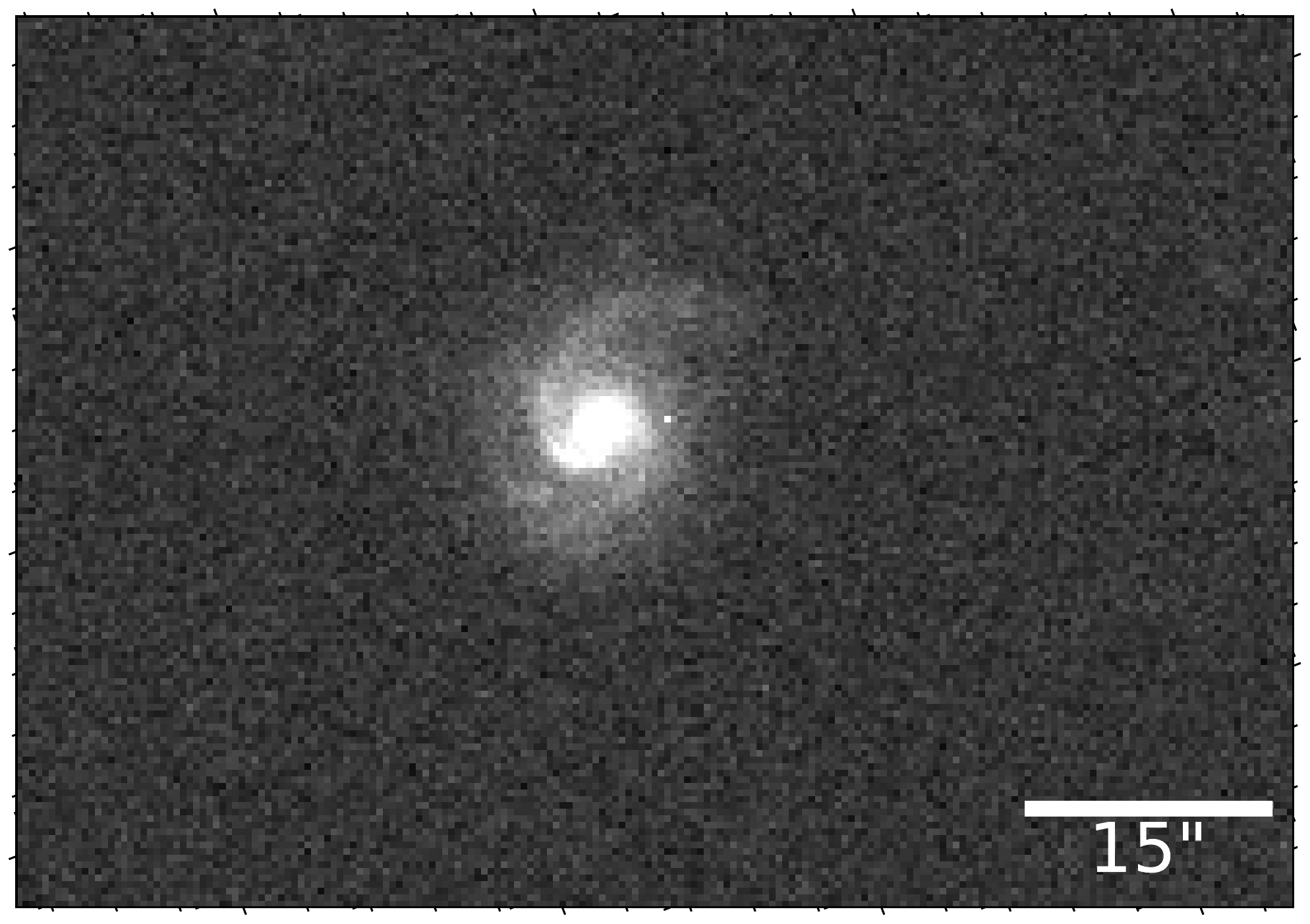}\label{217G2original}}
 \hspace{0.3cm}
\subfloat{\includegraphics[scale=0.2, bb=8 8 560 416]{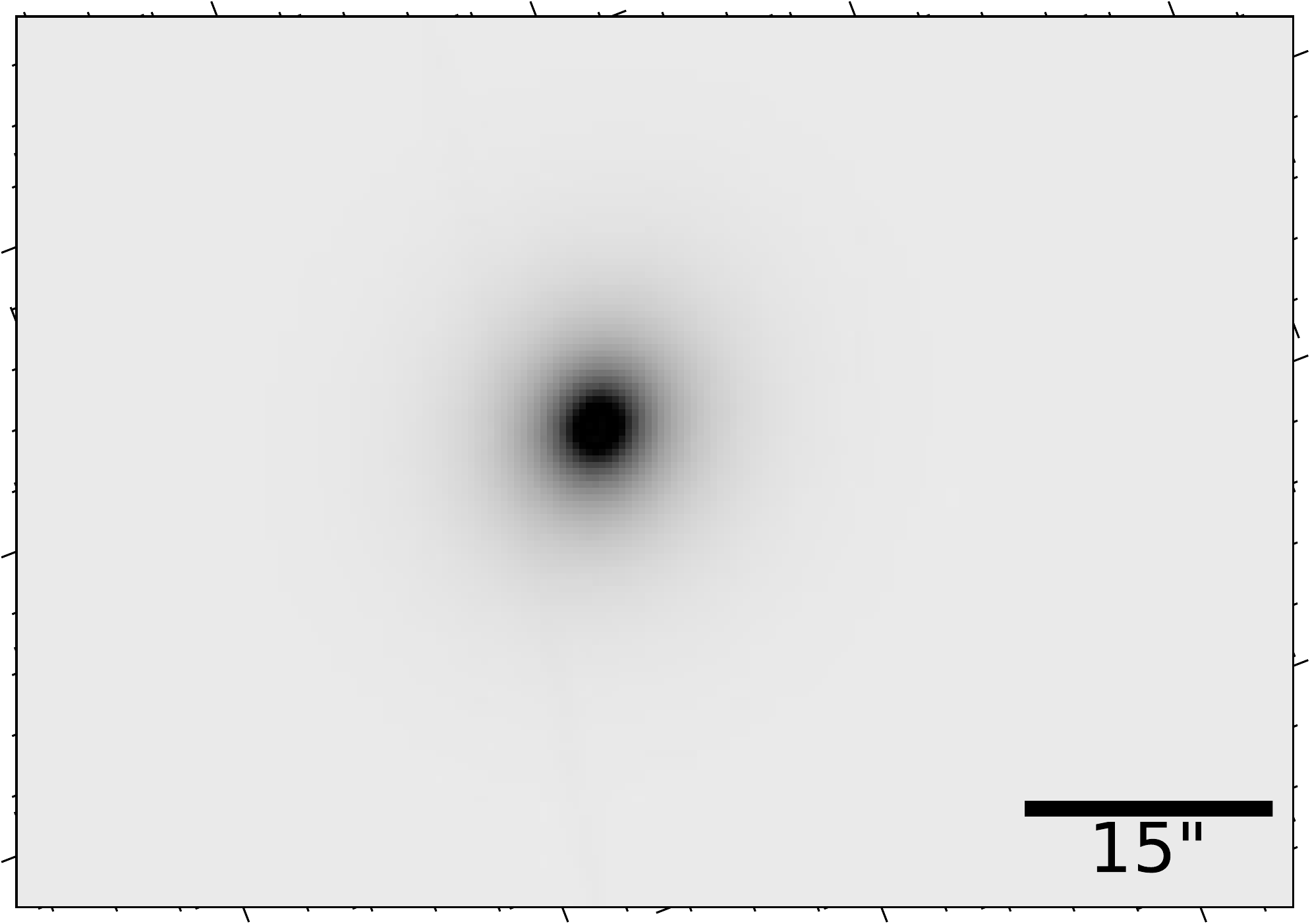} \label{217G2resid1}}
\hspace{0.3cm}
\subfloat{\includegraphics[scale=0.2, bb=8 8 560 416]{Figures_pdf/217G2_residue2-eps-converted-to.pdf} \label{217G2resid2}}\\
\subfloat{\includegraphics[scale=0.2, bb=8 8 560 416]{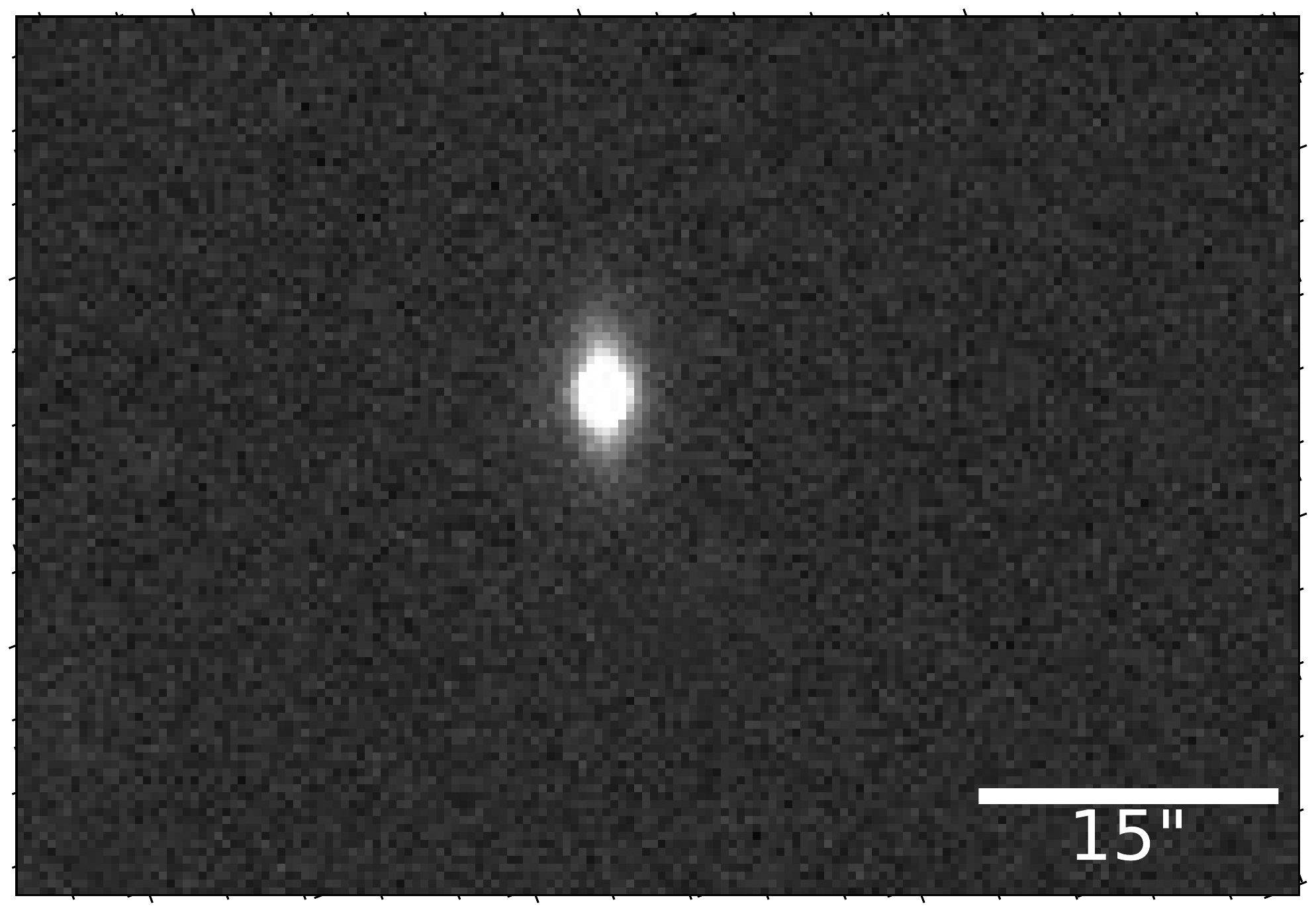}\label{217G3original}}
\hspace{0.3cm}
\subfloat{\includegraphics[scale=0.2, bb=8 8 560 416]{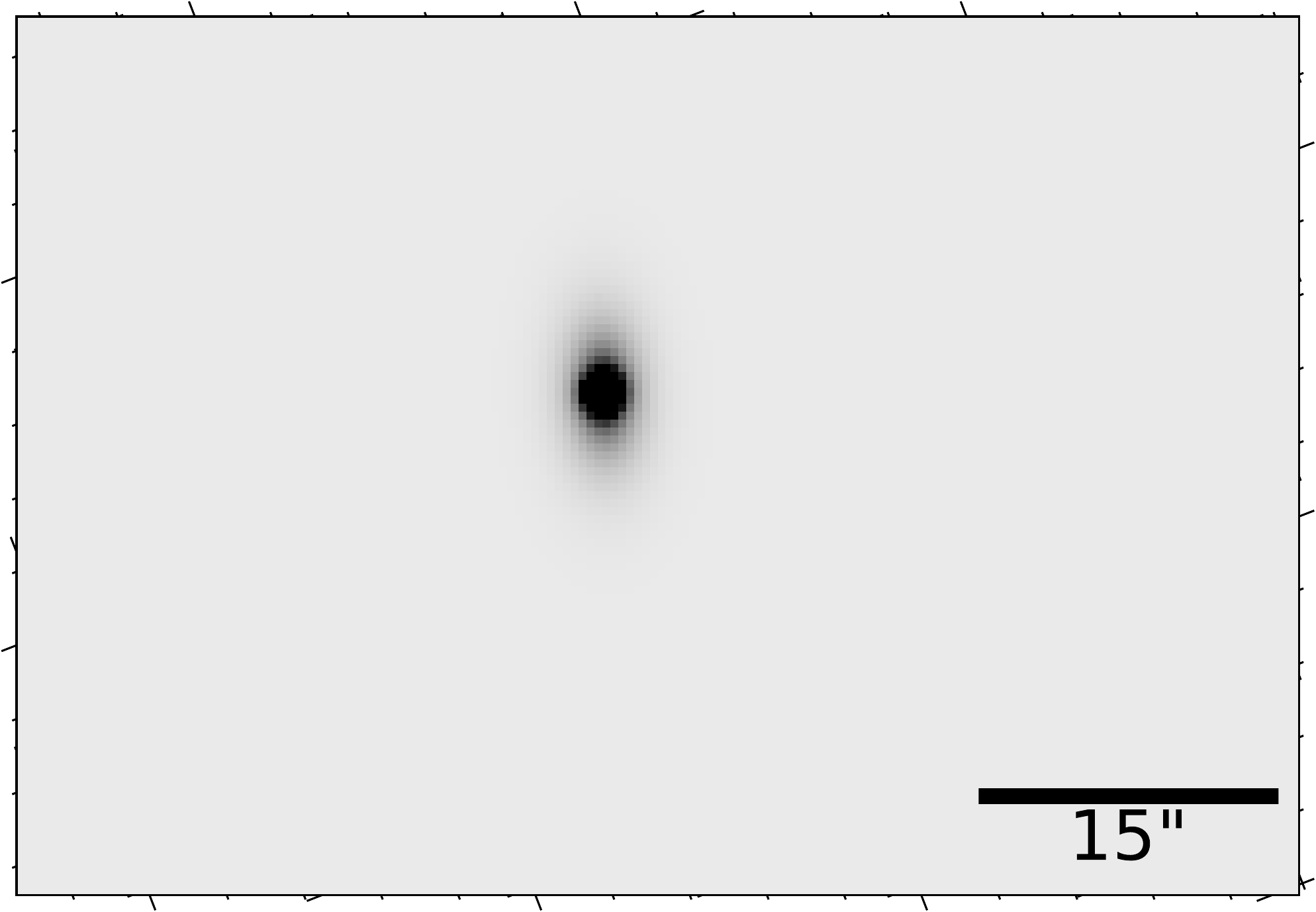} \label{217G3resid1}}
\hspace{0.3cm}
\subfloat{\includegraphics[scale=0.2, bb=8 8 560 416]{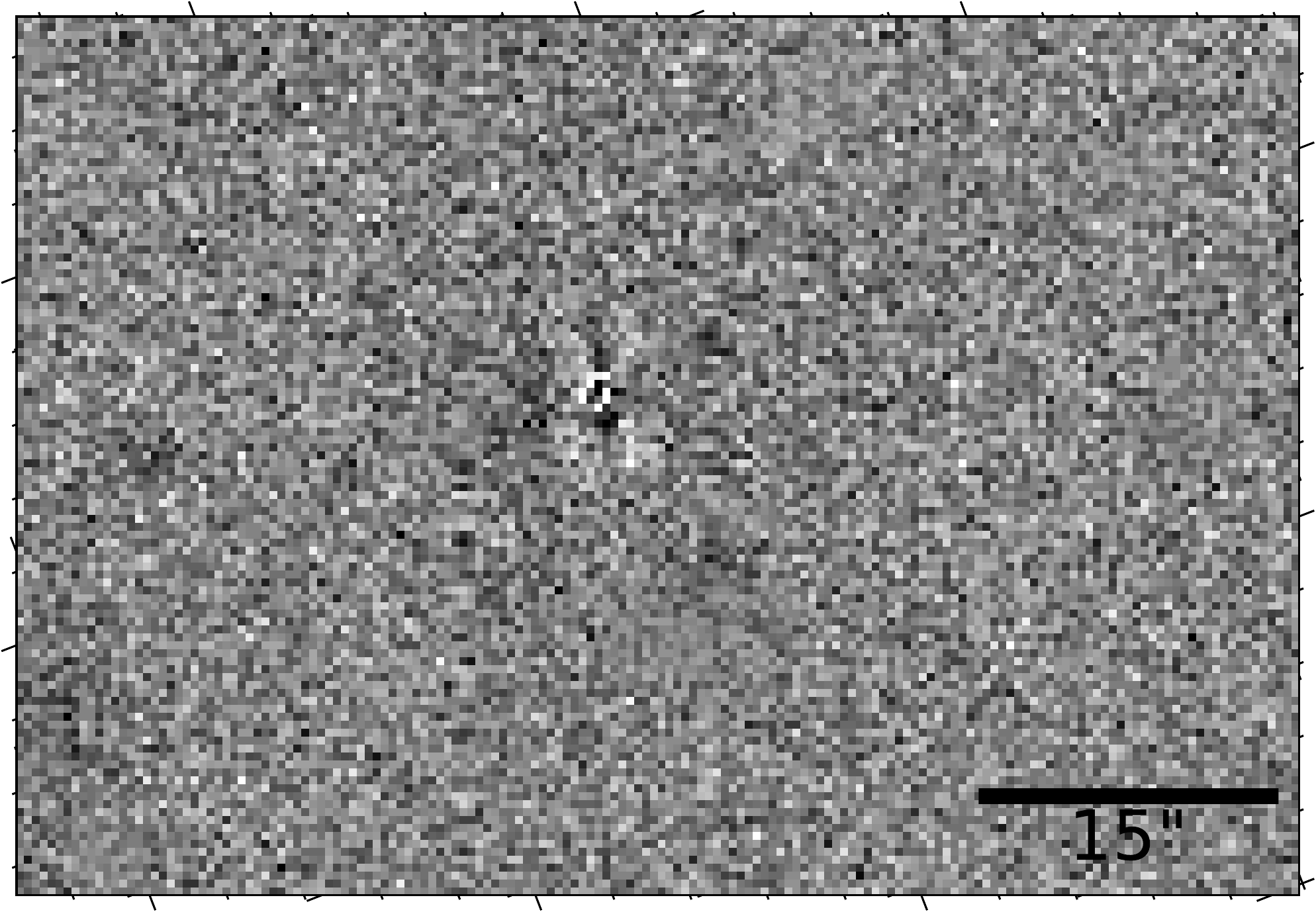} \label{217G3resid2}}\\

\caption{Photometric analysis of the triplet system SIT 217. See caption Fig.~\ref{fig:30analysis}}
\label{fig:217analysis}
\end{figure*}

This triplet system is composed of two elliptical galaxies (\textbf{G1} and \textbf{G3}) and one Sc galaxy (\textbf{G2}) \citep{Dobrycheva2013, Willett2013}. The optical image of \textbf{G1} presents a simple elliptical galaxy with a prominent bulge. A set of two anti-clockwise fuzzy spiral arms enclosing the core of \textbf{G2} are clearly observed. The northern arm appears somewhat fragmented and extended apart from the entire galaxy, while the southern arm seems fainter and bifurcated. \textbf{G3} has an apparent eye-shape enclosing a prominent bulge. This raises the possibility to categorize \textbf{G3} as a S0 or as an early type spiral (Figure~\ref{fig:SIT217}).

Figure~\ref{217_SB} clarifies the intensity profile of \textbf{G1} fitted by a single $Ser$ component with no breaks, while the middle panel of Figure~\ref{217_SB} presents the complicated profile of \textbf{G2} with multiple breaks. An extended excess hump, representing a bar, between 2$'' <$ SMA $<$ 3.6$''$ is well-defined.At SMA $<$ 9.5$''$, the intensity profile starts to bend downward causing a truncation in the exponential profile representing a type II.o. An SBc classification might be more appropriate for this galaxy.
The intensity profile of \textbf{G3} (Figure~\ref{217_SB}, right panel) presents type III-d profile with an anti-truncation at SMA $=$ 4.3$''$.  


Figures ~\ref{fig:217analysis} illustrates the decomposition of SIT 217. Using $Ser$ plus $Exp$ model for \textbf{G1} and \textbf{G2} reveals a low diffuse positive halo at the central region of both galaxies and an asymmetric arm in \textbf{G2} with $\chi^2_{\nu}$ of 0.9 and 1.17, respectively, (Figures~\ref{217G1resid2} and ~\ref{217G2resid2}).
The decomposition of \textbf{G3} using $Ser$ plus $Exp$ model fits nearly all its components  and a $\chi^2_{\nu}$ of 1.08 is obtained (Figure~\ref{217G3resid2}). The prominence of the bulge component from the 1D and 2D fitting match better an early-type spiral, more than the E/S0 initial classification.

\subsection{The triplet system SIT 263}   
 
\begin{figure*}
\centering
\subfloat{\includegraphics[scale=0.5, bb=130 -150 100 100]{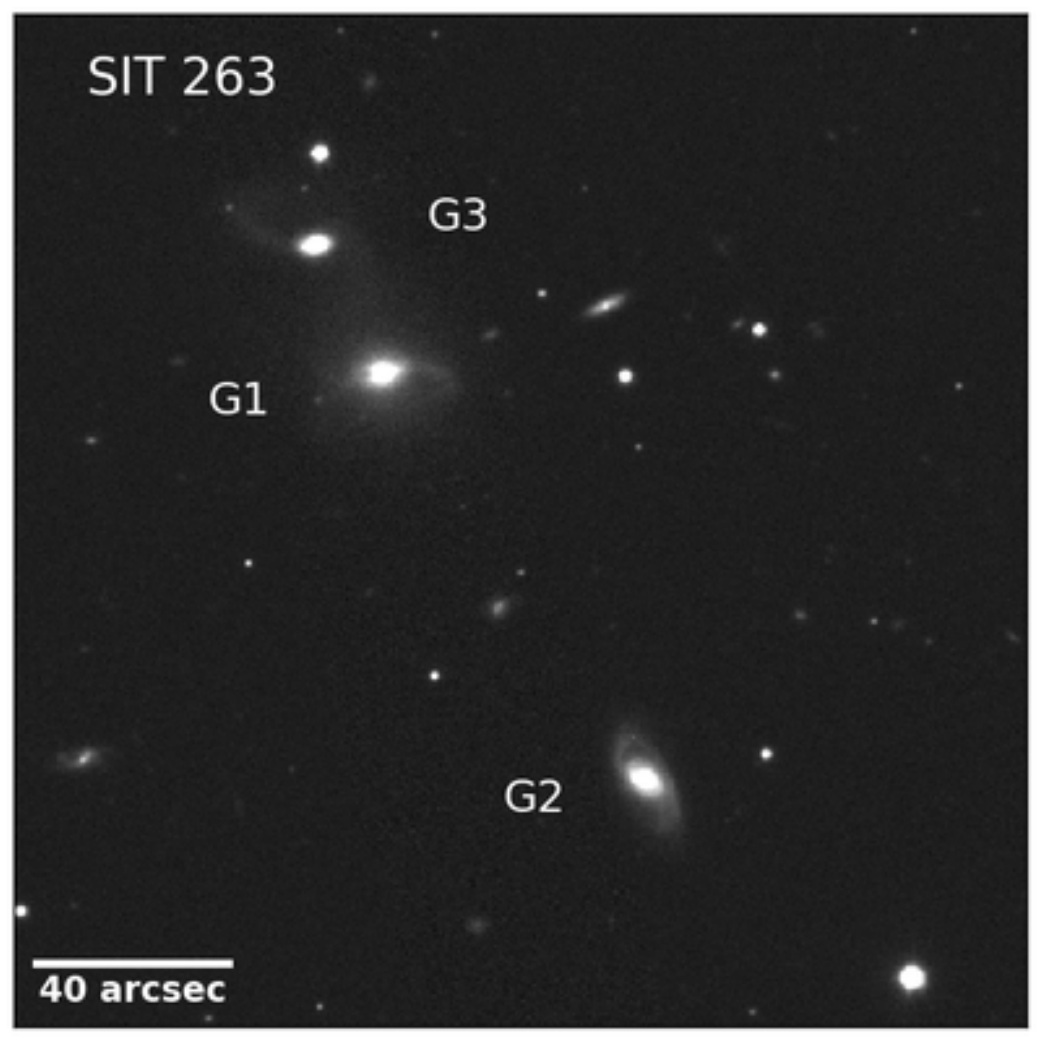}
 \label{fig:SIT263}}\\
\subfloat{\includegraphics[scale=0.7, bb=80 20 500 100]{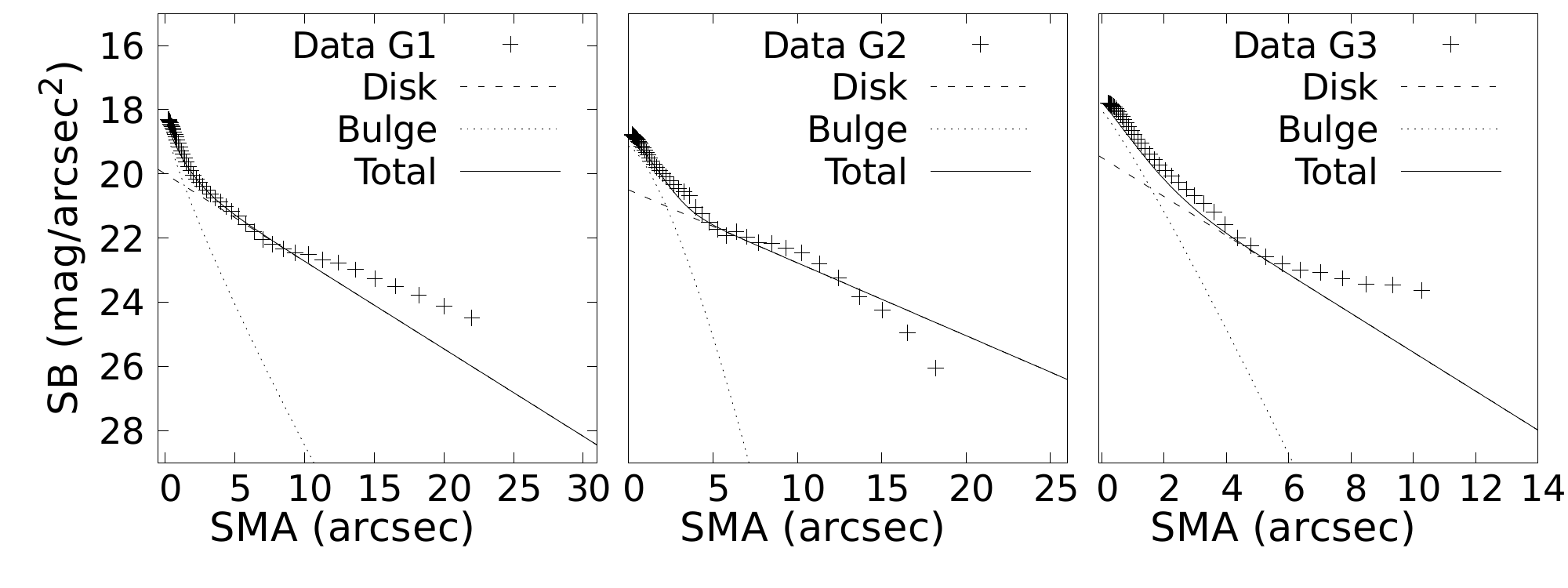}
\label{263_SB}}\\
\subfloat{\includegraphics[scale=0.2, bb=8 8 560 416]{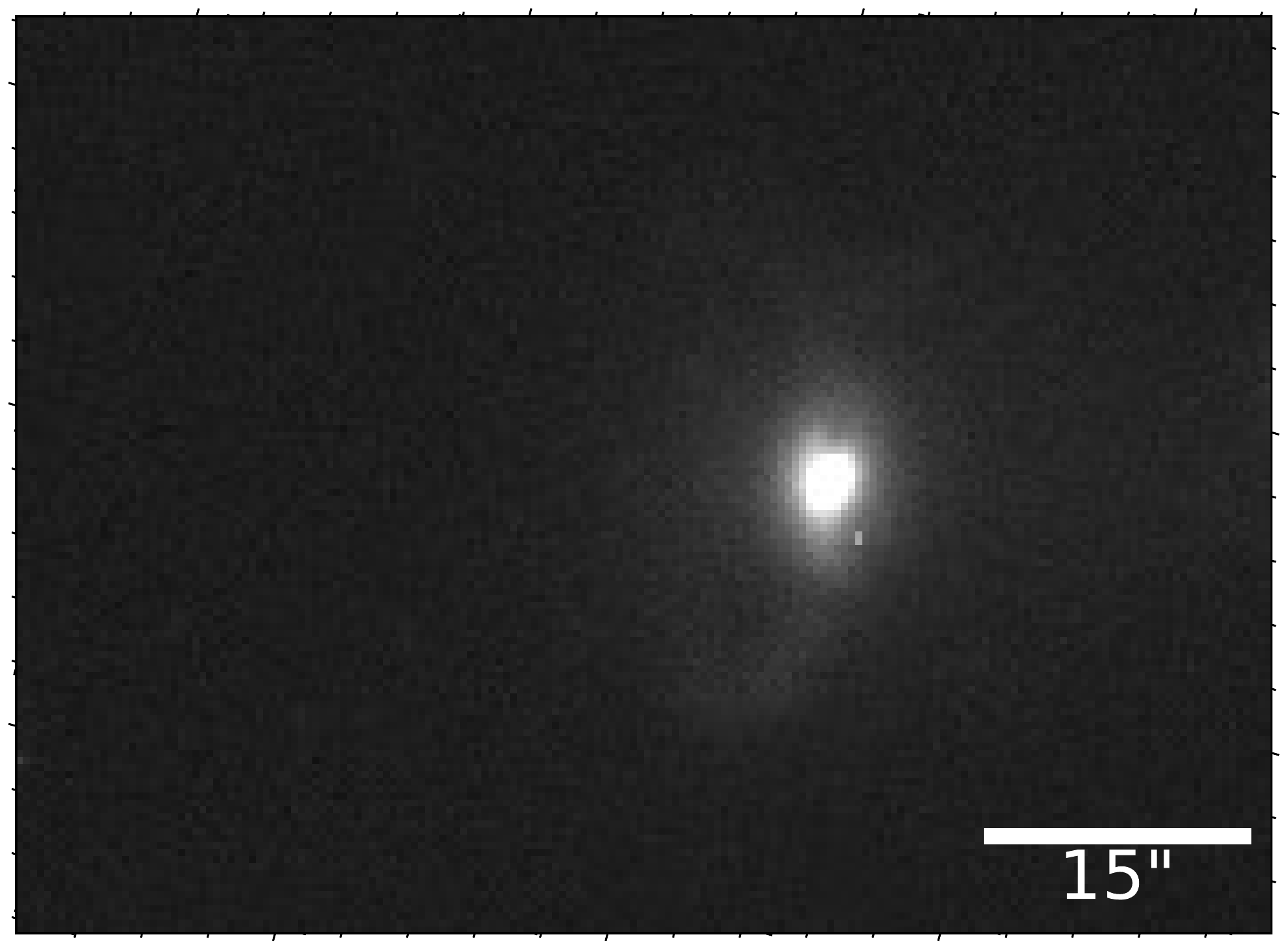}\label{263G1original}}
 \hspace{0.3cm}
\subfloat{\includegraphics[scale=0.2, bb=8 8 560 416]{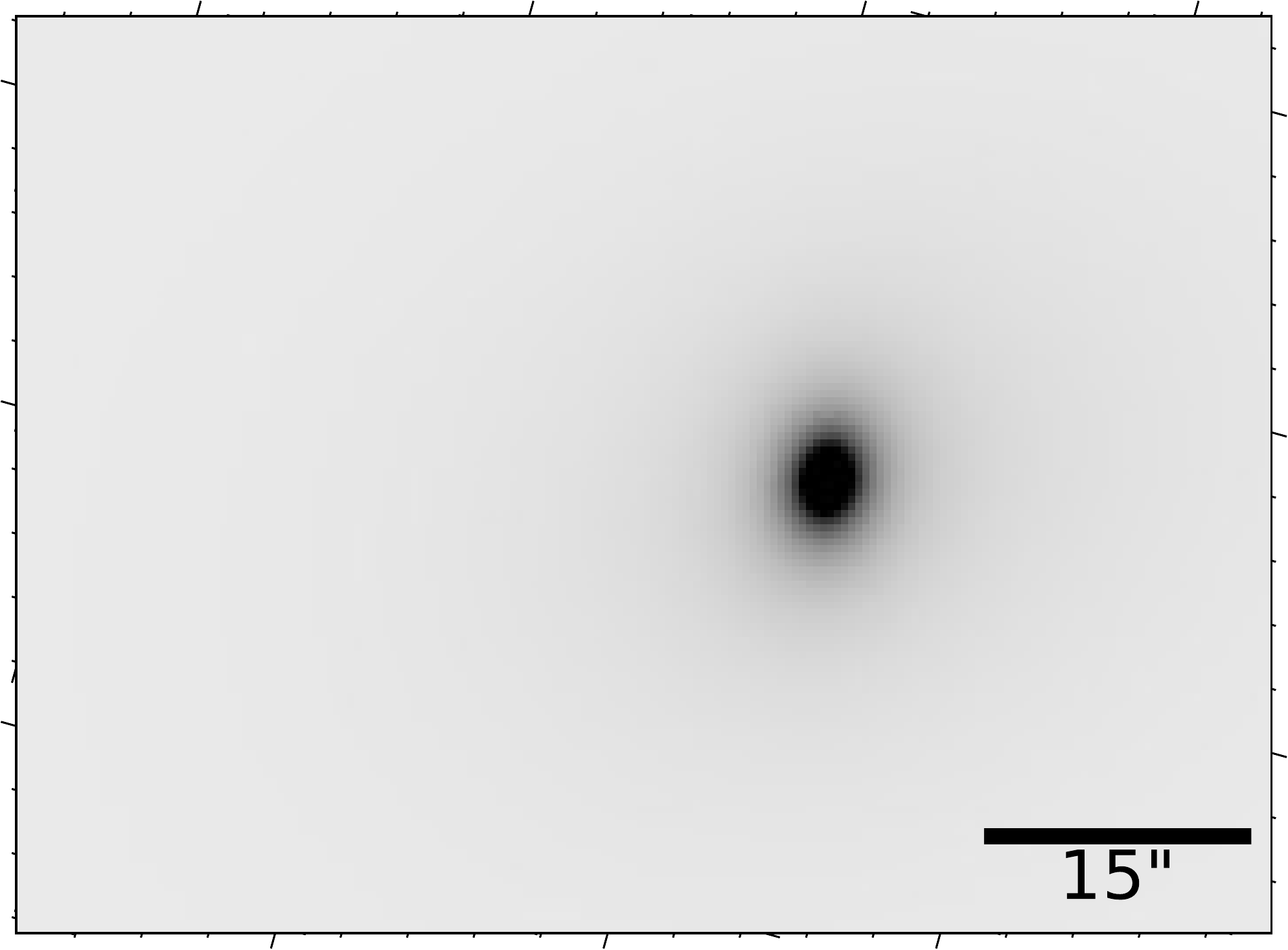} \label{263G1resid1}}
\hspace{0.3cm}
\subfloat{\includegraphics[scale=0.2, bb=8 8 560 416]{Figures_pdf/263G1_residue2-eps-converted-to.pdf} \label{263G1resid2}}\\
\subfloat{\includegraphics[scale=0.2, bb=4 4 570 266]{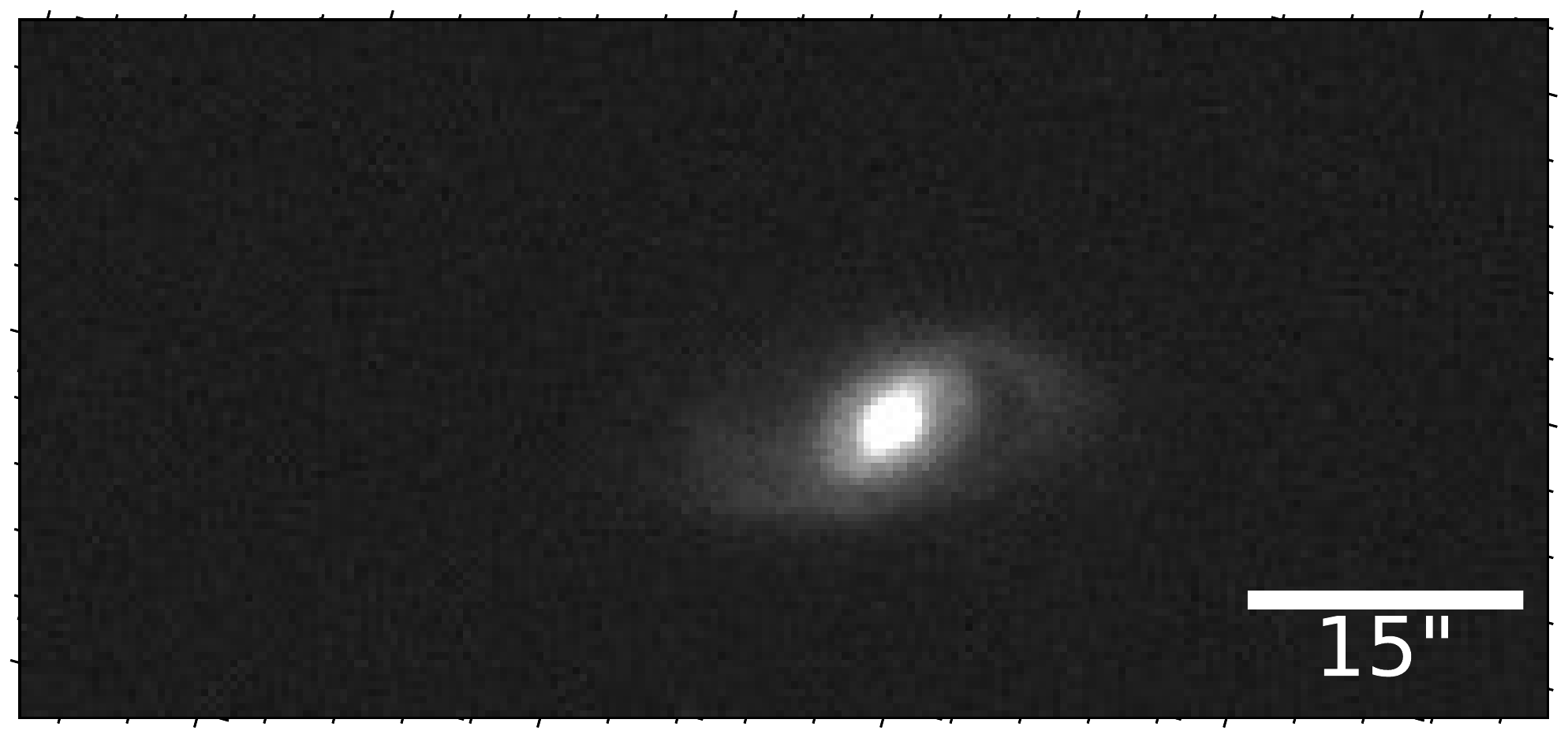}\label{263G2original}}
 \hspace{0.3cm}
\subfloat{\includegraphics[scale=0.2, bb=4 4 570 266]{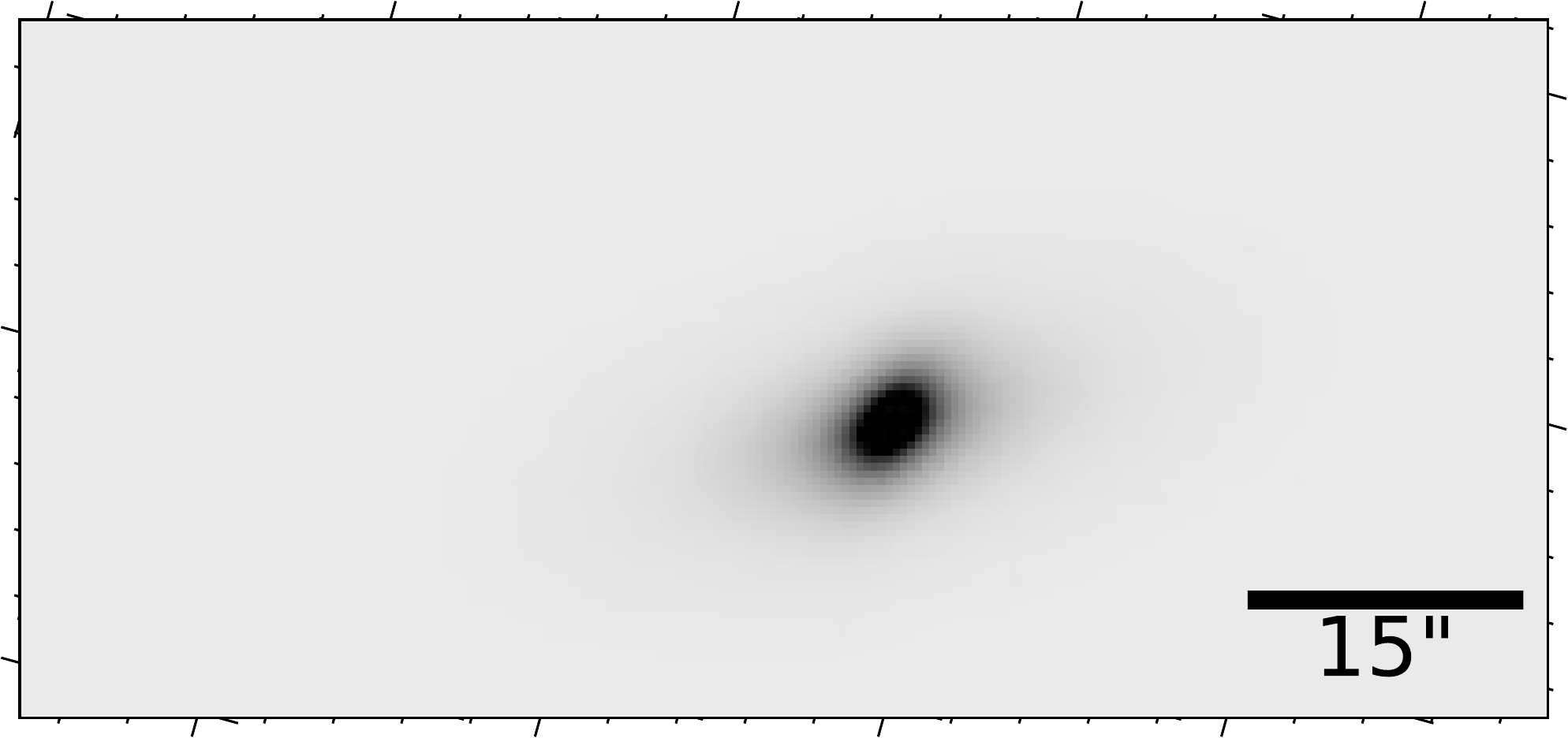} \label{263G2resid1}}
 \hspace{0.3cm}
\subfloat{\includegraphics[scale=0.2, bb=4 4 570 266]{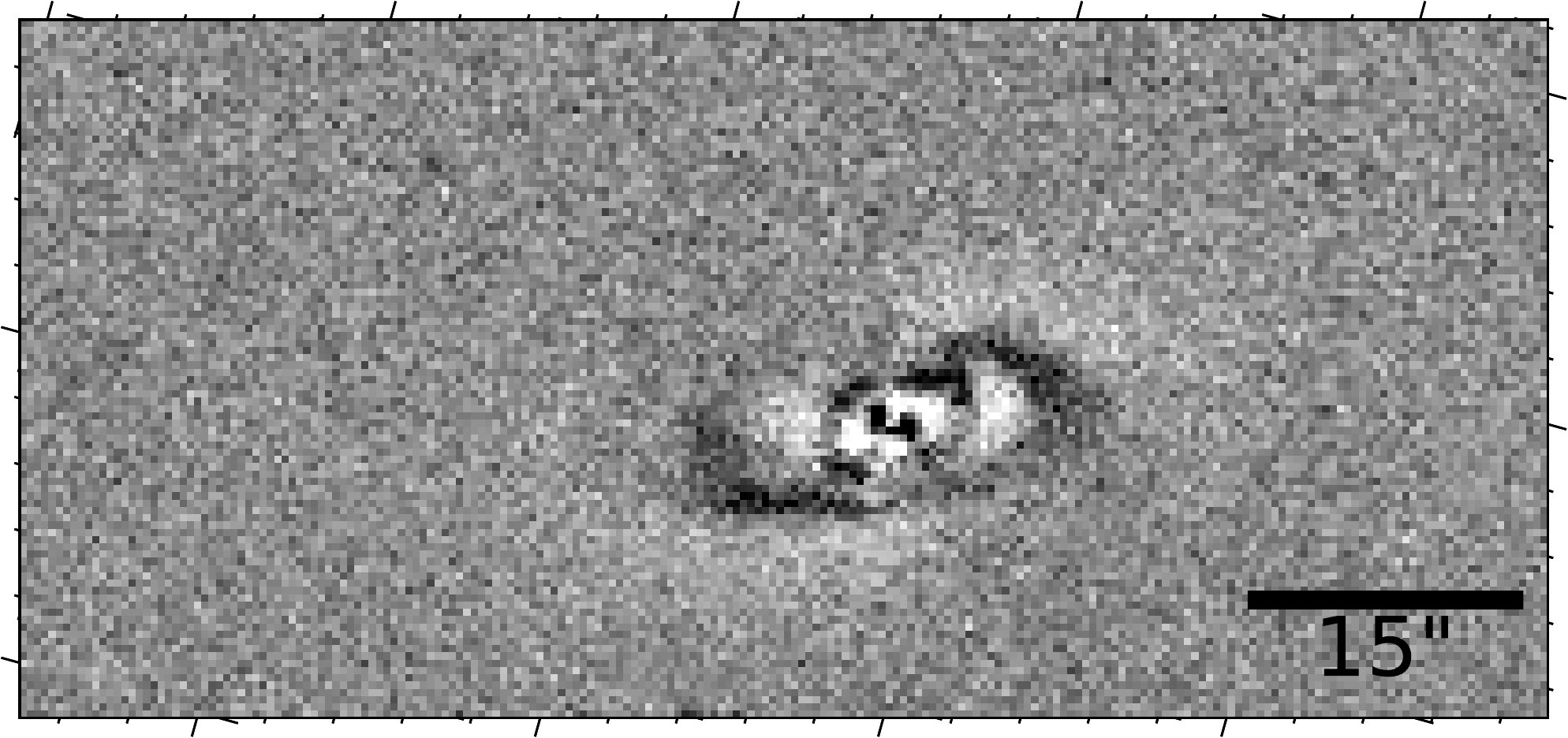} \label{263G2resid2}}\\
\subfloat{\includegraphics[scale=0.2, bb=6 6 570 488]{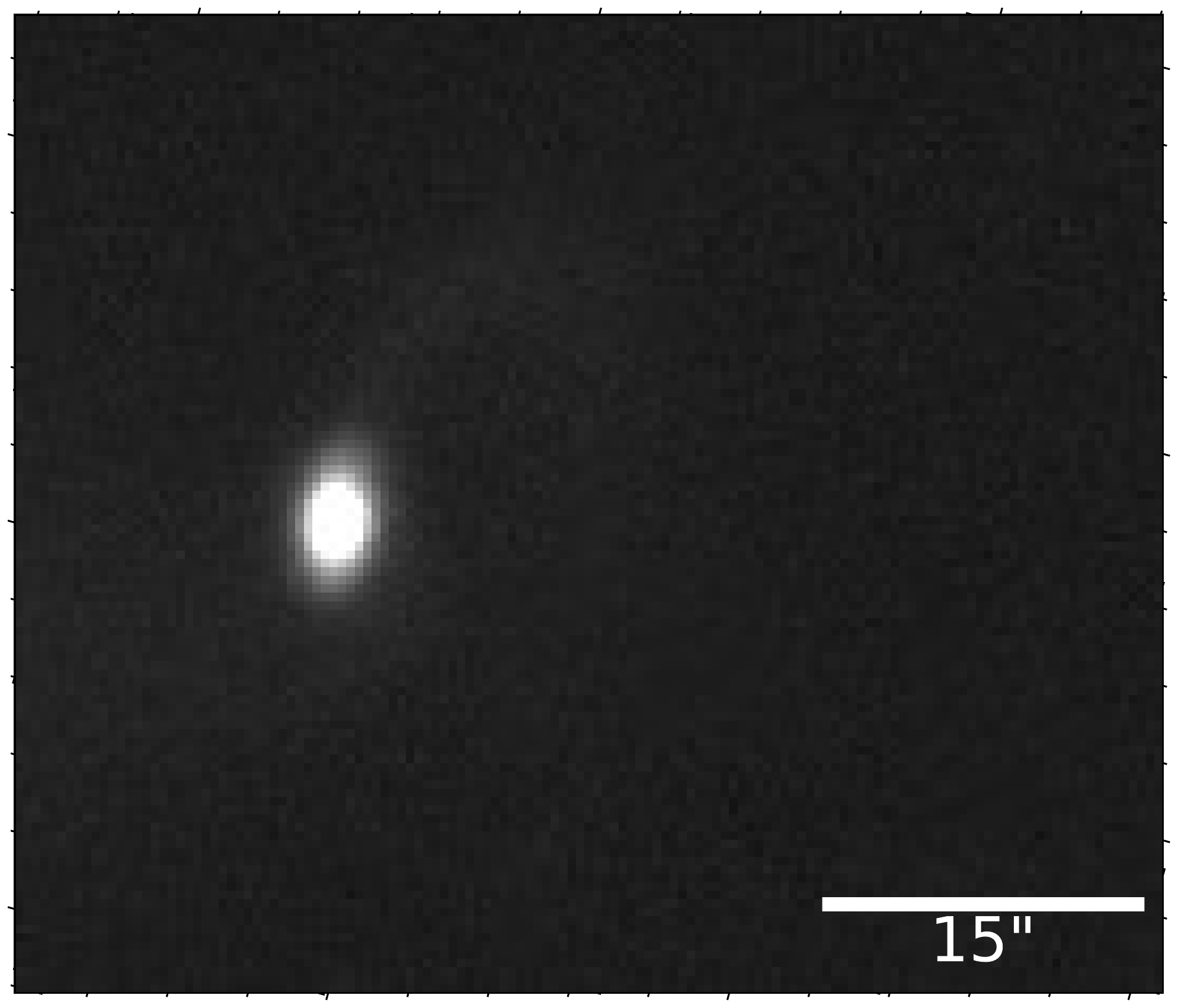}\label{263G3original}}
\hspace{0.3cm}
\subfloat{\includegraphics[scale=0.2, bb=6 6 570 488]{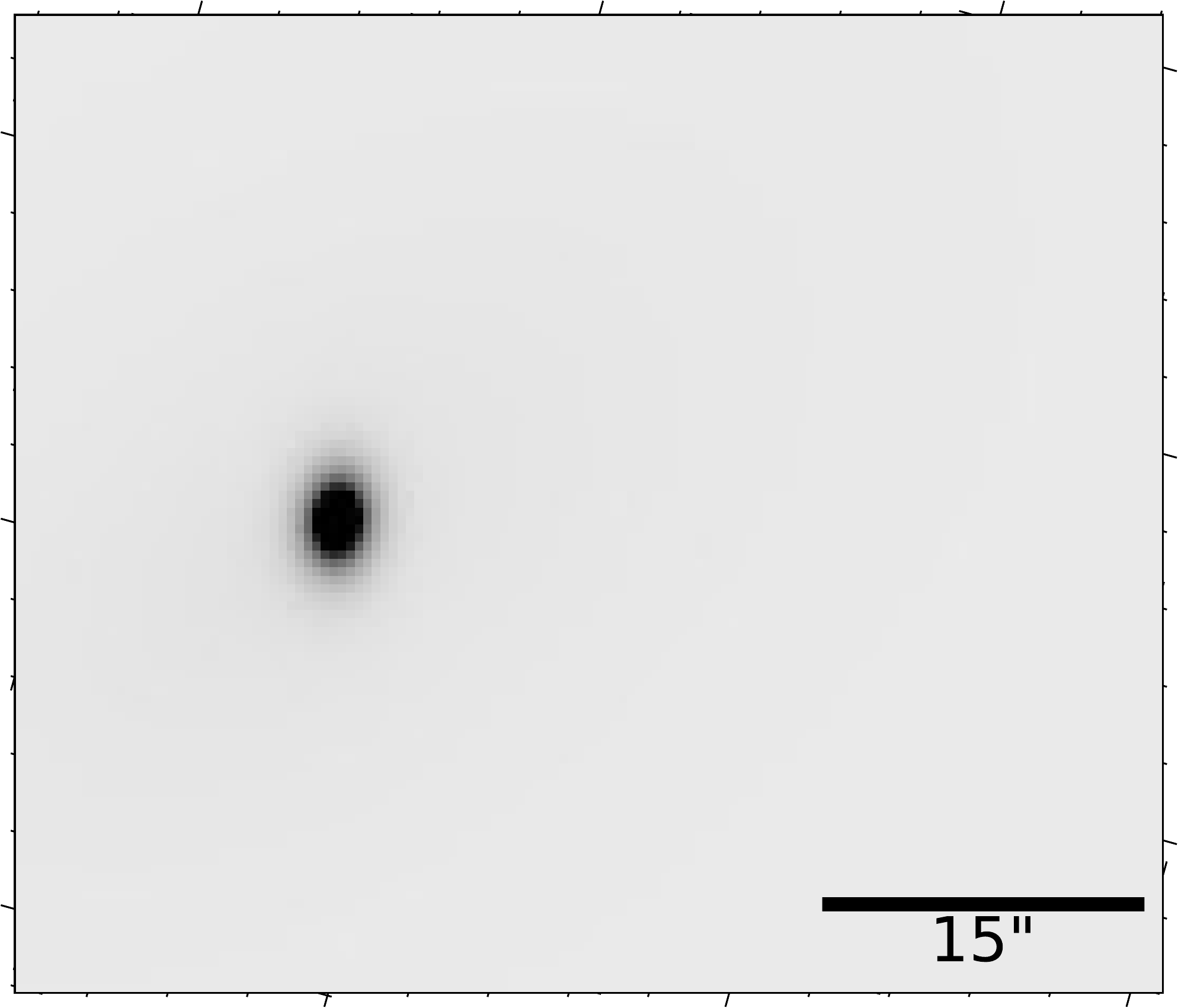} \label{263G3resid1}}
\hspace{0.3cm}
\subfloat{\includegraphics[scale=0.2, bb=6 6 570 488]{Figures_pdf/263G3_residue2-eps-converted-to.pdf} \label{263G3resid2}}

\caption{Photometric analysis of the triplet system SIT 263. See caption Fig.~\ref{fig:30analysis}}
\label{fig:263analysis}
\end{figure*}

The optical image of SIT 263 (Figure~\ref{fig:SIT263}) represents the most complicated triplet system in our sample. It comprises three luminous spiral galaxies, in which \textbf{G1} and \textbf{G3} show a strong evidence of interaction or a merger state evinced by the bridge between them with high bifurcated asymmetric spiral arms emerging from the center of both galaxies. \textbf{G2} is a spiral galaxy with a set of two tightly closed spiral arms and a cuspy core in the center. It is located at a distance of 1.61 arcmin from \textbf{G1}. 

The intensity profile of \textbf{G1} shows an anti-truncation of type III-d at SMA $\sim$ 9.5$''$ (Figure~\ref{263_SB}, left panel). The intensity profile of \textbf{G2} shows an inner break at SMA $\sim$ 6.5$''$ of type III profile (Figure~\ref{263_SB}, middle panel), similarly, the profile of \textbf{G3} is no doubt, considered as type III-d profile with a break at SMA $\sim$ 5.8$''$ (Figure~\ref{263_SB}, right panel).

Decomposition of \textbf{G1} using $Ser$ + $Exp$ reveals many residuals including an off-set bar, two distorted spiral arms, and a negative diffused portion that cause the arise of a ``Butterfly-Shape'' (Figure~\ref{263G1resid2}). 

Similarly, the decomposition of \textbf{G2} by $Ser$ + $Exp$ reveals many residuals with two over-subtracted arms and two other non-subtracted arms, negative and positive discontinuity portions are also remarked in the bulge region with a $\chi^2_{\nu}$ of 1.23 (Figure~\ref{263G2resid2}).

Decomposition of \textbf{G3} using $Ser$ + $Exp$ shows a ring and two asymmetric spiral arms as residuals with $\chi^2_{\nu}$ of 0.85 (Figure~\ref{263G3resid2}).

\subsection{The triplet system SIT 264}

\begin{figure*}
\centering
\subfloat{\includegraphics[scale=0.5, bb=130 -150 100 100]{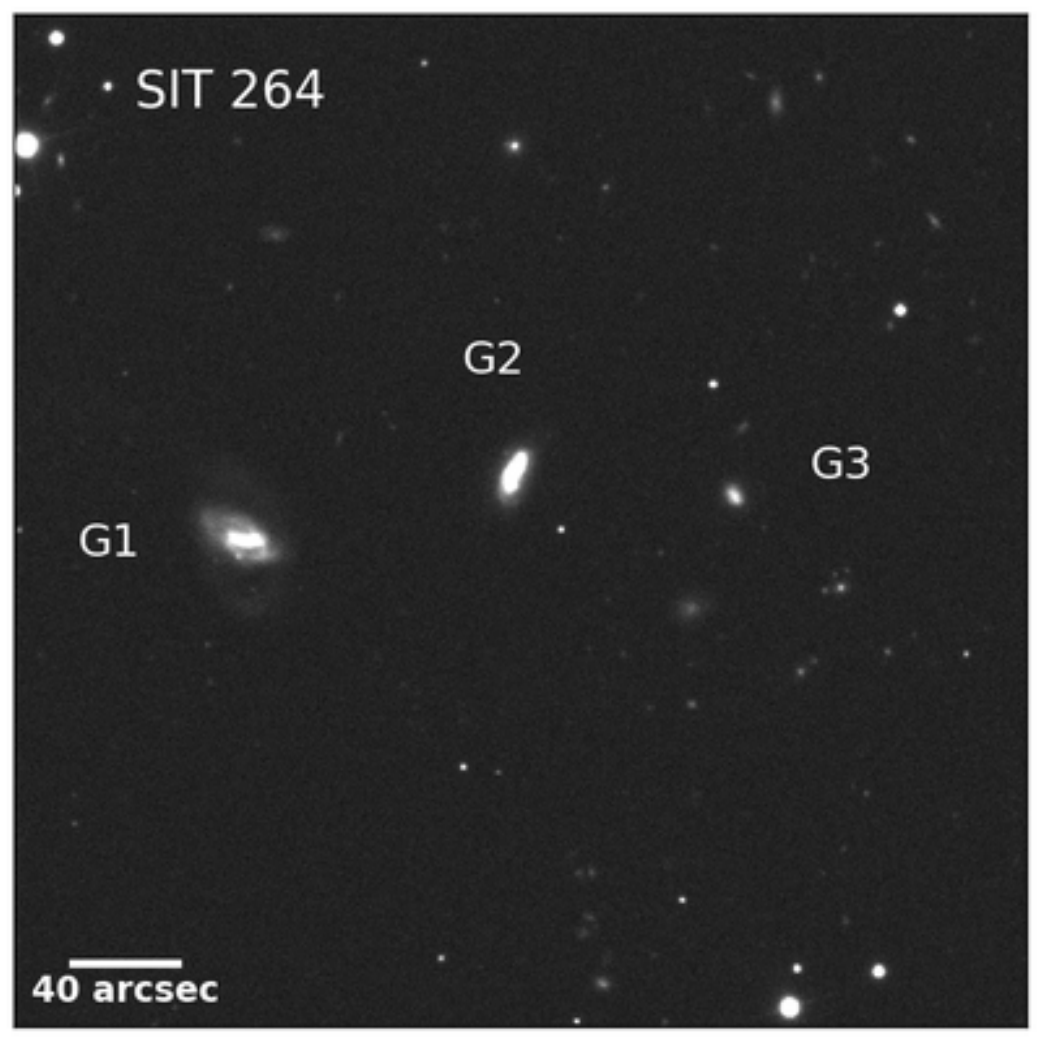}
 \label{fig:SIT264}}\\
\subfloat{\includegraphics[scale=0.7, bb=80 20 500 100]{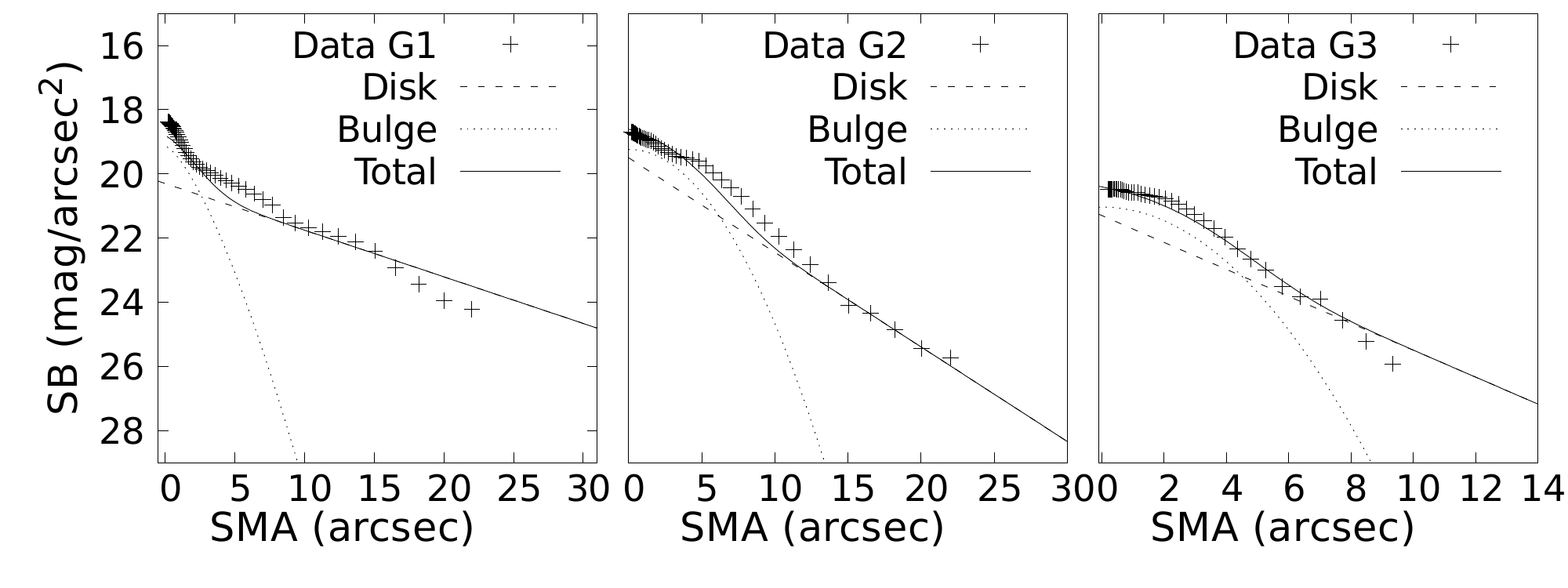}\label{264_SB}}\\
\subfloat{\includegraphics[scale=0.2, bb=8 8 560 416]{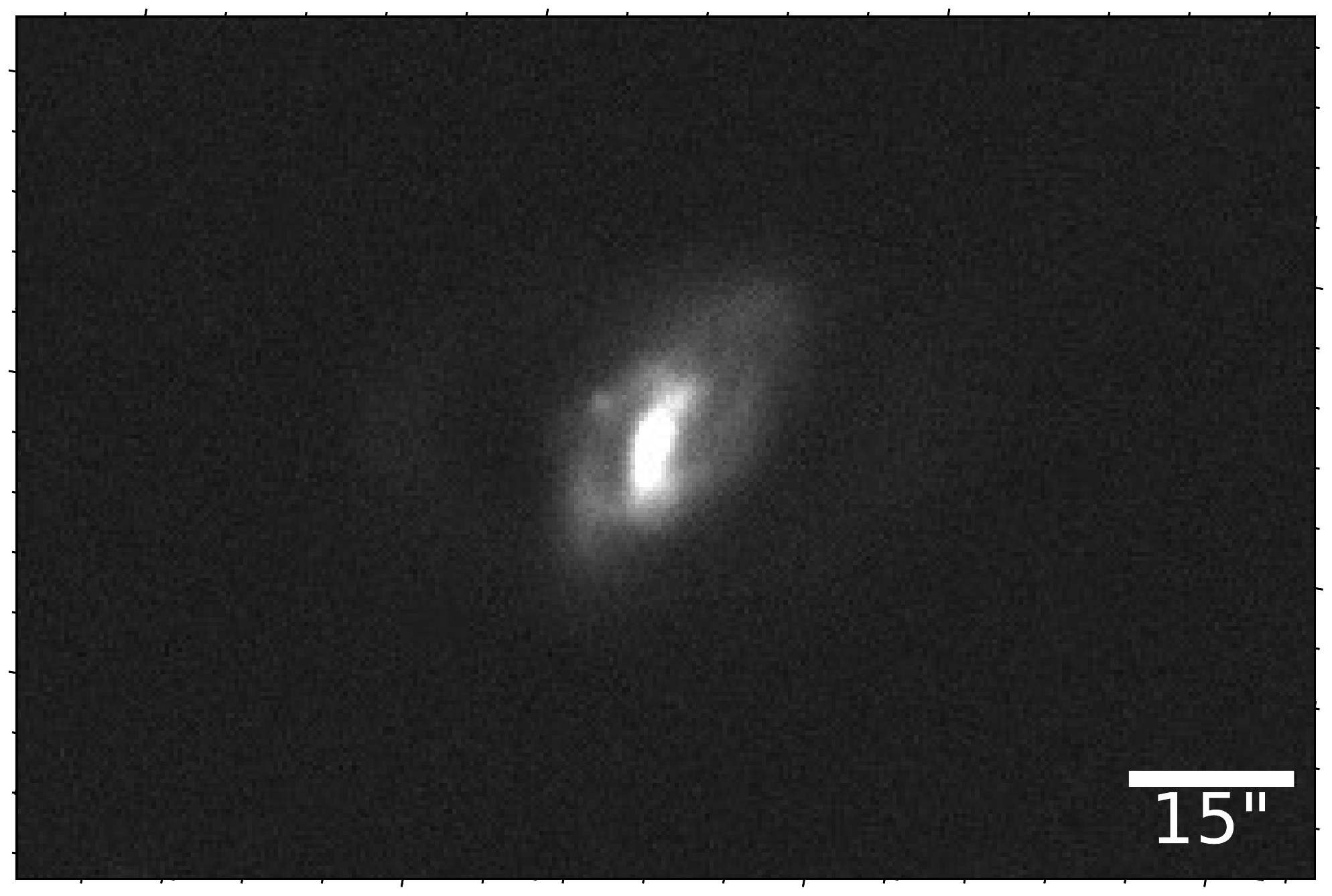}\label{264G1original}}
\hspace{0.3cm}
\subfloat{\includegraphics[scale=0.2, bb=8 8 560 416]{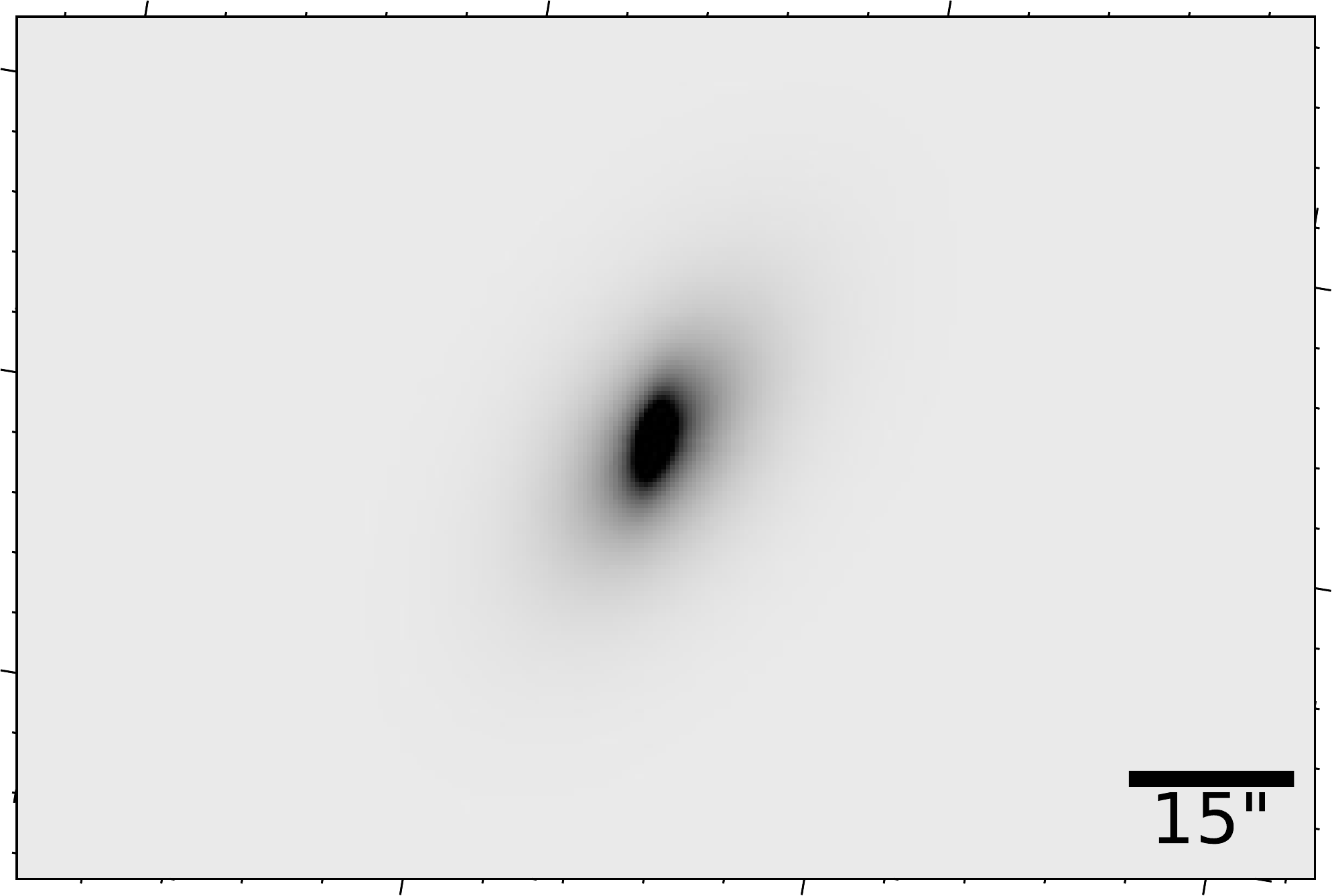} \label{264G1resid1}}
\hspace{0.3cm}
\subfloat{\includegraphics[scale=0.2, bb=8 8 560 416]{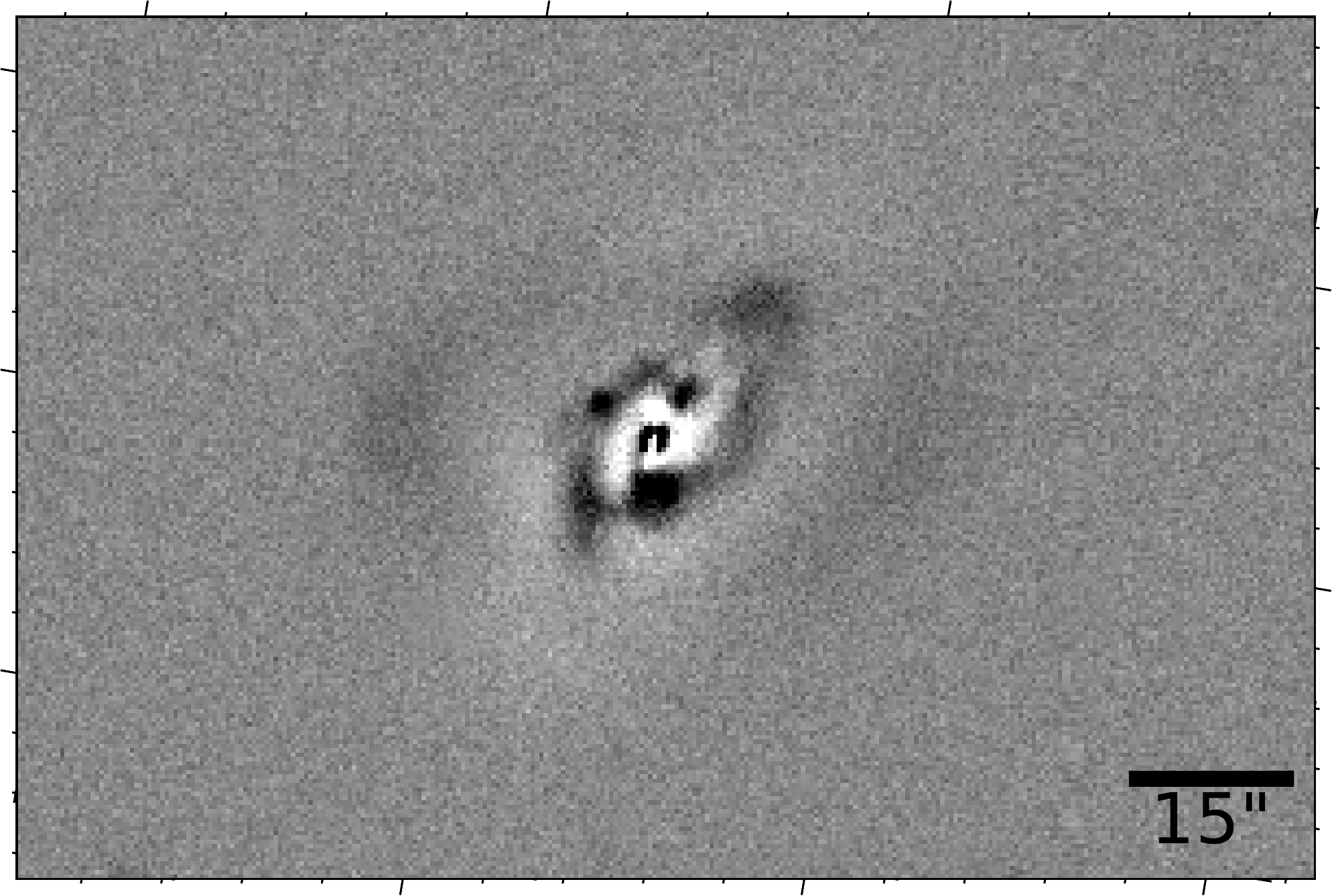} \label{264G1resid2}}\\
\subfloat{\includegraphics[scale=0.2, bb=8 8 560 416]{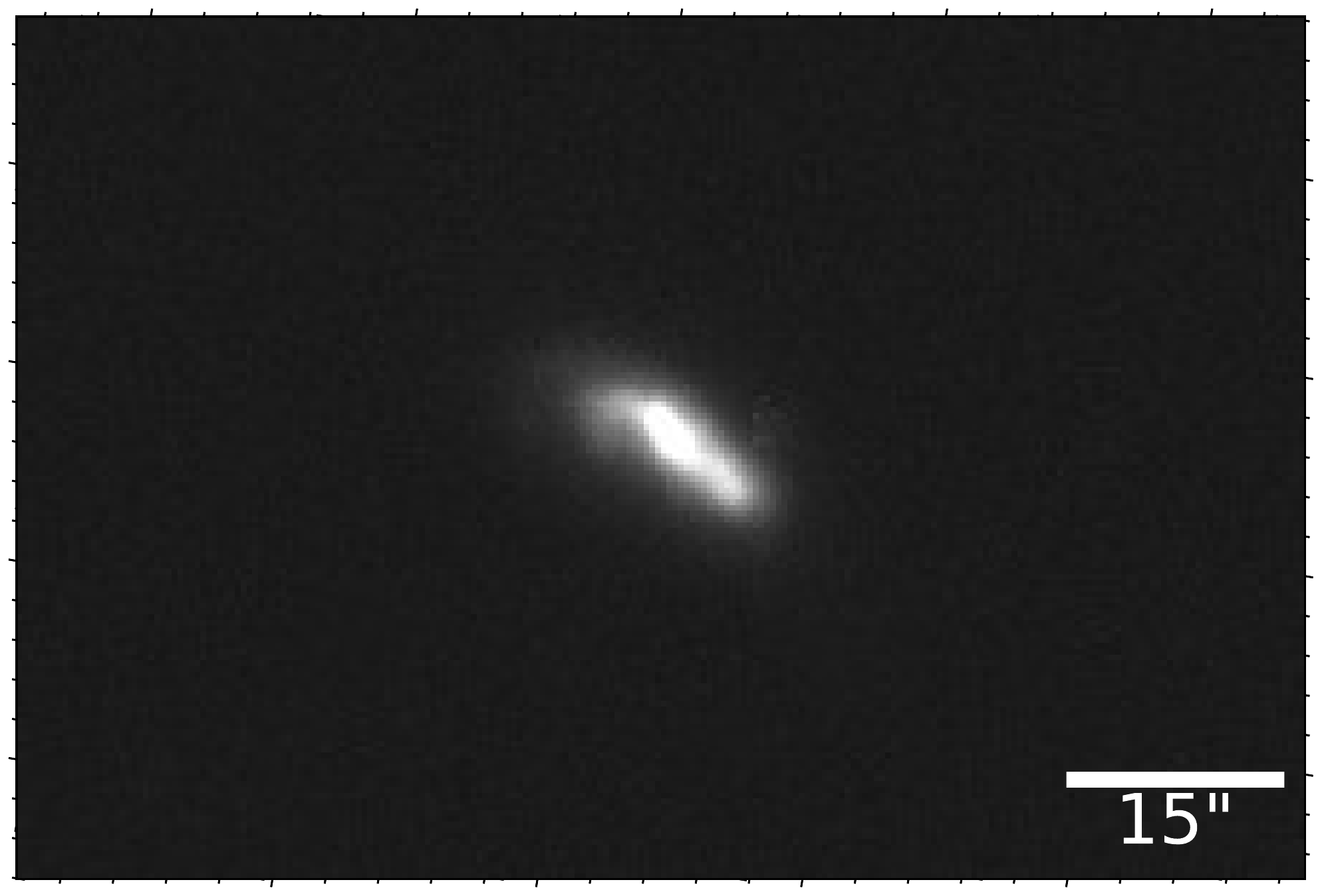}\label{264G2original}}
\hspace{0.3cm}
\subfloat{\includegraphics[scale=0.2, bb=8 8 560 416]{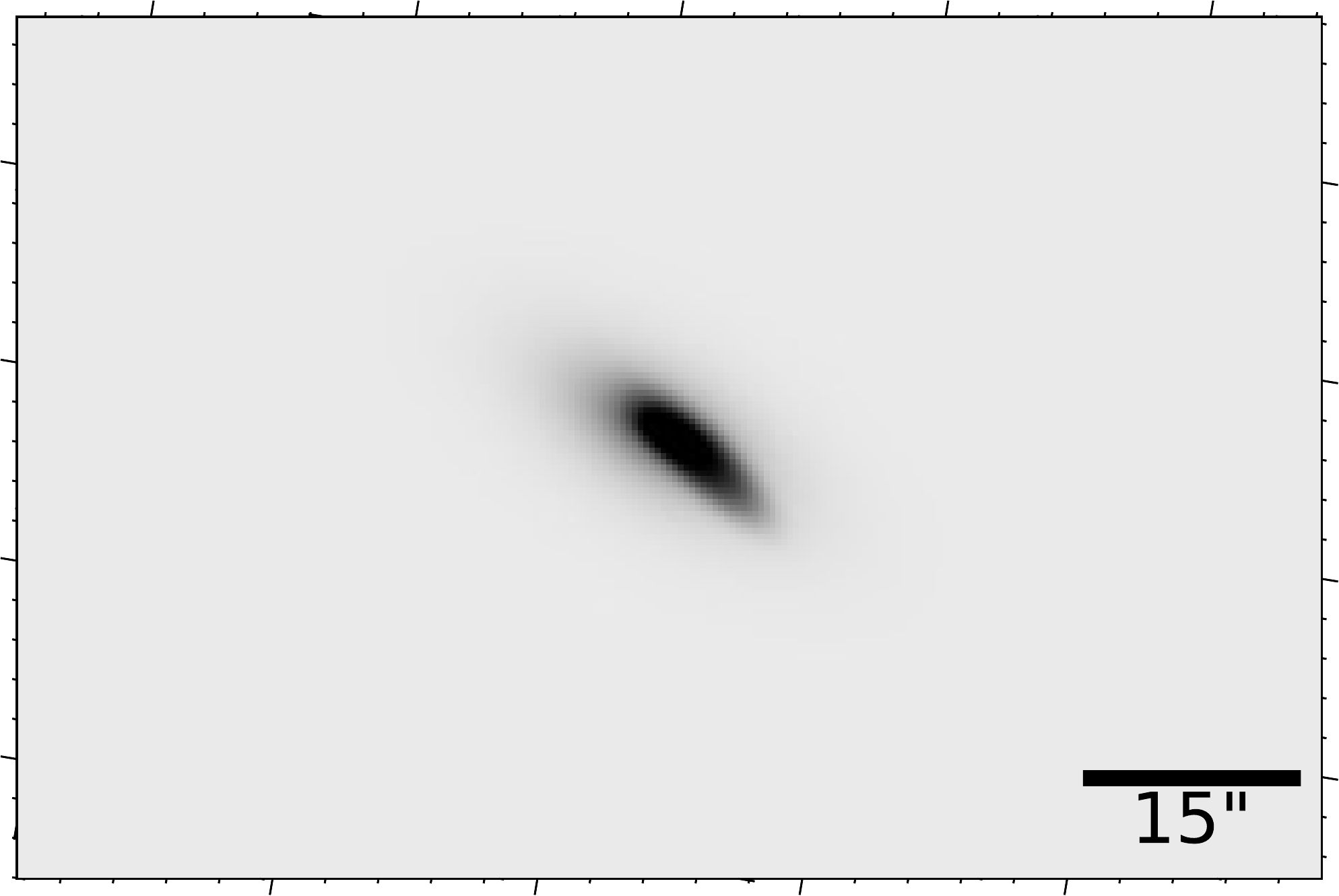} \label{264G2resid1}}
\hspace{0.3cm}
\subfloat{\includegraphics[scale=0.2, bb=8 8 560 416]{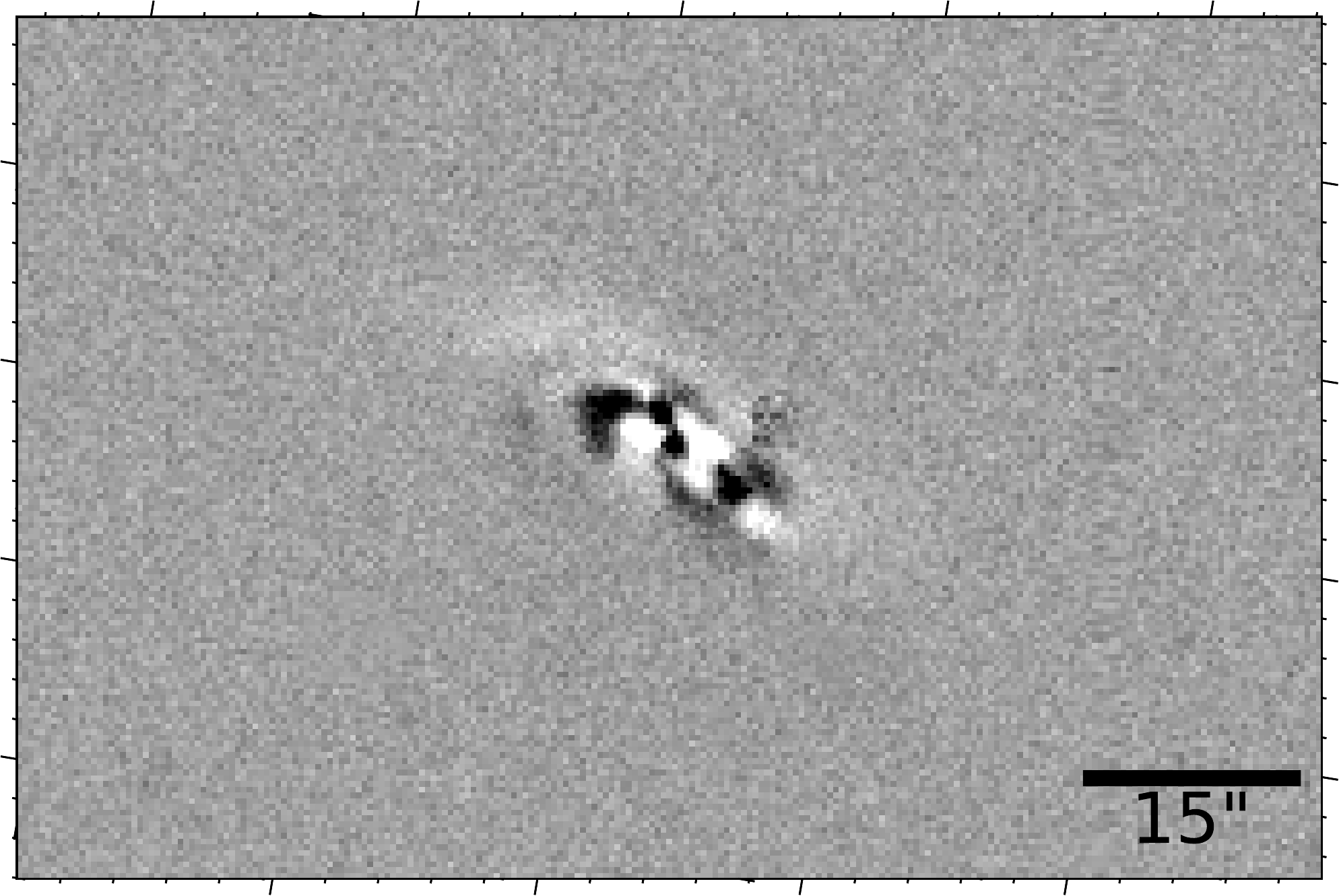} \label{264G2resid2}}
\hspace{0.3cm}\\
\subfloat{\includegraphics[scale=0.2, bb=8 8 560 416]{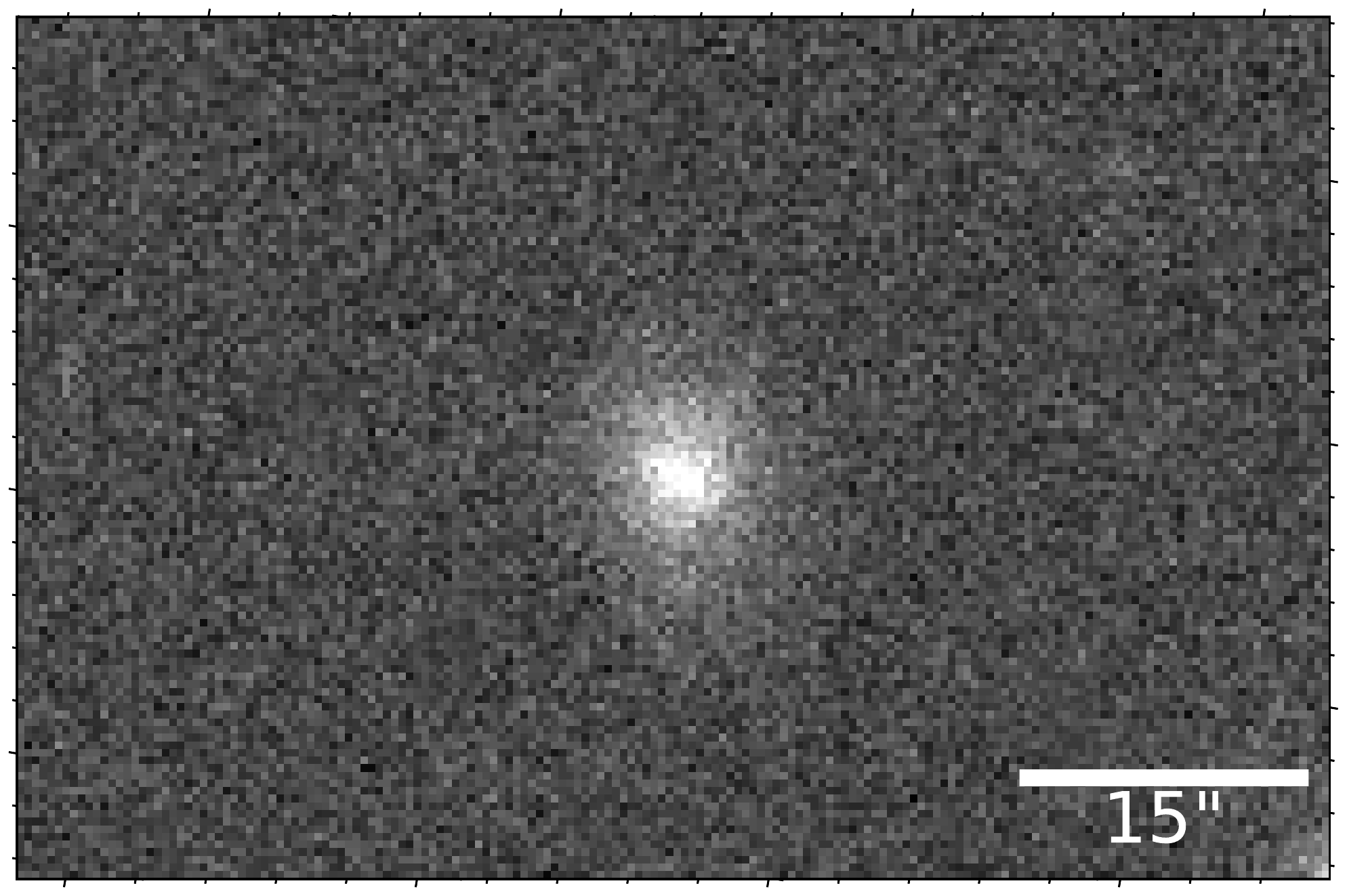}\label{264G3original}}
\hspace{0.3cm}
\subfloat{\includegraphics[scale=0.2, bb=8 8 560 416]{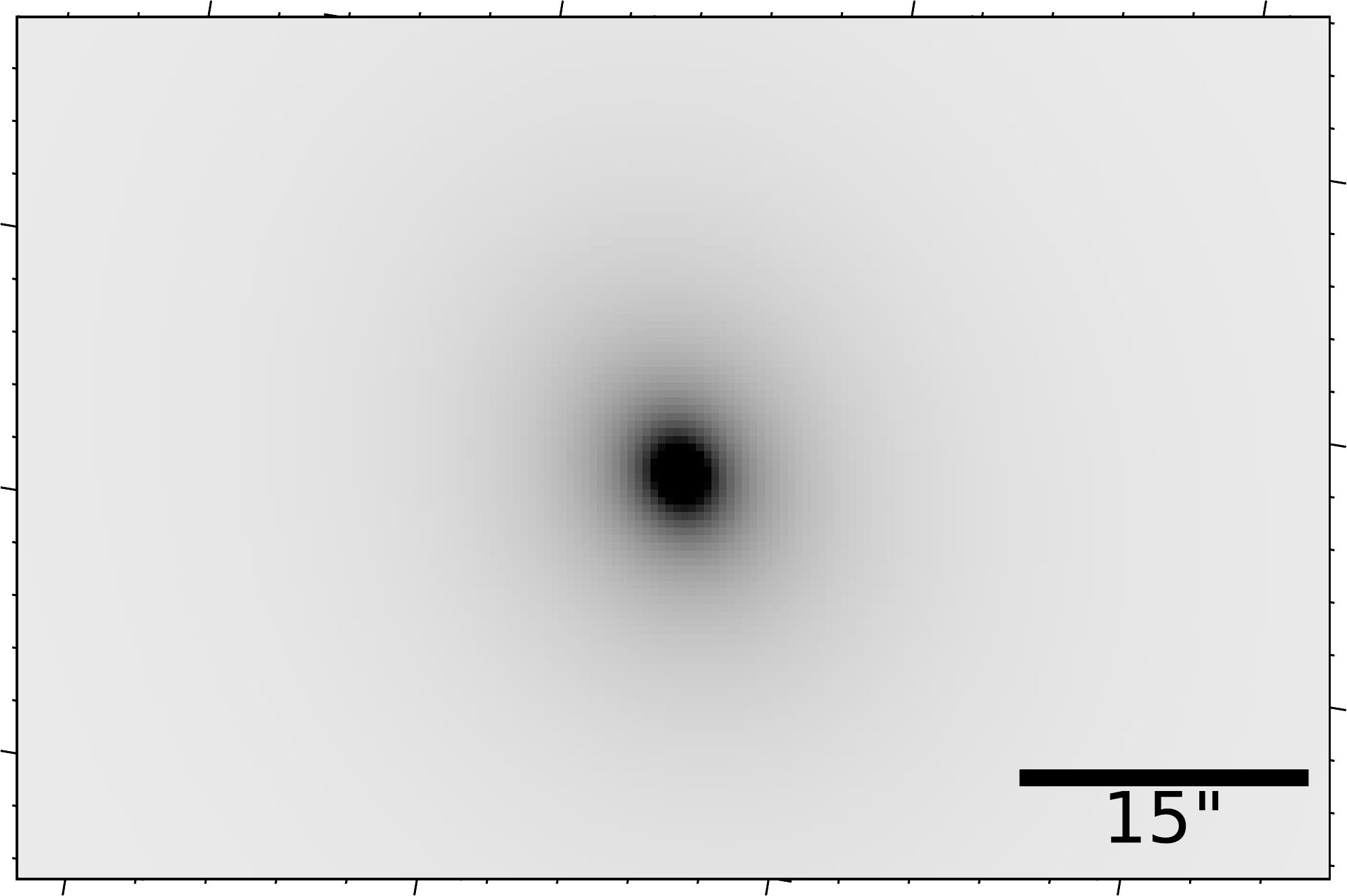} \label{264G2resid1}}
\hspace{0.3cm}
\subfloat{\includegraphics[scale=0.2, bb=8 8 560 416]{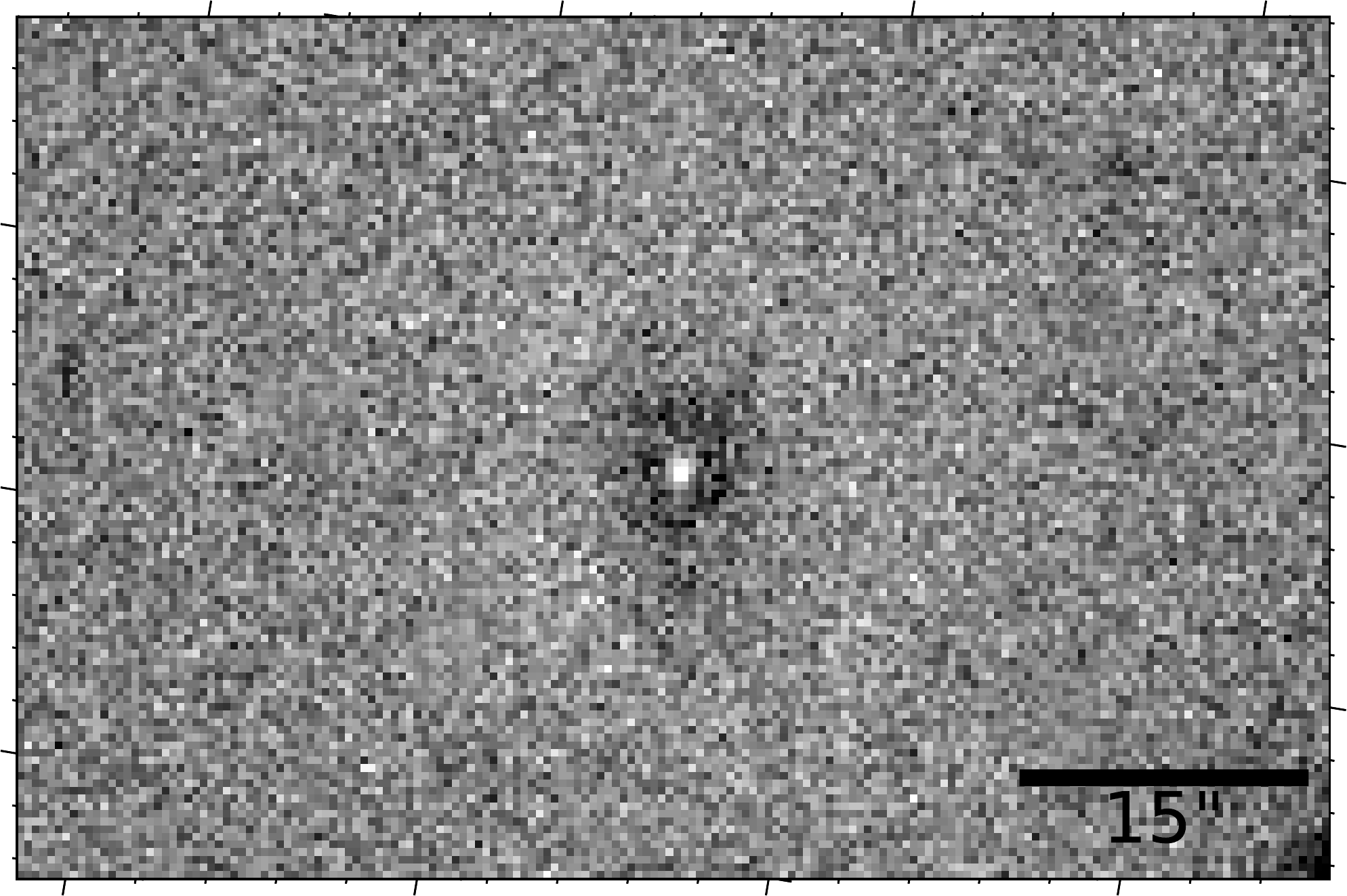} \label{264G3resid1}}\\
\caption{Photometric analysis of the triplet system SIT 264. See caption Fig.~\ref{fig:30analysis}}
\label{fig:264analysis}
\end{figure*}

This system is composed of three members of different morphological types; \textbf{G1} is an Sc galaxy while \textbf{G2} and \textbf{G3} are Irr and E, respectively (Figure~\ref{fig:SIT264}). The optical image of \textbf{G1} shows an elongated central component with a set of inner spiral arms that bifurcated as they wind out. In the outer region, another two asymmetric and fuzzy arms appear enclosing the galaxy in an anti-clockwise direction. These arms could be of a tidal origin. The irregular galaxy, \textbf{G2}, has an elongated structure without sub-components e.g. bar, arms, or disk components. The visual appearance of the elliptical galaxy \textbf{G3} resembles a faint early-type galaxy without additional components.

\textbf{G1} has a complex intensity profile (Figures~\ref{264_SB}, left panel) with two breaks at SMA of 4$''$ and 15$''$. The first break is of type III-d profile, while the second one is clearly type II.o.
The intensity profile of \textbf{G2} (Figure~\ref{264_SB}, middle panel) exhibits a strong evidence of an extremely bright ring located between 3$''$ $<$ SMA $<$ 6.5$''$ after which a hump of type III-d appears in the inner disk, while the outer region of the disk represents type I profile with no breaks.
The intensity profile of \textbf{G3} (Figure~\ref{264_SB}, right panel) shows a clear down-bending profile of type II.o at SMA $\sim$ 7.8$''$.

The decomposition of \textbf{G1} was not successful when combining a $Ser$ with an $Exp$ model ($\chi^2_{\nu}$ = 1.63). The residual image showed an inverse ``S-shape'' structure around the bulge of the galaxy that might be explained by a heavy dust lanes. In addition, an off-set in the center of the galaxy and two asymmetric spiral arms are observed (Figure~\ref{264G1resid2}). 

The irregular shape of \textbf{G2} makes its decomposition very difficult with a poor residual image. A two-component decomposition ($Ser$ + $Exp$) reveals an over-subtracted faint tail at the Eastern portion with some dark patches in the central region (Figure~\ref{264G2resid2}). Even by summing a $Dev$ model to the $Ser$ and $Exp$ models, some features remain unfitted in the residual image with $\chi^2_{\nu}$ of 1.41. The residual image of this galaxy match a distorted spiral galaxy more than an irregular galaxy. In addition, both $A$ and $S$ values of this galaxy confirm that it is a late type spiral galaxy as estimated by \cite{Conselice2003}. 

A single $Ser$ model with $n$ $=0.85$, was successfully capable to fit the simple exponential component of \textbf{G3} with a $\chi^2_{\nu}$ of 1.06 (Figure~\ref{264G3resid1}). Although this galaxy has been classified as S0 by \cite{Nair2010}, our classification agrees better with the one proposed by \cite{Park2005} who classify it as a spiral galaxy.

\subsection{The triplet system SIT 280}  

\begin{figure*}
\centering
\subfloat{\includegraphics[scale=0.5, bb=130 -150 100 100]{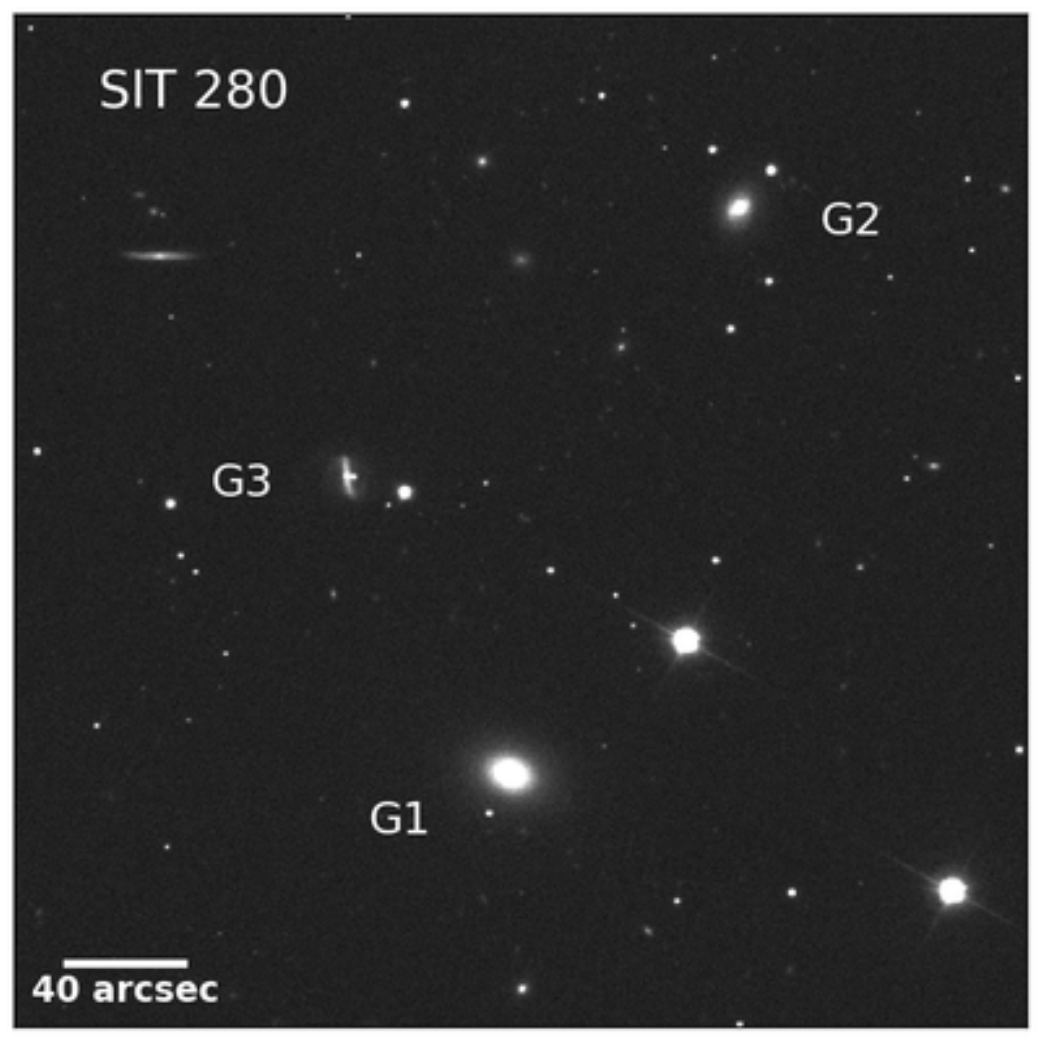}
 \label{fig:SIT280}}\\
\subfloat{\includegraphics[scale=0.7, bb=80 20 500 100]{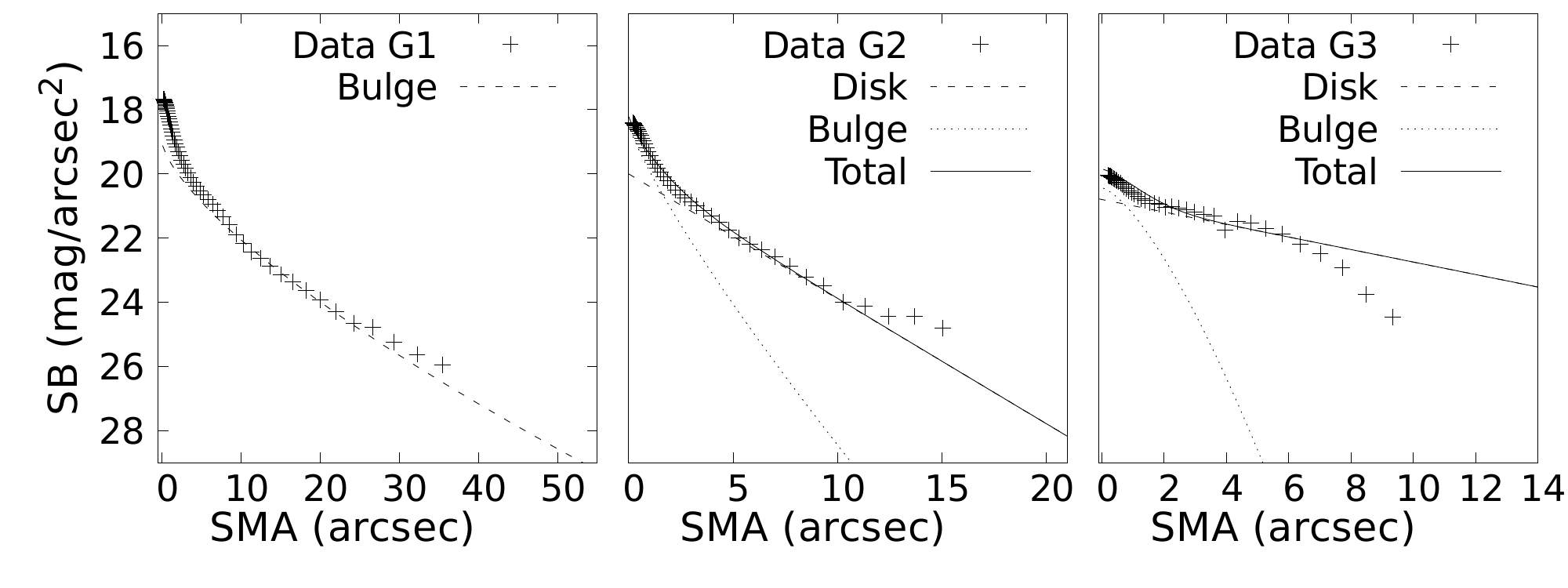}\label{280_SB}}\\
\subfloat{\includegraphics[scale=0.2, bb=7 5 560 360]{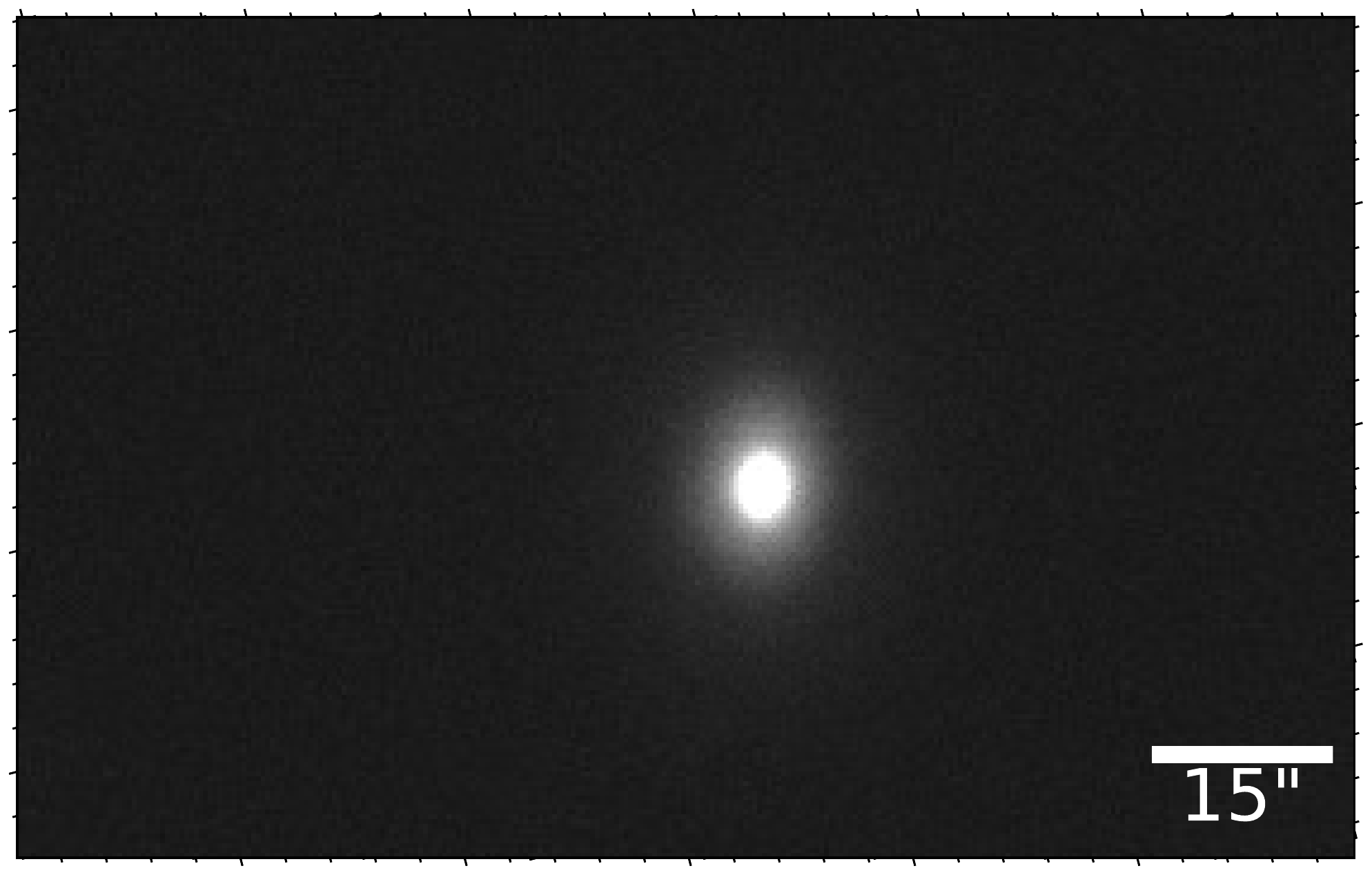}\label{280G1original}}
\hspace{0.3cm}
\subfloat{\includegraphics[scale=0.2, bb=7 5 560 360]{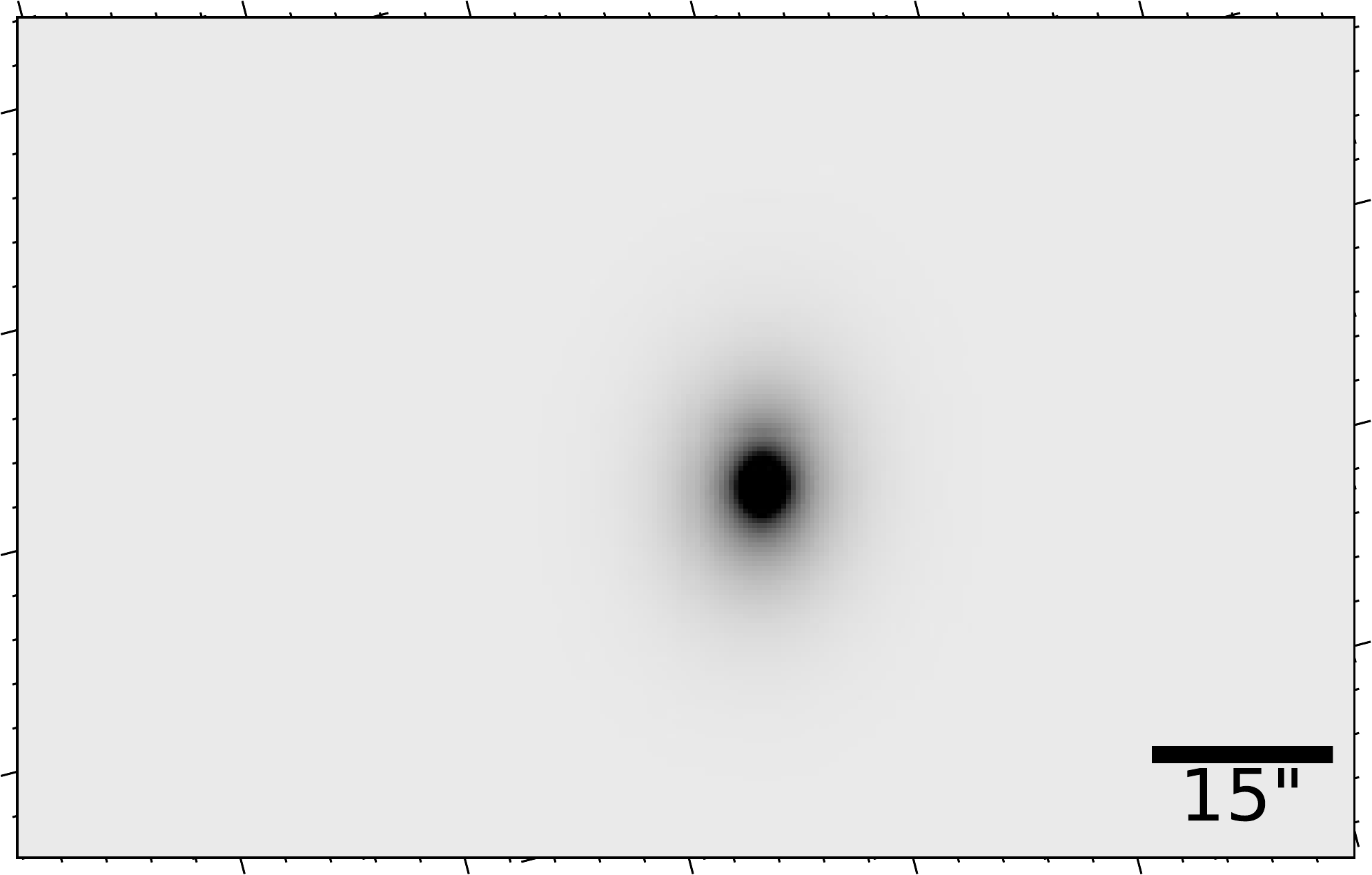} \label{280G1resid1}}
\hspace{0.3cm}
\subfloat{\includegraphics[scale=0.2, bb=7 5 560 360]{Figures_pdf/280G1_residue2-eps-converted-to.pdf} \label{280G1resid2}}\\
\subfloat{\includegraphics[scale=0.2, bb=4 4 560 270]{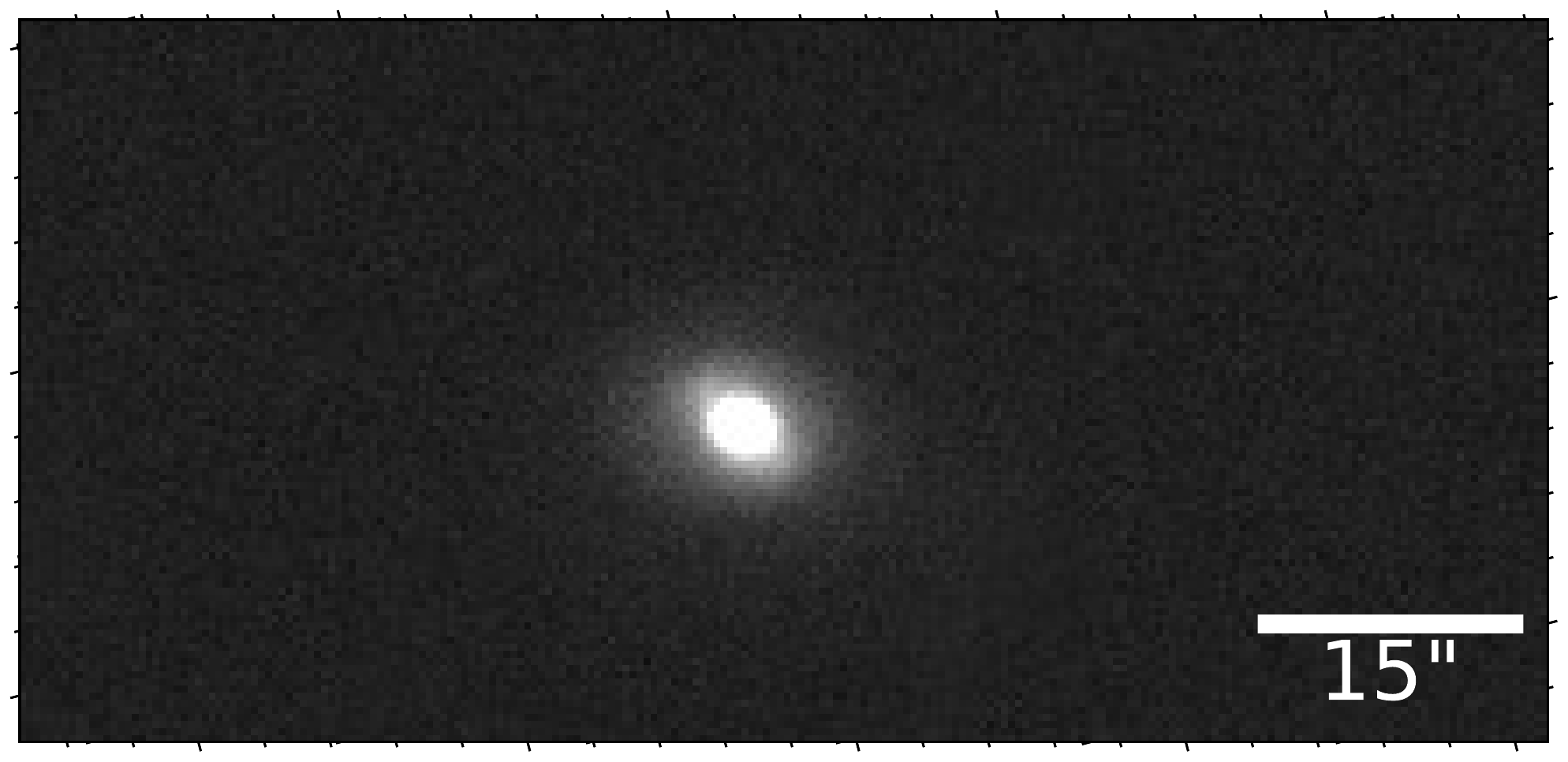}\label{280G2original}}
\hspace{0.3cm}
\subfloat{\includegraphics[scale=0.2, bb=4 4 560 270]{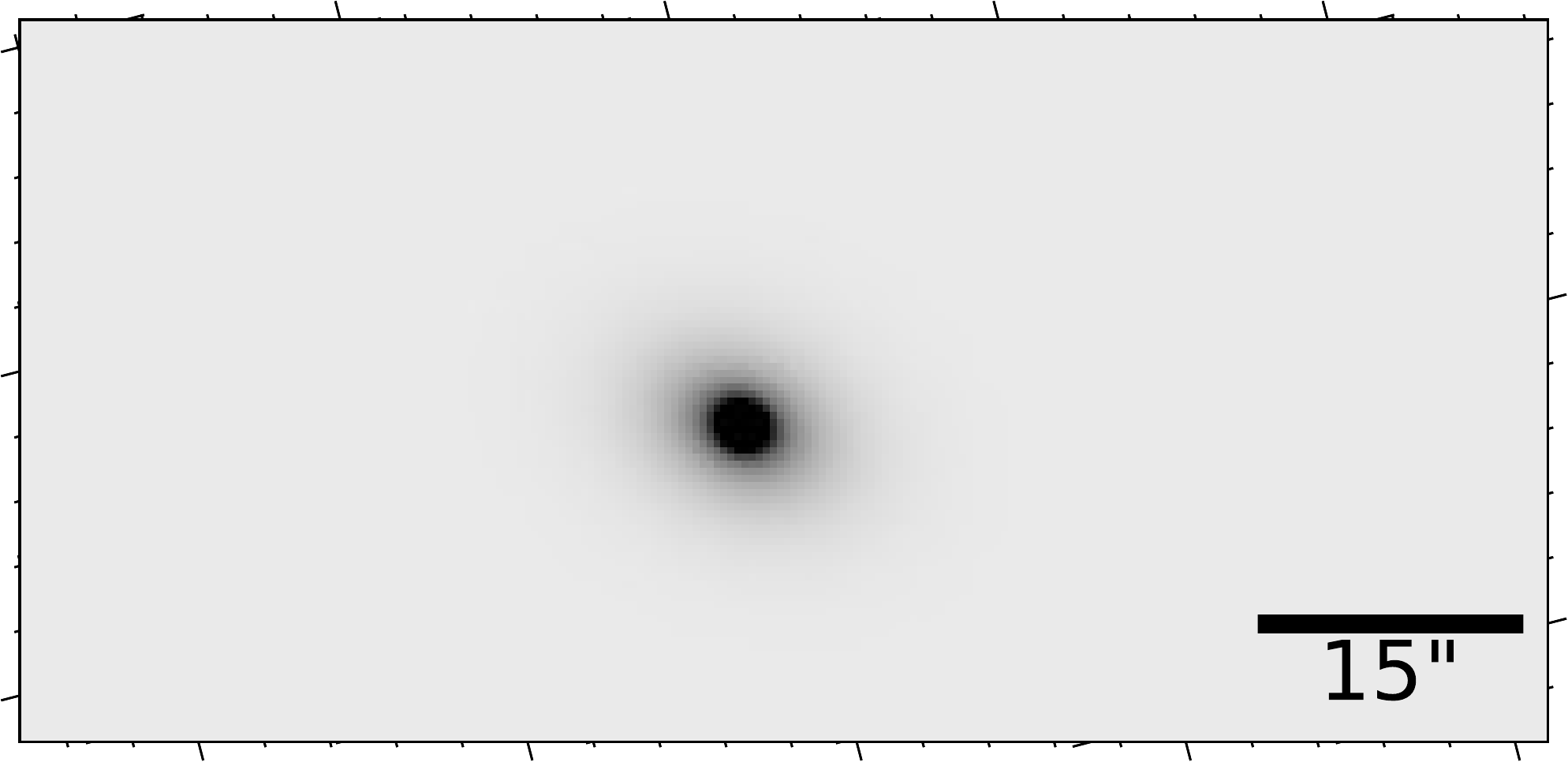} \label{280G2resid1}}
\hspace{0.3cm}
\subfloat{\includegraphics[scale=0.2, bb=4 4 560 270]{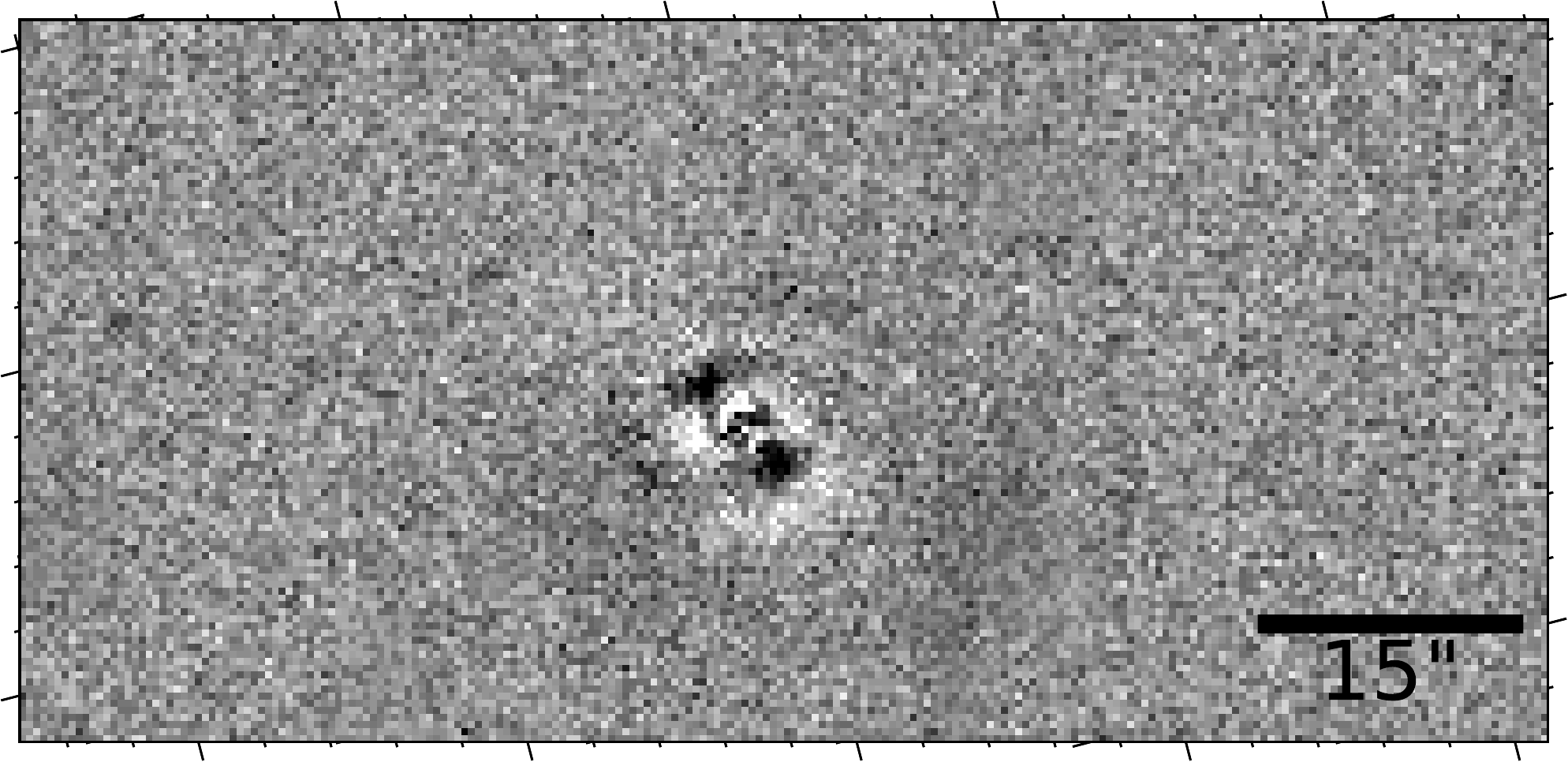} \label{280G2resid2}}\\
\subfloat{\includegraphics[scale=0.2, bb=4 4 560 260]{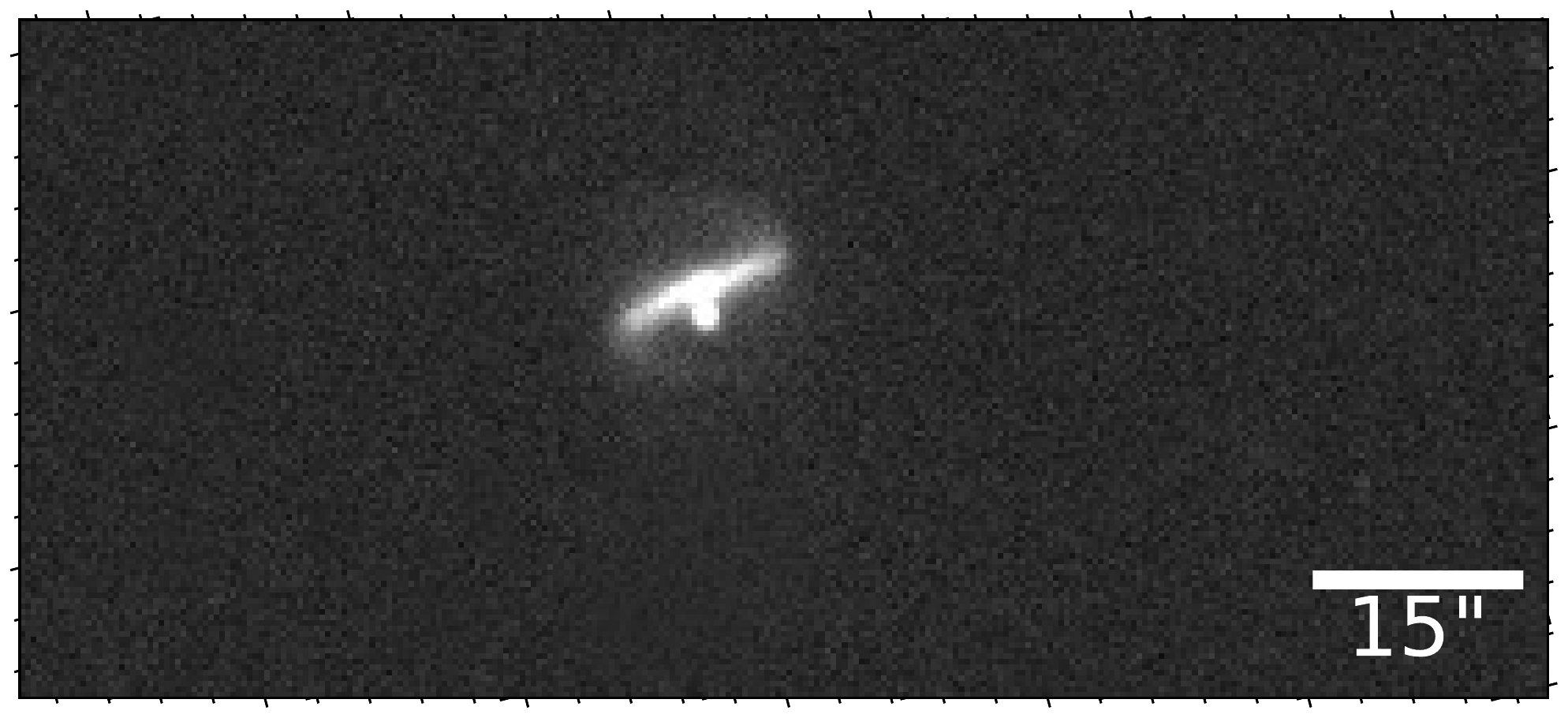}\label{280G3original}}
\hspace{0.3cm}
\subfloat{\includegraphics[scale=0.2, bb=4 4 560 260]{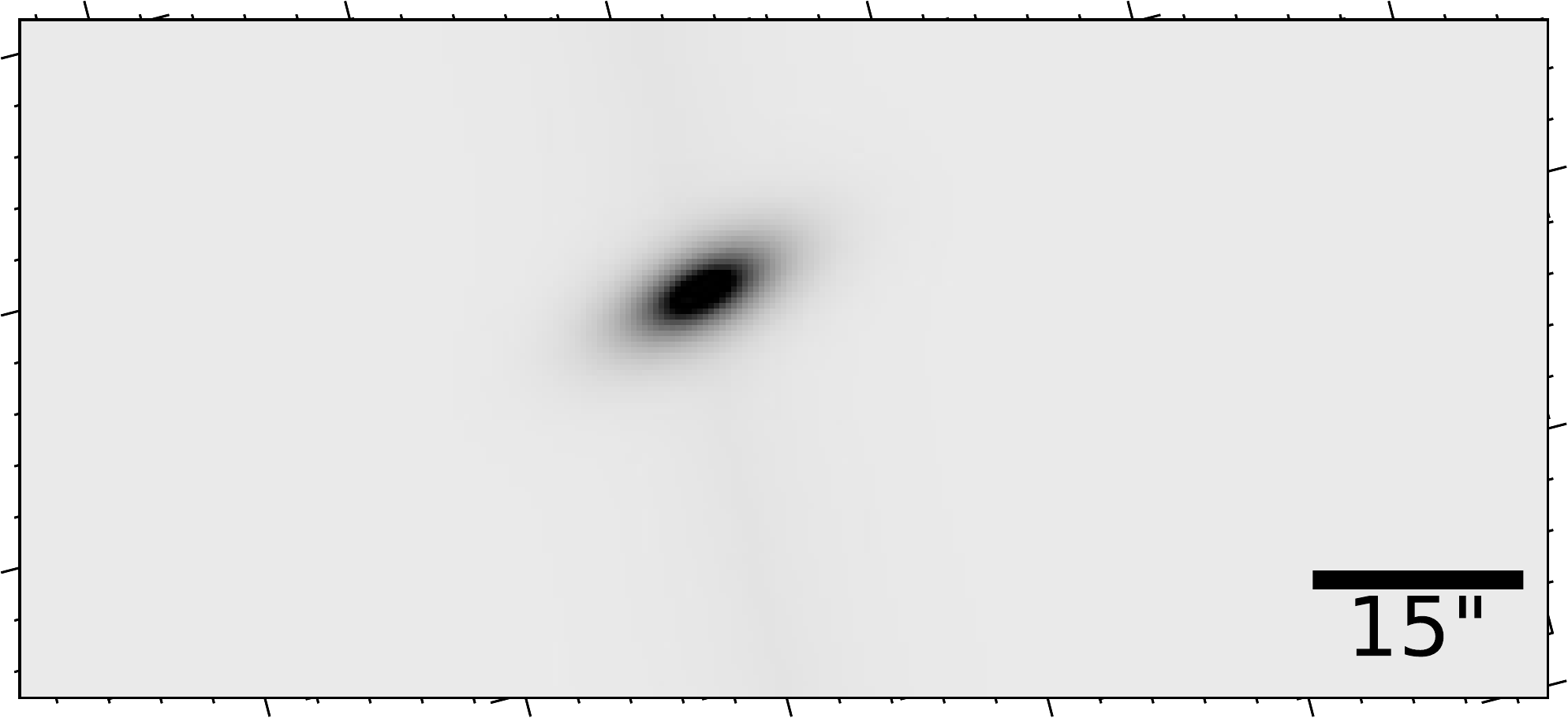} \label{280G3resid1}}
\hspace{0.3cm}
\subfloat{\includegraphics[scale=0.2, bb=4 4 560 260]{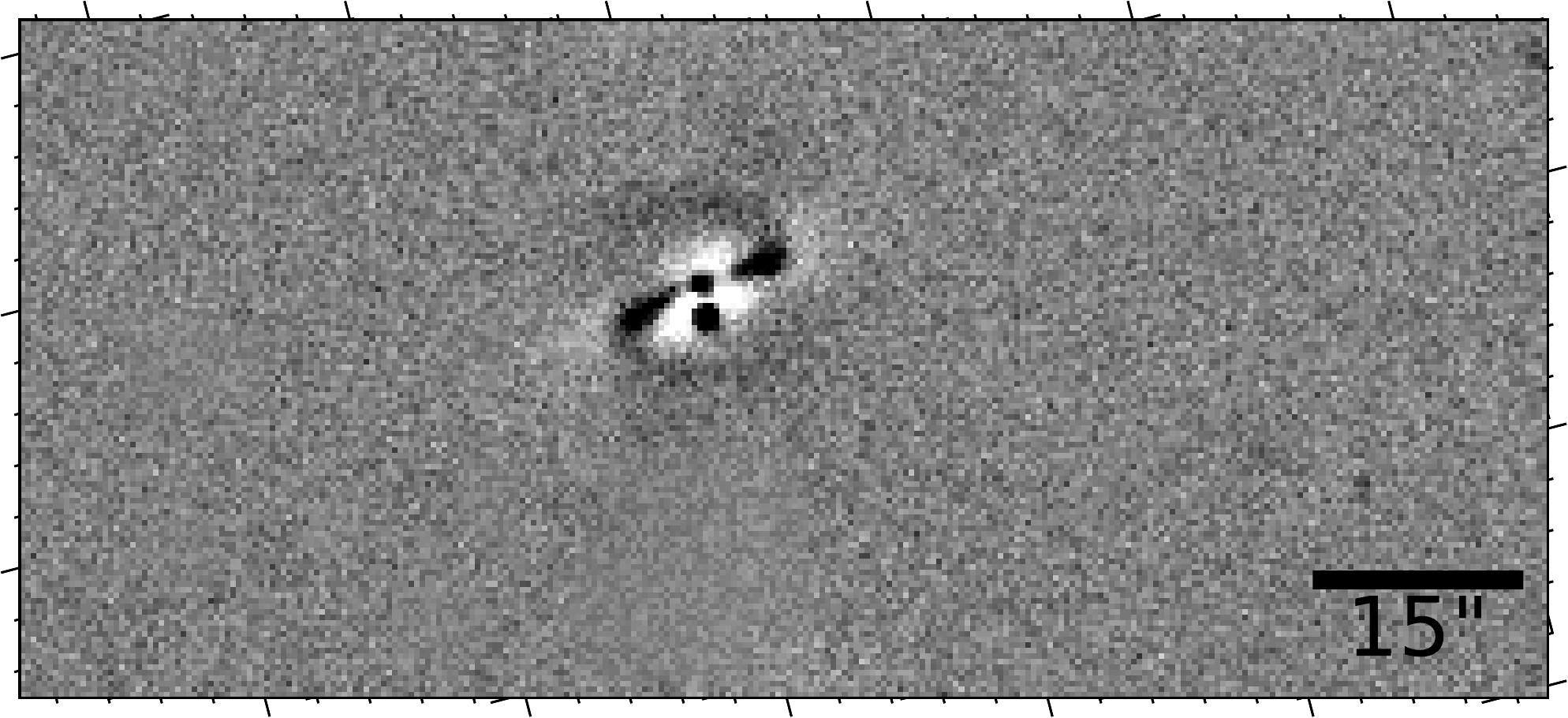} \label{280G3resid2}}\\

\caption{Photometric analysis of the triplet system SIT 280. See caption Fig.~\ref{fig:30analysis}}
\label{fig:280analysis}
\end{figure*}

SIT 280 is a triplet system of two early-type galaxies (\textbf{G1} and \textbf{G2}) and one late-type galaxy (\textbf{G3}) (Figure~\ref{fig:SIT280}). \textbf{G1} and \textbf{G2} are elliptical galaxies with a definite bulge surrounded by a faint halo. \textbf{G3} is an SBc galaxy with an elongated bar structure and two symmetric arms.

The intensity profile of \textbf{G1} (Figure~\ref{280_SB}, left panel) was fitted by a single $Ser$ profile that defines the pure bulge component of this elliptical galaxy. While, \textbf{G2} shows a break at SMA $=$ 10.8$''$ of type III-d, where most of the light dominated from the main disk of the galaxy (Figure~\ref{280_SB}, middle panel). This galaxy match well the classification of \cite{Nair2010} who classified it as S0 or Sa galaxy.

The SDSS image of \textbf{G3} is strongly affected by the extremely bright star projected at the south just beyond the galaxy's core, which could not be masked before applying our analysis. The projected output light of this star covers most of the central region of \textbf{G3}. Consequently, the intensity profile should be considered somewhat uncertain with a truncated profile at SMA $=$ 6.8 $''$ representing type II.o (Figure~\ref{280_SB}, right panel). 

Decomposition of \textbf{G1} by $Dev$ and $Ser$ model lefts a well defined ring in the central region of the galaxy that extends up to 4.44$''$ and a negative over-subtracted halo in the outer region of the galaxy (Figure~\ref{280G1resid2}).

Performing a decomposition of \textbf{G2} using $Ser$ and $Exp$ model reveals a residual image with a traces of spiral arms and a $\chi^2_{\nu}$ of 1.03 (Figure~\ref{280G2resid2}). This confirm the classification of this galaxy as S0 or Sa. 

Decomposition of \textbf{G3} represents one of the most complicated cases in our sample. The projected star is clearly recognized in the residual image of this galaxy  as shown in Figures~\ref{280G3resid1} and ~\ref{280G3resid2}. A clear mis-modeling of the core, as well as over-subtracted regions in the East and the West directions were observed while using $Ser$ and $Exp$ model. In addition, the bar and the two spiral arms could not be fitted and a $\chi^2_{\nu}$ value of 1.91 was obtained.

\end{document}